\documentclass[11pt,a4paper]{amsart}
\usepackage{amssymb}
\usepackage{mathptmx}
\usepackage{bm}
\usepackage{fullpage}
\usepackage{microtype,mathtools,enumitem,booktabs,array}
\usepackage{graphicx}
\usepackage{tikz}
\usepackage[hidelinks]{hyperref}
\usepackage[nameinlink,capitalize]{cleveref}
\allowdisplaybreaks
\newtheorem{theorem}{Theorem}[section]
\newtheorem{lemma}[theorem]{Lemma}
\newtheorem{proposition}[theorem]{Proposition}
\newtheorem{corollary}[theorem]{Corollary}
\newtheorem{fact}[theorem]{Fact}
\newtheorem*{restatedtheorem}{Restatement}
\theoremstyle{definition}

\newtheorem{example}[theorem]{Example}
\newtheorem{openproblem}{Open Problem}
\crefname{fact}{Fact}{Facts}
\crefname{example}{Example}{Examples}
\crefname{openproblem}{Open Problem}{Open Problems}
\Crefname{openproblem}{Open Problem}{Open Problems}
\AtBeginDocument{\AddToHook{env/lemma/begin}{\crefalias{theorem}{lemma}}\AddToHook{env/proposition/begin}{\crefalias{theorem}{proposition}}\AddToHook{env/corollary/begin}{\crefalias{theorem}{corollary}}\AddToHook{env/fact/begin}{\crefalias{theorem}{fact}}\AddToHook{env/definition/begin}{\crefalias{theorem}{definition}}\AddToHook{env/example/begin}{\crefalias{theorem}{example}}}
\newcommand{\Z}{\mathbb Z}
\newcommand{\Computes}{\mathsf{Computes}}
\newcommand{\R}{\mathbb R}
\newcommand{\Q}{\mathbb Q}
\newcommand{\N}{\mathbb N}
\newcommand{\MAX}{\operatorname{MAX}}
\newcommand{\relu}{\operatorname{ReLU}}
\newcommand{\Vone}{\mathbf V^1}
\newcommand{\FP}{\mathsf{FP}}
\newcommand{\Net}{\mathsf{Net}}
\newcommand{\Err}{\mathsf{Err}}
\newcommand{\Depth}{\mathsf{Depth}}
\title{\textbf{A Witnessing Duality for ReLU Network Depth}}
\author[Dimitrios Myrisiotis]{%
\texorpdfstring{Dimitrios Myrisiotis\\
Great Bay University\\
China}{Dimitrios Myrisiotis}}
\date{}
\hypersetup{pdftitle={A Witnessing Duality for ReLU Network Depth},pdfauthor={Dimitrios Myrisiotis}}
\begin{document}


\begin{abstract}
A depth lower bound for ReLU networks asserts that every sufficiently shallow network fails on some input.
What does a proof of such a bound compute?
We study this question for exact computation of the maximum of $n$ real inputs, with depth measured by the number of hidden layers.
For every separately fixed prime $q$ and depth bound $K$, we give a uniform $\mathsf{TC}^0$ refuter for rational networks with at most $K$ hidden layers and more than $q^K$ inputs, provided that the reduced denominators of their nonbias weights are not divisible by $q$.
Biases may be arbitrary rationals.
The refuter finds a rational input on which the network fails, and its correctness is provable in $\mathbf{VTC}^0$, the bounded-arithmetic theory of uniform constant-depth threshold computation.
The underlying numerical depth lower bound follows from Averkov, Hojny and Merkert (ICLR 2025), extending Haase, Hertrich and Loho (ICLR 2023); our contribution is an internal proof that makes its geometric argument into a feasible refuter.
On the construction side, $\mathbf{VTC}^0$ verifies locally checkable certificates for the balanced integer-weight maximum network of depth $\lceil\log_2 n\rceil$.
At $q=2$, the construction and refuter meet: for each fixed $K$, the theory verifies that integer-nonbias-weight networks with arbitrary rational biases can compute the maximum with at most $K$ hidden layers exactly when $n\le 2^K$.
We also give certified rational constructions of polynomial encoding length and depth $O(\log n/\log\log n)$.
These results instantiate a witnessing duality: bounded proofs of certified existence and of refutation correspond, respectively, to threshold-computable constructors and refuters whose guarantees the theory proves.
\end{abstract}
\space

\maketitle


\begingroup
\makeatletter
\def\@pnumwidth{2.5em}
\def\@tocrmarg{3.5em}
\def\@dotsep{4.5}
\def\l@section#1#2{%
\@dottedtocline{1}{0pt}{2.8em}{\bfseries\numberline{}#1}{#2}}
\def\l@subsection#1#2{%
\@dottedtocline{2}{2.8em}{3.3em}{\numberline{}#1}{#2}}
\renewcommand{\tocsection}[3]{%
\@ifnotempty{#2}{\llap{\hb@xt@\@tempdima{#2.\hfil}}}#3}
\let\tocsubsection\tocsection
\makeatother
\tableofcontents
\endgroup
\space


\section{Introduction}
\label{sec:intro}

A depth lower bound for neural networks guarantees that every shallower network fails on some input.
Such a guarantee need not say how to find a failing input, and a feasible algorithm that finds one is a refuter for the bound.
We study when a weak arithmetic theory can prove that a refuter succeeds, and compare this with its ability to certify explicit constructions.
This computational content of lower bounds also motivates work on constructive separations and refuter problems \cite{CJSW21,GC25}.

Our target is the exact representation of $\MAX_n(x)=\max\{x_1,\ldots,x_n\}$ by rational ReLU networks on $\mathbb R^n$.
Depth counts hidden layers, and the output is affine.
The nonbias weights are the entries of the weight matrices, including the output weights, as opposed to the biases.
For this target, the weight arithmetic determines whether depth is an obstruction.

With unrestricted rational weights, two hidden layers suffice \cite{BHRSY26,TR26}, so for such weights there is no growing-depth lower bound to witness.
By contrast, Haase, Hertrich and Loho proved that integer weights require logarithmic depth \cite{HHL23}, and Averkov, Hojny and Merkert extended the bound to weights whose denominators are powers of a fixed integer \cite{AHM25}.
The bound of Averkov, Hojny and Merkert requires that a fixed prime $q$ not divide the fixed integer, and the resulting depth bound is logarithmic to base $q$.
If the reduced denominators of all nonbias weights of a network omit $q$, then their product is not divisible by $q$ and every nonbias weight is an integer divided by a power of this product, so the bound applies to the network.
Thus this modular obstruction applies whenever the reduced nonbias denominators omit a fixed prime $q$, and rational biases remain unrestricted.

For the maximum, we show that $\mathbf{VTC}^0$, the bounded-arithmetic theory of uniform constant-depth threshold computation, can both prove a refuter correct and certify explicit constructions.
Our main technical result is on the refutation side: for every separately fixed prime $q$ and hidden-depth cap $K\ge1$, the theory proves the known modular depth lower bound together with the correctness of a specified refuter, which is itself a uniform constant-depth threshold computation.
The refuter handles networks with more than $q^K$ inputs, at most $K$ hidden layers, and nonbias weights whose reduced denominators omit $q$ (in particular, integer nonbias weights), with arbitrary rational biases.

On the construction side, uniformly in $n$, the theory verifies a certified balanced integer-weight construction, described below, with $\lceil\log_2 n\rceil$ hidden layers, which meets the refuter at $q=2$.
For each fixed $K$, the theory thus witnesses both sides of a sharp threshold: with integer nonbias weights and arbitrary rational biases, at most $K$ hidden layers suffice exactly when $n\le2^K$.
A second certified construction uses rational maximum blocks of slowly growing arity; the theory proves its correctness, verifies its complete rational evaluation histories (defined below), and proves a uniform depth inequality giving $O(\log n/\log\log n)$ hidden layers.
Because depth alone does not measure description size, we also require polynomial total encoding length, including every matrix entry and its precision, and both constructions, together with their certificates, meet this requirement.
Although two hidden layers suffice for unrestricted rational weights, none of the three known depth-two constructions, those of \cite{BHRSY26} and \cite{TR26} and the blocks built here from the dictionary of \cite{Zim26}, is known to meet it (\Cref{d2:constructions}).

We work in $\mathbf{VTC}^0$ as developed by Cook and Nguyen \cite{CookNguyen}, with Je\v r\'abek's internal integer arithmetic \cite{Jerabek22}.
The witnessing theorem for this theory turns a proof of a bounded existential assertion into a witnessing function computed by uniform constant-depth threshold circuits, an $\mathsf{FTC}^0$ function, whose guarantee the theory proves.
For real-input networks, applying it requires finite codes for networks and rational disagreements, explicit length bounds, and proofs that the finite tests express the intended all-real assertions (\Cref{app:witness} supplies the rational-witness bridge).

We apply witnessing in two directions, and call the resulting pair of correspondences the witnessing duality of the title.
Bounded existence of a network together with a checkable certificate of its correctness, described below, corresponds to a threshold-computable constructor for which the theory proves that the output passes the check.
A proof of the bounded assertion that every candidate network meeting the hypotheses has an error corresponds to a threshold-computable refuter for which the theory proves that the returned input is an error.
Conversely, a constructor or refuter gives the corresponding existence theorem when the theory can prove its guarantee.
The network-specific task is to supply sound checks, suitable bounds, and the internal proofs required by these correspondences.

The logical framework is standard, and the work lies in the concrete geometric and arithmetic constructions.
The numerical lower bound is known; on the refutation side, the contribution is its internal geometric proof and the specified refuter.

The geometric argument iterates an intersection product of piecewise-linear functions over polyhedral cones (\Cref{ssec:fixed-prime-depth}).
It must be made finite: cone operations require explicit coordinates, divisibility requires actual integer quotients, and the contradiction must yield bounded rational test points.
Every intermediate value is computed by an explicit parallel formula rather than by a loop over the input, so the refuter is a constant-depth threshold computation.

On the construction side, we give local certificates for assemblies of maximum gadgets, small fixed subnetworks that each compute the maximum of two values.
The certificate labels each gadget by the input coordinates it aggregates, and local union checks verify these labels.
Every gadget's value can then be computed directly from the original inputs.
An evaluation history records the rational value of every neuron, including the final output, and satisfies every equation between adjacent layers; with the labels, exact evaluation histories are available in parallel even when assembly depth grows.

In a related setting, Li, Li and Ren \cite{LLR26} study the reasoning needed for resolution lower bounds through the search for an error in a purported short proof.
Here the candidate is a real-input network and the error is a rational point; on the upper side, the witness is a network with a locally checkable certificate.

\subsection{Our Results}

\Cref{tab:results} summarizes the three main results.

\begin{table}[ht]
\centering
\caption{The three main results; lower-bound primes and depth caps are fixed outside the input.}
\label{tab:results}
\begin{tabular}{>{\raggedright\arraybackslash}p{.25\linewidth}p{.66\linewidth}}
\toprule
Result & Guarantee\\\midrule
Internal modular refutation & For each fixed prime $q$ and depth cap $K$, a specified $\mathsf{FTC}^0$ refuter has $\mathbf{VTC}^0$-provable correctness for $n>q^K$ when reduced nonbias denominators omit $q$. Arbitrary rational biases and unreduced encodings are allowed (\Cref{tc:refuter}).\\[3pt]
Certified integer construction & A uniform depth-$\lceil\log_2n\rceil$ integer-weight family has polynomial total codes and full histories verified in $\mathbf{VTC}^0$. Combined with the prime-two refuter, it gives the sharp threshold $n\le2^K$ for each fixed $K$ (\Cref{uc:balanced,uc:sharp}).\\[3pt]
Certified rational construction & Polynomial total network and certificate length and full histories, with a uniform internal depth inequality giving $O(\log n/\log\log n)$ hidden depth (\Cref{rc:upper} and \Cref{ir:rate}).\\
\bottomrule
\end{tabular}
\end{table}

The main comparison in \Cref{tab:results} pairs the modular refuter with the certified integer construction: at prime two they witness the two sides of the same sharp depth threshold in one theory.
The third row extends the construction side to rational weights through short block certificates and exact substitution of the blocks into a balanced assembly.
This gives a certified sublogarithmic-depth family whose correctness, histories and a uniform depth inequality are proved in the same theory.

\begin{samepage}
\paragraph{\emph{The internal refuter.}}
Fix a prime $q$ and an integer $K\ge1$, and write $d=q^K$.
A rational weight is $q$-free if its reduced denominator is not divisible by $q$.
The weights covered below are all matrix entries, including output weights; biases are exempt.
A defining graph is a formula defining a function's input-output relation.
A function given together with its specified defining graph is called \emph{chosen}, and a guarantee proved for it is a theorem about that graph.

\end{samepage}
\begin{theorem}[Informal; restated as \Cref{tc:refuter}]
\label{ithm:refuter}
For each separately fixed $q,K$, there is an $\mathsf{FTC}^0$ function that, given a rational network $N$ on $n\ge q^K+1$ inputs with at most $K$ hidden layers and $q$-free nonbias weights, returns a rational $n$-vector $x$ such that
\[
N(x)\ne\max_i x_i.
\]
One polynomial in the code length $|N|$ of $N$, independent of $q,K$, bounds the output length.
With the function introduced by its specified defining graph, and with a fixed-depth network evaluator that rejects malformed inputs, $\mathbf{VTC}^0$ proves the function's guarantee and the corresponding bounded refutation sentence, which asserts that every such network has such a rational error.
Stored rational fractions need not be reduced, and rational biases are arbitrary.
\end{theorem}
The function of \Cref{ithm:refuter} is chosen in this sense.
Its circuit depth and the exponent of its polynomial size bound may depend on the externally fixed $q,K$, so the statement does not give a polynomial-time algorithm when the depth cap is part of the input.
The output bound, by contrast, is one polynomial for all $q,K$; it comes from an explicit bound on the integer test points.
The intermediate arrays of the construction have polynomial lengths whose degrees may depend on $q,K$, but these lengths do not enter the coordinates of the returned point.
Widths and parameter magnitudes are unrestricted apart from the finite input encoding.

At $q=2$ the theorem covers integer nonbias weights, the setting of \cite{HHL23}, and more generally rational weights whose reduced denominators are odd.
For any fixed base $\beta$ coprime to $q$, it covers weights in $\mathbb Z[1/\beta]$, the setting of \cite[Theorem~2]{AHM25}.

\paragraph{\emph{Certified constructions and the two witnessing correspondences.}}
For the positive direction we use local certificates that specify the input-coordinate set aggregated by each maximum gadget.
The labels are supplied, rather than recovered from an arbitrary network.
We prove that passing the local checks certifies exact maximum computation.
The dimension $n$ is supplied to each constructor in unary.
\begin{theorem}[Informal; restated as \Cref{uc:assemblies}, \Cref{uc:balanced} and \Cref{uc:duality}]
\label{ithm:upper-duality}
There are chosen $\mathsf{FTC}^0$ functions constructing, uniformly over $n\ge1$, an integer-weight network of hidden depth $\lceil\log_2 n\rceil$ and an accepted maximum-assembly certificate.
The theory $\mathbf{VTC}^0$ proves acceptance and the certificate checker's soundness; supplied labels yield exact evaluation histories in $\mathsf{FTC}^0$ even when assembly depth grows.
More generally, a theorem of bounded existence of a network with an accepted certificate is equivalent to a chosen $\mathsf{FTC}^0$ constructor for which the theory proves that the output is accepted.
A theorem of bounded refutation, stated with a network evaluator defined in the theory, is likewise equivalent to a chosen $\mathsf{FTC}^0$ refuter whose error guarantee the theory proves.
\end{theorem}
The equivalences use standard witnessing and the conservative extension of $\mathbf{VTC}^0$ by chosen functions, with the specified checkers and bounds retained.
They do not extract a network from the bare assertion that some network is correct on all inputs, which has an extra universal quantifier over inputs (see the Discussion below).

The concrete upper family is the classical balanced construction (cf.~\cite[Lemma~D.3]{ABMM18}, whose two-input gadget uses weights $\pm1/2$) with the integer gadget $\max(a,b)=\relu(b-a)+\relu(a)-\relu(-a)$, where $\relu(t)=\max(t,0)$ is the ReLU.
It has hidden depth $\lceil\log_2 n\rceil$, zero biases and nonbias weights in $\{-1,0,1\}$.
At prime two, this construction and the refuter of \Cref{ithm:refuter} meet at the classical integer-depth boundary, whose lower bound is due to \cite{HHL23}.

In the corollary below, representation is expressed by the fixed bounded correctness predicate of \Cref{uc:sharp}, which tests rational inputs up to a length cutoff; agreement on those inputs also characterizes agreement on all real inputs (\Cref{app:witness}).
The existential quantifier of \Cref{uc:sharp} ranges over codes of length at most a fixed polynomial in $n$ whose nonbias weights are stored with denominator one, and the constructed network meets both conditions.
This length bound enters the equivalence of \Cref{uc:sharp}; the refuter of \Cref{ithm:refuter} itself applies to candidates of every length.
\begin{corollary}[Informal; restated as \Cref{uc:sharp}]
\label{icor:sharp}
For each separately fixed $K\ge1$, the theory $\mathbf{VTC}^0$ verifies both sides of the sharp threshold: $\MAX_n$ has an integer-nonbias-weight representation of hidden depth at most $K$ exactly when $n\le2^K$.
The positive side has the certified constructor of \Cref{ithm:upper-duality}; above the threshold, every such candidate, with arbitrary rational biases, has the bounded rational error returned by the refuter of \Cref{ithm:refuter} at $q=2$.
\end{corollary}
The constructor is uniform across dimensions, whereas the refuter and its internal proof retain the externally fixed cap $K$.

\paragraph{\emph{Certified rational compression.}}
The integer family has a rational counterpart using maximum blocks of slowly growing arity, adapted from the dictionary construction of \cite{Zim26}.
We construct certified blocks directly and assemble them in a balanced network.
\begin{theorem}[Informal; restated as \Cref{rc:upper} and \Cref{ir:rate}]
\label{ithm:rational}
There are chosen $\mathsf{FTC}^0$ constructors for rational networks and certificates of polynomial total length.
The theory $\mathbf{VTC}^0$ proves their acceptance, exact MAX correctness, and one uniform integer inequality giving $O(\log n/\log\log n)$ hidden depth.
For every rational input $x$, a chosen $\mathsf{FTC}^0$ function supplies a full evaluation history of length polynomial in $n+|x|$, where $|x|$ is the code length of $x$, and the theory proves every adjacent evaluation equation.
\end{theorem}
The theorem does not assert polynomial size at depth two, and it claims no improvement in semantic size-depth bounds.
Its contribution to the construction side is the certificate and history guarantee for this family, proved in \Cref{rc:appendix,ir:appendix}.

\paragraph{\emph{Quantified propositional proofs.}}
Because $\Vone$, the bounded-arithmetic theory of polynomial-time reasoning, proves every theorem of $\mathbf{VTC}^0$ (\Cref{lem:vtc-in-v1}), the classical translation theorem for $\Vone$ \cite[Theorem~7.57]{CookNguyen} turns the internal proof of \Cref{ithm:refuter} into finite proof families.
The tree-like quantified sequent calculus $\mathbf G_1^\star$, used here, restricts cut formulas to $\Sigma^q_1\cup\Pi^q_1$; in this standard notation the superscript $q$ stands for quantified and is unrelated to the prime $q$.
The classes $\Sigma^q_1$ and $\Pi^q_1$ are the smallest pair of classes that contain the quantifier-free formulas and are closed under conjunction and disjunction, with $\Sigma^q_1$ closed under existential and $\Pi^q_1$ under universal Boolean quantification, and with negation taking each class into the other.
\begin{samepage}
\begin{corollary}[Informal; restated as \Cref{tc:proofs}]
\label{icor:proofs}
For fixed $q,K$ and every input dimension and network-code length, the bounded assertion that each network meeting the hypotheses of \Cref{ithm:refuter} has the chosen refuter's rational disagreement has a polynomial-size $\mathbf G_1^\star$ proof.
A polynomial-time function generates these proofs from unary dimension and code-length parameters.
For each network meeting the hypotheses, the satisfying existential assignments of this assertion decode to the chosen rational disagreement.
\end{corollary}
\end{samepage}
This is a proof-size upper bound for quantified translations of the refutation assertion.
The conclusion of each proof still quantifies its counterexample existentially; no quantifier-free tautology encoding, bounded-depth threshold proof or proof-length lower bound is asserted.

The translation theorem is classical, and \Cref{sec:translation-cases} writes out the cases of its proof that the source leaves as exercises.
Its proof in Cook and Nguyen's book first puts the given proof in anchored form, in which every cut formula occurs in a non-logical axiom or in an induction instance.
We use this anchoring step as stated; its proof in the book leaves the construction of a countermodel as an exercise \cite[Exercise~6.44]{CookNguyen}.
The application of the translation theorem rests on the internal geometric proof and on an explicit formula, in the original language, that retains the refuter's output.

The certified balanced construction has the same propositional form: polynomial-size $\mathbf G_1^\star$ proofs, generated from $1^n$, of a bounded assertion whose satisfying existential assignments for $n\ge1$ retain the balanced network and its certificate (end of \Cref{sec:proofs}).

\paragraph{\emph{Unrestricted depth two.}}
The blocks behind \Cref{ithm:rational} also bear on the question of \Cref{app:depth-two}: how much arithmetic suffices to prove that two hidden layers compute the maximum when the weights are unrestricted rationals.
They answer it in $\mathbf{VTC}^0$ up to padding (\Cref{d2:padded}).
For every arity $b\ge2$ and every padding number $n\ge2^{771(b+1)^5}$ (here $n$ is a padding parameter, not the number of inputs), the theory proves that a chosen $\mathsf{FTC}^0$ function of $n$ and $b$ returns a network with $b$ inputs, exactly two hidden layers and zero biases, together with a certificate that the cone-certificate checker of \Cref{rc:cones} accepts.
It also proves that every evaluation history of this network at a rational input outputs the maximum of that input.

The padding number $n$ only makes room for the block: a threshold-computable function returns strings of length polynomial in its inputs, while the bounds we prove are $2^{512(b+1)^2}$ on the code of the block for $b\ge3$ (\Cref{ir:cones}; for $b=2$ the block has three neurons in each hidden layer) and $2^{771(b+1)^5}$ on the combined length of the block and its cone certificate (\Cref{ir:small}).
Adding the axiom $\mathrm{Exp}_A$, an instance of exponentiation stating that such an $n$ exists for every $b$, removes the padding: $\mathbf{VTC}^0+\mathrm{Exp}_A$ proves the collapse at every arity (\Cref{d2:exp}).
In the other direction, a proof of this unpadded collapse in a sound polynomial-bounded theory, such as $\mathbf{VTC}^0$ or the union $\bigcup_{i<\omega}\mathbf V^i$, would yield depth-two networks for $\MAX_b$ of total encoding length polynomial in $b$ (\Cref{d2:all-bounded} and the end of \Cref{d2:general-bound}).
None of the three known constructions is known to have such a size (\Cref{d2:constructions}), and we do not decide whether such networks exist.

Together with \Cref{ithm:refuter}, $\mathbf{VTC}^0$ thus proves a form of each half of the opening observation that the weight arithmetic determines whether depth is an obstruction.
For the lower bound, this is the bound with its refuter, for each fixed prime and depth, when the reduced nonbias denominators avoid the prime.
For the collapse, it is the padded depth-two collapse for unrestricted rational weights, with correctness proved inside the theory at rational inputs.

\subsection{Our Techniques}

The refuter rests on the cone product, which we first illustrate in the plane and then use to explain the obstruction at two hidden layers.
We then describe how the obstruction becomes a refuter and how the two certified constructions supply exact evaluation histories in parallel.

\paragraph{\emph{A planar view of the cone product.}}
The cone product $f\odot C$ of a continuous piecewise-linear function $f$ with a weighted polyhedral cycle $C$ is a cycle of one lower dimension, supported where $f$ fails to be linear on $C$ and weighted by the jumps of its gradient, as in tropical intersection theory.
A cycle is balanced if, around every face of codimension one, the weighted primitive normal vectors of the adjacent faces sum to zero modulo the span of that face.

\begin{figure}[ht]
\centering

\begin{tikzpicture}[x=1bp,y=1bp,>=latex,every node/.style={font=\normalsize}]
\path[use as bounding box] (0,0) rectangle (220,145);
\fill[black!10] (55,19) rectangle (103,67);
\fill[black!22] (55,67) -- (103,67) -- (151,115) -- (55,115) -- cycle;
\fill[black!4] (103,19) -- (151,19) -- (151,115) -- (103,67) -- cycle;
\draw[gray,dashed,->,line width=.5bp] (103,67) -- (157,67);
\draw[gray,dashed,->,line width=.5bp] (103,67) -- (103,121);
\draw[black,->,line width=1bp] (103,67) -- (55,67);
\draw[black,->,line width=1bp] (103,67) -- (103,19);
\draw[black,->,line width=1bp] (103,67) -- (151,115);
\fill[black] (103,67) circle (1bp);
\node at (80,43) {$g=0$};
\node at (138,43) {$g=x$};
\node at (79,98) {$g=y$};
\node[anchor=west,text=gray] at (159,67) {$x$};
\node[anchor=south,text=gray] at (103,123) {$y$};
\node[anchor=east] at (50,67) {$(-1,0)$};
\node[anchor=north west] at (108,20) {$(0,-1)$};
\node[anchor=west] at (155,118) {$(1,1)$};
\node[anchor=north west,inner sep=0pt] at (107,63) {$0$};
\end{tikzpicture}
\space
\caption{Gradient regions of $g(x,y)=\max(0,x,y)$ and the three rays supporting its first cone product. Primitive directions are marked; each displayed ray has lattice weight one in this example.}
\label{fig:planar-cone-product}
\end{figure}
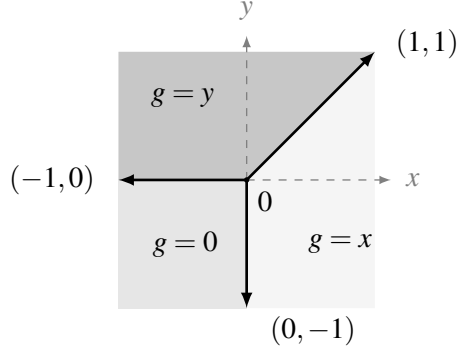

The function $g(x,y)=\max(0,x,y)$ gives a small example of the operation.
Its gradient is $0$, $e_1$ or $e_2$ in the three regions of \Cref{fig:planar-cone-product}.
Let $U$ be the planar unit cycle, with weight one on every two-dimensional chamber.
The first product $g\odot U$ records the gradient jumps of $g$.
Its nonzero rays have the primitive integer directions below (their coordinates are coprime), each with lattice weight one:
\[
v_1=(-1,0),\qquad v_2=(0,-1),\qquad v_3=(1,1).
\]
They balance because $v_1+v_2+v_3=0$.
The second product is a scalar at the origin:
\[
\left(g\odot(g\odot U)\right)(0)
=g(v_1)+g(v_2)+g(v_3)=1.
\]
The defining sum of the product subtracts one reference linear function evaluated at each ray, and these terms sum to zero because the rays balance.
For rational $a,b$, the ReLU of a single linear form, $h(x,y)=\max(0,ax+by)$, has its first product supported on the line $ax+by=0$, where $h=0$, so applying $h$ again gives zero.
When $a=b=0$ the product is zero directly.
These are planar instances of \Cref{lem:fp-product-engine}, \Cref{lem:fp-refinement-locality} and \Cref{lem:fp-max-normalization}.

\paragraph{\emph{The obstruction at two hidden layers.}}
The case $q=2$ and $K=2$ illustrates the invariant.
Vary four inputs and set the others to zero; there is at least one other input, since $n\ge q^K+1=5$, so the target is $\max(0,y_1,y_2,y_3,y_4)$.
For a bias-free integer-weight network, apply the cone product four times to the unit four-dimensional cycle.
The ReLU of a single linear form has zero repeated product after two applications.

Locality of rectification, explained below, and commutation of the cone product then force the fourth product of every second-layer activation, and of their integer-weighted output, to be even.
The fourth product of the target is one.
This is the known obstruction for integer weights \cite{HHL23}, here phrased through cone products.

The proof must construct its integer quotients and track integer coordinates through the cone operations; divisibility alone does not supply those finite witnesses.
The full identities are proved in \Cref{lem:fp-product-engine}, \Cref{lem:fp-refinement-locality}, \Cref{lem:fp-max-normalization} and \Cref{lem:fp-prime-quotients}.

For odd-denominator rational weights, clearing denominators scales the target by an odd integer, so its fourth product stays odd.
Bias domination, which scales the test points until the bounded change caused by deleting the biases cannot matter, and five agreement probes per chamber, one more than the four varied inputs (both under the next heading), then turn the obstruction into a disagreement for the original network.
For general $q,K$, prime-power binomial identities replace this parity calculation, and the construction must supply every quotient, integer coordinate and test point inside the arithmetic theory.
\paragraph{\emph{From the geometric obstruction to a refuter.}}
We first reduce the rational candidate to a bias-free integer network in the fixed dimension $d=q^K$.
A positive common denominator clears the nonbias weights, and deleting biases changes the original network by a computable uniform additive bound.
The target becomes $\max(0,x_1,\ldots,x_d)$ after fixing the remaining inputs to zero.

We encode the relevant geometry by a finite arrangement of integer hyperplanes.
Its cones have rational representatives obtained by a fixed number of perturbations of determinant-defined rays.
Since $d$ is fixed, all representatives can be enumerated in polynomial time in the arrangement length.
The enumeration has a completeness proof using bounded basis scans and two-dimensional sector calculations.
This supplies actual finite data, rather than an appeal to an unspecified cell decomposition.

On these cones, a continuous piecewise-linear function supplies compatible chamber gradients.
A cone product lowers dimension by one.
We prove balance, commutation and invariance under refinement by explicit coordinate identities, following the intersection construction of Allermann and Rau \cite{AR10}.
A family with integer gradients need not give integer weights on every intermediate rational cone.
Instead we maintain an \emph{integer twin}: the product of a cone weight and its chosen exterior orientation has specified integer coordinates.
At the final zero-dimensional cone the orientation is one, so the product value is an actual integer.
This is the integrality needed by the modular contradiction.

The network-specific identity is locality of rectification.
Writing $H=\max(F,0)$ for an activation family, $HC$ for the cone product $H\odot C$, and powers for repeated cone products, it gives
\[
H^r C=F^{r-1}(HC)
\]
for every balanced $k$-cycle $C$ and $1\le r\le k$, provided every nonzero chamber gradient of $F$ is a normal of the arrangement.

For each prime power $e=q^c$ with $c\ge1$, the mixed coefficients $\binom er$ are divisible by $q$ for $0<r<e$, the fact behind the additivity of normalized volume modulo $q$ in \cite[Proposition~11]{AHM25}.
Together these identities construct, layer by layer, the quotient by $q$ of each repeated cone product of an activation family.
For a weighted sum of activations we use a direct prefix sum of fixed product words, avoiding recursion through the lengths of previously constructed quotient arrays.
The output's $d$-fold product is consequently $qz$ for a specified integer $z$.

For the comparison with the maximum, let $D$ be the product of the nonbias denominators after canceling common factors of $q$ in each stored fraction (stored fractions need not be reduced), let $k\le K$ be the candidate's actual hidden depth, and let $a=D^{k+1}$ be the positive denominator-clearing scale.
The $d$-fold product of the maximum is one, so the scaled maximum has product $a^d$, which is nonzero modulo $q$ because $D$ is coprime to $q$.

Finally, $d+1$ integer probes per chamber recover a homogeneous output gradient.
Scale the probes large enough to dominate every bias error and the integer gap.
If all tests against the original network passed, every chamber gradient would agree with the scaled maximum, forcing the impossible equality $qz=a^d$.
The first failed test is the refuter's answer.
\Cref{ex:refuter} carries out this search for a three-input network with one hidden layer.

We specify every intermediate construction by a total defining graph and prove its output bound on all strings before using its correctness invariant.
We first specify these graphs through bounded scans in $\Vone$, then replace the scans by explicit parallel histories in $\mathbf{VTC}^0$.
Integer products and division use Je\v r\'abek's internal arithmetic theorems \cite{Jerabek22}.
A greedy basis scan keeps each vector independent of those kept before it, so it keeps at most $d$ vectors; we select them in $d$ fixed parallel stages and reconstruct every prefix of the scan from them.
Together with the direct quotient formulas, this removes the input-dependent sequential recursion from the refuter.

\paragraph{\emph{Labels instead of layer-by-layer evaluation.}}
Evaluating a network layer by layer composes a number of arithmetic steps that grows with its depth, so it does not give a threshold-computable evaluator for growing depth.
The certificate avoids this by labeling each maximum gadget with the set of input coordinates whose maximum it should compute; the checker verifies that each label is the union of its children's labels.
Given an input, all gadget values are then computed at once as maxima of input coordinates over their labels.

\Cref{uc:label-max} shows that the maximum over a union is the largest of the maxima over its parts.
This is the equation of each gadget, so every adjacent-layer equation can be checked in parallel, and the checker is sound (\Cref{uc:assemblies}).
For the balanced family the labels are dyadic blocks of padded leaves, each defined bit by bit from $n$ (\Cref{uc:balanced}); \Cref{uc:example} shows them for $n=3$.
The two correspondences of \Cref{uc:duality} apply the standard witnessing theorem with the checkers and output bounds kept inside the witnessed matrix.

\paragraph{\emph{Short certified blocks and parallel histories.}}
We use a finite integer-feature dictionary for two-layer MAX representations~\cite{Zim26}.
Finite support identities give the rational readout coefficients, whose precision is bounded using a common factorial denominator and integer sums over decreasing subset chains.
Small determinants bound the local geometric solves.
We also bound a complete cone certificate: splits cover both closed halfspaces, and nonnegative rational conic identities certify each activation and each winner region.
These are standard verification ingredients~\cite{IBZK22,Gok26,DISK25}.
For a block of arity $b$, the direct block and its complete certificate have length at most $2^{A(b+1)^{r_\ast}}$, where $A=771$ and $r_\ast=5$ are fixed constants of the block construction (\Cref{ir:small}).

The constructor selects the greatest arity whose block budget $2^{A(b+1)^{r_\ast}}$ is at most $n$, uses the binary family if no arity passes, and otherwise assembles the selected block by exact dense substitution.
Selecting maxima directly from the original input sets supplies the inputs and outputs of every block.
Their fixed two-layer histories are computed independently, and the proof checks every adjacent equation at the substituted boundaries.

The construction of each block starts with a signed average of explicit two-layer functions.
The error is the average minus the maximum, expressed as a signed sum of support functions of lower-dimensional polytopes.
The finite support identities express the error at one arity using local terms and two transformed copies of the error at the preceding arity.
After normalization, only two exceptional directions contribute these copies, and their extra segment terms cancel separately for every signed lower-arity term.
Thus a term in the full error expression is indexed by a local starting term and a binary word recording which transformation is used at each higher arity.

For each vertex, we concatenate these transformations with those defining its starting vertex and evaluate the resulting integral matrix product once.
The constructor can therefore print the data from these indices rather than compute each arity from the completed preceding array.
With all arrays fixed, the theory verifies their support identities by numerical induction over arities and predetermined transformed rational inputs (\Cref{ir:tiers}).
The arity selector converts this block budget into the uniform depth inequality of \Cref{ir:rate}.

\subsection{Other Related Work}

We group the related work by theme: depth lower bounds under weight restrictions, two hidden layers without weight restrictions, tropical intersections and finite arithmetic, logic, refuters and lower bounds, and exact constructions with certified verification.

\paragraph{\emph{Depth lower bounds under weight restrictions.}}
Haase, Hertrich and Loho \cite{HHL23} prove logarithmic depth lower bounds for integer-weight networks using lattice polytopes.
Averkov, Hojny and Merkert \cite[Theorem~2]{AHM25} extend the modular obstruction from integer weights to weights whose denominators are powers of a fixed base; in both results only the weights are restricted, and the biases are arbitrary real numbers.
We retain the modular strategy; the contribution is its finite cone computation, explicit prime quotients, and a specified refuter whose guarantee the theory proves.

With unrestricted real weights, Safran~\cite[Theorem~1]{Saf26} proves superlinear width lower bounds for exact maximum computation at every fixed number of hidden layers at least two, using extremal graph theory.
Our refuter concerns a depth obstruction under a denominator restriction, with width unrestricted.

\paragraph{\emph{Two hidden layers without weight restrictions.}}
Recent unrestricted representation results must be distinguished from these weight-restricted statements.
Brunck, Hertrich, Rue{\ss}, Stade and Yehudayoff \cite{BHRSY26} prove that two hidden ReLU layers suffice for exact representation without a fixed-base restriction.
Tallavarjula and Rastegar \cite{TR26} give another proof, with an explicit maximum network that has zero biases, integral hidden weights and output denominators dividing $(n-1)!$.

For six inputs, the author gave an explicit identity with a hand-checkable proof, which yields a two-hidden-layer maximum network with zero biases, rational weights and hidden widths $17$ and $41$ \cite{Myr26}.
The two-hidden-layer collapse rules out unrestricted growing-depth lower bounds as the target of a total refuter problem.
It does not invalidate a bound requiring a missing denominator prime: as \cite{TR26} also note, by \cite[Theorem~2]{AHM25}, a common denominator of any rational two-hidden-layer network for $\MAX_n$ is divisible by every prime $p$ with $p^2<n$.
Our theorem keeps that restriction throughout.
For the collapse itself, what we formalize is narrower: for the blocks of our rational construction, which follow Zimmermann's dictionary~\cite{Zim26}, $\mathbf{VTC}^0$ proves depth-two correctness at each arity under a padding hypothesis (\Cref{d2:padded}).
The constructions of \cite{BHRSY26} and \cite{TR26} are cited and not formalized here, and we claim neither a new numerical collapse nor a polynomial-size depth-two network.

\paragraph{\emph{Tropical intersections and finite arithmetic.}}
The cone product is modeled on the divisor-intersection construction of Allermann and Rau \cite[Definition~3.4 and Proposition~3.7]{AR10}.
Rather than import tropical intersection theory as a theorem inside bounded arithmetic, we prove the required fixed-dimensional identities from finite determinants and bounded coordinate scans.
Fu~\cite{Fu25} develops a toric-geometric description of unbiased rational ReLU networks using fans and Cartier divisors.
The rational-field and arithmetic interfaces use standard provable integer arithmetic \cite{CookNguyen,Jerabek15,Jerabek22}.
The short rational disagreement principle is familiar from neural verification \cite{SL21}; we include the specialization needed to connect the bounded predicate to all-real correctness.

\paragraph{\emph{Logic, refuters and lower bounds.}}
Witnessing and propositional translation are classical \cite{CookNguyen}.
Carmosino, Dang and Jackman \cite{CDJ26} extract refuters from selected Boolean gate-elimination lower bounds.
We establish the arithmetic correctness of a refuter for a modular geometric obstruction on real-input networks.
Zhang and B\"olcskei \cite{ZB26} study equational identification of deep ReLU networks through \L ukasiewicz logic, including a complete rational identification result under their hypotheses.
Our question concerns the strength and computational content of a proof that an entire restricted network class cannot represent a target.

For fixed input dimension and depth, verification under rational polyhedral input and output constraints is decidable in polynomial time by activation-cell enumeration and linear programming~\cite[Proposition~3.3 and Lemma~3.4]{FGHS26}.
This supplies an algorithmic comparison for the refuter problem; the lower-bound result here also proves a specified modular refuter's guarantee in $\mathbf{VTC}^0$.

Garl{\'i}k et al.~\cite{GGLST26} study bounded-arithmetic proofs of algebraic circuit lower bounds, formalizing rank arguments in $\mathbf{VNC}^2$ with dense coefficient encodings.
Their circuit model and encoding differ from the real-input ReLU networks considered here.

\paragraph{\emph{Exact constructions and certified verification.}}
{\widowpenalty=10000 Exact block construction and substitution have precedents in shallow ReLU representation.
Bakaev, Brunck, Hertrich, Stade and Yehudayoff~\cite{BBHSY26} construct maximum networks with $\lceil\log_3(n-2)\rceil+1$ hidden layers for $n\ge4$ and weights whose denominators are powers of two.
Zimmermann's integer-feature dictionary~\cite{Zim26} gives the finite span used here.
Wang and Basu~\cite{WB26} give explicit two-hidden-layer identities for $\MAX_5,\ldots,\MAX_8$.
The constructions in~\cite{R26} use exact linear-algebraic synthesis and structured substitution.

Proof-producing neural verification already uses case splitting and Farkas certificates~\cite{IBZK22}, rational certificate checking~\cite{Gok26}, and certified checker soundness~\cite{DISK25}.
Our construction result supplies the bounded-arithmetic guarantees: chosen conservative defining graphs, polynomial total dense encodings, and full rational histories for the constructed growing-depth family.
We do not claim these individual verification or synthesis methods as new.
\par}

\subsection{Discussion}

The duality concerns provability and provably correct witnessing, with construction and refutation as its two parts.
Every algorithm carries its chosen defining graph and a proof of its guarantee.
Ordinary algorithmic correctness alone does not supply that proof.
On the construction side, the certificate check is feasible and its soundness is a separate theorem, so correctness on all inputs is not assumed inside the check.

\Cref{uc:duality} states the correspondence in threshold arithmetic, and \Cref{prop:witnessing} records the polynomial-time interpretation of the refutation side.
The assertion that some network $N$ is correct on every input $x$, witnessed by an evaluation history $Z$, has the form $\exists N\,\forall x\,\exists Z$ at growing depth, and $\exists N\,\forall x$ at fixed depth, where the theory has an evaluator.
Witnessing extracts a constructor from neither form; it applies to the bounded existential certificate assertion instead.

{\widowpenalty=10000 The rational construction clarifies the denominator restriction on the refutation side.
For any family of rational networks computing \(\MAX_n\) exactly with hidden depth \(k(n)=o(\log n)\), every fixed prime \(q\) eventually divides the reduced denominator of some nonbias weight.
Indeed, put \(K(n)=\max\{1,k(n)\}\).
For fixed \(q\), the depth assumption gives \(K(n)\log q<\log n\) for all sufficiently large \(n\), hence \(n>q^{K(n)}\).
The known modular obstruction then excludes weights whose reduced denominators all omit \(q\)~\cite{AHM25}.
This is an external numerical consequence, not an internal asymptotic theorem.
\par}

Fixing the prime and depth is essential to the current argument.
The number of arrangement representatives and the product length depend on $d=q^K$.
The proof establishes a schema of separately fixed-depth theorems; it supplies no single polynomial-time procedure for the full logarithmic-depth regime.
The finite proof families for the lower bound have the same fixed-parameter scope; the certified upper family is uniform in $n$.

The depth-two results of \Cref{app:depth-two} relate provability of the unrestricted collapse to description size, as described under \emph{Unrestricted depth two} above.
Whether $\MAX_n$ has rational networks of depth at most two and polynomial total encoding length is not decided here.
A negative answer would show that no sound polynomial-bounded theory proves the unpadded collapse or the bounded depth-two assertion of \Cref{d2:all-bounded} (end of \Cref{d2:general-bound}).

\subsection{Open Problems}

The remaining uniform question concerns growing depth with integer-valued nonbias weights.
It is distinct from the question for unrestricted rational weights, where two hidden layers suffice \cite{BHRSY26,TR26}.
\paragraph{\emph{Uniform refutation at logarithmic integer depth.}}
The known semantic bound applies throughout the logarithmic range, while the current internal construction fixes the depth outside the input.

\begin{openproblem}[Uniform refutation at logarithmic integer depth]
\label{op:uniform}
Is there one polynomial-time refuter that, for every $n\ge3$, returns a polynomial-length rational disagreement for every rational network with integer-valued nonbias weights, arbitrary rational biases, and at most $\lfloor\log_2(n-1)\rfloor$ hidden layers?
\end{openproblem}
The semantic lower bound makes this a total search problem, but the separately fixed-depth algorithms do not give the required uniform running time.
A uniform refuter whose guarantee $\mathbf V^1$ proves would extend the present internal result over the full logarithmic integer regime.
The present cone enumeration does not supply that extension: the degrees of its polynomial bounds may depend on $d=2^K$, which exceeds $(n-1)/2$ at the largest depth of this range.
We name $\Vone$ here rather than $\mathbf{VTC}^0$ because at growing depth even the error check uses the layer-by-layer evaluator that \Cref{sec:foundations} specifies in $\Vone$; the threshold refuter uses only a fixed-depth evaluator.

\paragraph{\emph{Weak theories for the depth-two collapse.}}
For unrestricted rational weights, $\mathbf{VTC}^0$ proves the depth-two collapse for the maximum whenever the number sort contains a padding number exponential in a polynomial of the arity (\Cref{d2:padded}).
With an exponentiation axiom it also proves the collapse in the form that every arity $n\ge2$ has such a network (\Cref{d2:exp}).
The question is whether the second form needs exponentiation.

\begin{openproblem}[Weak theories for the unpadded depth-two collapse]
\label{op:depth-two-theory}
Is there a sound theory in the base two-sorted language, extending $\mathbf V^0$ and not proving the totality of exponentiation (the sentence $\mathrm{Exp}$ at the end of \Cref{d2:general-bound}), for instance a polynomial-bounded theory such as $\Vone$, that proves the conclusion of \Cref{d2:exp}, namely that for every $n\ge2$ the function $\MAX_n$ has a certified rational network with two hidden layers?
\end{openproblem}
For a sound polynomial-bounded theory, a positive answer would give rational depth-two networks for $\MAX_n$ of polynomial total encoding length (end of \Cref{d2:general-bound}), which are not known to exist.
For a theory whose provably total functions grow quasi-polynomially, the same Parikh-type argument is expected to give networks of quasi-polynomial size, but this extension is not proved here.
A positive answer by a sound polynomial-bounded theory would therefore give a polynomial upper bound on the size of depth-two networks for the maximum, and a superpolynomial lower bound on that size would rule out every such theory.


\phantomsection\addcontentsline{toc}{subsection}{AI Disclosure}
\begingroup
\makeatletter
\let\addcontentsline\@gobblethree
\makeatother
\subsection*{AI Disclosure}

This work was developed with substantial assistance from Codex and Claude Code.
These tools were used to discover almost all of the proofs in this work, find relevant literature, and write parts of the prose; the author wrote the rest.
The author posed the problem, guided the development, and is responsible for any errors.

\endgroup
\space

\subsection{Paper Organization}

\Cref{sec:foundations} fixes the codes, the theories and the arithmetic interfaces, including the inclusion of $\mathbf{VTC}^0$ in $\mathbf V^1$ (\Cref{lem:vtc-in-v1}).
\Cref{ssec:fixed-prime-depth,sec:locality} construct the finite cone calculus and its rectification and normalization identities.
\Cref{sec:quotients} constructs the prime quotients for network families.

\Cref{sec:probes,sec:graphs} supply the agreement probes and the defining graphs, and prove the internal theorem in $\mathbf V^1$ with its value-based normalization; \Cref{ex:refuter} runs the resulting refuter on a three-input network.
\Cref{tc:parallel-arithmetic} replaces the scans by parallel histories and proves the threshold-arithmetic theorem.

\Cref{uc:section} constructs the certified integer family, illustrated for $n=3$ in \Cref{uc:example}, states both witnessing correspondences, combines them at the sharp integer threshold, and states the certified rational construction.
\Cref{sec:proofs} records the general polynomial-time refuter interpretation and the classical quantified propositional consequences for both refuters and for the certified balanced constructor.

\Cref{app:witness} proves the rational witness bound used by the finite correctness predicate.
\Cref{rc:appendix} proves the certified rational assembly and its ordinary histories; \Cref{ir:appendix} opens with a roadmap of its steps and constructs the blocks directly.
That appendix then verifies the blocks on fixed arrays of rational points (\Cref{ir:tiers}), constructs their cone certificates (\Cref{ir:cones}), supplies them under the size guard (\Cref{ir:small}), and proves the uniform internal depth inequality (\Cref{ir:rate}).
\Cref{sec:translation-cases} writes out the cases of the propositional translation theorem \cite[Theorem~7.57]{CookNguyen} that its source leaves as exercises.
\Cref{app:depth-two} derives from these blocks the padded depth-two collapse in $\mathbf{VTC}^0$ and the unpadded collapse under one instance of exponentiation, compares the three known depth-two constructions, and proves the polynomial-size consequence of bounded provability in the base language of bounded arithmetic.
\space

\section{Preliminaries}
\label{sec:foundations}

This section fixes the formal setting of the paper: the arithmetic theories in which we reason, the codes for integers, rationals and networks, and the predicate stating that a network computes the maximum.
Every object in the internal proof is a finite binary code.
A number parameter is measured in unary, while network parameters and rational coordinates are binary strings.

We first specify the constructions of the paper sequentially, by bounded scans, in the standard two-sorted theory $\Vone$ of polynomial-time reasoning \cite[Chapter~6]{CookNguyen}.
A refuter for a depth bound is a function that receives a candidate network meeting the hypotheses of the bound and returns an input at which the value of the network differs from the maximum of its inputs.
The refuter of this paper, for a fixed prime $q$ and depth cap $K$, is specified sequentially in \Cref{thm:fixed-prime-depth} and by parallel formulas in \Cref{tc:refuter}.
\Cref{tc:parallel-arithmetic} later replaces the scans used by the refuter with explicit parallel histories and proves the refuter's guarantee already in the weaker theory $\mathbf{VTC}^0$.
The general limited-recursion lemma below is used only for the sequential specification.

The theory $\Vone$ has two sorts: numbers, which range over the natural numbers, and strings, which range over finite sets of numbers viewed as binary strings.
The number sort has $0,1,+,\cdot,\le$; the string sort has the bit relation $X(i)$, meaning that $i$ belongs to $X$, and the length $|X|$, which is one more than the largest element of $X$, or zero when $X$ is empty.
The standard model is the structure in which numbers are the natural numbers, strings are the finite subsets of $\N$, and every symbol of the language, including each function symbol adjoined later, has its intended meaning.

A bounded number quantifier ranges over numbers at most a displayed number term; a bounded string quantifier restricts the string length by such a term.
A $\Sigma^B_0$ formula has only bounded number quantifiers, and a $\Sigma^B_1$ formula is a $\Sigma^B_0$ formula preceded by a block of bounded existential string quantifiers.

A bounded formula is one in which every quantifier, of either sort, is bounded, and a bounded sentence is a bounded formula with no free variables.
When a formula is written as a displayed prefix of quantifiers followed by a formula $\psi$, we call $\psi$ its matrix; context distinguishes this logical sense from the matrices of numbers defined below.
For a string variable \(Y\), the notation \(Y\le t\) abbreviates \(|Y|\le t\).

Comprehension for a formula class asserts the following for each formula $\varphi(i)$ of the class and each bound $b$.
There is a string whose bit $i$ holds exactly when $\varphi(i)$ holds, for every $i<b$, and which has no bits at or above $b$.
The new string variable does not occur in $\varphi$.

Induction for a formula class asserts, for each formula $\varphi(i)$ of the class, that $\varphi(0)$ and $\forall i\,(\varphi(i)\to\varphi(i+1))$ imply $\forall i\,\varphi(i)$, with the other free variables of $\varphi$ as parameters.

The theory $\Vone$ consists of the basic axioms $\text{2-BASIC}$ for the two sorts \cite[Figure~2]{CookNguyen} together with $\Sigma^B_1$ comprehension \cite[Definition~5.3]{CookNguyen}.
Each conservative extension of $\Vone$ used below proves no new sentences of the original language, and proves induction for the polynomial-time predicates with fixed string parameters that occur in the construction.

We use $\forall i<t\,\theta(i)$ for a universal quantifier over numbers strictly below $t$.
The biconditional $P\leftrightarrow Q$ abbreviates $(\neg P\vee Q)\wedge(\neg Q\vee P)$.
The formula used to expand string equality $X=Y$ is
\[
|X|=|Y|\wedge\forall i<|X|\,\left(X(i)\leftrightarrow Y(i)\right).
\]
Equality substitution proves the forward implication, and the string-extensionality axiom of $\text{2-BASIC}$ proves the reverse implication.

\subsection{Defining Graphs, Arrays, and Bounded Scans}

We write $\mathsf{FP}$ for polynomial-time functions, with number arguments measured in unary and string arguments in binary.
A $\Pi^B_1$ formula is, dually to a $\Sigma^B_1$ formula, a $\Sigma^B_0$ formula preceded by a block of bounded universal string quantifiers.
A relation is $\Delta^B_1$ in a theory if it has $\Sigma^B_1$ and $\Pi^B_1$ definitions that the theory proves equivalent.
A defining graph is a formula specifying which output belongs to each input.
Totality asserts an output for every input, and uniqueness asserts that two outputs for the same input are equal.
A function given together with its specified defining graph is called \emph{chosen}, and a guarantee proved for it is a theorem about that graph.
The following classical interface justifies the chosen defining graphs used here.
\begin{fact}[Polynomial-time definitions and extensions]
\label{fact:fp-interface}
A function is in $\mathsf{FP}$ if and only if it has a $\Sigma^B_1$ graph whose totality and uniqueness $\Vone$ proves.
Polynomial-time relations have $\Delta^B_1$ definitions.
Adjoining such chosen defining functions gives a conservative extension of $\Vone$, retaining $\Sigma^B_1$ comprehension and bounded-number induction for its $\Delta^B_1$ predicates, with fixed string parameters \cite[Theorems~5.27 and~6.6, Corollaries~6.7--6.8 and~6.27]{CookNguyen}.
\end{fact}
We use this interface with explicit bounds and equations.
A function is admitted, that is, added to the language, together with three items: a chosen $\Sigma^B_1$ graph, a $\Vone$ proof of totality and uniqueness, and a proved polynomial bound on its output length.
Adjoining its symbol and graph equation is a definitional extension, so it proves no new theorems in the old language.

The correctness of an operation is never inferred from the existence of a polynomial-time implementation of it.
Instead, the graph equations of the operations, and the invariants proving their correctness, are part of the proof.
The definability and conservative-extension requirements are those of \Cref{fact:fp-interface}; the limited-recursion interface is stated in \Cref{lem:scan}.
Number functions use unary input size, so polynomially bounded numeric scans are feasible.

We code pairs of numbers by the pairing function
\[
\langle u,v\rangle=(u+v)(u+v+1)+2v.
\]
This function is injective: unequal sums $u+v$ give disjoint successive intervals, and at a fixed sum the value determines $v$ and then $u$.
It is strictly increasing in each argument, since increasing $u$ or $v$ by one increases $u+v$ and hence the value by at least $2(u+v+1)$.
It is also at least each argument and at most $(u+v+2)^2$.

Strings code arrays through this pairing.
Row $a$ of a string $X$ is the string $X^{[a]}$ given by $X^{[a]}(i)\leftrightarrow i<|X|\wedge X(\langle a,i\rangle)$.
A bit $i$ of a string $X$ is present when $X(i)$ holds, and an array is a string whose present bits all lie at values $\langle a,i\rangle$ of the pairing.

If rows $0,\ldots,v-1$ of an array have length at most $B$ and later rows are empty, then the array has length at most $\langle v,B\rangle$.
Indeed, every present bit has position $\langle a,i\rangle$ with $a<v$ and $i<B$, so by strict monotonicity $\langle a,i\rangle\le\langle v-1,B-1\rangle<\langle v,B\rangle$.
Conversely bounded comprehension forms the array with exactly the supplied rows, setting all other bits to zero; we call it the exact array with these rows.
Injectivity proves row recovery and uniqueness.
We call these facts (Arr).

A tuple of length $r$ is an array with empty row zero whose entry $a$, for $1\le a\le r$, is its row $a$.
A vector of dimension $n$ is a tuple of length $n$; a matrix is an array that stores its entry $(i,j)$ in array row $\langle i,j\rangle$.

A shape predicate, such as the statement that $X$ is a tuple of length $r$, asserts that $X$ equals the exact array whose rows are the rows of $X$ that the shape uses, so that $X$ has no other bits.
A tuple or array that satisfies its shape predicate is called exact.
We write $(X_1,\ldots,X_r)$ for the exact tuple of length $r$ whose entry $a$ is $X_a$, for $1\le a\le r$.
All these tests have only bounded number quantifiers after row symbols are admitted.

The graph of a bounded scan is built from the following recursion principle.
\begin{lemma}[Bounded scans and their invariants]
\label{lem:scan}
Let $u$ be a list of number and string parameters, let $G(u)$ and $H(i,u,Z)$ be string-valued functions with admitted graphs, where $i$ is a number and $Z$ is a string, and let $b(i,u)$ be a polynomial number term.
The limited-recursion theorem \cite[Definition~6.15 and Section~6B.2]{CookNguyen} supplies an admitted graph for the string-valued function $F$ with
\[
F(0,u)=G(u),\qquad
F(i+1,u)=\operatorname{Cut}\left(b(i,u),H(i,u,F(i,u))\right),
\]
where $\operatorname{Cut}(b,X)$ keeps the bits below $b$.
If $I(i,u,Z)$ is a polynomial-time predicate, $\Vone$ proves the base case $I(0,u,G(u))$, and $\Vone$ proves
\[
I(i,u,Z)\to I(i+1,u,H(i,u,Z)) \wedge |H(i,u,Z)|\le b(i,u),
\]
then $\Vone$ proves $I(i,u,F(i,u))$, and the cuts are idle on these states.
\end{lemma}
\begin{proof}
Limited recursion supplies the total unique graph and its displayed equations \cite[Definition~6.15 and Section~6B.2]{CookNguyen}.
For the invariant, use induction on the number $i$, with the input strings as parameters.
The base is the assumed property of $G$.
At a successor the transition invariant gives the next property and length bound.
That bound makes $\operatorname{Cut}$ equal to the uncut next state, by equality of bits and string extensionality.
This proves both assertions.
\end{proof}
We use this lemma for prefix sums, greedy basis scans, minima, quotient/remainder scans, and the first failing probe.
A minimum keeps one of the supplied entries rather than forming a sum.
A least-failure scan stores a bit recording whether a failure has been found, together with the index of the first failure; a malformed state, one not of this form, is replaced by a fixed default state.
For each use below we specify the transition, its length cap on all strings, the invariant satisfied by valid states, and the bounded number on which the induction runs.
Consequently no argument uses induction over a growing string length or over arbitrary cycle arrays (defined in \Cref{ssec:fixed-prime-depth}).

\subsection{Canonical Integers and Coded Rationals}

Before coding integers, we introduce the weaker theory used for the parallel constructions, because the integer laws below are imported from it.
A Boolean circuit is a finite directed acyclic graph whose gates compute bits from their incoming bits.
An AND gate outputs one when all its inputs are one, an OR gate when at least one is one, and a NOT gate reverses its single input bit.
The size is the number of gates, the depth is the largest number of gates along an input-to-output path, and a gate's fan-in is its number of incoming bits.

A threshold gate tests whether at least a specified number of its input bits are one.
A circuit family is a sequence of circuits $C_0,C_1,\ldots$ in which $C_\ell$ has $\ell$ input bits; its size and depth are the functions of $\ell$ giving the size and depth of $C_\ell$.
Polynomial size means size at most a polynomial in $\ell$, and constant depth means depth at most a constant independent of $\ell$.
A $\mathsf{TC}^0$ circuit family has polynomial size, constant depth, and unbounded-fan-in threshold, AND and OR gates, together with NOT gates.
We use DLOGTIME uniformity: from gate indices and the input length, gate types and wiring are decidable in logarithmic time.
The function class $\mathsf{FTC}^0$ permits polynomially many output bits computed by such a family.

The theory $\mathbf V^0$ has the same basic axioms as $\Vone$ and $\Sigma^B_0$ comprehension.
The theory $\mathbf{VTC}^0$ extends $\mathbf V^0$ by the counting axiom.
It states that for every string $X$ and number $n$ there is a sequence $c_0,\ldots,c_n$ with $c_0=0$, in which $c_{i+1}=c_i+1$ if $X(i)$ holds and $c_{i+1}=c_i$ otherwise.
Thus $c_i$ is the number of present bits of $X$ below $i$.
These are the standard uniform threshold conventions \cite[Section~2]{Jerabek22}.

A canonical signed integer has a sign bit and binary magnitude, with zero assigned its unique nonnegative zero code.
For definiteness, bit zero is the sign, present exactly when the value is negative, and magnitude bit $i$, the coefficient of $2^i$, occupies position $i+1$; leading zeros are omitted and the empty string codes zero.
Thus canonical integer one has length two.
Unsigned arithmetic, followed by this sign convention, gives canonical signed addition, multiplication, negation, comparison and absolute value.

A commutative ring has associative and commutative addition and multiplication, additive zero and inverses, multiplicative identity one, and distributivity.
An ordered ring has a total order preserved by addition, with $0<1$ and with the product of two positive elements positive.
It is discretely ordered when one is the least positive element.
An ordered field is an ordered ring in which every nonzero element has a multiplicative inverse.
The following imported arithmetic laws concern binary integers; the coded rationals below are built on them.
\begin{fact}[Provable binary arithmetic]
\label{fact:binary-arithmetic}
Binary addition and comparison have definitions in $\mathbf V^0$ satisfying their usual ordered-additive laws.
Binary multiplication and iterated addition have definitions in $\mathbf{VTC}^0$ satisfying the discretely ordered commutative-ring laws and the prefix-sum recurrence.
These laws extend to signed binary integers by a sign bit \cite[Section~9C.6]{CookNguyen}\cite[Section~2]{Jerabek15}.
\end{fact}
The integer laws are used inside $\Vone$ through the following inclusion.
The inclusion is standard: Cook and Nguyen develop $\mathbf{VTC}^0$ among the subtheories of their theory $\mathbf{VP}$ of polynomial-time reasoning \cite[Chapter~9, p.~255]{CookNguyen}, and $\Vone$ extends $\mathbf{VP}$ \cite[Theorem~8.3]{CookNguyen}.
Since we know of no single numbered statement of the inclusion, we include a short direct proof.
Because $\Vone$ proves every axiom of $\mathbf{VTC}^0$ (\Cref{lem:vtc-in-v1} below), adjoining the same defining axioms of these operations to $\Vone$ proves every law that they yield over $\mathbf{VTC}^0$, including the totality and uniqueness of the defining graphs.
\begin{lemma}[Counting in $\Vone$]
\label{lem:vtc-in-v1}
Every theorem of $\mathbf{VTC}^0$ is a theorem of $\Vone$.
\end{lemma}
\begin{proof}
The term $(Z)^u$ is the least $w<|Z|$ with $Z(\langle u,w\rangle)$, and is $|Z|$ when there is none \cite[Definition~5.56]{CookNguyen}; the pairing is the one defined above.
Let $\delta_{\mathrm{NUM}}(y,X,Z)$ state that $(Z)^0=0$ and, for every $u<y$, $(Z)^{u+1}=(Z)^u+1$ if $X(u)$ and $(Z)^{u+1}=(Z)^u$ otherwise.
The theory $\mathbf{VTC}^0$ is axiomatized by the axioms of $\mathbf V^0$ and the counting axiom, in the bounded form that the source notes is equivalent over $\mathbf V^0$ \cite[Definition~9.29 and the paragraph following it]{CookNguyen}
\[\forall X\,\forall y\,\exists Z\le1+\langle y,y\rangle\,\delta_{\mathrm{NUM}}(y,X,Z).\]

First, the terms $(Z)^u$ can be eliminated: a formula with bounded number quantifiers that uses such a term is equivalent in $\mathbf V^0$ to a $\Sigma^B_0$ formula.
Indeed, let $\mu(Z,u,w)$ be the $\Sigma^B_0$ formula $\left(w<|Z|\wedge Z(\langle u,w\rangle)\wedge\forall v<w\,\neg Z(\langle u,v\rangle)\right)\vee\left(w=|Z|\wedge\forall v<|Z|\,\neg Z(\langle u,v\rangle)\right)$.
The $\Sigma^B_0$ least number principle of $\mathbf V^0$ proves that exactly one $w\le|Z|$ satisfies $\mu(Z,u,w)$, and that $w$ is $(Z)^u$.
Hence an atomic formula $\theta((Z)^u)$ is equivalent to $\exists w\le|Z|\,(\mu(Z,u,w)\wedge\theta(w))$, and eliminating the occurrences one at a time gives a $\Sigma^B_0$ formula.

The theory $\Vone$ has the basic axioms of $\mathbf V^0$, and its $\Sigma^B_1$ comprehension contains $\Sigma^B_0$ comprehension, so it remains to prove the counting axiom in $\Vone$.
Fix $X$ and let $\varphi(y)$ be
\[
\exists Z\le1+\langle y,y\rangle\,\left(\delta_{\mathrm{NUM}}(y,X,Z)\wedge\forall u\le y\,(Z)^u\le u\right).
\]
This is a $\Sigma^B_1$ formula with parameter $X$, and $\Vone$ proves induction for it \cite[Corollary~6.1]{CookNguyen}.

For the base, $\Sigma^B_0$ comprehension gives the string $Z_0$ whose only bit is $0=\langle0,0\rangle$.
Then $|Z_0|=1=1+\langle0,0\rangle$ and $(Z_0)^0=0$, and the clause over $u<0$ is vacuous, so $\varphi(0)$ holds.

For the step, let $Z$ witness $\varphi(y)$.
Put $c=(Z)^y+1$ if $X(y)$ and $c=(Z)^y$ otherwise, so $c\le y+1$ by the invariant at $u=y$.
Simply adjoining the bit $\langle y+1,c\rangle$ to $Z$ need not work: $Z$ may have bits in rows above $y$, or empty rows whose value $(Z)^u=|Z|$ would change when the length grows.

Instead, $\Sigma^B_0$ comprehension below $1+\langle y+1,y+1\rangle$ forms the string $Z'$ whose bits are exactly the positions $\langle u,(Z)^u\rangle$ for $u\le y$ and the position $\langle y+1,c\rangle$.
Every such position has both coordinates at most $y+1$, by the invariant and $c\le y+1$, so it is at most $\langle y+1,y+1\rangle$ because the pairing is increasing in each argument; hence these bits lie below the comprehension bound and $|Z'|\le1+\langle y+1,y+1\rangle$.

By injectivity of the pairing, row $u\le y$ of $Z'$ contains exactly $(Z)^u$, and row $y+1$ contains exactly $c$.
Each of these entries is less than $|Z'|$, since the pairing is at least each argument and its position is a bit of $Z'$.
Hence $(Z')^u=(Z)^u$ for $u\le y$ and $(Z')^{y+1}=c$.
The clauses of $\delta_{\mathrm{NUM}}(y+1,X,Z')$ for $u<y$ are those of $\delta_{\mathrm{NUM}}(y,X,Z)$, the clause for $u=y$ holds by the choice of $c$, and $(Z')^0=(Z)^0=0$.
The invariant holds for $u\le y$ by that of $Z$ and for $u=y+1$ by $c\le y+1$.
Thus $\varphi(y+1)$ holds.

Induction gives $\forall y\,\varphi(y)$, and dropping the invariant gives the counting axiom.
\end{proof}
We use the signed operations of \Cref{fact:binary-arithmetic} with the canonical recoding just specified.
Using the shifted sign position is a bounded bit recoding of that signed arithmetic: deleting or adjoining the sign bit preserves magnitude addition and multiplication, and the definitions by signs commute with the recoding.
For opposite signs, the sum takes the sign of the operand with the larger magnitude, and its magnitude is the nonnegative difference of the two magnitudes, with the canonical zero when they are equal.
The defining equation of that difference is the unsigned addition equation stating that the smaller magnitude plus the difference equals the larger magnitude.
Hence signed equalities reduce to the unsigned ring equations by writing each value as its positive part minus its negative part.

Order is the corresponding signed order, with reversed magnitude order on negative values.
The least positive value is one, as follows from the first present magnitude bit.
We use these imported ring laws, not any unbounded integer induction scheme.

For canonical signed integers of lengths at most $L,L'$, negation and absolute value do not increase length, while the code lengths of sums and products satisfy
\[
|a+b|\le\max(L,L')+2,\qquad |ab|\le L+L'+1.
\]
Both inequalities follow by binary carries: a sum gains at most one magnitude bit, and a product has at most the sum of the two magnitude lengths; the extra displayed allowances include the sign position.
For a sum of $v\ge1$ entries, sequential addition gives the bound $\max_i|a_i|+2v$.
The empty sum is the canonical integer zero, of length zero.
The same prefix recursions prove these length bounds internally using \Cref{lem:scan}.

A rational code, or valid rational pair, is the exact tuple $(a,b)$ of length two, with $a$ canonical signed and $b$ canonical positive.
It denotes $a/b$, with no coprimality requirement.
Let $0_\Q=(0,1)$ and $1_\Q=(1,1)$.
Equality and order are cross-product tests:
\[
(a,b)=_\Q(c,e) \leftrightarrow ae=cb,
\qquad (a,b)\le_\Q(c,e) \leftrightarrow ae\le cb.
\]

Define the value operations without reducing fractions by
\[
(a,b)+(c,e)=(ae+bc,be),\quad
(a,b)(c,e)=(ac,be),\quad -(a,b)=(-a,b).
\]
The absolute value of a valid pair $(a,b)$ is $(|a|,b)$, where $|a|$ here means the canonical integer absolute value, rather than the length of its code.
Since $b>0$, this pair represents the absolute value of $a/b$ and does not increase either component length.
For a nonzero numerator the guarded inverse exchanges numerator and denominator, normalizing the denominator sign; the inverse at zero is $0_\Q$.
A malformed rational read returns $0_\Q$.
These definitions give the following interpretation used by the geometric argument.
\begin{lemma}[The rational-field interface]
\label{lem:code-field}
The theory $\Vone$ proves that $=_\Q$ is an equivalence relation respected by the coded operations and order, and that the quotient of rational codes satisfies the ordered-field laws.
If both components of two inputs have length at most $L$, each component of their sum or product has length at most $\operatorname{inc}(L)=2L+3$.
Negation, absolute value and the guarded inverse preserve their common component-length bound.
\end{lemma}
\begin{proof}
Positivity of denominators is preserved by multiplication.
Reflexivity and symmetry of cross-product equality are immediate.
For transitivity, let $(a,b)=_\Q(c,e)$ and $(c,e)=_\Q(d,f)$; then $ae=cb$ and $cf=de$ imply $aef=cbf=db e$; cancel the positive factor $e$ in the ordered integer ring to get $af=db$.

For pairs equal under $=_\Q$, multiplying their cross-product equations proves that the product respects $=_\Q$, and negating both sides of such an equation proves the same for negation; the sum is treated at the end of this paragraph.
For the absolute value and the guarded inverse, let $(a,b)=_\Q(a',b')$, so that $ab'=a'b$ with $b,b'>0$.
Then $a$ and $a'$ have the same sign, and $|a|b'=|a'|b$.
This is the cross-product equation for the absolute values.
When $a\ne0$, it is also the cross-product equation for the inverses $(\pm b,|a|)$ and $(\pm b',|a'|)$, and when $a=a'=0$ both inverses are $0_\Q$.
For the sum, $(ae+bc)/(be)$ is the usual common-denominator expression.
Replacing $(a,b)$ by the equal pair $(a',b')$ gives the two sides $(ae+bc)(b'e)=ab'e^2+bb'ce$ and $(a'e+b'c)(be)=a'be^2+bb'ce$ of the cross-product test; they share the term $bb'ce$, and $ab'e^2=a'be^2$ since $ab'=a'b$.
Replacing $(c,e)$ by a pair $(c',e')$ with $ce'=c'e$ gives the two sides $abee'+b^2ce'$ and $abee'+b^2c'e$, which are equal for the same reason, and replacing both pairs follows by transitivity.

Associativity and distributivity follow after taking the product of all participating denominators: the resulting numerator equations are the integer associativity and distributivity laws.
The additive zero is $(0,1)$ and the multiplicative identity is $(1,1)$.
The negative gives numerator $ab-ab=0$ on the common denominator; a nonzero inverse gives equal numerator and denominator up to the sign normalization.
No zero divisor remains, since a positive denominator is nonzero and the integer ring is ordered.

Cross-product order is total; multiplication by a positive common denominator transfers transitivity and compatibility with addition and positive multiplication to the integer order.
The order is respected by $=_\Q$: if $(a,b)=_\Q(c,e)$, then $ae=cb$ gives both $(a,b)\le_\Q(c,e)$ and $(c,e)\le_\Q(a,b)$, so transitivity carries every order relation of a code to each code equal to it.
This proves the ordered-field laws modulo $=_\Q$.

The component bounds follow directly from the preceding integer length inequalities: $ae+bc$ has length at most $2L+3$, and $be,ac$ at most $2L+1$.
The other operations only exchange, negate, or choose existing components, or return zero.
\end{proof}
The lemma bounds one operation at a time, so a fixed finite formula built from these operations has a component cap given by finitely many compositions of $\operatorname{inc}$.
Sums of a variable number of rationals need a separate bound, which we derive next.
For a variable sum of $v$ rationals, a common-denominator computation gives component cap $(v+1)(\langle3,2L\rangle+3)$ when input components have cap $L$.

To prove this cap, compute the common denominator and each cleared numerator as an iterated product of $v$ components, starting from canonical integer one, whose length is two.
The integer multiplication bound therefore gives denominator length at most $v(L+1)+2$, including $v=0$.

Each cleared numerator is a product of $v$ components and has the same bound; adding $v$ such numerators adds at most $2v$.
Thus $v(L+3)+2$ bounds both components.
Since $\langle3,2L\rangle+3\ge L+3$ and $L+3\ge2$, this is at most the displayed cap.

The products and sums are prefix scans with precisely these caps, so \Cref{lem:scan} proves the same statements internally.
Permuting or regrouping a bounded sum preserves value by induction using the ring laws; equality here means rational value equality, not equality of unreduced output strings.

Vector addition and scalar multiplication act coordinatewise; a linear combination of vectors $v_j$ is a sum $\sum_j a_jv_j$ with scalar coefficients $a_j$.
For a matrix $B$ and a column vector $x$, $(Bx)_i=\sum_jB_{ij}x_j$; for compatible matrices, $(AB)_{ij}=\sum_kA_{ik}B_{kj}$.
The transpose $B^T$ has entries $(B^T)_{ji}=B_{ij}$.
A coefficient vector $a$ represents the linear form $x\mapsto a\cdot x=\sum_i a_ix_i$.

For a nonnegative integer $r$, let $S_r$ consist of the maps from $\{1,\ldots,r\}$ to itself that use every output value exactly once; these are its permutations.
For $\pi\in S_r$, let $\operatorname{inv}(\pi)$ count pairs $i<j$ with $\pi(i)>\pi(j)$.
For an $r\times r$ matrix $A$, its determinant is the finite sum
\[
\det A=\sum_{\pi\in S_r}(-1)^{\operatorname{inv}(\pi)}
\prod_{i=1}^r A_{i,\pi(i)}.
\]
The determinant of the empty matrix is one.

A minor is the determinant of a square submatrix.
The cofactor of entry $(i,j)$ is $(-1)^{i+j}$ times the minor obtained by deleting row $i$ and column $j$.
The adjugate is the transpose of the cofactor matrix: its $(j,i)$ entry is that cofactor.
Multilinearity means linearity in each row separately when the other rows are held fixed.
Writing $I_r$ for the matrix with diagonal entries one and other entries zero, the adjugate identity is
\[
A\operatorname{adj}(A)=\operatorname{adj}(A)A=(\det A)I_r.
\]
The same definitions apply to coded rationals, with equality interpreted by $=_\Q$.

Fixed-order determinants require no extra algebraic axiom.
\begin{fact}[Fixed determinant identities]
\label{fact:dt-det}
For any externally fixed matrix order, Leibniz determinants satisfy multilinearity, sign change under a row interchange, vanishing with repeated rows, the adjugate identity, and the product formula $\det(AB)=\det A\det B$ for square matrices over the coded rational field.
All these identities have $\Vone$ proofs with the admitted operations.
\end{fact}
\begin{proof}
Fix the externally specified order $r$.
At $r=0$ the determinant and empty products are one, the adjugate identity has no entries, and the other assertions have no row choices.
Assume henceforth that $r\ge1$.

Expanding the Leibniz sum and grouping by the entry selected in a row proves linearity in that row, since every term uses exactly one of its entries.
Swapping two adjacent positions in a permutation changes its inversion count by one.
Swapping positions $i<j$ is a sequence of $2(j-i)-1$ adjacent swaps, so it reverses the sign $(-1)^{\operatorname{inv}(\pi)}$.
Pair the terms before and after a row interchange by swapping the two assigned column values.
Their products agree and their signs are opposite, proving sign change.
If the rows are equal, the same pairing is an involution without fixed points on the terms of one determinant, so they cancel.

Group the terms with $\pi(i)=j$ to expand along row $i$.
If $s$ of the positions before $i$ have values below $j$, removing position $i$ and value $j$ removes
\[
(i-1-s)+(j-1-s)=i+j-2-2s
\]
inversions.
Thus the sign factor relative to the remaining permutation is $(-1)^{i+j}$, and the grouped sum is exactly the cofactor expansion.
The entry $(i,j)$ of $A\operatorname{adj}(A)$ is the cofactor expansion along row $j$ after replacing that row by row $i$.
It is $\det A$ when $i=j$, and zero when $i\ne j$ by the repeated-row cancellation.

Reindexing the Leibniz sum by the inverse permutation gives $\det(A^T)=\det A$.
Indeed, an inversion $(i,j)$ of $\pi$ corresponds to the inversion $(\pi(j),\pi(i))$ of its inverse, so their inversion counts agree.
Transposing the deleted submatrices therefore gives $\operatorname{adj}(A^T)=\operatorname{adj}(A)^T$.
Apply the already proved identity to $A^T$ and transpose it to obtain $\operatorname{adj}(A)A=(\det A)I_r$.

For the product formula, let $\alpha$ range over all maps from $\{1,\ldots,r\}$ to itself and put
\[
T_\alpha=\sum_{\pi\in S_r}(-1)^{\operatorname{inv}(\pi)}
\prod_{i=1}^r B_{\alpha(i),\pi(i)}.
\]
Expanding every selected entry of $AB$ and grouping by its intermediate indices gives
\[
\det(AB)=\sum_\alpha\left(\prod_{i=1}^r A_{i,\alpha(i)}\right)T_\alpha.
\]
If $\alpha(i)=\alpha(j)$ at two distinct positions, swap the values $\pi(i)$ and $\pi(j)$.
This is a fixed-point-free pairing of equal products with opposite signs, so $T_\alpha=0$.
Otherwise $\alpha$ uses $r$ distinct values in an $r$-element range and hence is a permutation, say $\tau$.

Each permutation $\tau$ can be sorted by swapping adjacent inverted values, one inversion removed per swap; an unsorted permutation has an adjacent descent.
Consequently it is a composition of $\operatorname{inv}(\tau)$ adjacent transpositions.
For $\pi(i)=\nu(\tau(i))$, applying these position swaps gives
\[
(-1)^{\operatorname{inv}(\pi)}
=(-1)^{\operatorname{inv}(\nu)}(-1)^{\operatorname{inv}(\tau)}.
\]
Also $\prod_i B_{\tau(i),\nu(\tau(i))}=\prod_{a=1}^rB_{a,\nu(a)}$, by reindexing with $\tau$.
As $\pi$ runs over $S_r$, so does $\nu$, and hence $T_\tau=(-1)^{\operatorname{inv}(\tau)}\det B$.
Substituting the two cases for $\alpha$ gives
\[
\det(AB)
=\sum_{\tau\in S_r}(-1)^{\operatorname{inv}(\tau)}
\left(\prod_{i=1}^rA_{i,\tau(i)}\right)\det B
=(\det A)\det B.
\]

Every sum and permutation list in this proof has an externally fixed length.
The calculations are therefore finite deductions from the commutative-ring laws of \Cref{lem:code-field}, using the admitted finite sums and products.
They assert no variable-order determinant algorithm.
\end{proof}
The same principle transfers any fixed ordered-field deduction in the geometric lemmas: substitute the interpretation of \Cref{lem:code-field} into its finite proof.
It does not transfer arbitrary true sentences about real numbers.
A second, more limited observation is useful for closed finite objects such as the fixed braid arrangement defined in \Cref{sec:locality}.
A numeral is one of the closed terms $0$, $1$, $1+1$, $(1+1)+1$, and so on; we write $\underline c$ for the numeral with value $c$.
For a finite set $S$ of natural numbers, the string constant $\underline S$ is the string formed by $\Sigma^B_0$ comprehension from the formula $\bigvee_{s\in S}i=\underline s$.
The comprehension bound is $\underline{1+\max S}$, or $0$ when $S$ is empty.
Adjoining it is a definitional extension, and $\Vone$ proves $\underline S(i)\leftrightarrow\bigvee_{s\in S}i=\underline s$ and $|\underline S|=\underline{1+\max S}$, or $|\underline S|=0$ when $S$ is empty.
\begin{lemma}[Closed bounded calculations]
\label{lem:sigma1-complete}
A true closed bounded sentence of the admitted polynomial-time language whose constants are numerals and string constants has a $\Vone$ proof.
\end{lemma}
\begin{proof}
After the outer variables of the sentence have been replaced by constants, a bounded number quantifier with bound $\underline b$ ranges over the numerals $\underline0,\ldots,\underline b$.
Indeed, the basic order axioms prove $x\le y+1\leftrightarrow x\le y\vee x=y+1$, and $b$ applications of it, ending with $x\le0\to x=0$, prove $x\le\underline b\to\bigvee_{c\le b}x=\underline c$.
A bounded string quantifier of bound $b$ ranges over the fixed finite list of strings of length at most $b$.
This list is exhaustive in $\Vone$: split on the $b$ possible bit positions by excluded middle, and in each case use the length bound and string extensionality to identify the string with the constant having those bits.

Existential quantifiers are proved by one successful value; universal quantifiers are proved by all values in their respective finite lists.
Repeat this expansion through the fixed formula, evaluating each bound after its outer values have been fixed.

Every closed admitted function value has a finite witnessing computation for its chosen graph; its defining equation and uniqueness prove the returned value.
The basic axioms decide the resulting numeral equalities, inequalities and bit tests.
The true finite Boolean combination therefore has a finite proof.
This construction chooses a finite proof externally.
It asserts no feasible bound uniform in the numeral constants or string bounds.
\end{proof}
Neither of these two transfers supplies an induction principle for arrangements given by a string variable, such as those built from an arbitrary candidate network.
Those arrangements are handled by the explicit bounded scans and coordinate invariants below.

\subsection{Network and Correctness Conventions}

For \(n\ge1\), write \([n]=\{1,\ldots,n\}\).
For an input vector \(x=(x_1,\ldots,x_n)\), put \(\MAX_n(x)=\max_{1\le i\le n}x_i\).
An affine map has coordinates \(y_i=\sum_t W_{it}x_t+\eta_i\), with matrix \(W\) and bias vector \(\eta\).
A scalar rational ReLU network is a sequence of affine maps with coordinatewise $\relu(t)=\max(0,t)$ at its hidden layers and an affine scalar output.
Depth counts hidden layers; a depth-zero network is affine.
The width of a hidden layer is its number of coordinates, or neurons, and all widths are positive.

A network $N$ on $n$ inputs computes the function $x\mapsto N(x)$ on $\R^n$ obtained by applying its affine maps in order and applying $\relu$ coordinatewise to the output of every affine map except the last.
The value of a hidden coordinate before $\relu$ is applied is its preactivation.
An activation pattern chooses, for each hidden ReLU, either its linear branch or its zero branch; at a zero preactivation either choice agrees with the value.
We write $N\ne\MAX_n$ on $\R^n$ when $N(x)\ne\MAX_n(x)$ for some $x\in\R^n$.
Weights are the nonbias matrix entries; biases are stored separately and may be arbitrary rationals.

A network is coded by an exact array that contains metadata followed by the flattened parameter list, in fixed layer, row and column order.
The metadata records the input and layer dimensions, the hidden depth, the total hidden-neuron count and the parameter count, each as a unary string (the number $r$ is stored as the string of $r$ ones), together with their consistency equations.
Each parameter is a rational pair as above; matrices are reconstructed by their declared dimensions and consecutive offsets in the list.
This explicitly stores every zero entry too.

The validity predicate $\Net(N,n)$, specified with the fixed record below, checks all metadata, rational entries, and unused bits.
Consequently for length $m=|N|$ the input dimension, hidden depth, total hidden-neuron count, and parameter count are at most $m$.
Every integer component also has length at most $m$.
These bounds follow from the unary metadata rows, because by the definition of rows no row of an array, and hence no nested entry, is longer than the array itself.

We record a length principle for normalizations, which later rewrite the stored parameters of a network.
Suppose a normalization changes no metadata, introduces no new parameter rows and does not increase any component length.
Then the whole network's encoded length does not increase.
Since the enclosing pairings are nondecreasing, every new bit position is bounded by the position of the old last bit of its corresponding nonempty component.
A zero component remains the canonical zero.


\paragraph{\emph{A Fixed Network Record.}}
We fix the following refinement of the network encoding, and use it for every network constructor and history checker in the paper.
Write \(\operatorname{Un}(r)\) for the string whose first \(r\) bits are one and which has no other bits; thus its length is \(r\), and \(\operatorname{Un}(0)=\varnothing\).
We also write \(1^r\) for \(\operatorname{Un}(r)\), as is usual in complexity theory.
This unary field is distinct from a canonical signed binary integer.

Matrix neuron and column indices in this record description are zero-based.
Input column \(i\) corresponds to the main text's mathematical coordinate \(x_{i+1}\).
An exact tuple stores its entry \(a\) in row \(a\), for \(a\ge1\), and has empty row zero, as in (Arr); below, the positions of a tuple are called its entries.

Consider a network with input dimension \(d_0=n\), hidden depth \(k\), positive hidden widths \(d_1,\ldots,d_k\), and scalar output \(d_{k+1}=1\).
Its hidden-neuron count is \(q_{\mathrm{hid}}=\sum_{j=1}^k d_j\).
For affine layer \(1\le j\le k+1\), let $p_j=d_jd_{j-1}+d_j$ count its matrix and bias parameters, and let $M=\sum_{j=1}^{k+1}p_j$ be the total count.
Define offsets $O_j=\sum_{1\le\ell<j}p_\ell$ for $1\le j\le k+2$; in particular $O_1=0$ and $O_{k+2}=M$.

The exact dimension tuple \(\mathsf{Dim}=(\operatorname{Un}(d_1),\ldots,\operatorname{Un}(d_{k+1}))\) lists in unary the dimensions after the input.
The tuple \(\mathsf{Meta}\) stores this tuple and the input dimension, hidden depth, hidden-neuron count and parameter count.
The tuple \(\mathsf{Par}\) lists every weight and bias in the order specified next.
The exact record is
\[
N=(\mathsf{Meta},\mathsf{Par}),\qquad
\mathsf{Meta}=(\operatorname{Un}(n),\mathsf{Dim},
\operatorname{Un}(k),\operatorname{Un}(q_{\mathrm{hid}}),\operatorname{Un}(M)).
\]

The exact tuple \(\mathsf{Par}\) has \(M\) rational entries.
Its entry \(O_j+i d_{j-1}+t+1\) is the weight \(W_j[i,t]\), for \(i<d_j\) and \(t<d_{j-1}\).
The bias \(\eta_j[i]\), for \(i<d_j\), is the entry \(O_j+d_jd_{j-1}+i+1\) of \(\mathsf{Par}\).
Thus, for each layer \(1\le j\le k+1\), the weight entries are exactly \(O_j+1,\ldots,O_j+d_jd_{j-1}\) and the bias entries are exactly \(O_j+d_jd_{j-1}+1,\ldots,O_j+p_j=O_{j+1}\).
Hence each entry holds exactly one weight or bias, and each weight and bias has exactly one entry.
In particular, each full matrix precedes its bias vector, and every zero entry is present in this ordering.

For example, a network with two inputs, one hidden layer of width two and scalar output has \(p_1=6\), \(p_2=3\), \(M=9\), \(O_1=0\) and \(O_2=6\).
Its entries \(1,\ldots,4\) hold the matrix \(W_1\) row by row, entries \(5\) and \(6\) the bias vector \(\eta_1\), entries \(7\) and \(8\) the matrix \(W_2\), and entry \(9\) the bias \(\eta_2[0]\).

The predicate \(\Net(N,n)\) checks this exact two-entry record, every unary field, the positivity of all dimensions, \(d_0=n\ge1\), \(d_{k+1}=1\), the equations for \(q_{\mathrm{hid}}\) and \(M\), that \(\mathsf{Par}\) has exactly \(M\) entries, every rational entry, and the absence of unused bits.
It validates dimensions and offsets before using them.
A failed read has the previously specified rational-zero default.
The prefix counts and offsets are formed by the admitted aggregate arithmetic, with bounds from the input-code length.
No allocation uses the value of an unchecked binary field.

The affine identity on one input is the particular record with \(k=q_{\mathrm{hid}}=0\), \(\mathsf{Dim}=(\operatorname{Un}(1))\), \(M=2\), and \(\mathsf{Par}=(1_{\Q},0_{\Q})\).
Denote it by \(\mathsf{Id}_1\).

For later reference, we record the history vector of a network, which stores the value of every hidden neuron and the output.
It has \(q_{\mathrm{hid}}+1\) rational entries.
For \(1\le j\le k\), the value of hidden neuron \(i<d_j\) of layer \(j\) is entry \(1+\sum_{1\le\ell<j}d_\ell+i\), and the scalar output is entry \(q_{\mathrm{hid}}+1\).

A valid rational input \(n\)-vector is an exact tuple of \(n\) valid rational pairs in the coordinate order specified above.
The predicate \(\mathsf{LEval}(N,n,x,Z)\) first requires \(\Net(N,n)\).
It then checks that \(Z\) is this exact history vector and \(x\) is a valid rational input tuple, and checks all adjacent affine/ReLU equations and the final affine equation.
At depth zero its single equation applies the output affine map directly to \(x\).
All equations compare rational values by cross-products, rather than comparing unreduced pair codes.
\space

\paragraph{\emph{The General-Depth Evaluator.}}
The general-depth network evaluator specified in this subsection computes every layer value and the output value as an integer numerator over a positive common denominator, using canonical integer arithmetic and no division.
On arbitrary input strings it first checks complete network and input validity, returning the empty string on failure.
On a valid network of depth bounded by a separately fixed $K$, its layer recursion is a fixed composition of bounded array scans.

The evaluator uses common denominators.
Let $m=|N|$, let $s=|x|$ for a valid input vector, let $D$ be the positive product of all parameter denominators, and let $E$ be the positive product of all input denominators.
For a parameter $a_\lambda/b_\lambda$, define its cleared integer by
\[
\widehat a_\lambda=a_\lambda\prod_{\mu\ne\lambda}b_\mu.
\]
Thus $\widehat a_\lambda=D(a_\lambda/b_\lambda)$; no division is used.

Use zero-based input indices $0\le i<n$, as in the network record: $\alpha_i/e_i$ denotes the mathematical coordinate $x_{i+1}$.
Put $V_{0,i}=\alpha_i\prod_{0\le t<n, t\ne i}e_t$, so $V_{0,i}/E=x_{i+1}$.
Write $\widehat W_j[i,t]$ and $\widehat\eta_j[i]$ for the cleared weight and bias parameters.

At affine layer $j\ge1$ form the integer
\[
\bar V_{j,i}=\sum_{t<d_{j-1}}\widehat W_j[i,t]V_{j-1,t}
+\widehat\eta_j[i]D^{j-1}E.
\]
At a hidden layer set $V_{j,i}=\max(0,\bar V_{j,i})$; at the output layer set $V_{j,0}=\bar V_{j,0}$.
The denominator at that layer is exactly $D^jE$.

Induction on $j$ shows that $V_{j,i}/(D^jE)$ is the value of coordinate $i$ at layer $j$.
The base $j=0$ is the displayed input identity.
For the step, dividing the displayed recurrence by $D^jE$ gives the required affine combination of the preceding values, and positivity of $D^jE$ makes taking the maximum with zero agree with ReLU.
At the last layer no ReLU is taken.
This proves every adjacent value equation.

We now give explicit polynomial component bounds.
Put $B=4(m+1)^2$, $L_0=(m+1)(s+2)+2$, and $R=2B+2m+8$.
There are at most $m$ parameter denominators, each of length at most $m$; hence $D$ and every cleared parameter have length at most $B$.
There are at most $m$ input coordinates, each component of length at most $s$, so $E$ and every $V_{0,i}$ have length at most $L_0$.

We bound the numerators layer by layer.
Suppose the preceding numerators have length at most $L_0+(j-1)R$.
Each weighted term then has length at most $L_0+(j-1)R+B+1$.
The scaled bias has length at most
\[
L_0+j(B+1)+2\le L_0+(j-1)R+B+3.
\]
Adding at most $m+1$ terms increases this latter bound by at most $2(m+1)$.
Since $B+3+2(m+1)\le R$, the next numerator has length at most $L_0+jR$; replacing a negative numerator by zero does not increase length.
The positive denominator $D^jE$ has length at most $L_0+j(B+1)+1\le L_0+jR$ for $j\ge1$.

There are at most $m$ affine layers, since each affine layer has at least two parameters and the parameter count is at most $m$; so $L_0+mR$ bounds every component.
The number of stored entries is polynomial in $m$, and exact tuple reconstruction has fixed nesting depth, giving a polynomial bound on the full history in $m+s$.

The products, cleared parameters and layer transitions are bounded scans with these caps.
\Cref{lem:scan} therefore supplies their chosen $\Vone$ graphs and proves, by numerical induction on the layer, both the caps and the displayed value equations.
Malformed networks or input vectors return the empty string before any layer allocation.
This gives the stated polynomial-time rational-value evaluator without reducing fractions.
This evaluator is a general-depth construction in $\Vone$; only the separately specified fixed-depth evaluator $\mathsf{Eval}_K$ of \Cref{tc:parallel-arithmetic} is used by the parallel refuter of \Cref{tc:refuter}.

\paragraph{\emph{Correctness and Denominator Conditions.}}
The denominator conditions at the end of this subsection use the following arithmetic notions.
For integers $a,b$, $a$ divides $b$ when $b=ac$ for an integer $c$.
A prime is an integer $q\ge2$ whose only positive divisors are one and itself.
Integers are coprime when their only common positive divisor is one.
Congruence modulo $q$ means that $q$ divides the difference; a residue of $a$ modulo $q$ is an integer $r$ with $0\le r<q$ congruent to $a$.
A denominator has nonzero residue precisely when $q$ does not divide it.

We now define the correctness predicate.
Recall that $\Net(N,n)$ is complete validity with input dimension $n$.
Write $\Depth(N,K)$ for the conjunction of $\Net(N,n)$, for the input dimension $n$ recorded in $N$, with the condition that the hidden depth is at most $K$.
Let $\Err(N,n,x)$ say that $x$ is a valid rational $n$-vector and the general-depth evaluator at $N,x$ returns a nonempty valid rational output whose value differs from $\MAX_n(x)$.
A failed network or input guard counts as no error.
Fix a polynomial $p_0$ with nonnegative integer coefficients supplied by \Cref{lem:short}, and put
\[
p(m)=p_0(m)+2^{100}(m+2)^{40}.
\]
This one polynomial is used for every fixed prime and depth; its second term bounds the probes, the test inputs that the refuter evaluates and from which it returns its output (\Cref{sec:probes}).
Set
\[
\Computes(N,n)\equiv\mathsf{Net}(N,n)\wedge
\forall x\,\left(|x|\le p(|N|)\rightarrow\neg\Err(N,n,x)\right).
\]
The quantifier is over binary strings; strings that are not valid rational $n$-vectors do not satisfy $\Err$.
In the standard model this predicate means correctness on all real inputs by \Cref{lem:short}.
That equivalence is an ordinary mathematical statement; no real sort is added to $\Vone$.

For a fixed prime $q$, the raw denominator predicate $\mathsf{GoodDen}_q(N,n)$ is complete validity together with nonzero residue modulo $q$ for every stored nonbias denominator.
The actual residue scan is specified in \Cref{sec:graphs}.
A value-based version $\mathsf{ValGood}_q(N,n)$ first cancels common factors $q$ in each numerator/denominator pair and requires success at every nonbias entry; its graph and value-preservation proof are given with \Cref{cor:fixed-prime-depth-values}.
No restriction is imposed on bias denominators.

For a real number $t$, $\lfloor t\rfloor$ and $\lceil t\rceil$ are its greatest integer lower bound and least integer upper bound, respectively.
For $a>0$ and a base $b>1$, $\log_b a$ is the real exponent $u$ satisfying $b^u=a$.

For a positive integer \(\beta\), write \(\Z[1/\beta]=\{a/\beta^t:a\in\Z, t\in\N\}\).
\begin{fact}[Known fixed-base threshold, {\cite[Theorem~2]{AHM25}}]
\label{fact:fixed-base-depth}
If $q$ is prime and does not divide a fixed positive integer $\beta$, a network that represents $\max(0,x_1,\ldots,x_d)$ and whose nonbias weights belong to $\mathbb Z[1/\beta]$ requires at least $\lceil\log_q(d+1)\rceil$ hidden layers.
\end{fact}
The cited model permits arbitrary real biases.
Setting the last input of a network that computes $\MAX_n$, with nonbias weights in $\Z[1/\beta]$, to zero gives a network of the same depth that computes $\max(0,x_1,\ldots,x_{n-1})$ and whose nonbias weights are among the original ones.
The fact with $d=n-1$ then shows that such a network has at least $\lceil\log_q n\rceil$ hidden layers; in particular, $K$ hidden layers suffice only when $n\le q^K$.
For a finite family of nonbias denominators not divisible by $q$, their product $\beta$ is not divisible by $q$, since $q$ is prime, and every nonbias weight with one of these denominators lies in $\Z[1/\beta]$, so \Cref{fact:fixed-base-depth} applies with this $\beta$.

The same semantic consequence also follows from the case of \Cref{fact:fixed-base-depth} for networks without biases, by passing to the recession function $x\mapsto\lim_{t\to\infty}N(tx)/t$.
By positive homogeneity of $\relu$, $N(tx)/t$ is the value at $x$ of the same network with every bias divided by $t$.
By continuity of the network function in its biases, the limit therefore exists and is computed by the same network with every bias set to zero, which has the same nonbias weights.
If $N$ computes $\MAX_n$, so does this network, since $\MAX_n(tx)/t=\MAX_n(x)$.
It has the depth and the nonbias weights of $N$, so the preceding argument, applied to it, gives the same bound for $N$ while using \Cref{fact:fixed-base-depth} only for networks without biases.
Our proof below instead establishes a bounded-arithmetic version of this bound, with a specified refuter, and does not infer provability from the imported semantic fact.
\space

\section{A Finite Cone Calculus}
\label{ssec:fixed-prime-depth}

This section and the four that follow prove the lower bound of \Cref{thm:fixed-prime-depth} inside $\Vone$.
The proof combines an intersection calculation in a fixed dimension, in which piecewise-linear functions act on weighted cones of a hyperplane arrangement, with a divisibility argument for the functions computed by a candidate network.
For each externally fixed prime $q$ and hidden depth $K\ge1$, the relevant dimension is $d=q^K$.
We construct the finite cone calculus in any separately fixed $d\ge2$.

The product defined below is modeled on the cycle intersection construction of \cite[Definition~3.4, Remark~3.5 and Proposition~3.7]{AR10}; the modular obstruction follows \cite{HHL23,AHM25}.
Every geometric identity required by the internal argument is proved below from the coded-rational and bounded-scan interfaces in \Cref{sec:foundations}.
No general tropical intersection theorem is assumed to hold inside the arithmetic theory.
The dimensions and product lengths are fixed numerals, while the network widths and arrangement lengths are inputs.

\paragraph{\emph{The fixed-dimensional product.}}
Fix an integer $d\ge2$, and write $[d]=\{1,\ldots,d\}$.
For vectors \(u,v\in\Q^d\), their dot product is \(u\cdot v=\sum_{c=1}^d u_cv_c\), and \(\operatorname{sgn}(t)\) is \(-1,0,1\) according as \(t<0,t=0,t>0\).
The unit vector \(e_c\) has coordinate one at \(c\) and zero elsewhere.
The span of a list consists of its rational linear combinations; the list is independent if only the zero coefficient list has combination zero.
A basis is an independent spanning list, and dimension is its length.
For a matrix \(B\), its kernel \(\ker B\) is the space of vectors \(x\) with \(Bx=0\).
A linear subspace contains zero and is closed under rational linear combinations.
A linear flat is a linear subspace; its codimension in \(\Q^d\) is \(d\) minus its dimension.
We order finite coordinate sets by their increasing coordinate lists.
Lexicographic order compares two distinct equal-length lists at their first differing entry.

An arrangement $\mathcal H=(h_0,\ldots,h_{s-1})$ is a list of nonzero integer $d$-vectors, called its normals or rows, that begins with the $d$ units in increasing order, that is, $h_{c-1}=e_c$ for $c\in[d]$.
Repeated normals are allowed.
A sign vector has entries in $\{-1,0,1\}$.

For $x\in\Q^d$, the covector of $x$ is the sign vector $\operatorname{cov}(x)$ of length $s$ with entries $\operatorname{cov}(x)_i=\operatorname{sgn}(h_i\cdot x)$ for $i<s$.
For a sign vector $X$ of length $s$, let $Z(X)=\{i<s:X_i=0\}$ denote its zero-sign index set.

A vector $h$ kills a vector $x$ when $h\cdot x=0$, and kills a set when it kills each of its elements.
The realized sign cone of $X$ is the set of rational points $x$ for which $\operatorname{cov}(x)=X$; only nonempty such sets are called cones, and $X$ is a realized covector when its realized sign cone is nonempty.
The zero normals of $X$ are the normals $h_i$ with $i\in Z(X)$.

For sign vectors, $X\sqsubseteq Y$ means $X_i\in\{0,Y_i\}$ for every index, and $X\circ Y$ retains the nonzero entries of $X$ and uses $Y$ elsewhere.
We write $Y\sqsupseteq X$ for $X\sqsubseteq Y$, and $X\sqsubset Y$ when $X\sqsubseteq Y$ and $X\ne Y$.
For realized covectors with $X\sqsubseteq Y$, we say that $Y$ lies above $X$.
All scalar equalities of rational codes mean equality of values.

The rank of a matrix is the largest order of a nonzero square minor, where a minor is the determinant of a square submatrix.
A full-row minor of a matrix with $r$ rows is a minor of order $r$, so that its submatrix uses every row.
For a square matrix $B$, the adjugate $\operatorname{adj}(B)$ is the transpose of its cofactor matrix.

For a sign vector $X$, scan its zero normals $h_i$ in increasing order of $i$, and retain $h_i$ when appending it as a row to the rows retained so far raises the rank.
Let $B_X$ be the retained matrix, $r_X$ its row count, and
\[
E(X)=\ker B_X,\qquad \dim X=d-r_X.
\]
We call $E(X)$ the associated linear flat of $X$ and $\dim X$ its dimension; \Cref{lem:fp-flats-complete} shows that $E(X)$ is the set of points killed by every zero normal of $X$.
A cone of dimension $d$ is a chamber, and a cone of dimension $d-1$ is a ridge.

The basis $\mathcal N(X)$ of $E(X)=\ker B_X$ is fixed as follows.
If $r_X=0$, use the units; if $r_X=d$, use the empty list.
Otherwise, list the $r_X$-element column sets $J$ in lexicographic order.
For such a $J$, let $B_J$ be the square submatrix of $B_X$ formed by its columns in $J$.
Choose the first $J$ with $\det B_J\ne0$, and put $\delta=\det B_J$.
For each free coordinate $f\notin J$, in increasing order of $f$, use the vector
\[
v^f_f=\delta,\quad v^f_{f'}=0\text{ for }f'\notin J\cup\{f\},\quad
v^f_J=-\operatorname{adj}(B_J)B_f.
\]
Here $B_f$ is column $f$ of the retained row matrix $B_X$.
Every determinant in this geometric engine has order at most the externally fixed $d$.
Each is evaluated by its fixed Leibniz expression.

For ordered $(d-1)$-tuples of arrangement rows define the cross by
\[
\operatorname{cr}(h_{g_1},\ldots,h_{g_{d-1}})_c
=(-1)^{d+c}\det(h_{g_1},\ldots,h_{g_{d-1}})_{[d]\setminus\{c\}}.
\]
Here $(h_{g_1},\ldots,h_{g_{d-1}})_{[d]\setminus\{c\}}$ is the square matrix with rows $h_{g_1},\ldots,h_{g_{d-1}}$ restricted to the columns in $[d]\setminus\{c\}$.
The dot product of the cross with $x$ is the determinant of the $d\times d$ matrix with rows $h_{g_1},\ldots,h_{g_{d-1}},x$.
The rays are the entries of the following list, indexed from zero: an explicit zero vector at index zero, followed by both signs of every such ordered cross.
For a positive ray index $a$, write $a-1=2g+\varepsilon$ with $\varepsilon\in\{0,1\}$, and read the $d-1$ base-$s$ digits of $g$ from least to most significant as $g_1,\ldots,g_{d-1}$.
This index denotes $(-1)^\varepsilon\operatorname{cr}(h_{g_1},\ldots,h_{g_{d-1}})$.
Thus
\[
R=1+2s^{d-1},\qquad S=R^d.
\]
Here $R$ is the length of the ray list and $S$ is the number of ordered $d$-tuples of ray indices.

We also need a perturbation by which the signs of a vector $w$ break the ties in those of an integer vector $u$.
For a rational $d$-vector $w$ and an integer $d$-vector $u$ put
\[
\Delta(w)=2\max\left(\{|h_i\cdot w|:i<s\}\cup\{0\}\right)+1,
\qquad u\boxplus w=\Delta(w)u+w.
\]
Decode $\sigma<S$ into its $d$ base-$R$ digits from least to most significant; these are its ordered ray indices.
The integer representative $\varpi(\sigma)$ is computed from $u=0$ by replacing $u$ with $u\boxplus w$, for the ray $w$ of each of these indices in this order.
Since $\Delta(0)=1$, the zero ray leaves $u$ unchanged, so ray index zero makes its perturbation step idle; the proof of \Cref{lem:fp-flats-complete} uses this to represent points of dimension below $d$ by indices whose last rays are zero.
The explicit zero ray is needed at $d=2$: for $d\ge3$ a tuple with a repeated row already gives a zero cross, but at $d=2$ a cross is formed from the single nonzero row $h_{g_1}$ and is never zero.

The covector of an index $\sigma<S$ is $\operatorname{cov}(\varpi(\sigma))$, and $\sigma$ matches a point $x$ when its covector is $\operatorname{cov}(x)$.
A cone index is an index $\sigma<S$ that is least among the indices with its covector.
Explicitly, the canonical-index predicate is
\[
\mathsf{Can}(i) \leftrightarrow i<S \wedge
\forall j<i\,\left(\operatorname{cov}(\varpi(j))\ne\operatorname{cov}(\varpi(i))\right).
\]

When order, cover or chamber predicates below take numerical indices, they first require \(\mathsf{Can}\) for every such index and then test its covector.
The arrays of families and cycles defined below are indexed by the indices below $S$ and have zero entries at every noncanonical index.
Thus summing over all indices below \(S\) counts each cone once; the remaining slots are padding.
The function $\operatorname{idx}(x)$ is the least matching index and uses sentinel $S$ when there is none.
Before \Cref{lem:fp-flats-complete} proves that every rational point has a matching index, $\operatorname{idx}$ and the other definitions above remain total guarded definitions.

\begin{lemma}[Flats, completeness and perturbation]
\label{lem:fp-flats-complete}
For every sign vector $X$ of length $s$, the retained zero normals of $X$ are independent and span every zero normal of $X$.
Consequently $E(X)$ is the common zero flat of those normals.
Every rational point has a matching cone index.
For realized covectors $X\sqsubseteq Y$ with $X\ne Y$, one has $\dim X<\dim Y$.
For an integer vector $u$ and a rational vector $w$, one has $\operatorname{cov}(u\boxplus w)=\operatorname{cov}(u)\circ\operatorname{cov}(w)$.
\end{lemma}

\begin{proof}
We first record the fixed-dimensional algebra used in the scan.
Appending a row to a matrix of rank $r$ raises rank by at most one: expansion of a new minor of order at least $r+2$ along the appended row uses only old minors of order at least $r+1$, all zero.
A list is independent exactly when it has a nonzero full-row minor.
If such a minor $B_J$ exists, the adjugate identity shows that $B_J^{\mathsf T}c=0$ forces $c=0$.

Conversely, in a greedy scan of an independent list every row raises the rank.
Indeed, the rows retained so far each raised the rank by exactly one, so their rank equals their number and they have a nonzero full-row minor.
A row that did not raise the rank would then be written by the spanning step below as a combination of them, contradicting independence.

Deleting a row from an independent list preserves independence, since a vanishing combination of the shorter list, extended by a zero coefficient on the deleted row, is a vanishing combination of the original list, so all its coefficients are zero.
We call the following argument the spanning step.
If $B$ has a nonzero full-row minor and appending $v$ does not raise rank, solve $B_J^{\mathsf T}c=v_J$ on its first such minor.
With $b_1,\ldots,b_r$ the rows of $B$, the residual $z=v-\sum_jc_jb_j$ vanishes on the coordinates in $J$.
Each augmented minor on $J\cup\{f\}$ vanishes and, after subtracting that row combination, equals $\pm z_f\det B_J$.
Thus the residual is zero.
The rank-zero and full-rank cases say respectively that $v=0$ and that there are no remaining coordinates.

The prescribed kernel vectors satisfy $Bv^f=0$ by the adjugate identity and are independent by their free-coordinate matrix $\delta I_{d-r_X}$.
If $Bx=0$, subtract $\sum_f(x_f/\delta)v^f$; the residual has zero free coordinates and satisfies the invertible homogeneous system on $J$, so it is zero.

Next comes a fixed determinant proof that an independent list cannot lie in the span of fewer vectors than its length.
Suppose an independent list $a_1,\ldots,a_r$ lies in the span of $t<r$ vectors $y_1,\ldots,y_t$, say $a_k=\sum_{l=1}^t\mathsf X_{kl}y_l$.
Let $A$ and $\mathsf Y$ be the matrices with rows $a_1,\ldots,a_r$ and $y_1,\ldots,y_t$, so that $A=\mathsf X\mathsf Y$.
Choose a column set $J$ with $\det A_J\ne0$.
Then $A_J=\mathsf X\mathsf Y_J$, with $\mathsf X$ of size $r\times t$ and $\mathsf Y_J$ of size $t\times r$.
Pad $\mathsf X$ with $r-t$ zero columns and $\mathsf Y_J$ with $r-t$ zero rows, producing square matrices $\mathsf X',\mathsf Y'_J$ with $A_J=\mathsf X'\mathsf Y'_J$.
The product determinant formula gives $\det A_J=\det\mathsf X'\det\mathsf Y'_J=0$, since $\mathsf X'$ has a zero column, contradicting the chosen minor.

Let one span lie inside another of the same dimension; we show that the two are equal.
Let $T$ be the change-of-basis matrix whose rows express the basis of the contained span in the basis of the containing span.
The matrix $T$ is invertible.
Indeed, suppose instead that $\det T=0$, and scan the rows of $T$ in order; some row does not raise the rank, since otherwise the rank would equal the order of $T$ and its only minor of that order, $\det T$, would be nonzero.
The rows before the first such row each raised the rank by exactly one, so their rank equals their number and they have a nonzero full-row minor.
The spanning step therefore writes that row as a combination of the earlier rows.
This gives a nonzero $c$ with $c^{\mathsf T}T=0$, and hence a nonzero vanishing combination of the basis of the contained span, contradicting its independence.
Invertibility of $T$ gives the reverse inclusion.

The preceding assertions are finite families of ordered-field sentences for this fixed $d$.
Expand their determinants and cases and replace each guarded inverse by its inverse equation.
The general-ring identities of \Cref{fact:dt-det} prove their algebraic assertions.
Substitute the coded-rational interpretation of \Cref{lem:code-field} into these finite proofs to obtain their coded versions, as in the interpretation procedure of \Cref{sec:foundations}.
This step concerns fixed coordinate lists, not the variable-length arrangement scan.

For the scan itself, induct on the processed normal index.
An accepted row preserves independence and the old spanning equations.
A zero-sign normal rejected by the rank test has unchanged rank and the computed Cramer spanning equation just proved.
At full rank the retained rows span every vector.
The invariant uses only bounded numeric indices and the fixed-coordinate coefficient functions, with the input strings as parameters.
The scan state stores at most $d$ retained indices.
For a list of length $v$ it has the explicit array bound $\langle d+1,v+d+2\rangle$, given with the defining graph of the scan in \Cref{sec:graphs}, so \Cref{lem:scan} proves that cutting each state to this bound is idle.
This establishes the common-zero description of $E(X)$.

For realized $X$, a normal vanishes on $E(X)$ exactly when its $X$-sign is zero.
To see that vanishing on $E(X)$ forces a zero $X$-sign, evaluate at a realizing point; the converse is the common-zero description just proved.
If $X\sqsubseteq Y$, then $Z(Y)\subseteq Z(X)$, so $E(X)\subseteq E(Y)$.
If their dimensions agree, these flats agree, and the preceding characterization gives equal zero sets.
Every remaining sign already agrees by $X\sqsubseteq Y$, so $X=Y$.
Thus proper inclusion of realized covectors raises dimension.

If $u$ is integer, a nonzero $h_i\cdot u$ has magnitude at least one, and multiplication by $\Delta(w)$ makes it dominate $h_i\cdot w$.
At a zero evaluation the new sign is supplied by $w$.
Hence
\[
\operatorname{cov}(u\boxplus w)=\operatorname{cov}(u)\circ\operatorname{cov}(w).
\]

It remains to show that every rational point has a matching cone index.
Call $\dim\operatorname{cov}(x)$ the dimension of a point $x$, and call an index $\sigma<S$ that matches $x$ a representation of $x$; the rays of this representation are the $d$ rays denoted by the $d$ ray indices that make up $\sigma$, taken in order.
Prove separately, for each $j=0,\ldots,d$, that every point of dimension at most $j$ has a representation whose last $d-j$ rays are zero.
The dimension-zero point is zero.
For a point of dimension less than $j$, the assertion for $j-1$ applies, and a representation whose last $d-j+1$ rays are zero has its last $d-j$ rays zero.

For a point $x$ of dimension $j>0$, put $X=\operatorname{cov}(x)$ and extend $B_X$ greedily with the arrangement normals.
The units force a final independent $d$-list.
The cross $\varrho$ of its first $d-1$ rows is nonzero and lies in $E(X)$.
Some normal $h_i$ has nonzero evaluation on $\varrho$.
This normal is not a zero normal of $X$, because every zero normal of $X$ kills $\varrho$, so $h_i\cdot x$ is nonzero as well.
Choose a sign $\varepsilon$ so that the nonzero evaluations $h_i\cdot x$ and $\varepsilon h_i\cdot\varrho$ have the same sign.

Take the minimum positive ratio
\[
t^*=\min\left\{\frac{h_i\cdot x}{\varepsilon h_i\cdot\varrho}:
(h_i\cdot x)(\varepsilon h_i\cdot\varrho)>0\right\}.
\]
The list is nonempty and its minimum is positive.
Set $y=x-t^*\varepsilon\varrho$.
A zero evaluation stays zero.
Evaluations of opposite signs move strictly within the old sign; evaluations of the same sign remain weakly within it by the minimum property.
At a genuine minimum-attaining index a new zero occurs.
Thus $\operatorname{cov}(y)\sqsubset X$, its dimension is at most $j-1$, and
\[
X=\operatorname{cov}(y)\circ\operatorname{cov}(\varepsilon\varrho).
\]

Since $y$ has dimension at most $j-1$, the assertion for $j-1$ gives a representation of $y$ whose last $d-j+1$ rays are zero; replace the first of these zero rays by $\varepsilon\varrho$.
The perturbation identity gives the desired covector, and later zero rays are idle.
In the defining graph of this minimum (\Cref{sec:graphs}), failed ratio tests duplicate a selected genuine positive candidate, so they never insert a new default smaller value.
There are $d+1$ separate deductions, not induction on a growing dimension.
\end{proof}

From now on we identify a realized covector, its sign cone, and its least representative index whenever the argument position makes the type unambiguous.
In particular $\varpi(X)$ uses that canonical numeric index, and order and dimension tests use its covector.
Write $q_X=\varpi(X)$ for the integer representative of $X$; its cone subscript distinguishes it from the externally fixed prime $q$.

The cone $\bar X$ is $\operatorname{idx}\left((\cdots((q_X\boxplus e_1)\boxplus e_2)\cdots)\boxplus e_d\right)$, obtained by perturbing $q_X$ by the units one after another in increasing order of their index.
Every intermediate vector is integer, since each $\Delta(e_c)$ is an integer, so the perturbation identity applies at each step and the covector of $\bar X$ is $X\circ\operatorname{cov}(e_1)\circ\cdots\circ\operatorname{cov}(e_d)$.
A zero sign remaining in this covector would make its normal kill all units, which no normal does, since every normal is nonzero; so $\bar X$ has dimension $d$, that is, it is a chamber, and it lies above $X$.
For a chamber $M$, $\bar M=M$.
The unique zero-dimensional cone is $\mathsf o=\operatorname{idx}(0)$.

Define $\tau\lessdot\sigma$, read as $\sigma$ covers $\tau$, to mean $\tau\sqsubseteq\sigma$ and $\dim\sigma=\dim\tau+1$.
In that case the representative $q_\sigma$ lies outside $E(\tau)$.
Otherwise every zero normal of $\tau$ would vanish at $q_\sigma$, giving $Z(\sigma)=Z(\tau)$ and hence $\sigma=\tau$.
So $q_\sigma$ extends $\mathcal N(\tau)$ to a basis of $E(\sigma)$.

\begin{lemma}[Cones above a cone]
\label{lem:fp-stars-diamonds}
Let $\tau$ be a cone and let $w\in\Q^d\setminus E(\tau)$.
If a cone $\sigma\sqsupseteq\tau$ with $\sigma\ne\tau$ has $q_\sigma=t+\alpha' w$ with $t\in E(\tau)$, then $\alpha'\ne0$ and $\sigma=\operatorname{idx}(q_\tau\boxplus(\operatorname{sgn}(\alpha')w))$.
A ridge $\tau$ has exactly two chambers above it, namely $\operatorname{idx}(q_\tau\boxplus w)$ and $\operatorname{idx}(q_\tau\boxplus(-w))$.
If $\tau\sqsubseteq\sigma$ and $\dim\sigma-\dim\tau=2$, there are exactly two intermediate cones $\rho_1,\rho_2$, that is, cones $\rho$ with $\tau\sqsubseteq\rho\sqsubseteq\sigma$ and $\rho\ne\tau,\sigma$.
Their representatives extend $\mathcal N(\tau)$ to a basis of $E(\sigma)$, and
\[
q_\sigma=t+\alpha_1q_{\rho_1}+\alpha_2q_{\rho_2},
\qquad t\in E(\tau),\quad\alpha_1,\alpha_2>0.
\]
\end{lemma}

\begin{proof}
For $w\notin E(\tau)$, define the two perturbed cones by
\[
\sigma_\pm=\operatorname{idx}(q_\tau\boxplus(\pm w)).
\]
Their zero normals kill both $E(\tau)$ and $w$.
Whenever their dimension is $\dim\tau+1$, their flat is therefore $E(\tau)+\Q w$ and
\[
q_{\sigma_\pm}=t_\pm+\alpha'_\pm(\pm w),
\qquad t_\pm\in E(\tau),\quad\alpha'_\pm>0.
\]
The coefficient is nonzero by proper inclusion, and it is positive because a zero normal of $\tau$ that is nonzero at $w$ fixes its sign.

Conversely, let $\sigma\sqsupseteq\tau$ with $\sigma\ne\tau$ have representative $q_\sigma=t+\alpha' w$ with $t\in E(\tau)$.
For $i\in Z(\tau)$ the normal $h_i$ kills $t$, so $\sigma_i=\operatorname{sgn}(\alpha' h_i\cdot w)$.
For $i\notin Z(\tau)$ we have $\sigma_i=\tau_i$ by $\tau\sqsubseteq\sigma$.
If $\alpha'=0$, this gives $\sigma=\tau$, so $\alpha'\ne0$.
By the perturbation identity, $\sigma=\operatorname{cov}(q_\tau\boxplus(\operatorname{sgn}(\alpha')w))$, whose least index is $\operatorname{idx}(q_\tau\boxplus(\operatorname{sgn}(\alpha')w))$.

For a ridge $\tau$, let $w\notin E(\tau)$, for instance a unit outside its flat.
Every zero normal of the ridge is a nonzero multiple of its single retained normal, which is nonzero at $w$, so neither perturbed cone has a zero sign.
Both perturbed cones then have full dimension.
They are distinct, since at the index of the retained normal $h$ their signs are $\operatorname{sgn}(h\cdot w)$ and $-\operatorname{sgn}(h\cdot w)$.
A chamber $M$ above $\tau$ differs from $\tau$.
Since $\mathcal N(\tau),w$ is a basis of $\Q^d$, the representative $q_M$ of the chamber $M$ is $t+\alpha' w$ with $t\in E(\tau)$.
The converse classification then makes $M$ one of the two perturbed cones.

We need a two-dimensional fact, the sector classification, for the assertion on intermediate cones: under the hypotheses below, the nonzero feasible boundary points lie on exactly two rays.
Let finitely many nonzero linear forms $\psi_i$ on $\Q^2$ be given, each identified with its coefficient pair in $\Q^2$, and suppose they include an independent pair.
Let $\varepsilon_i\in\{-1,1\}$ be signs and $y^\circ\in\Q^2$ a point with $\varepsilon_i\psi_i(y^\circ)>0$ for every $i$.
A point $y\in\Q^2$ is feasible when $\varepsilon_i\psi_i(y)\ge0$ for every $i$, and it is a boundary point when moreover $\psi_i(y)=0$ for some $i$.

Choose the first independent pair and solve for $v$ so that the oriented evaluations $\varepsilon_i\psi_i(v)$ of that pair are $-1$ and $1$.
Define
\[
T_+=\min_{\varepsilon_i\psi_i(v)<0}
\frac{\varepsilon_i\psi_i(y^\circ)}{-\varepsilon_i\psi_i(v)},
\qquad
T_-=\min_{\varepsilon_i\psi_i(v)>0}
\frac{\varepsilon_i\psi_i(y^\circ)}{\varepsilon_i\psi_i(v)}.
\]
Both lists are nonempty, since the chosen pair contributes one form to each, and both minima are positive.

The endpoints $y_+=y^\circ+T_+v$ and $y_-=y^\circ-T_-v$ are feasible and each has a genuinely vanishing form.
Indeed, write $n_i=\varepsilon_i\psi_i(y^\circ)>0$ and $c_i=\varepsilon_i\psi_i(v)$, so that the oriented evaluation at $y^\circ+\theta v$ is $n_i+\theta c_i$.
For $\theta\ge0$ it stays positive when $c_i\ge0$, and for $c_i<0$ it stays nonnegative up to $\theta=n_i/(-c_i)$; the negative direction is identical with the signs reversed.
The selected independent forms show that $y^\circ,v$ are independent, since proportionality would force the proportionality coefficient to have both signs.
Hence
\[
\det(y_+,y_-)=-(T_++T_-)\det(y^\circ,v)\ne0,
\quad
(T_++T_-)y^\circ=T_-y_++T_+y_-.
\]

Every nonzero feasible boundary point is a positive multiple of one endpoint.
Indeed, write it as $\mu y_++\nu y_-$.
If both coefficients are positive, a vanishing nonzero form would vanish at both independent endpoints.
If both are negative, feasibility of the point and its negative forces the independent pair to vanish there, so the point is zero.
If the coefficients have opposite signs, a form vanishing at the endpoint with positive coefficient is strictly positive at the other endpoint and makes the combination infeasible.
If one coefficient is zero and the other negative, the endpoint and its negative are both feasible, again impossible for a nonzero vector.
These cases leave exactly the two positive endpoint rays.

We now prove the assertion on intermediate cones, with $\tau\sqsubseteq\sigma$, $\dim\tau=k-2$ and $\dim\sigma=k$.
Extend $\mathcal N(\tau)$ greedily by vectors of $\mathcal N(\sigma)$ to a basis of $E(\sigma)$; the two added vectors are the quotient vectors $u_1,u_2$.

For $x\in E(\sigma)$ there is a unique decomposition $x=t+\chi_1u_1+\chi_2u_2$ with $t\in E(\tau)$.
Write $[x]=(\chi_1,\chi_2)$ for these quotient coordinates and $\widehat y=y_1u_1+y_2u_2$.
Define
\[
J=Z(\tau)\setminus Z(\sigma),\qquad
\psi_i=(h_i\cdot u_1,h_i\cdot u_2)\quad\text{for }i\in J.
\]
We call the forms $\psi_i$ with $i\in J$ genuine.
Each form is nonzero: otherwise its normal would kill $E(\sigma)$ while having nonzero sign on $\sigma$.
They have no common nonzero kernel, because such a quotient vector would be killed by every zero normal of $\tau$ and thus lie in its flat.
An independent pair exists; otherwise all forms would be multiples of one nonzero form $(a,b)$ and kill $(-b,a)\ne0$, contrary to the previous sentence.

Take $y^\circ=[q_\sigma]$ and $\varepsilon_i=\sigma_i$; then every oriented evaluation $\varepsilon_i\psi_i(y^\circ)$ is positive.
The sector computation takes one form with its sign at each arrangement index $i<s$; at every index outside $J$ we place a copy of the first genuine independent form, with its sign.
This adds neither a zero form nor a false boundary condition.

For the two sector endpoints set
\[
\rho_\pm=\operatorname{idx}(q_\tau\boxplus\widehat y_\pm).
\]
Outside $Z(\tau)$ signs are retained; on $Z(\sigma)$ they remain zero; on $J$ they are zero or their $\sigma$ sign.
Thus these cones lie between $\tau$ and $\sigma$.

At each endpoint some form $\psi_i$ with $i\in J$ vanishes, since a form placed at an index outside $J$ is a copy of a form with index in $J$.
The normal $h_i$ of such a genuine form is a zero normal of $\rho_\pm$, because $i\in Z(\tau)$ gives $h_i\cdot q_\tau=0$ and $h_i\cdot\widehat y_\pm=\psi_i(y_\pm)=0$.
It is not in the span of the rows of $B_\sigma$, because those rows kill $u_1,u_2\in E(\sigma)$ while $\psi_i\ne0$.
So the zero normals of $\rho_\pm$ include $d-k+1$ independent vectors, and the dimensions of $\rho_\pm$ are at most $k-1$.
The independent list $\mathcal N(\tau),\widehat y_\pm$ lies in the flats of $\rho_\pm$, so their dimensions are at least $k-1$.
The endpoints give distinct cones, since a nonzero form vanishing at one cannot vanish at both independent endpoints.

Every intermediate cone $\rho$ has a nonzero feasible quotient, since otherwise $\rho=\tau$, and a genuine zero form, since otherwise $\rho=\sigma$.
The sector classification makes $\rho$ exactly one of the two cones.
Their quotient representatives are independent, proving the basis assertion.
The positive endpoint decomposition of $[q_\sigma]$ gives the displayed positive coefficients.
The case $k=2$, where $\tau$ is the origin and the lower basis is empty, is included.
\end{proof}

\paragraph{\emph{Exterior coordinates and normalization.}}
For $0\le r\le d$, a degree-$r$ exterior array has one coordinate for each increasing $r$-element subset of $[d]$.
The wedge of $r$ vectors records the corresponding $r$-row determinants.
For an independent list it is nonzero and encodes its oriented span; for a dependent list it is zero.
Wedge multiplication raises the degree by one, while contraction with a linear form lowers it by one.
For $v\in\Q^d$, a linear form $\varphi\in\Q^d$, a degree-$r$ exterior array $\Omega$, an $(r+1)$-element set $I\subseteq[d]$ and, when $r\ge1$, an $(r-1)$-element set $J\subseteq[d]$, define the wedge $v\wedge\Omega$, of degree $r+1$, and the contraction $\operatorname{ct}_\varphi\Omega$, of degree $r-1$, by
\[
(v\wedge\Omega)_I
=\sum_{i\in I}(-1)^{|\{j\in I:j<i\}|}v_i\Omega_{I\setminus\{i\}},
\]
\[
(\operatorname{ct}_\varphi\Omega)_J
=\sum_{i\notin J}(-1)^{|\{j\in J:j<i\}|}
\varphi_i\Omega_{J\cup\{i\}}.
\]
The empty wedge is one and contraction at degree zero is zero.
Wedge multiplication at degree $d$ is zero, represented by the empty array of degree $d+1$.

For vectors $v_1,\ldots,v_r\in\Q^d$, let $\operatorname{wedge}(v_1,\ldots,v_r)=v_1\wedge\operatorname{wedge}(v_2,\ldots,v_r)$, with $\operatorname{wedge}()=1$, and write $v_1\wedge\cdots\wedge v_r$ for it.
By expansion along the first row, its coordinate at $I$ is the determinant of the matrix with rows $v_1,\ldots,v_r$ restricted to the columns in $I$.
Let $O_X=\operatorname{wedge}(\mathcal N(X))$.
Its coordinates are minors of an integer matrix whose rows are independent, so it is a nonzero integer exterior array.
A basis change by $T$ multiplies a basis wedge by $\det T$: expand multilinearly, remove repeated basis vectors and group the surviving permutations.

For a cover $\tau\lessdot\sigma$ and a direction $v\in E(\sigma)\setminus E(\tau)$, let $I$ be the first coordinate set, in lexicographic order, with $(v\wedge O_\tau)_I\ne0$.
Define the nonzero coordinate ratio $\lambda(\sigma,\tau;v)=(O_\sigma)_I/(v\wedge O_\tau)_I$.
Such an $I$ exists, since $v,\mathcal N(\tau)$ is independent and its wedge is therefore nonzero; the ratio uses the guarded inverse of \Cref{lem:code-field}, whose guard succeeds because $(v\wedge O_\tau)_I\ne0$.
The list $v,\mathcal N(\tau)$ and the list $\mathcal N(\sigma)$ are both bases of $E(\sigma)$, so the basis-change rule above gives
\[
O_\sigma=\lambda(\sigma,\tau;v)(v\wedge O_\tau),
\qquad \kappa(\sigma,\tau;v)=|\lambda(\sigma,\tau;v)|.
\]
We call the displayed identity the orientation equation of the cover.
When the direction is omitted, as in $\lambda(\sigma,\tau)$ and $\kappa(\sigma,\tau)$, it is $v=q_\sigma$.

The coordinate definitions give
\[
\operatorname{ct}_\varphi(v\wedge\Omega)
=(\varphi\cdot v)\Omega-v\wedge\operatorname{ct}_\varphi\Omega.
\]
Equal contributed and contracted indices supply the dot product; interchanging distinct indices supplies the minus sign.
Induction on the number of basis vectors, starting from zero contraction at degree zero, shows that contraction of a basis wedge vanishes when the form kills its flat.
Also $t\wedge O_\tau=0$ for $t\in E(\tau)$, by expanding $t$ in $\mathcal N(\tau)$ and using vanishing with a repeated row.

The following identities are the normal-change rules, written with $\sigma$ and $\tau$ suppressed from $\lambda$ and $\kappa$.
If $v'=\alpha v+t$ with $\alpha>0$ and $t\in E(\tau)$, then
\[
\lambda(v')=\lambda(v)/\alpha,\qquad
\kappa(v')\varphi(v')=\kappa(v)\varphi(v)
\]
for every form killing $E(\tau)$, and $\kappa(v')v'-\kappa(v)v\in E(\tau)$.
At a chamber and at the origin, $O$ has a single coordinate, and it equals one.
These are fixed-coordinate identities, transferred to coded rationals in the same way as the finite families of ordered-field sentences for the fixed $d$ stated earlier.

\paragraph{\emph{Families and balanced cycles.}}
A family $f$ is an array that supplies a gradient $g^f_M\in\Q^d$ at the index of each chamber $M$.
Here a gradient is the coefficient vector of the linear function assigned to that chamber.
It is admissible when $g^f_M-g^f_{\bar X}$ kills $E(X)$ whenever $X\sqsubseteq M$.
Write $f[X]=g^f_{\bar X}\cdot q_X$, and call $g^f_{\bar X}$ the selected gradient of $f$ at $X$; an integer family has denominator-one gradients.
For an admissible family, $f[X]=g^f_M\cdot q_X$ for every chamber $M$ above $X$.
Indeed, $g^f_M-g^f_{\bar X}$ kills $E(X)$, and $q_X\in E(X)$.
Hence $f[X]$ does not depend on which chamber above $X$ is selected.
A family $f$ represents a function $\phi$ on $\Q^d$ when $\phi(x)=g^f_M\cdot x$ for every chamber $M$ and every point $x$ of its sign cone.

For $0\le k\le d$, a $k$-cone is a cone of dimension $k$.
A $k$-cycle $C$ is an exact array of rational entries $C(\sigma)$, one for each index $\sigma<S$, with $C(\sigma)=0$ unless $\sigma$ is a $k$-cone; its degree is $k$.
For $k\ge1$, define at $(k-1)$-cones
\[
(f\odot C)(\tau)
=\sum_{\sigma:\tau\lessdot\sigma}
C(\sigma)\kappa(\sigma,\tau)
(g^f_{\bar\sigma}-g^f_{\bar\tau})\cdot q_\sigma,
\]
and zero elsewhere.
Every such sum runs over all indices $\sigma$ below $S$, and its summand is multiplied by one if $\tau\lessdot\sigma$ and by zero otherwise.
In $(f\odot C)(\tau)$ we call $\tau$ the output cone, and a cover $\sigma$ of $\tau$ contributes when $C(\sigma)\ne0$.

Put
\[
V_C(\tau)=\sum_{\sigma:\tau\lessdot\sigma}
C(\sigma)\kappa(\sigma,\tau)q_\sigma.
\]
A $k$-cycle with $k\ge1$ is balanced when $V_C(\tau)\in E(\tau)$ at every $(k-1)$-cone $\tau$; zero-cycles are balanced by definition.
The unit cycle $U$ is the $d$-cycle with $U(M)=1$ at every chamber $M$.

An integer twin of a $k$-cycle $C$ is an array $\widetilde C$ that assigns to each index $X<S$ a degree-$k$ exterior array $\widetilde C(X)$ with integer coordinates, which is zero unless $X$ is a $k$-cone.
Writing $(a,1)$ for the rational code with numerator $a$ and denominator one, the twin satisfies
\[
C(X)O_X=(\widetilde C(X),1)
\]
coordinatewise.
The twin records integral exterior coordinates without asserting that each rational cycle weight is an integer.

\begin{lemma}[Balance, commutation and integer twins]
\label{lem:fp-product-engine}
The unit is balanced.
Products by admissible families preserve balance, and admissible families commute on balanced inputs of degree at least two.
If the family is integer and admissible and $C$ has a specified integer twin, the product has an explicitly constructed integer twin.
\end{lemma}

\begin{proof}
For a balanced $k$-cycle $C$, a $(k-1)$-cone $\tau$ and any form $\varphi$ killing $E(\tau)$, the definition of balance gives
\[
\sum_{\sigma:\tau\lessdot\sigma}
C(\sigma)\kappa(\sigma,\tau)\varphi(q_\sigma)=0.
\]
Thus, in $(f\odot C)(\tau)$, the lower gradient $g^f_{\bar\tau}$ may be changed by any such form without changing the value of the sum.

The unit is balanced because the two chambers above a ridge are the perturbed cones obtained from opposite transverse directions $w$ and $-w$.
Normal change identifies their contributions modulo the ridge flat with $\kappa(w)w$ and $\kappa(-w)(-w)$.
The orientation $O$ equals one at both chambers, so the absolute factors $\kappa(w)$ and $\kappa(-w)$ agree, and the contributions $\kappa(w)w$ and $\kappa(-w)(-w)$ cancel.

Let $f$ and $f'$ be admissible families and $C$ a balanced $k$-cycle with $k\ge2$.
Fix a $(k-2)$-cone $\tau$ and subtract from each family its own gradient at $\bar\tau$ throughout.
Products do not change, since only gradient differences enter them, and the shifted gradients above $\tau$ kill $E(\tau)$.
Write $\phi(X)=f[X]$ and $\gamma(X)=f'[X]$ for these shifted evaluations.
For each $k$-cone $\sigma$ above $\tau$, let $\rho_1,\rho_2$ be its two intermediate cones between $\tau$ and $\sigma$.
Let $\widehat\kappa(\sigma)$ be the absolute ratio of $O_\sigma$ to
\[
q_{\rho_1}\wedge q_{\rho_2}\wedge O_\tau.
\]

For a cover $\rho$ of $\tau$ and each cover $\sigma$ of $\rho$, write $\rho'$ for the intermediate cone between $\tau$ and $\sigma$ other than $\rho$; it depends on $\sigma$.
Set
\[
V'(\rho)=\sum_{\sigma:\rho\lessdot\sigma}
C(\sigma)\kappa(\sigma,\rho;q_{\rho'})q_{\rho'}.
\]
The positive decomposition of \Cref{lem:fp-stars-diamonds} gives $q_\sigma=\alpha' q_{\rho'}+t$ with $\alpha'>0$ and $t\in E(\rho)$.
Normal change gives $\kappa(\sigma,\rho;q_{\rho'})=\alpha'\kappa(\sigma,\rho)$.
Thus the difference between the alternative and original direction summands is $-\kappa(\sigma,\rho)t\in E(\rho)$, and multiplying by $C(\sigma)$ and summing preserves this membership.
Hence $V'(\rho)\in E(\rho)$, and its Cramer coordinates in the basis $\mathcal N(\tau),q_\rho$ of $E(\rho)$ give
\[
V'(\rho)=\zeta_\rho q_\rho+t_\rho,
\qquad t_\rho\in E(\tau).
\]

Admissibility and normal change give
\[
(f'\odot C)(\rho)
=\sum_{\sigma:\rho\lessdot\sigma}
C(\sigma)\kappa(\sigma,\rho;q_{\rho'})\gamma(\rho')
-\zeta_\rho\gamma(\rho).
\]
The omitted residual term is the shifted gradient applied to $t_\rho$, which is zero because the shifted gradients above $\tau$ kill $E(\tau)$.

The orientation equations give
\[
\kappa(\sigma,\rho;q_{\rho'})\kappa(\rho,\tau)
=\widehat\kappa(\sigma),
\]
since the two denominator wedges differ only by sign.
Substitute and exchange the bounded sums to obtain
\[
(f\odot(f'\odot C))(\tau)
={}\sum_{\sigma\sqsupseteq\tau, \dim\sigma=k}
C(\sigma)\widehat\kappa(\sigma)
\left(\phi(\rho_1)\gamma(\rho_2)+\phi(\rho_2)\gamma(\rho_1)\right)
-\sum_{\rho:\tau\lessdot\rho}
\zeta_\rho\kappa(\rho,\tau)\phi(\rho)\gamma(\rho).
\]
This expression is symmetric under exchanging $f$ and $f'$, that is, $\phi$ and $\gamma$, and so proves commutation.

For balance of $f'\odot C$ at the $(k-2)$-cone $\tau$, let $h$ be any zero normal of $\tau$ and substitute the scalar $h\cdot q_\rho$ for $\phi(\rho)$.
Direct expansion of $h\cdot V_{f'\odot C}(\tau)$ is exactly this scalar substitution; no fictitious admissible family is required.
Regrouping by the index carrying $\gamma$ gives
\[
\sum_{\rho:\tau\lessdot\rho}
\kappa(\rho,\tau)\gamma(\rho)
h\cdot\left(V'(\rho)-\zeta_\rho q_\rho\right)=0.
\]
The common-zero description of $E(\tau)$ proves balance.
For $k=1$ the output is a zero-cycle and is balanced by convention.

For an integer family define its twin product by
\[
(f\circledast\widetilde C)(\tau)
=\sum_{\sigma:\tau\lessdot\sigma}
\operatorname{sgn}\lambda(\sigma,\tau)
\operatorname{ct}_{g^f_{\bar\sigma}-g^f_{\bar\tau}}
\widetilde C(\sigma).
\]
The gradient difference kills $E(\tau)$, so contraction of $O_\tau$ vanishes.
The contraction identity applied to the orientation equation gives each product summand times $O_\tau$; multiplying by $\operatorname{sgn}\lambda$ replaces $\lambda$ by its absolute value.
Summing proves the twin assertion.
Every contraction and sum is integer, the unit has its top-coordinate-one twin, and integer combinations preserve twins.
At the origin $O_{\mathsf o}=1$, so the cycle value is the integer empty-coordinate twin.
No individual rational cycle weight is assumed integer.
\end{proof}
\space

\section{Refinement, Rectification, and the Maximum}
\label{sec:locality}

Refinement lets the same product be computed on the successive arrangements needed by a network.
The rectification identity then controls repeated products by a ReLU activation.

\paragraph{\emph{Refinement and locality.}}
Let $\mathcal H_0$ be an initial segment of the list $\mathcal H$ that contains the units; we call $\mathcal H_0$ the base arrangement.
A superscript or subscript zero, or the word base, denotes the objects of $\mathcal H_0$, and the words refined and ambient denote those of $\mathcal H$.
Put $\check X=\operatorname{idx}_0(q_X)$, the restriction of the covector of $X$ to $\mathcal H_0$.
Restriction preserves order and composition.
It maps chambers to chambers, since a point with no zero signs on $\mathcal H$ has no zero signs on $\mathcal H_0$.
The common-zero description gives $E(X)\subseteq E_0(\check X)$ and $\dim X\le\dim\check X$.
When these dimensions agree the two flats agree, so there is a unique nonzero rational $\vartheta_X$ with
\[
O^0_{\check X}=\vartheta_X O_X.
\]
Define $\bar\kappa(X)=|\vartheta_X|$ in this case.

Lift an admissible family by $g^{\widehat f}_M=g^f_{\check M}$, and lift a $k$-cycle by
\[
(\mathcal L C)(X)=
\begin{cases}
\bar\kappa(X)C(\check X),&\dim X=\dim\check X=k,\\
0,&\text{otherwise}.
\end{cases}
\]
Admissibility of the lifted family follows because its relevant gradient difference kills the larger flat $E_0(\check X)$.
An integer family remains integer.

For a family $F$, its rectification $F^+$ is the family whose gradient on a chamber $M$ is $g^F_M$ when $F[M]>0$ and zero otherwise, where $F[M]=g^F_M\cdot q_M$.
For a family $A$ and a cycle $C$, write $AC$ for the cone product $A\odot C$.
Powers denote repeated cone products, not pointwise powers: $A^0C=C$ and $A^rC=A(A^{r-1}C)$.
For a $d$-cycle $C$, the degree-$d$ product $A^dC$ has $d$ factors and is a zero-cycle; we call its value at the origin the degree-$d$ value.

\begin{lemma}[Refinement and rectification]
\label{lem:fp-refinement-locality}
For an admissible base family $f$ and a balanced positive-degree base cycle $C$,
\[
\widehat f\odot\mathcal L C=\mathcal L(f\odot_0 C).
\]
Consequently, for admissible base families $f_1,\ldots,f_d$, the value at the origin of $\widehat f_d\odot(\cdots\odot(\widehat f_1\odot U)\cdots)$ equals the value at the origin of $f_d\odot_0(\cdots\odot_0(f_1\odot_0U_0)\cdots)$, where $U_0$ is the unit cycle of $\mathcal H_0$.
If a base $k$-cycle $C$ has an integer twin $\widetilde C$, then $\mathcal L C$ has the integer twin whose value at $X$ is $\operatorname{sgn}(\vartheta_X)\widetilde C(\check X)$ when $\dim X=\dim\check X=k$ and is zero otherwise.
If every nonzero chamber gradient of an admissible family $F$ is an arrangement normal, its rectification $H=F^+$ is admissible.
For such a family $F$, every balanced $k$-cycle $C$ and fixed $1\le r\le k$,
\[
H^r C=F^{r-1}(HC).
\]
\end{lemma}

\begin{proof}
The proof treats, in this order, the lifting identity, the value at the origin, the lifting of integer twins, the admissibility of $H$, and the power identity.
For the lifting identity, let $k\ge1$ be the degree of $C$, fix a $(k-1)$-cone $\tau$, and write $\nu_0=\check\tau$.
There are three cases: $\dim\nu_0=k-1$, $\dim\nu_0=k$, and $\dim\nu_0\ge k+1$.
Suppose first that $\dim\nu_0=k-1$.
The two flats at $\tau,\nu_0$ are equal.
For each base cover $\nu_0\lessdot\nu$, set
\[
\sigma_\nu=\operatorname{idx}(q_\tau\boxplus q^0_\nu).
\]
The flat of this cone contains $E(\tau)$ and the independent direction $q^0_\nu$.
All zero normals of $\nu$ kill both summands, so this flat is exactly $E_0(\nu)$ and has dimension $k$.
Its restriction is $\nu$.

Conversely a refined cover contributing a nonzero lifted weight has a base restriction of dimension $k$.
That restriction covers $\nu_0$, since restriction preserves order, so $\nu_0\sqsubseteq\check\sigma$, and the two dimensions are $k-1$ and $k$.
Its flat is the corresponding $E_0(\nu)$.
Since $E_0(\nu)$ is spanned by $E(\tau)$ and $q^0_\nu$, any representative is a flat vector plus a multiple of $q^0_\nu$, and the converse classification of \Cref{lem:fp-stars-diamonds} makes that multiple nonzero.
The multiple cannot be negative, since a base normal vanishing at $\nu_0$ and nonzero at $\nu$ fixes its sign.
Thus these covers are exactly the $\sigma_\nu$.

Comparing the two exterior orientation equations gives
\[
\bar\kappa(\sigma_\nu)\kappa(\sigma_\nu,\tau;q^0_\nu)
=\bar\kappa(\tau)\kappa_0(\nu,\nu_0).
\]

Normal change permits $q^0_\nu$ in each summand.
Admissibility replaces the selected upper chamber gradient by the base chamber gradient above $\nu$.
The lower selected gradient may differ from the selected gradient above $\nu_0$.
Their difference kills $E_0(\nu_0)$; base balance therefore cancels this difference after the whole cover sum.
The remaining expression is $\bar\kappa(\tau)(f\odot_0 C)(\nu_0)$, as required.

If $\dim\nu_0=k$, the right side is zero.
Every potentially nonzero lifted cover must restrict to the same $\nu_0$: its restriction contains $\nu_0$ and has the same dimension.
Both selected base chamber gradients lie above $\nu_0$, so their difference kills its flat, which contains the cover representative.
Each left summand is zero.

If $\dim\nu_0\ge k+1$, every refined cover has restriction of dimension greater than $k$.
Its lifted weight is zero, and the right side is zero too.
These cases are exhaustive since $\dim\nu_0\ge\dim\tau=k-1$.

On chambers both orientations are the standard full exterior coordinate, so $\bar\kappa=1$ and the lifted unit is the ambient unit.
Apply the proved product identity successively, using the already proved balance of actual products from the unit on the base arrangement.
Hence each ambient product $\widehat f_j\odot(\cdots\odot(\widehat f_1\odot U)\cdots)$, which starts from the ambient unit $U=\mathcal LU_0$, is the lift of the corresponding base product $f_j\odot_0(\cdots\odot_0(f_1\odot_0U_0)\cdots)$; the $j$th step has an input of positive degree $d-j+1$.
We do not need, and do not claim, that lifting preserves balance for every balanced cycle.
At the origin both orientations equal one and $\bar\kappa=1$, so a lift has the value of the base cycle there; this proves the assertion on the value at the origin.

When a base cycle has integer twin $\widetilde C$, its lifted twin on matching dimensions is $\operatorname{sgn}(\vartheta_X)\widetilde C(\check X)$.
Indeed, coordinatewise,
\[
(\mathcal L C)(X)O_X
=|\vartheta_X|C(\check X)O_X
=\operatorname{sgn}(\vartheta_X)C(\check X)O^0_{\check X}
=(\operatorname{sgn}(\vartheta_X)\widetilde C(\check X),1).
\]

For locality, define $\operatorname{sgn}_X F=\operatorname{sgn}(F[X])$, and call a cone $X$ positive, zero or negative according as $\operatorname{sgn}_XF$ is $1$, $0$ or $-1$.
A cycle has nonnegative support when it vanishes at every negative cone.
If $X\sqsubseteq Y$ and $F[X]\ne0$, every chamber gradient above $Y$ evaluates at $q_X$ to the nonzero value $F[X]$, by admissibility.
By hypothesis this nonzero gradient is an arrangement normal, so its nonzero sign at $X$ persists at $Y$.
Hence the sign of $F$ persists.
If $F[X]=0$, each gradient above $X$ is either zero or a normal vanishing on $X$, and therefore kills $E(X)$.

By definition, $H$ retains the chamber gradient where $F[M]>0$ and is zero elsewhere.
In the positive and negative sign cases, sign persistence proves admissibility of this rectified assignment.
In the zero case both possible gradients kill $E(X)$, proving it there as well.
The value $H[X]$ is $\max(F[X],0)$; every nonzero chamber gradient of $H$ is still an arrangement normal, and $H$ is integer when $F$ is.

The product $H\odot C$ has nonnegative support: at a negative output cone every chamber above it is negative, so both rectified gradients in each summand vanish.

Now assume $C$ is balanced and supported only on cones where $F[X]\ge0$; this refers to the sign of $F$, not to the cycle weights.
At a positive contributing cover, the upper gradient is unchanged by rectification.
At a zero cover the upper gradient evaluates to zero on its representative whichever rectification choice is made.

At a negative output cone there is no contributing nonnegative cover, by sign persistence.
At a positive output cone the lower gradients agree.
At a zero output cone both lower gradients kill its flat, and balance cancels their difference across the entire sum.
Thus $H\odot C=F\odot C$ for every such $C$.

Starting with $D_0=H\odot C$, \Cref{lem:fp-product-engine} makes $D_0$ balanced and the support argument makes its support lie where $F[X]\ge0$.
Replace each of the next $r-1$ copies of $H$ by $F$, successively applying that equality.
Each intermediate product is balanced by \Cref{lem:fp-product-engine} and stays supported where $F[X]\ge0$, because it equals the corresponding product by $H$.
All inputs have positive degree since $r\le k$.
This gives the displayed power identity.
\end{proof}

\paragraph{\emph{The fixed maximum family.}}
Use the arrangement $\mathcal H_0$, called the braid arrangement, which lists the units $e_1,\ldots,e_d$, in this order, followed by the vectors $e_a-e_b$ for $1\le a<b\le d$, in lexicographic order of the pairs $(a,b)$.
It has $c_d=d(d+1)/2$ rows.

Introduce labels $0,1,\ldots,d$, with $y_0=0$ and $y_c=x_c$ for $c\ge1$.
An ordered partition is a list of nonempty, pairwise disjoint blocks whose union is the label set.
A realized covector of this arrangement records an ordered partition $(P_1,\ldots,P_r)$ of these labels, whose blocks are the classes of labels with equal values, listed in decreasing order of value.
Every such partition is realized: assign value $r-t$ to labels of $P_t$, then subtract the assigned value of label zero from every label.
Distinct partitions differ in the sign of some arrangement row.

For a block $P$, define its vector by the following formula, in which a bracket around a statement is its zero-one indicator:
\[
(b_P)_c=[c\in P]-[0\in P]\qquad\text{for }1\le c\le d.
\]
Thus $b_P$ is the indicator vector of $P\setminus\{0\}$ when $0\notin P$, and minus the indicator vector of $[d]\setminus P$ when $0\in P$.
The flat of the realized covector $(P_1,\ldots,P_r)$ has basis $b_{P_2},\ldots,b_{P_r}$.
Indeed these vectors assign constant label values on each block.

For a vector in the flat, give its blocks their common values $a_P$, with $a_P=0$ on the block containing label zero.
Writing $P(c)$ for the block containing label $c$, its coordinate $c$ equals $a_{P(c)}-a_{P(0)}$, so the vector is $\sum_P a_P b_P$.
Since $\sum_P b_P=0$, subtracting the coefficient of $P_1$ from every coefficient removes that block without changing the vector.

For independence, adjoin a zero coefficient on the omitted block.
If a combination of these vectors is zero, then the coefficient of the block containing any label equals the coefficient of the block containing label zero.
The omitted zero coefficient forces all coefficients to vanish.
Thus dimension is $r-1$.

Every pair of labels is compared by an arrangement row, so $\tau\sqsubseteq\sigma$ means that labels equal in $\sigma$ are equal in $\tau$ and that every strict comparison in $\tau$ holds in the same direction in $\sigma$.
Hence the blocks of $\sigma$ refine those of $\tau$ in the same order, and a cover, which has exactly one more block, splits one block of $\tau$ into two consecutive nonempty blocks.
Chambers are precisely the partitions into singletons.

Write $e_0=0$.
The leading label of a chamber is the label of its first block, and the selected leading label at a cone $X$ is the leading label of $\bar X$.
The maximum family $\Upsilon$ has gradient $e_a$ on the chamber whose leading label is $a$.
For a cone, all possible leading labels lie in its first block.
Their gradient differences vanish on its flat, since those labels have equal value there.
Hence $\Upsilon$ is an integer admissible family representing $\max(0,x_1,\ldots,x_d)$.

\begin{lemma}[Normalization of the maximum family]
\label{lem:fp-max-normalization}
The degree-$d$ product $\Upsilon^dU_0$ of the maximum family with the base unit $U_0$ has value one at the origin.
The same is true for its lifted family on every arrangement beginning with $\mathcal H_0$.
\end{lemma}

\begin{proof}
Put $\omega_X=b_{P_2}\wedge\cdots\wedge b_{P_r}$, with empty wedge one.
Both $O^0_X$ and $\omega_X$ are wedges of bases of the same flat, so the basis-change rule gives $O^0_X=\xi_X\omega_X$ for a nonzero rational $\xi_X$.
Define the lattice weight of a cycle by
\[
\operatorname{wt}_C(X)=C(X)|\xi_X|.
\]
At the origin $\xi_X=1$.
On a chamber $|\xi_X|=1$ as well.
The basis $b_{P_2},\ldots,b_{P_{d+1}}$ of a chamber is obtained by omitting one vector from the list $-\sum_ce_c,e_1,\ldots,e_d$, and the base basis of a chamber is the list of units.
If the negative sum is omitted, the remaining vectors are the units in some order, so its determinant is $\pm1$.
If a unit is omitted, expansion along the negative-sum row leaves only that missing unit term, again of determinant absolute value one.

For a cover $\tau\lessdot\sigma$, a positive pointing quotient direction is a vector $u=\alpha q_\sigma+z$ with $\alpha>0$ and $z\in E(\tau)$.
Every cover $\tau\lessdot\sigma$ of the braid arrangement has such a direction $u_\sigma$ with $u_\sigma\wedge\omega_\tau=\pm\omega_\sigma$.
Indeed, let $\sigma$ split a block $P$ of $\tau$ into consecutive blocks $P',P''$.
Let $b=b_{P''}$ if $P$ is the first block of $\tau$, and $b=b_{P'}$ otherwise.

If $P$ is the first block, then $P''$ is the second block of $\sigma$ and $\omega_\sigma=b_{P''}\wedge\omega_\tau$.
Otherwise $b_P$ is a factor of $\omega_\tau$.
Since $b_P=b_{P'}+b_{P''}$, we have $b_{P'}\wedge b_P=b_{P'}\wedge b_{P''}$.
Moving $b_{P'}$ into its position then gives $b\wedge\omega_\tau=\pm\omega_\sigma$.
The label values of $b$ are constant on the blocks of $\sigma$ and differ on $P'$ and $P''$, so $b\in E(\sigma)\setminus E(\tau)$.
Since $E(\sigma)=E(\tau)+\Q q_\sigma$, we have $b=\alpha q_\sigma+z$ with $z\in E(\tau)$ and $\alpha\ne0$, and $u_\sigma=\operatorname{sgn}(\alpha)b$ is a positive pointing quotient direction.

For a cover $\tau\lessdot\sigma$ and a positive pointing quotient direction $u$ with $u\wedge\omega_\tau=\pm\omega_\sigma$, the orientation equations give
\[
\kappa(\sigma,\tau;u)=\frac{|\xi_\sigma|}{|\xi_\tau|}.
\]
For an admissible family $f$, normal change and multiplication by $|\xi_\tau|$ therefore express the product as
\[
\operatorname{wt}_{f\odot C}(\tau)
=\sum_{\sigma:\tau\lessdot\sigma}
\operatorname{wt}_C(\sigma)
(g^f_{\bar\sigma}-g^f_{\bar\tau})\cdot u_\sigma.
\]
This scalar identity does not assume balance.

For each separately fixed $t=0,\ldots,d$, let $C_t=\Upsilon^tU_0$.
We prove that the weight $\operatorname{wt}_{C_t}$ on a cone of dimension $d-t$ is one if the first block has $t+1$ labels and every later block is a singleton, and is zero otherwise.
At $t=0$ these are exactly the chambers, and the unit weights are one.

For the next step take a cone $\tau$ of dimension $d-t-1$, and compute $\operatorname{wt}_{C_{t+1}}(\tau)$ from the weights of $C_t$ at the covers of $\tau$, which we call the upper weights.
Let $\ell$ be the first-block size of $\tau$.
Its partition has $d-t$ blocks; counting the labels outside the first gives $\ell\le t+2$.
If $\ell\le t$, no cover can have first block of size $t+1$: splitting cannot enlarge a first block.
Every upper weight is therefore zero.

If $\ell=t+1$, a contributing cover can only split a later block.
The two leading chamber labels remain in the unchanged first block, and their gradient difference vanishes on the cover flat.
Thus each contribution is zero.

If $\ell=t+2$, every later block is a singleton, by the same label count.
Let $B=P_1(\tau)$ be the first block of $\tau$, and let $\{c_1\},\{c_2\},\ldots$ be its later blocks, in order.
The contributing covers are exactly
\[
\sigma_a=(B\setminus\{a\},\{a\},\{c_1\},\ldots)\qquad\text{for }a\in B.
\]
For each take $u_a=-b_{\{a\}}$.
The prescribed bases give $u_a\wedge\omega_\tau=-\omega_{\sigma_a}$.
This direction splits off label $a$ below all other labels in $B$.
Indeed, let $a'\in B\setminus\{a\}$.
The recorded normal $e_{a'}-e_a$, or its stored negative, vanishes at $\tau$.
It has the same nonzero sign at $u_a$ and on $\sigma_a$, since the form $y_{a'}-y_a$ equals one at $u_a$ and is positive on $\sigma_a$.

Its label values are constant on every block of $\sigma_a$, so $u_a\in E(\sigma_a)=E(\tau)+\Q q_{\sigma_a}$ and $u_a=\alpha q_{\sigma_a}+z$ with $z\in E(\tau)$.
That normal kills $z$, so its value at $u_a$ is $\alpha$ times its value at $q_{\sigma_a}$.
Both values are nonzero and of the same sign, hence $\alpha>0$ and $u_a$ is a positive pointing quotient direction.

For each $a\in B$, let $c\in B\setminus\{a\}$ be the selected leading label at the upper cover $\sigma_a$, and let $c'\in B$ be the selected leading label at $\tau$.
Since $(e_c-e_{c'})\cdot(-b_{\{a\}})=[c'=a]$, each summand of the scalar identity for $\operatorname{wt}_{C_{t+1}}(\tau)$ equals $[c'=a]$, because the weights of $C_t$ at the covers $\sigma_a$ are one.
Hence the sum over $a\in B$ is one.
The formula also holds when a selected label is zero, because $e_0=0$ and the common zero-label indicator cancels in the difference.
This proves the next weight formula in every case.

At $t=d$ the only cone is the origin, whose first block contains all $d+1$ labels and whose $\xi$ is one.
Its cycle value is consequently one.
Refinement gives the second assertion.
\end{proof}

For this fixed $d$, the braid arrangement, maximum family and their arrays are closed objects.
Their index bound is the fixed numeral $(1+2c_d^{d-1})^d$.
After their graphs have been admitted, \Cref{lem:sigma1-complete} gives $\Vone$ proofs of the following three closed assertions, which are true by the arguments above: that this fixed list is an arrangement in the sense of \Cref{ssec:fixed-prime-depth}, that its fixed integer maximum family is admissible, and that its final value is one.
This use of closed completeness does not concern a network or an arrangement given by a string variable.
For such arrangements the conclusion follows instead from the uniform refinement proof.
\space

\section{Network Families and Explicit Prime Quotients}
\label{sec:quotients}

We now apply the finite calculus to the functions computed by a candidate network.
Denominator clearing, as in \cite[Lemma~16]{AHM25}, makes the relevant families integer, and an explicit quotient induction supplies the divisibility obstruction, in the role that \cite[Theorem~14]{AHM25} plays for normalized volumes.

\paragraph{\emph{Network families at a fixed depth.}}
Fix a prime $q$, an integer $K\ge1$, and $d=q^K$ externally.
Let $N$ be a valid network with input dimension $n\ge d+1$ and actual hidden depth $k\le K$.
Write $m=|N|$ for the length of its code.

For now suppose every stored nonbias denominator has nonzero residue modulo $q$.
Let $D$ be their product.
For each stored nonbias entry $e$, let $u_e$ be its stored numerator, let $P_e$ be the product of all these denominators except that of entry $e$, and let $\widehat u_e=u_eP_e$.
The empty product is integer one.
The cleared network $\bar N$ retains the original input dimension, every width and every column, replaces each nonbias weight by $(\widehat u_e,1)$, and sets every bias to zero.

Set the $n-d$ input coordinates after the first $d$ to zero.
Let $F$ be the resulting bias-free restriction of the original network to its first $d$ inputs, and let $\bar N^\circ$ be the corresponding restriction of $\bar N$.
The integer $a=D^{k+1}$ satisfies
\[
\bar N^\circ(y)=aF(y).
\]
Indeed every hidden affine sum is multiplied by one further positive factor $D$; rectification commutes with multiplication by a positive integer, and the final affine output contributes the last factor.

For $k\ge1$, let $\widehat w_i$ be the first $d$ cleared weights of hidden neuron $i$ in layer one.
Start $\mathcal H_1$ with the braid arrangement, then append all $\widehat w_i$ in increasing order of $i$, substituting $e_1$ for a zero vector.

For each separately fixed $\ell=2,\ldots,k$, start $\mathcal H_\ell$ with all rows of $\mathcal H_{\ell-1}$ in their existing order, and then proceed as follows.
For every index $\sigma$ below the bound $S_{\ell-1}$ computed from $\mathcal H_{\ell-1}$ (defined below), take the representative $\varpi(\sigma)$ and form the preactivation gradients of layer $\ell$ at it.
To do so, compute at that index the preactivation gradients of the earlier layers successively by their weighted sums, and at each activation retain the gradient of its preactivation exactly when the dot product of that gradient with the representative is positive.

Taking the indices in increasing order, at a chamber index append the layer-$\ell$ gradients in increasing order of the neuron, with $e_1$ in place of a zero gradient; at every other index append $e_1$.
This defines $\mathcal H_\ell$ before the completeness assertion of \Cref{lem:fp-flats-complete} is used.

If $s_\ell$ is the row count of $\mathcal H_\ell$ and $S_\ell=(1+2s_\ell^{d-1})^d$, then
\[
s_1\le c_d+m,\qquad s_{\ell+1}\le s_\ell+mS_\ell.
\]
There are only $K$ separately fixed constructions, so these are polynomial number terms in $m$.
Use the final arrangement $\mathcal H_k$ for the products.
A family is global linear when it assigns the same gradient to every chamber.
For $k=0$, use just the braid arrangement and the global linear output family, whose gradient is the vector of the first $d$ cleared output weights.

For $1\le\ell\le k$ and each neuron $i$ of layer $\ell$, write $F_{\ell i}$ for its preactivation family and $H_{\ell i}=F_{\ell i}^+$ for its activation family, both families on the final arrangement $\mathcal H_k$.
On the final chambers the layer-one preactivation gradients are $\widehat w_i$, later ones are the actual integer weighted sums of preceding activation gradients, and the output is the corresponding weighted sum without a rectifier.
These are integer admissible families.

To justify these claims layer by layer, we begin with a global linear family, which is admissible.
A chamber of a later arrangement restricts to a chamber of every earlier arrangement.
By induction on $j$, its preactivation signs at layers $1,\ldots,j$ are those of the chamber of $\mathcal H_j$ to which it restricts, since every nonzero such gradient was inserted as a normal and restriction keeps the signs of the earlier normals.
Consequently its layer-$\ell$ preactivation gradient equals the gradient listed at the corresponding chamber of $\mathcal H_{\ell-1}$, and occurs as a final normal when nonzero.

\Cref{lem:fp-refinement-locality} proves admissibility of the rectification of each such preactivation family, and integer weighted sums preserve admissibility and integrality.
The same induction through actual sums and signs proves that the output family evaluates to $\bar N^\circ$ at every point of every chamber, including zero preactivations.
No enumeration of all Boolean activation patterns is used.

\paragraph{\emph{Explicit quotients by the prime.}}
Recall that $AC=A\odot C$ and that $A^rC$ denotes $r$ consecutive cone products with $C$, with $A^0C=C$.
All degrees and exponents in the following construction are fixed numerals.
For a nonnegative integer $t$, the factorial $t!$ is the product $1\cdots t$, with $0!=1$.
For integers $e\ge0$ and $0\le r\le e$, the binomial coefficient $\binom er$ counts the $r$-element subsets of an $e$-element set.
In a product of $e$ copies of $X+Y$, choosing the $r$ positions contributing $Y$ gives exactly those subsets, so this is also the coefficient of $X^{e-r}Y^r$.
Ordering the chosen and unchosen positions gives $r!(e-r)!$ permutations per subset, hence $r!(e-r)!\binom er=e!$.

For each externally fixed integer $b$ with $1\le b\le K$, set $e=q^b$ and, for $1\le r<e$, put
\[
b_{e,r}=\binom er/q.
\]
These are integer constants, as the following elementary argument shows.

The argument shows that $q$ divides $\binom er$ for $1\le r<e=q^b$, and it works with integer coefficients modulo the fixed prime $q$.
Two integers are congruent when their difference is an integer multiple of $q$, and two polynomials with integer coefficients are congruent, written $\equiv$, when their corresponding coefficients are congruent.
For $0<r<q$, the identity $r!(q-r)!\binom qr=q!$ has a right side divisible by $q$, while neither factorial on the left is divisible by $q$.
Since $q$ is prime, a product of integers not divisible by $q$ is not divisible by $q$; therefore $\binom qr$ is divisible by $q$.

Expand a product of $q$ copies of $X+Y$, where $X,Y$ are formal symbols.
Its coefficients are these binomial integers.
Modulo $q$ only the two pure terms remain, giving $(X+Y)^q\equiv X^q+Y^q$.
Repeating this finite coefficient identity $b$ times gives $(X+Y)^{q^b}\equiv X^{q^b}+Y^{q^b}$.
Comparing the intermediate coefficients of $(X+Y)^{q^b}$ with those of $X^{q^b}+Y^{q^b}$ shows that $q$ divides $\binom er$ for $1\le r<e=q^b$, so $b_{e,r}$ is an integer.
Here $q,b$ are externally fixed; these conclusions are finite integer equalities for the displayed constants, not an internal variable-prime theorem.

\begin{lemma}[Explicit prime quotients]
\label{lem:fp-prime-quotients}
For each $1\le\ell\le k$ and each neuron $i$ of layer $\ell$ there are specified rational-cycle operations $Q_{\ell i},Z_{\ell i}$, that is, operations from rational cycles to rational cycles.
Their domains are balanced inputs of degree at least $q^{\ell-1}$ and $q^\ell$, respectively, and they lower degree by those respective amounts.
On these domains,
\[
F_{\ell i}^{q^{\ell-1}}C=qQ_{\ell i}(C),\qquad
H_{\ell i}^{q^\ell}C=qZ_{\ell i}(C).
\]
They preserve balance and have specified integer-twin operations whenever a twin of $C$ is supplied.
The degree-$d$ output product with the unit has an integer twin divisible by $q$.
\end{lemma}

\begin{proof}
A global linear family has equal chamber gradients, so its product with any positive-degree cycle is zero.
At layer one, where the preactivation families are global linear, put $Q_{1i}(C)=0$.
Suppose $F_{\ell i}^{e}C=qQ_{\ell i}(C)$ has been constructed, where $e=q^{\ell-1}$.
Define, on inputs of degree at least $qe$,
\[
Z_{\ell i}(C)
=F_{\ell i}^{(q-1)e-1}Q_{\ell i}(H_{\ell i}C).
\]
The exponent is nonnegative, including $q=2,e=1$.
The input $H_{\ell i}C$ has degree at least $qe-1\ge e$.
\Cref{lem:fp-refinement-locality} and the already constructed identity give
\[
H_{\ell i}^{qe}C
=F_{\ell i}^{qe-1}(H_{\ell i}C)
=qF_{\ell i}^{(q-1)e-1}Q_{\ell i}(H_{\ell i}C).
\]
Every factor has positive input degree.
Balance follows from product balance, linearity, and the already constructed quotient operations, without division of a cycle by $q$.
Use the same fixed words, integer coefficients and integer-twin products to construct the twin of $Z_{\ell i}$.

To pass to the layer-$\ell$ preactivations for $\ell\ge2$, write $H_j=H_{\ell-1,j}$ and $Z_j=Z_{\ell-1,j}$ for the activation families of layer $\ell-1$ and their quotients, which satisfy $H_j^eC=qZ_j(C)$ for $e=q^{\ell-1}$.
Index these activation families as $H_1,\ldots,H_u$.
For integer weights $v_j$ and balanced input $C$ of degree at least $e$, let $P_t=\sum_{j=1}^t v_jH_j$ for $0\le t\le u$.
The quotient $Q_t$, which is to satisfy $P_t^eC=qQ_t(C)$, is defined directly by
\[
Q_t(C)=\sum_{j=1}^t\left(
v_j^e Z_j(C)+
\sum_{r=1}^{e-1} b_{e,r}v_j^r
P_{j-1}^{e-r}H_j^r C
\right).
\]

For $t=0$, both sides of $P_t^eC=qQ_t(C)$ vanish, since $e\ge1$.
For the step, expand $(P_{t-1}+v_tH_t)^eC$ by bilinearity.
Commutation on balanced suffixes makes every word with $r$ occurrences of $H_t$ have the same value.
Indeed, at most $e(e-1)/2$ adjacent interchanges reorder a word.
Each interchange acts on a balanced cycle of degree at least two, since the input has degree at least $e$ and each word has $e$ factors.
There are $\binom er$ such words.
The pure $P_{t-1}$ term is the previous prefix value, the pure $H_t$ term is $qv_t^eZ_t(C)$, and each mixed coefficient is $qb_{e,r}v_t^r$.
This gives exactly $qQ_t(C)$.

Induction is on the numeric prefix, with $C$ and its supplied twin fixed string parameters; its equations quantify only bounded coordinates of the specified arrays.
It does not put a universal cycle-string quantifier inside the formula on which induction is performed.
All mixed words act on balanced cycles and preserve balance.
The explicit sum defining $Q_t$ uses integer coefficients, so the same sum with every cycle replaced by its integer twin is an integer twin of $Q_t(C)$.
Here the twin of $Z_j(C)$ is supplied by the preceding layer, the twins of the words $P_{j-1}^{e-r}H_j^rC$ are the twin products of \Cref{lem:fp-product-engine} for the integer admissible families $P_{j-1}$ and $H_j$, and integer combinations preserve twins.
For neuron $i$ of layer $\ell$, whose preactivation family is $F_{\ell i}=P_u$ for its integer weights $v_j$, this construction supplies $Q_{\ell i}=Q_u$.
There is no recursion on the preceding quotient's code length.

For actual depth $k=K$, apply this prefix construction to the output with exponent $d=q^K$ and the quotients $Z_{Ki}$.
For $1\le k<K$, replace them by
\[
\widehat Z_{ki}(C)=H_{ki}^{d-q^k}Z_{ki}(C).
\]
Then $H_{ki}^dC=q\widehat Z_{ki}(C)$, with balanced output and the specified twin.
Use the same output prefix formula with exponent $d$.
At degree $d$ the final cycle is at the origin and its orientation is one, so the computed twin is an integer $z$ and the output value is $qz$.
At depth zero the output is global linear; its first product is zero and hence its degree-$d$ value is zero, with quotient $z=0$.
\end{proof}
\space

\section{From Agreement Probes to a Refuter}
\label{sec:probes}

The product calculation becomes an error-finding algorithm by testing finitely many rational points derived from the candidate.
Bias domination transfers those tests to the homogeneous integer network.

\paragraph{\emph{Agreement probes.}}
In the following scalar formulas, write $w_{\ell ij}$ and $\eta_{\ell i}$ for the candidate weights and biases, with neuron positions numbered from one; the stored matrix indices are shifted by one as in \Cref{sec:foundations}.
Let $\operatorname{NumAbs}(w)$ be the magnitude of the canonical signed numerator of a stored rational, irrespective of cancellation.
For the first hidden layer put $E_{1i}=\operatorname{NumAbs}(\eta_{1i})$.
For every later affine layer, including the final output, put
\[
E_{\ell i}=\operatorname{NumAbs}(\eta_{\ell i})+
\sum_{j=1}^{d_{\ell-1}}\operatorname{NumAbs}(w_{\ell ij})E_{\ell-1,j},
\]
where $d_{\ell-1}$ is the width of layer $\ell-1$ in the layer-dimension notation $d_0=n,d_1,\ldots,d_{k+1}=1$ of the fixed network record (\Cref{sec:foundations}); these subscripted widths are unrelated to the dimension $d=q^K$.
At depth zero put $E=\operatorname{NumAbs}(\eta_{11})$; otherwise let $E$ be the final output entry.
All entries are nonnegative integers.

For every rational input $x$, deleting the biases changes each layer value by at most its corresponding $E_{\ell i}$.
At layer one this follows from the absolute bias bound.
For the step, the difference of affine sums is at most the bias magnitude plus the sum of weight magnitudes times preceding differences.
Every stored positive denominator is at least one, so those magnitudes are bounded by the numerator magnitudes in the formula.

The bound passes through a hidden rectifier because $|\max(0,u)-\max(0,v)|\le|u-v|$, which holds by cases.
If $u,v\ge0$, the two sides are equal; if $u,v\le0$, the left side is zero.
If $u\ge0\ge v$, the left side is $u\le u-v=|u-v|$; if $v\ge0\ge u$, it is $v\le v-u=|u-v|$.
At the output there is no rectification.
These deductions are induction on bounded neuron coordinates at each separately fixed layer.

With the clearing factor $a=D^{k+1}$ of the cleared network, set $\bar a=a(E+1)>0$ and define, for $y\in\Z^d$, the input point with integer coordinates
\[
x(y)=(\bar ay_1,\ldots,\bar ay_d,0,\ldots,0)\in\Q^n.
\]
For integer $y$, the bias-free restriction is positively homogeneous, so $F(\bar ay)=\bar aF(y)$.
If $N$ agrees with $\MAX_n$ at $x(y)$, the bias bound gives $|\bar aF(y)-\bar a\max(0,y_1,\ldots,y_d)|\le E$.
Multiplying by $a/\bar a$ and using $\bar N^\circ=aF$ yields
\[
|\bar N^\circ(y)-a\max(0,y_1,\ldots,y_d)|
\le \frac{aE}{\bar a}=\frac E{E+1}<1.
\]
Here the zero among the input coordinates is available because $n\ge d+1$.
Both terms inside the absolute value are integers, and a nonzero integer has magnitude at least one; hence their difference is zero.
The integer fact used here follows by trichotomy.

For every final chamber $\Gamma$ test $y=q_\Gamma$ and the $d$ points $y=q_\Gamma\boxplus e_c$.
They are integer points in the same chamber, by the perturbation identity of \Cref{lem:fp-flats-complete}, since the covector of $q_\Gamma$ has no zero entry.

If agreement holds at all of their $x(y)$, let $g$ be the difference between the output gradient and $a$ times the maximum gradient on this chamber.
Agreement at $y=q_\Gamma$ gives $g\cdot q_\Gamma=0$, and agreement at $y=q_\Gamma\boxplus e_c$ gives $\Delta(e_c)g\cdot q_\Gamma+g_c=0$.
Thus every $g_c=0$.

Thus on every final chamber the output family and $a$ times the lifted maximum family $\widehat\Upsilon$ of \Cref{lem:fp-max-normalization} have the same gradient, so their degree-$d$ products with the unit agree.
Linearity of the product in the family and \Cref{lem:fp-max-normalization} then give output degree-$d$ value $a^d$.
\Cref{lem:fp-prime-quotients} makes the same value $qz$ for an actual computed integer twin $z$.
Under the standing assumption of \Cref{sec:quotients}, every stored nonbias denominator has nonzero residue modulo $q$, so their product $D$ is not divisible by the prime $q$, and neither is $a^d=D^{d(k+1)}$.
Hence $a^d\ne qz$, and the residue calculation of \Cref{sec:graphs} proves this inequality inside the theory.

\paragraph{\emph{One common output cutoff.}}
The output cutoff $p$ of this paper is one polynomial shared by all externally fixed $q,K$; the calculation below shows that the probes stay within it.
The following calculation bounds the actual returned probes, not the much larger auxiliary cycle arrays.
Put $\bar m=(m+2)^2$.
On a candidate that passes the refuter's initial test of the complete antecedent of the depth bound in \Cref{thm:fixed-prime-depth}, $n\ge d+1$ and complete network validity give $d\le n\le m$ and at most $m$ stored parameters, each of component length at most $m$.
Hence the sum of the component lengths of the stored nonbias denominators is at most $m^2\le \bar m$.

From here to the end of this section, $|\cdot|$ applied to a number denotes the length of its code, not its absolute value.
Multiplication of integers adds at most their lengths plus one, and addition adds at most two to the maximum length.
For the product of all nonbias denominators, its omitted-entry products, and cleared numerators this gives
\[
|D|,|P_e|\le2\bar m+2,\qquad |\widehat u_e|\le3\bar m+3.
\]
An integer sum of at most $m$ terms adds at most $2m\le2\bar m$ to their maximum length.
Let $L_j$ be the largest coordinate length of a layer-$j$ preactivation gradient, the output counting as layer $k+1$.
Starting with $L_1\le3\bar m+3$, these gradient caps thus satisfy $L_{j+1}\le L_j+5\bar m+4$, and consequently $L_j\le7j\bar m$.
Since $K\le d$, every arrangement normal has coordinate length at most $\ell_H=7d\bar m$.
The units, braid differences, and replacement units obey this bound as well.

Compute each cross determinant by its fixed Leibniz expression, adding its monomials in a balanced binary tree.
With $h=\lceil\log_2((d-1)!)\rceil\le d^2$, a ray coordinate has length at most
\[
\ell_r=(d-1)\ell_H+d-2+2h\le8d^2\bar m.
\]
For integer coordinates bounded by $L_u,L_w$, the perturbation obeys
\[
|\Delta(w)|\le\ell_H+L_w+2d+5,\qquad
|(u\boxplus w)_c|\le L_u+L_w+\ell_H+2d+8.
\]
The first perturbation from zero returns its ray.
After $d$ perturbations the representative and the subsequent unit-perturbed probe therefore have caps
\[
L_q\le d\ell_r+(d-1)(\ell_H+2d+8)\le12d^3\bar m,
\quad L_y\le L_q+\ell_H+2d+10\le15d^3\bar m.
\]
Here $\bar m\ge25$, $d\ge2$, and $\ell_H+2d+8\le8d\bar m$ verify the displayed loose constants.

For the bias majorant, each layer adds at most $3\bar m+3$ to the preceding integer length cap.
As $k+1\le K+1\le d$, this gives
\[
|E+1|\le5d\bar m,\qquad |a|\le3d\bar m,\qquad |\bar a|\le9d\bar m.
\]

Every scaled numerator has length at most $9d\bar m+15d^3\bar m+1\le18d^3\bar m$.
All denominators are canonical integer one of length two, including at padded zero coordinates.
Let $S_*$ be the sum of the numerator and denominator component lengths over all coordinates of $x(y)$; it satisfies
\[
S_*\le n(18d^3\bar m+2)\le19nd^3\bar m\le19(m+2)^6.
\]

For the pairing of \Cref{sec:foundations}, $\langle u,v\rangle\le(u+v+2)^2$.
Put $r=m+2$; then $S_*+5\le20r^6$, so the inner rational pair has bound at most $400r^{12}$.
As $n+3\le r+1\le r^{12}$, the outer vector bound is at most
\[
|x(y)|\le\langle n+1,\langle3,S_*\rangle\rangle
\le(401r^{12})^2<2^{22}r^{24}<2^{100}r^{40}\le p(m).
\]

The polynomial $p$ is fixed in \Cref{sec:foundations}; it includes an independently fixed short-disagreement bound.
The common bound does not follow merely from separate polynomial running times: it follows from this explicit probe calculation using $d,K\le m$.
The degrees of the algorithms' running times and internal-array bounds may still depend on $q,K$.
\space

\section{Defining Graphs and the Internal Theorem}
\label{sec:graphs}

The preceding identities describe both geometry and arithmetic computations.
To obtain a theorem of $\Vone$, we now assign total defining graphs to these computations and verify the bounded invariants used by the proof.
The guards and caps below ensure totality even on malformed inputs; the recurrence invariants recover the intended geometry on valid inputs.
The next section replaces these histories by threshold-computable graphs with the same required equations.

\paragraph{\emph{Defining graphs and their bounds.}}
We specify the computation underlying these identities.
All new symbols carry a suffix determined by the externally fixed $q,K$; they are distinct from the foundational symbols.
An integer read retains a canonical signed integer and otherwise returns zero.
A rational read retains a valid rational with positive denominator and otherwise returns $0_\Q$.
An integer-family read retains a denominator-one rational entry and otherwise returns the integer zero, rather than taking an arbitrary rational numerator.

Every network construction used only for \Cref{thm:fixed-prime-depth} first checks that $\Net(N,n)$ holds, that $n\ge d+1$, that the actual depth is at most $K$, and that the raw denominator predicate $\mathsf{GoodDen}_q(N,n)$ of \Cref{sec:foundations} holds.
The residue graph used by $\mathsf{GoodDen}_q$ is given below.
Failure returns the empty failure string before forming any cleared array.
Standalone geometric graphs instead use guarded reads independently of any network promise.
The value-normalization constructor at the end is exempt from the raw denominator guard: it checks complete network validity, and its composed refuter additionally checks the stated dimension and depth bounds.

For component bounds define
\[
\operatorname{inc}(L)=2L+3,\quad
P(L)=\langle3,2L\rangle,\quad
\operatorname{Tot}(v,L)=(v+1)(P(L)+3).
\]
A rational is component-bounded by $L$ when both canonical integer components have length at most $L$.
The established length laws bound a sum of $v$ such rationals by $\operatorname{Tot}(v,L)$ and an integer sum of $v$ entries of length at most $L$ by $L+2v$.
A guarded inverse normalizes the denominator sign and exchanges components, so it does not enlarge their common bound.
A minimum returns an actual list entry.
All other fixed scalar sums, products, differences and absolute values have bounds obtained by composing $\operatorname{inc}$.

We fix this composition count explicitly, rather than asserting a numerical exponent.
Form a finite registry $\mathcal E_d$ of the scalar expressions used above.
For linear algebra, include dot products with at most $2d$ entries, every Leibniz determinant of orders $0,\ldots,2d$, and the adjugate coordinates and Cramer quotients of those orders.
For cone geometry, include cross and perturbation coordinates, the two-dimensional sector direction, ratios, endpoints and lifted coordinates, wedge and contraction coordinates in degrees $0,\ldots,d$, and a nonzero orientation-coordinate ratio and its absolute value.

For products, include the per-cover product, balance-vector, shifted-gradient, commutation double-sum, residual and integer-twin summands.
Also include refinement orientation ratios, lifted weights and signed twin coordinates, and weighted family, cycle and twin summands.
Finally include powers $v^r$ and constants $b_{e,r}v^r$ for $r,e\le d$, and the affine-clearing, bias-majorant and scaled-probe scalar summands.

In this registry expand every fixed-index determinant, dot, wedge, contraction, power and sum into its displayed scalar formula.
A guarded read, selected index, or result of a variable-length sum or minimum is an input leaf with its own bound.
A branch is the finite case tree of its scalar branches; guarded division is a product with the guarded inverse.
There is no variable-length loop inside a registry expression.

Define $\operatorname{count}(T)$ for $T\in\mathcal E_d$ by counting input and literal leaves as zero, a unary operation as one plus its child count, a binary operation as one plus its two child counts, and a finite branch as one plus the sum of its branch counts.
Write $\operatorname{inc}^t$ for the $t$-fold composition of $\operatorname{inc}$.
Put
\[
t_d=1+\sum_{T\in\mathcal E_d}\operatorname{count}(T),\qquad
\mathcal F(L)=\operatorname{inc}^{t_d}(L+d+3).
\]

For each fixed $d$, this definition is an explicit finite prescription for the numeral $t_d$.
Every registry literal has component length at most $d+3$: binomial coefficients are at most $2^d$, their quotients are smaller, and all other scalar literals are $0,1,-1,2$.
Structural induction on the finite expressions proves their component bound $\mathcal F(L)$ when all leaves have bound $L$.
The same bound covers integer expressions by their integer length laws.
The induction is an external finite derivation selecting the numeral; it does not iterate $\operatorname{inc}$ an input-dependent number of times.
Variable arrays and sums use the separate rules just stated.

For exact arrays use (Arr), with rows $0,\ldots,v-1$ and bound $\langle v,B\rangle$ when each row has bound $B$.
Define
\[
\operatorname{Vec}_d(L)=\langle d+1,P(L)\rangle,\quad
\operatorname{Ext}_d(L)=\langle2^d+1,P(L)\rangle,\quad
\operatorname{Ext}^{\Z}_d(L)=\langle2^d+1,L\rangle.
\]
Vector row zero and rows above $d$ are empty.
For exterior degree $r$ use the fixed lexicographic list of increasing $r$-subsets as rows $1,\ldots,\binom dr$, all other rows empty.
The number of representative indices of \Cref{ssec:fixed-prime-depth}, the ordered $d$-tuples of ray indices, is $S=R^d$; the geometric setup below guards it by its barred bound $\overline S$.
For a cycle the cone rows are $0,\ldots,S-1$ and may include row zero; they are rational zero away from cones of its degree.
A twin has the same cone rows, each an integer exterior array of the appropriate degree, and zero exterior rows off those cones.
For a cycle or a twin, shape means equality of its string to the bitwise reconstruction of the array from the rows its shape uses, that is, the shape predicate of \Cref{sec:foundations}, not just validity of named entries.
So the string has no present bit in a row that the shape does not use, and none at a position that is not a value of the pairing.
The twin predicate is this pair of shapes together with its bounded coordinate equations.

Families have $S$ rows of $d$-vectors, with zero vectors at nonchambers.
Wrong degrees return the exact zero array of degree zero, and product calls require positive input degree.
Every fixed word tests that its length is at most the input degree.
All these predicates have only bounded numeric quantifiers after the chosen arithmetic graphs are added.

For the geometric setup put $\bar s=|\mathcal H|$, and let $c_{\mathrm{br}}$ be the length of the fixed braid code.
Read $s$ as the least empty normal-row index, bounded by $\bar s$.
Set
\[
\overline R=1+2\bar s^{d-1},\quad\overline S=\overline R^d,\quad
B_N=\mathcal F(\bar s+c_{\mathrm{br}}+2),\quad B_O=\mathcal F(B_N+2).
\]
Normal rows are guarded integer vectors of component length at most $\bar s+2$.
A malformed setup need not be geometrically complete; all arithmetic constructors below remain total on it.
Use $R=1+2s^{d-1}$ and $S=R^d$, guarded by their barred bounds.
Use the signed crosses and the ray and index order of \Cref{ssec:fixed-prime-depth}.
There a positive ray index $a$ is read through $a-1=2g+\varepsilon$ as a sign bit $\varepsilon$ and the $d-1$ base-$s$ digits of $g$, and an index $\sigma<S$ as its $d$ base-$R$ digits, all least significant first.
Successive number quotients recover the sign bit and these least-significant-first digits.
For number quotient and remainder, use $u=vb+r$, $r<b$, $v,r\le u$ when $b>0$, and $v=r=0$ when $b=0$.
Existence follows by choosing the least $v\le u$ with $u<(v+1)b$; failure at its predecessor gives $vb\le u$.
Uniqueness follows since a quotient change changes $vb$ by at least $b$.
Thus fixed digit decoding is bounded before geometric correctness is used.

The following list binds every remaining geometric graph to its actual scalar formulas and array bounds.
\begin{enumerate}[leftmargin=2em]
\item Fixed minors, rank on lists of at most $2d$ vectors, adjugates, Cramer coordinates and kernel bases use the finite registry formulas, selecting the first nonzero minor by a bounded least index.
Missing pivots return zero coordinates; rank zero uses units and full rank the empty kernel.
Guard every division before evaluating its branch.
Kernel coordinates have bound $B_N$ and their fixed wedges have bound $B_O$.
A Cramer coefficient with input caps $L',L''$ has cap $\mathcal F(L'+L''+2)$.
Materialized scalar matrices with $r$ rows and $c$ columns use the flat (Arr) entry reconstruction, which stores entry $(i,j)$ in row $\langle i,j\rangle$ as in \Cref{sec:foundations}.
Such a matrix has row-pair bound $\langle r+1,c+1\rangle$ and its actual entry cap, and it is not identified with a nested vector array.

\item A greedy scan of a list of length $v$ has state count $1^r$ in row zero and unary indices $1^{j+1}$ in rows $1,\ldots,r$, with $r\le d$ and strictly increasing $j<i$ at stage $i$.
All other rows and unused bits are empty.
A malformed state or $i\ge v$ returns empty.
Otherwise read the retained original vectors and the current vector.
Append $i$ exactly when the admission test holds, $r<d$, and rank rises; the admission test is a zero sign at $i$ for the zero-normal scan, and no condition for the completion and quotient-extension scans.
In all other cases return the validated reconstruction.
Every branch has state bound $\langle d+1,v+d+2\rangle$ independently of the old state's length.
Limited recursion with this clipping bound is total before its spanning invariant is proved.
That invariant proves clipping and failure idle on valid scans.
The zero-normal, completion and quotient-extension scans have respectively $v=s,s+d,2d$ as sufficient list bounds.
Their list vectors are separately materialized at their (Arr) bounds.

\item Crosses use the balanced integer tree specified in \Cref{sec:probes}.
Their all-string cap is
\[
B_r=(d-1)(\bar s+2)+d-2+2h,
\]
including the zero ray, where $h=\lceil\log_2((d-1)!)\rceil$ as in \Cref{sec:probes}.
The $d$ integer perturbations have representative cap
\[
B_q=2+d(B_r+\bar s+2d+10).
\]
This bound follows directly from the integer dot/sum/update lengths on arbitrary guarded normal rows, without completeness.
Covectors are exact $s$-row lists of the three fixed sign codes.
The least-index lookup scans all $S$ representatives and returns $S$ when no covector matches.
A sentinel reads the zero representative before any sign or chamber test.
Completion by $d$ units has cap $B_q+d(\bar s+2d+12)$, and a subsequent lookup again returns a representative bounded by $B_q$.
Order, composition, dimension, cover and chamber tests use these sign lists, ranks and bounded indices.

\item In completeness, for a supplied point of component cap $L$, each positive ratio has cap $\mathcal F(\bar s+L+B_r+2)$.
Select a first genuine candidate before filling failed tests and the appended row with copies of it.
If none exists return zero residuals and sentinels.
The residual coordinate cap is $\mathcal F(L+B_r+\mathcal F(\bar s+L+B_r+2)+2)$.
The $d+1$ separate dimension deductions compose these bounds only a fixed number of times.
A perturbed cone $\operatorname{idx}(q_\tau\boxplus w)$ of \Cref{lem:fp-stars-diamonds}, with direction $w$ of arbitrary cap $L_w$, has temporary cap $\mathcal F(\bar s+B_q+L_w+2)$; its lookup has global representative cap $B_q$.

\item A standalone sector takes a list of $v$ forms and first validates that they are nonzero, with signs $\pm1$, an independent pair and strictly positive oriented values at its input point.
For forms of cap $L_\psi$ and point cap $L_y$, use successively
\[
V_{\mathrm{sec}}=\mathcal F(L_\psi+2),\quad
R_{\mathrm{sec}}=\mathcal F(L_\psi+L_y+V_{\mathrm{sec}}+2),\quad
E_{\mathrm{sec}}=\mathcal F(L_y+R_{\mathrm{sec}}+V_{\mathrm{sec}}+2).
\]
Direction, ratio and endpoint formulas are the ones in the sector proof.
Their variable ratio lists have $v+1$ rows with bound $\langle v+1,P(R_{\mathrm{sec}})\rangle$; failed tests duplicate a genuine positive candidate.
Invalid inputs return zero endpoints and sentinels.
For the two intermediate cones of \Cref{lem:fp-stars-diamonds} use quotient coordinates of cap $\mathcal F(B_N+B_q+2)$ and forms of cap $\mathcal F(\bar s+B_N+2)$, duplicate a genuine form outside $J$, then use the displayed sector caps.
Lift an endpoint of cap $E_{\mathrm{sec}}$ with cap $\mathcal F(B_N+E_{\mathrm{sec}}+2)$ and apply the preceding perturbed-cone bound.
Missing bases or pairs return the stated zero objects.

\item Orientation ratios use the first nonzero coordinate of $v\wedge O_\tau$, with all other cases returning zero.
For direction cap $L$ use
\[
\Lambda(L)=\mathcal F(B_O+\mathcal F(L+B_O+2)+2).
\]
The two-direction ratio in commutation has cap $\Lambda_2=\mathcal F(B_O+\mathcal F(2B_q+B_O+2)+2)$.

Shifts of a family of gradient cap $A$ have cap $\mathcal F(A+2)$ and evaluations have cap $\mathcal F(A+B_q+2)$.
Weighted sums of $v$ families or cycles, with scalar caps $L_w,L_g$, have cap $\operatorname{Tot}(v,\mathcal F(L_w+L_g+2))$.
Integer weighted sums and twins have cap $\mathcal F(L_w+L_g+2)+2v$ per integer coordinate.
They are direct coordinate sums, not a recursion on preceding array lengths.

\item A per-cover rational product summand is exactly
$C(\sigma)\kappa(\sigma,\tau)(g_{\bar\sigma}-g_{\bar\tau})\cdot q_\sigma$, enabled by the cover/degree tests.
An integer-twin summand is the earlier signed contraction, with genuine integer-family reads.
For a family cap $A$ and an input-cycle coordinate cap $L$ define
\[
\mathsf R_A(L)=\operatorname{Tot}(\overline S,\mathcal F(A+L+\Lambda(B_q)+B_q+2)),
\qquad
\mathsf I_A(L)=\mathcal F(A+L+2)+2\overline S.
\]
The output arrays have bounds $\langle S,P(\mathsf R_A(L))\rangle$ and $\langle S,\operatorname{Ext}^{\Z}_d(\mathsf I_A(L))\rangle$.

The unit and its twin have coordinate cap two.
A balance vector has cap $B_V(L)=\operatorname{Tot}(\overline S,\mathcal F(L+\Lambda(B_q)+B_q+2))$.
The same cap applies to $V'(\rho)$, whose alternative direction is another global representative.
The Cramer coefficient $\zeta_\rho$ of $V'(\rho)$ along $q_\rho$, from the proof of \Cref{lem:fp-product-engine}, has cap $Z_V=\mathcal F(B_N+B_q+B_V(L)+2)$.
A commutation double sum has cap
\[
\operatorname{Tot}(\overline S^2,\mathcal F(A+L+\Lambda_2+B_q+Z_V+2)).
\]
These are bounds for the actual bounded arrays used in the proof identities.

The fixed-word caps $\mathsf R_A^{[t]},\mathsf I_A^{[t]}$ are obtained by $t$ finite compositions, for each separately fixed $0\le t\le d$.

\item Refinement reads both arrangements and uses the same setup with input cap $\bar s+|\mathcal H_0|+c_{\mathrm{br}}+2$.
Restriction is its bounded lookup on the base covector.
Its orientation ratio is the first nonzero-coordinate ratio of the two equal-dimensional orientations; a failed dimension/pivot test returns zero.
A rational lift has cap $\mathcal F(L+L_{\bar\kappa}+2)$ when its ratio cap is $L_{\bar\kappa}$.
Its integer twin lift uses the sign of the ratio and the guarded integer twin entries.
Fixed braid arrays use their exact, externally fixed codes.
\end{enumerate}

Every standalone graph chooses raw caps equal to the sum of its string argument lengths plus two.
The preceding formulas, with those caps, bound its output on all strings.
The same actual graph equations prove the tighter component caps on valid callers; no tighter caller promise is substituted into a graph's only totality bound.
All finite searches select existing entries or stated sentinels, and all materialized arrays use their actual declared bounds.
Each recursion has its clipping bound before its invariant is used.

Thus the construction order is acyclic: scalar graphs, fixed linear algebra, scans, rays and lookups, geometric auxiliaries, exterior arrays, families and products, and finally the finite compositions.
It yields total unique graphs in $\Vone$ from (Arr) and limited recursion.
The proofs of correctness use those very defining equations and their stated invariants.

For the network constructors, invalid networks exit first.
As in \Cref{sec:quotients}, $m=|N|$ is the length of the network code.
On valid networks take $B_U=3(m+2)^2+3$, set $L_1=\mathcal F(B_U+2)$ and
\[
L_{j+1}=\operatorname{Tot}(m,\mathcal F(B_U+L_j+2))\quad\text{for }1\le j\le K,
\qquad A=2+\sum_{j=1}^{K+1}L_j.
\]
The cap $L_j$ is for the layer-$j$ gradients, the output of a network of actual depth $k\le K$ counting as layer $k+1$; it covers the output and every output prefix, as well as hidden gradients.

To materialize successive arrangements, use the normal counts $s_1\le c_d+m$ and $s_{\ell+1}\le s_\ell+mS_\ell$ displayed in \Cref{sec:quotients}, together with (Arr) and these gradient caps; then use the actual final arrangement length $\bar s$ in the geometric list above.
The number of such constructions is fixed by $K$.
Clearing uses a scan below $m$ skipping nonentries; each excluded product skips its selected entry too.
A coarse product bound obtained by multiplying at most $m$ entries of length at most $m+2$, namely $2+m(m+3)$, bounds all-string product histories after the network guard.
On valid networks these product histories have the bound $2\bar m+2$, where $\bar m=(m+2)^2$, proved in \Cref{sec:probes}.

For bias-majorant histories use the component caps $B^E_0=m+2$ and $B^E_{j+1}=B^E_j+3m+6$ for the separately fixed $0\le j\le K$, including the final output.
A row product adds at most $m+3$, its at-most-$m$-term sum adds $2m$, and the bias addition fits the remaining allowance.
The histories are exact arrays of at most $K+2$ rows, each an integer vector with at most $m+1$ entries, hence (Arr) bounds from $B^E_{K+1}$.

Writing $B_D=2+m(m+3)$, a coarse scale bound is
\[
B_a=2+(K+1)(B_D+1),\qquad B_{\bar a}=B_a+B^E_{K+1}+3.
\]
An arbitrary input direction with component cap $L_y$ gives scaled numerator cap $B_{\bar a}+L_y+1$ and padded vector bound $\langle n+1,\langle3,B_{\bar a}+L_y+3\rangle\rangle$.
Scale, padded points and the bounded least-error scan then have polynomial bounds by the same rules.

We also give explicit bounds for the direct quotient constructors.
Put $C_w=\mathcal F(B_U+d+3)$, a cap for the scalar coefficients $v_j^e$ and $b_{e,r}v_j^r$.
Abbreviate fixed-word composition by $\mathsf R^{[t]},\mathsf I^{[t]}$ with family cap $A$, writing $\mathsf R=\mathsf R^{[1]}=\mathsf R_A$ and $\mathsf I=\mathsf I^{[1]}=\mathsf I_A$.
Superscripts $R$ and $I$ mark caps of rational cycles and of their integer twins, as functions of the input cap $L$.
At layer one choose the zero quotient graphs directly, with caps $Z^R_1(L)=2$, $Z^I_1(L)=0$; the already proved global-linear calculation validates this specialization.
For $j\ge2$, write $r=q^{j-1}$ and set
\[
\begin{aligned}
Q^R_j(L)&=\operatorname{Tot}(mr,\mathcal F(C_w+Z^R_{j-1}(L)+\mathsf R^{[r]}(L)+2)),\\
Q^I_j(L)&=\mathcal F(C_w+Z^I_{j-1}(L)+\mathsf I^{[r]}(L)+2)+2mr,\\
Z^R_j(L)&=\mathsf R^{[(q-1)r-1]}(Q^R_j(\mathsf R(L))),\\
Z^I_j(L)&=\mathsf I^{[(q-1)r-1]}(Q^I_j(\mathsf I(L))).
\end{aligned}
\]
The direct prefix contains at most $mr$ summands, each a weighted previous quotient or a fixed mixed word, so the first two bounds follow from the actual rational and integer sum laws.
The last two are the specified rectification quotient words, not recurrences over neuron prefixes.

On branch $k<K$ first replace $Z^R_k$ and $Z^I_k$ by their $d-q^k$ fixed-word extensions $\mathsf R^{[d-q^k]}(Z^R_k(L))$ and $\mathsf I^{[d-q^k]}(Z^I_k(L))$, the caps of the quotients $\widehat Z_{ki}=H_{ki}^{d-q^k}Z_{ki}$ of \Cref{lem:fp-prime-quotients} and of their twins.
At the output use the first two formulas with $r=d$ and this last activation quotient.
All wrong-degree or malformed-cycle branches return their exact zero arrays.

When these constructors are called standalone, use their raw argument-length caps in place of the valid-network caps and materialize their family lists at their own declared bounds.
On a supplied valid network, the tighter $B_U,A$ bounds follow from the already stated caller invariants.
All compositions have fixed depth at most $K$ and fixed word length at most $d$, so they expand to polynomial number terms.
These constructors satisfy the identities proved in \Cref{lem:fp-prime-quotients} and its proof: balance of their output cycles, the integer-twin equations, and the prefix equations $P_t^eC=qQ_t(C)$.
Coordinate predicates for these identities have only bounded numeric quantifiers, with input cycle and twin strings retained as parameters.
This is why the numeric prefix induction of that proof is permitted, and why the symbols adjoined for these constructors can be eliminated in the proof of \Cref{thm:fixed-prime-depth}.

\paragraph{\emph{The fixed-prime residue predicate.}}
We give the actual arithmetic graph of this predicate, rather than infer its correctness from a polynomial-time remainder algorithm.
The graph performs binary long division of the magnitude of the integer by $q$, reading one bit per stage.
Let $u$ be a canonical signed integer code, and write $|u|$ for the length of the code $u$.
Let $L=\max(|u|-1,0)$ be its number of available magnitude positions, and process these positions from $L$ down to one.
Maintain a nonnegative integer quotient code $Q$ and a residue $r\in\{0,\ldots,q-1\}$, initially zero.
Encode the residue by the single bit at position $r$.
At stage $t<L$ read $b=u(L-t)$, choose the fixed finite table entries
\[
c=\left\lfloor\frac{2r+b}{q}\right\rfloor\in\{0,1\},\qquad
r'=2r+b-qc,
\]
and put $Q'=2Q+c$ by the admitted canonical integer operations.
The table depends only on the fixed numeral $q$.

Reject a malformed preceding state or input integer by returning the fixed zero state.
Store the quotient and residue in tuple rows one and two, row zero empty, with bound $\langle3,|u|+3+q\rangle$; clip the quotient to $|u|+3$ and the residue to $q$.
The update is bounded on every guarded state, and limited recursion defines it before its invariant.

Let $Q_t,r_t$ be the quotient and residue after $t$ stages, and let $A_t$ be the integer magnitude prefix given by $A_0=0$ and $A_{t+1}=2A_t+b$, with $b$ the bit read at stage $t$.
Induction on $t$ proves
\[
A_t=qQ_t+r_t,\qquad 0\le r_t<q.
\]
At a step substitute $A_{t+1}=2A_t+b$ and the finite table equation.
The prefix is nonnegative and has at most $t+1$ signed magnitude positions, so $0\le Q_t\le A_t$ gives the quotient length bound and makes clipping idle.
Binary reconstruction of the magnitude by its shift/add recurrence proves that $A_L$ is the magnitude of $u$, using the actual signed-code value laws.

For a negative input with final $r=0$, change $Q$ to $-Q$.
For a negative input with $r>0$, change $Q$ to $-Q-1$ and $r$ to $q-r$.
For a nonnegative input retain both.
In the next display, an overline denotes the canonical nonnegative integer code for one of these finitely many residues.
The resulting signed quotient $\widehat Q_q(u)$ and residue $\operatorname{Res}_q(u)$ satisfy
\[
u=q\widehat Q_q(u)+\overline{\operatorname{Res}_q(u)},\qquad
0\le\operatorname{Res}_q(u)<q.
\]
The sign adjustment adds at most two to the quotient length and is covered by the stated cap.
Invalid signed strings have residue zero by definition.
The residue number is obtained from its one-hot state by a bounded least-index decode.

These decompositions are unique.
Indeed if $qQ+r=qQ'+r'$ with $0\le r,r'<q$, then $q(Q-Q')=r'-r$ has magnitude less than $q$.
A positive integer $Q-Q'$ is at least one, and a negative integer is at most minus one; either would give magnitude at least $q$ by integer order.
Thus $Q=Q'$ and then $r=r'$.

For valid integer $u,v$, write $u=qQ+r$ and $v=qQ'+s$ for their decompositions with $0\le r,s<q$, and substitute them into their actual canonical sum.
Here, for an integer $w\ge0$, $w\bmod q$ denotes the remainder of $w$ on division by $q$.
The fixed finite carry table for $r+s$ supplies a decomposition of the sum with remainder $(r+s)\bmod q$.
Uniqueness gives
\[
\operatorname{Res}_q(u+v)
=(\operatorname{Res}_q(u)+\operatorname{Res}_q(v))\bmod q.
\]

For the product, use the same decompositions and distribute
\[
(qQ+r)(qQ'+s)=q(qQQ'+Qs+Q'r)+rs.
\]
The finite table decomposing $rs<q^2$ and uniqueness similarly give
\[
\operatorname{Res}_q(uv)
=(\operatorname{Res}_q(u)\operatorname{Res}_q(v))\bmod q.
\]

These deductions use proved ring/order equations for the admitted canonical operations, so they bind the residue graph to those operations.
Every table assertion is a finite true statement about numbers below $q^2$ and is provable by fixed numeral arithmetic.
Primality of the externally fixed $q$ makes a product of two nonzero residues nonzero; its finite multiplication-table assertion has such a proof too.
Induction through the actual denominator-product scan therefore proves that its product and every fixed power have nonzero residue when every selected denominator does.
Also $\operatorname{Res}_q(qz)=0$ for every signed integer $z$.

Recall that $\mathsf{GoodDen}_q(N,n)$ is $\Net(N,n)$ together with nonzero residue of every raw stored nonbias denominator; its graph uses the residue predicate just defined.
Bias denominators are unrestricted.
This is a polynomial-time-checkable bounded predicate with its specified graph.
It tests the stored denominators without reducing the entries; equivalent encodings can receive different answers, as $q/q$ and $1/1$ illustrate.
Under $\mathsf{GoodDen}_q$, the cleared product $D$ is positive with nonzero residue.
Thus $a=D^{k+1}$ and $a^d$ have nonzero residue, and the equation $qz=a^d$ obtained from agreement at every probe is impossible.

\paragraph{\emph{The internal statement and its chosen refuter.}}
In the standard model, the depth threshold of the next theorem is the known fixed-base bound of \Cref{fact:fixed-base-depth}, applied as explained after that fact.
The theorem adds a proof inside $\Vone$ and a specified refuter, and the theory proves the refuter's guarantee with the fixed cutoff $p$.
The theorem gives no single polynomial-time procedure that works uniformly as $K$ grows.

\begin{theorem}[A fixed prime and a fixed depth]
\label{thm:fixed-prime-depth}
Externally fix a prime $q$ and $K\ge1$, and put $d=q^K$.
With the fixed evaluation and disagreement definitions, $\Vone$ proves
\[
\forall n\ge d+1\,\forall N\,
\left(\mathsf{GoodDen}_q(N,n)\wedge\Depth(N,K)
\rightarrow \neg\Computes(N,n)\right).
\]
There is a specified polynomial-time refuter $\mathcal A_{q,K}$, including its particular defining graph, whose guarantee the theory proves: on every such candidate it returns a rational $n$-vector $x$ with $|x|\le p(|N|)$ and $\Err(N,n,x)$.
\end{theorem}

\begin{proof}
Define $\mathcal A_{q,K}$ to test the complete antecedent first, returning empty on failure.
On the actual depth branch $k\in\{0,\ldots,K\}$, form the cleared weights, bias bound, final arrangement and scale as specified above.
Enumerate the indices below $S$ and, at every chamber, its unperturbed probe followed by its $d$ unit-perturbed probes.

Evaluate the original network at the padded scaled $n$-vectors and compare with their coordinate maximum.
Return the point at the least failing pair of indices in this order.
If no test fails return empty.

The loop bound is $(d+1)S$, a polynomial number term for fixed $q,K$.
All its constructors, evaluations and least-index search have the total graphs and bounds just proved.
Its definition uses the original network code in the evaluation and disagreement tests, never the cleared one.

Suppose the antecedent holds.
If every test passed, the bias and integer-gap argument would give equality at each homogeneous integer probe.
The chamber-gradient argument would identify the output family with $a$ times the lifted maximum family.
Maximum normalization makes its degree-$d$ value $a^d$, while the explicitly computed prime quotient and its integer twin make it $qz$.
The residue identities prove these unequal.
Therefore the least-error search finds a genuine failure, and the encoded bound in \Cref{sec:probes} places its output within the fixed $p(|N|)$.
This proves the refuter's guarantee and contradicts $\Computes$.
The argument handles depth zero by the global-linear zero quotient and uses the shorter-depth quotients on every $k<K$ branch.

It remains to place the argument inside $\Vone$.
All coordinate invariants, shape predicates, scans, finite sums and graph equations used in this proof are formulas of the two-sorted language of $\Vone$ with its admitted function symbols.
Rational codes are strings, and no sort for the rational field is introduced; we call such formulas field-sort-free.
The fixed coordinate identities invoked in the geometric engine have already been transferred to coded rationals before the variable scans are used.

To eliminate the new symbols, use the field-sort-free $\Sigma^B_1$ representation of failure of the fixed evaluation and disagreement tests.
Totality and uniqueness of their chosen graphs make the extension conservative over the fixed evaluation and disagreement definitions; symbol removal gives the displayed old-language lower bound.
For the algorithm assertion retain the specified graph for $\mathcal A_{q,K}$, remove the other auxiliary symbols, and obtain the same stated guarantee in its definitional extension.
No assertion is made for an arbitrary implementation without a proved comparison to this graph.
\end{proof}

The smallest case $q=2$, $K=1$ already illustrates the search.
\begin{example}[The refuter on a three-input network]
\label{ex:refuter}
Take $q=2$ and $K=1$, so $d=2$, and let $N$ be the network on $n=3$ inputs whose one hidden layer has the three neurons
\[
\relu(x_1-x_2),\qquad\relu(x_2),\qquad\relu(-x_2),
\]
with output weights $1,1,-1$, all biases zero, and every parameter stored with denominator one.
Since $\relu(x_2)-\relu(-x_2)=x_2$, it computes $N(x)=x_2+\relu(x_1-x_2)=\max(x_1,x_2)$ and ignores $x_3$.
The antecedent of \Cref{thm:fixed-prime-depth} holds: $n=3\ge d+1$, the actual depth is $k=1$, and every stored denominator is one, which is odd.
Here the nonbias denominator product $D$ of \Cref{sec:quotients} is one, so the cleared network is $N$ itself and $a=D^{k+1}=1$.
All biases are zero, so the bias majorant $E$ of \Cref{sec:probes} is zero, $\bar a=a(E+1)=1$ and $x(y)=(y_1,y_2,0)$.

The final arrangement $\mathcal H_1$ lists the braid rows $h_0=e_1$, $h_1=e_2$, $h_2=e_1-e_2$, followed by the first-layer rows $h_3=(1,-1)$, $h_4=(0,1)$ and $h_5=(0,-1)$.
Thus $s=6$, $R=1+2s=13$ and $S=R^2=169$.
For $d=2$ the cross of a single row $(u,v)$ is $(-v,u)$.
Write $r_j$ for the ray of index $j$.
We have $r_0=0$, and the indices $1,\ldots,6$ denote $(0,1)$, $(0,-1)$, $(-1,0)$, $(1,0)$, $(1,1)$ and $(-1,-1)$, the two signs of the crosses of $h_0$, $h_1$ and $h_2$.
Write $\sigma=\sigma_0+13\sigma_1$ with base-$13$ digits $\sigma_0,\sigma_1$.
The first perturbation from zero returns its ray, so $\varpi(\sigma)=r_{\sigma_0}\boxplus r_{\sigma_1}=\Delta(r_{\sigma_1})r_{\sigma_0}+r_{\sigma_1}$.

For $\sigma<13$ the second ray is $r_0=0$ and $\Delta(0)=1$, so $\varpi(\sigma)=r_{\sigma_0}$.
Every ray is orthogonal to the row whose cross it is, so none of these representatives lies in a chamber.
For $\sigma_1=1$ the ray $r_1=(0,1)$ has $|h_i\cdot r_1|\le1$ for every $i$, with equality at $i=1,\ldots,5$, so $\Delta(r_1)=3$ and $\varpi(\sigma)=3r_{\sigma_0}+(0,1)$.
The indices $13$, $14$ and $15$ give $(0,1)$, $(0,4)$ and $(0,-2)$, on which $h_0$ vanishes.
For $\sigma_1=1$ and $\sigma_0=3,4,5,6$, that is, at the indices $16$, $17$, $18$ and $19$, the representatives are $(-3,1)$, $(3,1)$, $(3,4)$ and $(-3,-2)$.
Their values on $h_0,\ldots,h_5$ are
\[
(-3,1,-4,-4,1,-1),\quad(3,1,2,2,1,-1),\quad(3,4,-1,-1,4,-4),\quad(-3,-2,-1,-1,-2,2).
\]
None of these values is zero and the four sign vectors are pairwise distinct, so these are the first four chambers, each at its canonical index.

The unit directions have $\Delta(e_1)=\Delta(e_2)=3$, since every row entry has magnitude at most one and $h_0\cdot e_1=h_1\cdot e_2=1$.
The probes of the first three chambers are therefore
\[
(-3,1),(-8,3),(-9,4);\qquad(3,1),(10,3),(9,4);\qquad(3,4),(10,12),(9,13).
\]
Each probe $y$ has a positive coordinate, so $N(x(y))=\max(y_1,y_2)=\max(y_1,y_2,0)=\MAX_3(x(y))$, and all nine tests pass.
At index $19$ the unperturbed probe $y=(-3,-2)$ gives $x(y)=(-3,-2,0)$, and
\[
N(-3,-2,0)=\relu(-1)+\relu(-2)-\relu(2)=-2\ne0=\MAX_3(-3,-2,0).
\]
The least failing pair is therefore index $19$ with its unperturbed probe, and $\mathcal A_{2,1}$ returns $x=(-3,-2,0)$, with every coordinate stored over denominator one.
The threshold refuter of \Cref{tc:refuter} returns the same point: its normalization leaves these denominator-one entries unchanged, and it tests the same probes in the same order.

The returned point is consistent with the divisibility obstruction.
On the final chambers the output family is $H_{11}+(H_{12}-H_{13})$, where $H_{12}-H_{13}$ is the global linear family with gradient $e_2$ and $H_{11}=F_{11}^+$ is the rectification of the global linear family $F_{11}$ with gradient $h_2$.
A global linear family has zero product with every cycle of positive degree, and \Cref{lem:fp-refinement-locality} with $r=2$ gives $H_{11}^2U=F_{11}(H_{11}U)=0$ for the unit cycle $U$, which is balanced of degree two.
By linearity of the product in the family and in the cycle, the degree-two output value is $0=qz$ with $z=0$, whereas the scaled maximum has value $a^2=1$.
\end{example}

For a fixed positive integer $\beta$, a fixed-base weight with base $\beta$ is a nonbias weight whose value lies in $\Z[1/\beta]$, that is, equals $a/\beta^t$ for integers $a$ and $t\ge0$.
\Cref{cor:fixed-prime-depth-values} below needs power graphs for fixed-base weights, which we specify first.
For a fixed integer $\beta\ge2$, the canonical power graph of $\beta^t$ starts with integer one and iterates multiplication by the fixed positive integer $\beta$, clipping at $2+(t+1)(|\beta|+2)$, where $|\beta|$ is the length of the canonical code of $\beta$.
Prefix induction proves the actual power equations, the displayed length bound and idle clipping.
At $\beta=1$ the graph is constantly one.

\paragraph{\emph{Removing a representation restriction.}}
The raw theorem also yields a value-based promise by an explicit cancellation procedure.
This bridge is needed if a fixed-base weight is supplied in an unreduced encoding whose stored denominator contains $q$.
For one valid entry $(u,v)$, $v>0$, repeatedly do the following for at most $m$ stages, where $m=|N|$.
If $\operatorname{Res}_q(v)\ne0$, keep the pair unchanged.
If both residues are zero, replace both integers by their exact quotients $\widehat Q_q(u),\widehat Q_q(v)$ from the residue graph.
If the denominator residue is zero and the numerator residue is nonzero, set a failure flag and thereafter keep that failed state.

The exact state tuple is $(u,v,\operatorname{Un}(f))$: its first two entries are canonical signed integers, with $v>0$ and component caps $m+3$, and its third entry is the unary one-bit flag $f\in\{0,1\}$.
Thus the zero flag is empty and the one flag has its bit at position zero; it is distinct from the canonical signed integer one.
A malformed state returns $(0,1,\operatorname{Un}(1))$, abbreviated $(0,1,1)$ below.
Use exact tuple bound $\langle4,2(m+3)+2\rangle$ and limited recursion with this cap.

In an exact-division branch the positive denominator quotient is at least one and its signed length drops by at least one, since $q\ge2$.
The numerator's signed length does not increase, for either sign.
Thus these caps hold and clipping is idle on a valid starting pair.

Every nonfailed update preserves the rational value: a division stage replaces $(u,v)$ by $(u',v')=(\widehat Q_q(u),\widehat Q_q(v))$, and $u=qu'$ and $v=qv'$ imply $uv'=u'v$ with both denominators positive.
After $m$ stages every nonfailed pair has a denominator of nonzero residue.
Suppose instead that the denominator of some nonfailed pair has zero residue after $m$ stages.
Then all preceding stages were divisions, and the initial positive denominator of that pair, of length at most $m$, would have undergone $m$ strict length drops.
A positive denominator has length at least two, contradicting that bound.

Apply these independent entry scans to all nonbias entries, keeping the metadata and every bias code unchanged.
Call the resulting code $N^\sharp$, and let $\mathsf{ValGood}_q(N,n)$ mean complete network validity and absence of every entry failure flag.
This is a polynomial-time-checkable predicate with the specified bounded graph.
On it $N^\sharp$ is valid, has the same widths, depth and input dimension, and satisfies $\mathsf{GoodDen}_q(N^\sharp,n)$.
Its actual length satisfies $|N^\sharp|\le|N|$, by the length principle for normalizations in \Cref{sec:foundations}: the metadata and every bias code are unchanged, no parameter row is added, and, by the caps just proved, no integer component length increases.

Induction through the actual layer evaluation, with its bounded coordinate equations, shows that $N^\sharp(x)=N(x)$ for every rational input $x$, by the equal-weight cross-product identities at each step.
This equality is one of values, not identity of unreduced evaluation strings.

\begin{corollary}[Value-based denominator promise]
\label{cor:fixed-prime-depth-values}
The theorem remains true with $\mathsf{ValGood}_q$ in place of $\mathsf{GoodDen}_q$.
Its specified refuter first forms $N^\sharp$, then applies $\mathcal A_{q,K}$ to that code and the original $n$.
In particular it applies to fixed-base weights with base $\beta$ coprime to $q$, irrespective of their unreduced stored representation.
\end{corollary}

\begin{proof}
The returned point has length at most $p(|N^\sharp|)\le p(|N|)$, by the proved length comparison and the nonnegative coefficients of the fixed $p$.
The equality of evaluations makes its disagreement a disagreement for $N$ itself.
All normalization graphs have just been specified with their totality bounds and invariants; retaining their chosen composition graph gives the provable refuter guarantee, and the same field-sort-free symbol removal gives the old-language lower bound.

For the fixed-base assertion, fix $\beta\ge1$ with $q\nmid\beta$.
For a stored nonbias fraction $u/v$, suppose $a,t$ are supplied integer/power representation witnesses satisfying
\[
u\beta^t=va,\qquad v>0,
\]
using the canonical power graph described above.
The residue-product identity and induction through this power graph give nonzero residue of $\beta^t$.
At a normalization stage where $q$ divides the denominator, reducing the displayed equality modulo $q$ forces residue zero for the numerator, so the failure branch is impossible.

After dividing numerator and denominator by $q$, cancellation in the integer equation preserves the same representation witness $t,a$.
Induction through the at-most-$m$ stages therefore proves success.
The witness strings and $t$ are parameters; the invariant uses their actual operation graphs and bounded coordinate equations, and no bound on their lengths by $m$ is needed.
This proves the implication for every such representation witness, hence for the assertion that each stored weight denotes some $a/\beta^t$.
For $\beta=1$ the power graph is constantly one and the same proof applies.
Ordinary fixed-base weights have precisely these witnesses, so the standard-model consequence has the full value-based interpretation as well.
\end{proof}
\space

\section{Parallel Arithmetic and the Threshold Refuter}
\label{tc:parallel-arithmetic}

The refuter of \Cref{thm:fixed-prime-depth} was specified through sequential scans whose correctness is proved in $\Vone$.
This section replaces every such scan by an explicit parallel computation and verifies the same equations in the weaker theory $\mathbf{VTC}^0$ of \Cref{sec:foundations}, so that the refuter becomes a threshold-computable function whose guarantee $\mathbf{VTC}^0$ proves (\Cref{tc:refuter}).
The conservative function language of $\mathbf{VTC}^0$ supplies the bounded parallel operations needed by the refuter.
When this section speaks of the original or old construction, or of original caps, guards and equations, it means those of the sequential specification in \Cref{sec:foundations,ssec:fixed-prime-depth,sec:locality,sec:quotients,sec:probes,sec:graphs}.
Throughout, the original language of a theory means its base two-sorted language, without chosen function symbols.

Fix the prime $q$, the hidden-depth cap $K$, and $d=q^K$ externally.
Throughout this section $\mathbf{T}$ is the conservative universal extension of $\mathbf{VTC}^0$ by its chosen $\mathsf{FTC}^0$ functions.
We state the imported interfaces before using them.
\begin{fact}[Threshold arithmetic, closure, and witnessing]
\label{fact:tc-interface}
The theory $\mathbf{VTC}^0$ has a base-language axiomatization by bounded formulas \cite[Definition~9.29 and the following paragraph]{CookNguyen}.
It has a universal conservative extension by chosen $\mathsf{FTC}^0$ functions.
Every formula of this extension with bounded number quantifiers and no quantified strings has a quantifier-free equivalent \cite[Lemma~8.19]{CookNguyen}.
The extension proves comprehension for these formulas \cite[Theorem~9.34(c)]{CookNguyen}, and hence number induction and minimization for them \cite[Corollary~5.8]{CookNguyen}.

Let $F$ be a chosen function, and let $t$ be a nondecreasing polynomial term with $|F(Y)|\le t(|Y|)$.
The aggregate $F^\star(b,X)$ is the exact array whose row $i$ is $F(X^{[i]})$ for every $i<b$, with no other bits.
Since $|X^{[i]}|\le|X|$, its bit at position $z$ is given by the bounded formula $\exists i<b\,\exists j<t(|X|)\,\left(z=\langle i,j\rangle\wedge F(X^{[i]})(j)\right)$.
Comprehension for this formula supplies the array, and (Arr) proves its row equations.
Adjoining $F^\star$ with this bit definition is then a conservative extension \cite[Corollary~5.39]{CookNguyen}\cite[Theorem~2.1]{Jerabek22}.

In this extension, a proved assertion $\forall u\,\forall X\,\exists Y\,\varphi(u,X,Y)$ with such a matrix has an $\mathsf{FTC}^0$ witnessing function whose guarantee the extension proves; explicit output bounds can be retained in the matrix \cite[Lemma~8.19 and Theorem~8.20]{CookNguyen}.

Iterated multiplication of nonnegative binary integers supplies all interval products with their recurrence and empty product one.
Division of a nonnegative binary integer $A$ by a positive binary integer $B$, written DIV, supplies $Q,R$ with $A=BQ+R$ and $0\le R<B$ \cite[Theorem~8.6 and Corollary~8.8]{Jerabek22}.
\end{fact}
Together with \Cref{fact:binary-arithmetic}, these facts justify the equations specified below; no equation is postulated merely because the function satisfying it is computable in $\mathsf{FTC}^0$.
The signed-code convention is the bounded recoding in \Cref{sec:foundations}.

All induction in this section is bounded-number induction for formulas without quantified strings in this extension, with the input arrays as parameters.
In particular, the generic $\FP$ limited-recursion lemma is not imported into $\mathbf{T}$.

\paragraph{\emph{Exact arrays, reads and parallel selection.}}
Use exactly the pairing, row convention and reconstruction of (Arr).
Integer and rational reads apply the original canonicality and positive-denominator guards and return integer zero or $(0,1)$ respectively on failure.
Every scalar cap mentioned below is measured on these guarded inputs.
An exact array with $v$ rows of bit bound $B$ is formed by bounded comprehension below $\langle v,B\rangle$, putting no bits at non-pair positions or in unused rows.
Row recovery and uniqueness follow from injectivity of the pairing.
Thus no promise about unnamed bits is implicit.

If $P(i)$ is a bounded predicate using chosen functions, define
\[
j=\min(P;v)\quad\leftrightarrow\quad
\left(j<v\wedge P(j)\wedge\forall h<j\,\neg P(h)\right)
\vee \left(j=v\wedge\forall h<v\,\neg P(h)\right).
\]
The number $j$ is unique and bounded by $v$.
Its bits, or its unary code, are given by a bounded formula in the function language, hence belong to its $\mathsf{FTC}^0$ closure.
The same construction applied to the reversed order gives a maximum.
For a numerical minimum of supplied rationals, compare every entry with every other entry by cross-products, then choose the first minimizing index.
A different fixed tie convention is handled by ordering the indices accordingly.
This selects an existing code; it does not manufacture a reduced rational.
Every no-candidate branch returns its specified sentinel or zero before any guarded division.

\begin{lemma}[Parallel arithmetic histories]
\label{tc:arithmetic}
The integer prefix sums, integer interval and excluded-entry products, and rational prefix sums used in the finite cone calculus have chosen total $\mathsf{FTC}^0$ graphs.
On their stated valid domains $\mathbf{T}$ proves the precise prefix equations, canonical integer and tuple-code identities, the original component caps, and idle clipping.
Their all-string bounds use actual argument lengths and list lengths; no unrestricted recursion is needed.
\end{lemma}
\begin{proof}
For signed integer sums separate nonnegative and negative magnitudes, use two unsigned iterated sums, and subtract in the ordered ring with its canonical signed output.
At each prefix $t$ the resulting integer is the signed sum of the first $t$ entries.
The iterated-addition recurrence and the sign cases give the signed addition recurrence.
Canonical signed representations of one value are unique, so these are string equations.
The empty prefix is the unique zero code.
If input lengths are at most $L$, induction on $t$ gives length at most $L+2t$.

Write IMUL for the chosen iterated multiplication function of \Cref{fact:tc-interface}, including its interval-product outputs.
For products use IMUL on magnitudes.
If one magnitude is zero the result is the unique nonnegative zero code.
Otherwise its sign is the parity of the number of negative factors, obtained by counting and division by two.
The empty product is integer one.
The unsigned product recurrence and the two sign cases prove the signed recurrence, again as canonical string equality.

Aggregate closure forms all prefix and interval products simultaneously.
Replacing excluded or nonentry positions by one gives every excluded-entry product without dividing by a possibly zero factor.
For a list of integer factors $a_0,a_1,\ldots$ and an excluded position $e$, write $\Pi_t$ for the product of the first $t$ factors, at positions $0,\ldots,t-1$, and $\Pi^{(e)}_t$ for the same product with the factor at position $e$ replaced by one.

Bounded number induction on $t$ proves $\Pi_t=\Pi^{(e)}_tc_t$, where $c_t=a_e$ when $e<t$ and $c_t=1$ otherwise; this formula has no quantified strings.
The base is $1=1\cdot1$.
At a step with $t\ne e$ both prefixes gain the same factor and $c_{t+1}=c_t$; at $t=e$ the full prefix gains $a_e$, the excluded prefix gains one and $c_{t+1}=a_e$.
Commutativity and associativity close both cases, and at the last position this gives the full product as $a_e$ times the excluded-entry product, the identity behind the cleared entries $\widehat u_e$.

For participating factors $a_i$, the product starts from integer one, whose code has length two, and each factor $a_i$ adds at most $|a_i|+1$.
Hence the product has length at most $2+\sum_i(|a_i|+1)$, including the empty product.
These are the same bounds obtained by prefix multiplication in the original argument; bounded induction on the explicit prefixes proves them in $\mathbf{T}$.

For rational entries $(a_i,b_i)$, $i<v$, with positive $b_i$, set for $t\le v$
\[
B_t=\prod_{i<t}b_i,\qquad
A_t=\sum_{i<t}a_i\prod_{\substack{j<t\\j\ne i}}b_j,
\qquad R_t=(A_t,B_t).
\]
Every product and sum on the right has just been defined in parallel.
The empty pair is $(0,1)$.
For $t>0$, a zero numerator does \emph{not} change the denominator to one.
Splitting off the last factor in each product, and separating the last summand, gives
\[
B_{t+1}=B_tb_t,\qquad A_{t+1}=A_tb_t+a_tB_t.
\]
These are exactly the unreduced rational-addition equations, not only equalities of rational values.
Canonical integer uniqueness and exact tuple reconstruction prove equality of the pairs with the corresponding cross-product prefix history.
When all components have length at most $L$, the product $B_t$ and each cleared numerator product have length at most $t(L+1)+2$; signed addition adds at most $2t$.
The extra two accounts for the empty product $B_0=1$.
For $t\le v$, the resulting $t(L+3)+2$ bound is at most $\operatorname{Tot}(v,L)$ of \Cref{sec:graphs}, by the calculation in \Cref{sec:foundations}.
The same proof supplies every prefix bound, so clipping at the original cap is idle on these valid histories.

Besides the prefix convention $R_t$ just defined, a second convention fixes the full denominator $B_v$ first and then sums cleared numerators: its stage-$t$ integer numerator is $\sum_{i<t}a_i\prod_{j<v,j\ne i}b_j$, and its denominator is $B_v$ at every stage.
The preceding arithmetic proves its exact integer-addition recurrence and the same cap.
Each graph is defined with one of the two conventions, and the two are not asserted to have equal intermediate strings.

At every stage the rational values of the two conventions are equal by cross-product multiplication.
At the final stage $t=v$ they give the identical pair $(A_v,B_v)$, since both numerators are then $\sum_{i<v}a_i\prod_{j<v,j\ne i}b_j$ and both denominators are $B_v$.
For the bounded sums of \Cref{sec:foundations} we choose this full-denominator convention; $R_t$ above supplies the separate cross-product-prefix history when it is requested.

For arbitrary strings apply the guarded reads first.
Use $L$ equal to the sum of actual string argument lengths plus two, rather than an unproved caller cap.
The formulas above and the bounds for at most $v$ factors give total outputs within the displayed polynomial caps.
A signature requiring a globally valid exact list can instead test that shape first and return its stipulated zero on failure.
All branches are now bounded before invoking a valid-state invariant.
On arbitrary malformed states the result is not asserted to agree with the clipped sequential recursion of \Cref{sec:graphs}.
\end{proof}

The remaining basis scans depend on previously accepted vectors.
The fixed rank bound of these scans allows us to compute every accepted index in a fixed number of stages.
\begin{lemma}[The exact fixed-rank greedy history]
\label{tc:greedy}
For the zero-normal, completion and quotient-extension lists, the specified greedy basis history has a total $\mathsf{FTC}^0$ replacement.
The theory $\mathbf{T}$ proves that, for every $t\le v$, the stage-$t$ states satisfy, bit for bit, the initial equation and the update equation of the sequential greedy scan of \Cref{sec:graphs}, including the count row and all unused bits.
The theory $\mathbf{T}$ also proves the spanning invariant of that scan for these states.
\end{lemma}
\begin{proof}
All ambient vector dimensions are bounded by the fixed $d$, and each rank test is the finite disjunction of the fixed-order nonzero minors.
Let $A(i)$ be the list's admission predicate after its original guards; it depends on the supplied list and covector, not on an evolving code.

Start with an empty retained list.
Having chosen $j_1<\cdots<j_r<v$, choose $j_{r+1}$ to be the least $i$ with $j_r<i<v$ such that $A(i)$ holds and appending row $i$ to the retained rows increases their rank.
For $r=0$ this is the first admitted nonzero row.
If there is no such index, choose $v$ and keep all later indices equal to $v$.

Make these choices for the externally fixed $r=0,\ldots,d-1$.
Each is the parallel selection already defined and a fixed-size determinant calculation.
Consequently their composition has fixed depth.

For each $t\le v$ let $r_t$ count the chosen indices less than $t$.
Put $1^{r_t}$ in row zero and $1^{j_a+1}$ in row $a$, for $1\le a\le r_t$, and put zero bits everywhere else.
Each index is less than $v$, so this exact state has the original bound $\langle d+1,v+d+2\rangle$.
It is correctly shaped, independent and increasing.
Between two consecutive chosen indices every admitted row fails rank increase, by the definition of the minimum; at the next chosen index it succeeds.
When $d$ rows are retained, the old $r<d$ condition forbids further appends.

These observations prove, by bounded number induction on $t$, the exact old update equation and the initial equation.
They also prove the spanning invariant using the fixed-dimensional Cramer calculation already supplied in the cone proof.
A malformed input vector uses the same guarded read; if the signature rejects a malformed whole list, it returns its stipulated failure object before these choices.
The transition's reset on an externally supplied malformed state is retained, although that state is never reached by this valid history.
The list bounds $s$, $s+d$, and $2d$ are respectively the bounds already declared for the three uses.
\end{proof}

\paragraph{\emph{The remaining geometric functions.}}
We first define total functions for the geometric arrays, including their failure branches, and then verify the geometric invariants on valid inputs.
Begin with the scalar expressions in the fixed registry: determinants, adjugates, Cramer coordinates, dot products, wedges, contractions, fixed powers and guarded inverses.
Each is a finite expression of depth bounded by the fixed registry count $t_d$.
Structural induction on these expressions proves the cap $\mathcal F(L)$ using $|a+b|\le\max(|a|,|b|)+2$ and $|ab|\le|a|+|b|+1$.
Every inversion is preceded by its guard, and a missing pivot returns the prescribed zero.

Next use \Cref{tc:greedy} to form bases and kernel vectors.
Form every cross ray independently from its fixed Leibniz determinant, adding its monomials in the balanced binary tree of \Cref{sec:probes}, so that the original all-string cap $B_r$ applies.
Its coordinates are integers, and canonical signed integer codes are unique, so their final codes do not depend on the order of summation.

Recall from \Cref{ssec:fixed-prime-depth} the tie-breaking factor $\Delta(w)=2\max\left(\{|h_i\cdot w|:i<s\}\cup\{0\}\right)+1$, where $h_0,\ldots,h_{s-1}$ are the $s$ normals of the arrangement, and the perturbation $u\boxplus w=\Delta(w)u+w$.
For $\Delta(w)$ compare all $s$ absolute dot products and retain their maximum; then perform the specified $d$ perturbations.
Thus all $S=(1+2s^{d-1})^d$ representatives are formed in fixed composition depth.
Every covector is its exact $s$-row sign list.

Compare a requested covector with the sign lists of all $S$ representatives and take the least equal index, or $S$ if none.
The sentinel index $S$ reads zero before subsequent tests.
Chamber completion is another fixed list of $d$ perturbations and one such lookup.

In the completeness and sector constructions, first select a genuine positive ratio candidate or the first independent pair.
When none exists return exactly the stipulated residual/endpoints and sentinel.
Otherwise fill every failed ratio test with the selected genuine candidate before taking the minimum.
No zero default enters a nonempty positive minimum.
The ratios, endpoints, quotient coordinates and their lifts are the original fixed scalar expressions.
Orientation ratios use the first nonzero denominator coordinate and the same failure branch.
Restriction and refinement are the preceding lookups on the smaller arrangement.
This defines all graphs before geometric completeness is used.

The original caps $B_N,B_O,B_r,B_q$, the sector caps $V_{\mathrm{sec}},R_{\mathrm{sec}},E_{\mathrm{sec}}$, and $\Lambda,\Lambda_2$ follow from exactly those scalar expressions and selection of actual entries.
No cap changes merely because independent entries are computed in parallel.
Define a family/cycle weighted sum by \Cref{tc:arithmetic}; define a cone-product entry as its original cover-bracketed scalar summand summed over all $S$ indices.
Define the twin by its signed integer contraction sum.
Set all off-degree rows to their exact zero arrays and reject malformed shape or insufficient degree before a fixed word.
This gives precisely the caps $\mathsf R_A,\mathsf I_A,B_V,Z_V$ and the double-sum cap of \Cref{sec:graphs}.
A double sum has $S^2$ entries.
Every word has length at most the fixed $d$.

For network-specific graphs test complete validity, $n\ge d+1$, depth at most $K$, and the relevant denominator promise before constructing their arrays.
On failure return the stipulated failure string.
Compute the product $D$, each omitted product $P_e$ and $\widehat u_e=u_eP_e$ by the integer graphs.
Compute each layer's gradients and its neuron prefixes by coordinatewise sums; apply the activation sign at each representative; append the original gradient or replacement $e_1$ under the original condition.
There are only $K$ arrangement constructions.
Hence $s_{\ell+1}\le s_\ell+m(1+2s_\ell^{d-1})^d$ expands to a polynomial term for fixed $d,K$.
The actual final arrangement length supplies $\bar s$ in the geometric caps.
Bias majorants have $K+1$ such integer sum layers and the existing $B^E_j$ caps.

For the quotient constructors, form every prefix family $P_{j-1}$ of \Cref{lem:fp-prime-quotients}, a weighted sum of activation families rather than an omitted denominator product, independently by \Cref{tc:arithmetic}.
For each summand of the displayed direct $Q_t$ formula, form its fixed mixed word, coefficient and previous-layer quotient.
Then sum over $j\le t$ and $1\le r<e$, where $e=q^{\ell-1}$ is the exponent of that formula, not an entry index.
This has at most $me$ summands with $e\le d$, not a recursion through a previous quotient array.
The rectification quotient is the stated fixed word applied to this result.

Make only $K$ layer compositions and at most $d$ products per word.
These operations give the original $Q^R,Q^I,Z^R,Z^I$ cap formulas.
The shorter-depth extension is the same fixed additional word.
Standalone calls use raw input-length caps and their own materialized family lists; network calls satisfy the tighter caller caps by the just-proved arithmetic inequalities.

All shape, balance, twin, admissibility, coordinate-identity and prefix predicates in this construction use bounded number quantifiers only in $\mathbf{T}$'s language.
Their supplied cycle and twin strings remain parameters.
For example, balance is the bounded list of zero-normal dot-product equations for $V_C$, and the twin equation is equality of the bounded orientation coordinates.

Fixed determinant/ordered-field deductions transfer through the rational-field interpretation, whose ring and order proof is already valid in $\mathbf{T}$.
Bounded sums can be exchanged using the explicit prefix equations and bounded number induction.

Bounded sums can also be reindexed along an injection, as in the cover sums of refinement and the regrouping in commutation.
Let $P$ be a bounded predicate, and let $\rho$ be a chosen function that is injective, with values below $S$, on the indices $\nu<v$ satisfying $P$.
Let $a_\sigma$ vanish whenever no such $\nu$ has $\rho(\nu)=\sigma$.
Bounded number induction for $0\le t\le v$ proves
\[
\sum_{\substack{\nu<t\\P(\nu)}}a_{\rho(\nu)}
=\sum_{\sigma<S}\left[\exists\nu<t\,\left(P(\nu)\wedge\rho(\nu)=\sigma\right)\right]a_\sigma.
\]
The restricted sum reads $a_{\rho(\nu)}$ only when $P(\nu)$ holds, so its index is below $S$; the displayed quantifier in the bracket ranges over bounded numbers.
At a step with $P(t)$, injectivity makes $\rho(t)$ the only newly admitted index on the right, and a second bounded induction on the prefix of the right sum shows that admitting one index adds exactly its entry.
At $t=v$ the vanishing hypothesis turns the right side into the full sum.

Consequently, after this replacement, the following proofs use only deductions available in $\mathbf{T}$: flat completeness (\Cref{lem:fp-flats-complete}), the perturbed and intermediate cones of \Cref{lem:fp-stars-diamonds}, product balance, commutation and integer twins (\Cref{lem:fp-product-engine}), refinement and rectification (\Cref{lem:fp-refinement-locality}), maximum normalization (\Cref{lem:fp-max-normalization}), network-family admissibility, the direct prime-quotient identities (\Cref{lem:fp-prime-quotients}), and the bias-domination and chamber-gradient arguments of \Cref{sec:probes}.
Every induction there over a layer, a degree, an exponent or a dimension ranges below the fixed $d$ or $K$ and is a finite chain of separate deductions, not an internal induction.
This verifies the arithmetic requirements of those deductions.

The fixed braid calculation is a closed calculation of these chosen functions: expand its fixed number bounds, prove each closed function value by its defining equations, and combine the resulting finite proofs.
This is the same finite proof as in \Cref{lem:sigma1-complete}, now using only the chosen threshold-computable graphs.

\begin{lemma}[Residues and the complete cancellation history]
\label{tc:cancel}
The fixed-$q$ residue and normalization graphs have total $\mathsf{FTC}^0$ replacements, with exact canonical histories on valid inputs, the original caps, rational value preservation on successful branches, and the original whole-network length comparison.
\end{lemma}
\begin{proof}
For a valid signed input $u$ let $A_t$ be the number formed by its $t$ most significant magnitude bits, starting with $A_0=0$.
Bit slicing gives all $A_t$ in parallel.
Let $Q_t$ and $r_t$ be the unique DIV output with $A_t=qQ_t+r_t$ and $0\le r_t<q$.
Writing $A_{t+1}=2A_t+b$, where $b$ is the next magnitude bit, proves that its next remainder and quotient are exactly the fixed carry-table update, by uniqueness of the decomposition.
Encode $r_t$ one-hot as in \Cref{sec:graphs}.
Since $Q_t\le A_t$ and $r_t<q$, the old component and tuple caps hold.
This gives the old unsigned history as actual canonical strings.

For negative $u$, let $Q$ and $r$ be the final quotient and remainder of its magnitude, and apply the original final cases.
If $r=0$, negate $Q$.
Otherwise replace $Q$ by $-Q-1$ and $r$ by $q-r$.
This proves the signed decomposition and its cap.
Malformed input returns the same fixed zero residue state.
Uniqueness and ring distribution now give the original residue sum and product identities, including the finite nonzero-product table for fixed prime $q$.

For normalization let $m=|N|$.
Reject an invalid entry, a nonpositive denominator, or a component longer than $m$ with the stipulated failed state.
Otherwise write the entry as $(u,v)$, with $v>0$ and both component lengths at most $m$.
Compute $q^t$ for every $t\le m$ by IMUL on $t$ copies of $q$.
Each has length at most $2+t(|q|+1)$ and has the proved prefix-power equation.

Let $\nu_v$ be the greatest $t\le m$ whose DIV remainder on dividing $v$ by $q^t$ is zero.
For nonzero $u$ let $\nu_u$ be the analogous greatest exponent for the magnitude of $u$; for $u=0$ put $\nu_u=m$.
Positivity and $q^m\ge2^m>v$ imply $\nu_v<m$.
If $t<\nu_v$, divisibility of $v$ by $q^{t+1}$ gives an exact next quotient; at $t=\nu_v$ the quotient of $v$ by $q^{\nu_v}$ has nonzero residue modulo $q$, since another exact division would contradict maximality.
The same reasoning applies to nonzero $u$.

For every stage $i\le m$ put
\[
a_i=\min(i,\nu_u,\nu_v),\qquad
u_i=u/q^{a_i},\quad v_i=v/q^{a_i},\qquad
f_i=[\nu_u<\nu_v \wedge i>\nu_u].
\]
The reconstructed state is exactly $(u_i,v_i,\operatorname{Un}(f_i))$, using the normalization flag convention above.
Here each displayed quotient is the exact canonical signed quotient; for $u=0$ it is the unique zero code.

If $i<\min(\nu_u,\nu_v)$ the next stage divides both components by $q$.
If $\nu_v\le\nu_u$ and $i\ge\nu_v$, the denominator residue is nonzero and the state is unchanged with flag zero.
If $\nu_u<\nu_v$, stage $\nu_u+1$ first sets the failure flag, keeping the pair reached after $\nu_u$ divisions, and every later state is unchanged.
These exhaust the cases and prove the exact sequential update.
For zero numerator the first case continues until $\nu_v$, then the second applies; there is no failure.
Initial malformed entry input returns $(0,1,1)$; malformed whole networks return the earlier stipulated failure string.
The formulas therefore define total graphs independently of an old $\FP$ recursion.

Every exact quotient magnitude is no larger than the magnitude it divides; $v_i$ is positive.
Thus the original $m+3$ component caps and $\langle4,2(m+3)+2\rangle$ tuple cap bound every state, and clipping is idle on valid entry starts.

A successful division preserves the value by $u=qu'$, $v=qv'$ and cross multiplication.
At the final nonfailed stage $\nu_v<m$ proves that the denominator has nonzero residue.

Apply this construction independently to all nonbias entries and leave metadata and bias bits unchanged.
No component length increases, so the length principle for normalizations in \Cref{sec:foundations} proves $|N^\sharp|\le|N|$; its bit-position argument uses only monotonicity of the pairing and is valid in $\mathbf{T}$.
The rational numerator zero is canonicalized as an integer only; its positive denominator is retained and is not forced to one.
\end{proof}

\paragraph{\emph{A guarded evaluator and its actual equations.}}
Define $\mathsf{Eval}_K(N,x)$ to return the empty string unless both codes are valid and the network depth $k$ is at most $K$.
Put $m=|N|$ after the validity guard.
Let $D$ be the product of \emph{all} parameter denominators, including biases, and $E$ the product of input denominators.
This $D$ is used only for the guarded evaluator and its bounds; unlike the nonbias denominator product $D$ of the network-specific graphs above, it includes the bias denominators.
Likewise this $E$ is the input-denominator product of the general-depth evaluator of \Cref{sec:foundations}, not the bias majorant $E$ of \Cref{sec:probes}.

Index the parameter entries by $\lambda$, write their stored fractions as $u_\lambda/b_\lambda$, and set $U_\lambda=u_\lambda\prod_{\mu\ne\lambda}b_\mu$.
From here to the definition of $\mathsf{Err}_K$, neuron indices are one-based: $U_{jih}$ is the cleared integer for record weight $W_j[i-1,h-1]$, and $U_{ji,\mathrm{bias}}$ is that for bias $\eta_j[i-1]$.
For $1\le i\le n$, write input $x_i=p_i/\epsilon_i$ and put $A_{0i}=p_i\prod_{1\le h\le n, h\ne i}\epsilon_h$.
Set $B_j=D^jE$ for $0\le j\le K+1$.

For each actual affine layer use the exact canonical integer equation
\[
\bar A_{ji}=\sum_{h=1}^{d_{j-1}} U_{jih}A_{j-1,h}+U_{ji,\mathrm{bias}}D^{j-1}E.
\]
At a hidden layer put $A_{ji}=\max(0,\bar A_{ji})$; at the scalar output put $A_{k+1,1}=\bar A_{k+1,1}$.
Return exactly $(A_{k+1,1},B_{k+1})$.

The evaluator has $K+1$ fixed stages, with each width sum given by \Cref{tc:arithmetic}.
Here $D,E>0$, so every $B_j$ is positive, and the excluded-entry identity of \Cref{tc:arithmetic} gives $U_\lambda=D\cdot(u_\lambda/b_\lambda)$ and $A_{0i}=E\cdot(p_i/\epsilon_i)$ as rational values.

For an actual affine layer $j$, dividing the displayed equation for $\bar A_{ji}$ by $B_j=D\cdot D^{j-1}E$ therefore gives $\bar A_{ji}/B_j=\sum_{h}W_j[i-1,h-1]\cdot(A_{j-1,h}/B_{j-1})+\eta_j[i-1]$, the affine row of that layer at the preceding values.
At a hidden layer, positivity of $B_j$ gives $\max(0,\bar A_{ji})/B_j=\max(0,\bar A_{ji}/B_j)$, in either branch at zero; at the output no rectification is applied.
Thus the rational values $A_{ji}/B_j$, in layer order, satisfy every adjacent affine/ReLU equation and the final affine equation of $\mathsf{LEval}$; this is a finite chain of $k+1\le K+1$ deductions, not an internal induction.

We now give explicit all-string bounds after the validity guard.
Put $z=|x|$ and define the following lengths, where $\kappa_D$ and $\kappa_E$ bound the component lengths of the products $D$ and $E$, and $L_j$ bounds the component lengths of the integers $A_{ji}$ and $\bar A_{ji}$ at layer $j$:
\[
\kappa_D=2+m(m+3),\quad \kappa_E=2+z(z+3),\quad
L_0=m+z+\kappa_D+\kappa_E+10,
\]
\[
\bar L=(K+3)(\kappa_D+\kappa_E+m+z+10)+2(m+1),\qquad L_j=L_0+j\bar L.
\]

The products $D,E$ have component lengths at most $\kappa_D,\kappa_E$.
The cleared parameters have component length at most $m+\kappa_D+1$, the initial $A_{0i}$ at most $z+\kappa_E+1$, and $B_j$ at most $j(\kappa_D+1)+\kappa_E+1$.
A weighted previous numerator adds at most $m+\kappa_D+2$ to its length.
Summing at most $m$ terms adds at most $2m$.
The bias term has length at most $m+\kappa_D+1+(j-1)(\kappa_D+1)+\kappa_E+2$.
The displayed $\bar L$ dominates these increases for every $j\le K+1$, so $L_j$ bounds $A_{ji},\bar A_{ji}$ by finite induction.

Exact (Arr) reconstruction bounds the at most $K+2$ history rows, each with at most $m+1$ entries.
Invalid inputs return before allocation.
Thus this is a total chosen $\mathsf{FTC}^0$ graph with a proved polynomial bound.

If one retains the bounded existential computation graph of a general-depth evaluator, construct its depth-$K$ history by these very equations, with the required unused rows empty or final-state padding as specified by that graph.
For a graph using the $D^jE$ convention this gives canonical equality of every named coordinate by uniqueness of its integer recurrence.
A different unreduced representation requires an explicit comparison; cross-product value equality suffices for disagreement.
We do not adjoin any arbitrary general-depth $\FP$ evaluator to $\mathbf{T}$.

Let $\mathsf{Err}_K(N,n,x)$ be the error predicate.
It states that $x$ is a valid rational $n$-vector, and that the output of this guarded evaluator at $N$ and $x$ is nonempty and has a rational value that differs from $\MAX_n(x)$ under cross-product comparison.
It is $\Err(N,n,x)$ of \Cref{sec:foundations} with $\mathsf{Eval}_K$ as the evaluator, a failed guard counting as no error.

\begin{samepage}
\begin{theorem}[A threshold-computable refuter; restatement of \Cref{ithm:refuter}]
\label{tc:refuter}
For each fixed $q,K$ there is a total chosen $\mathsf{FTC}^0$ function $\mathcal A^{\mathsf{TC}}_{q,K}$ whose guarantee $\mathbf{VTC}^0$ proves in its conservative definitional extension.
Here $\mathsf{ValGood}_q(N,n)$ is evaluated by the normalization graphs of \Cref{tc:cancel}.
On valid networks with $n\ge q^K+1$, depth at most $K$, and $\mathsf{ValGood}_q(N,n)$, the output of $\mathcal A^{\mathsf{TC}}_{q,K}$ is a rational $n$-vector of length at most the same common polynomial $p(|N|)$.
This output satisfies $\mathsf{Err}_K$.
The polynomial bounding the auxiliary arrays of $\mathcal A^{\mathsf{TC}}_{q,K}$ may depend on $q,K$.
\end{theorem}
\end{samepage}
\begin{proof}
The refuter evaluates a fixed list of chamber probes in parallel and returns the first probe at which the network differs from the maximum; the work is to show that some probe fails.
First test complete network validity, dimension and depth.
On failure return empty.
Normalize all nonbias entries by \Cref{tc:cancel}; on a failure flag return empty.
All these tests are bounded predicates of chosen functions.

On success form the cleared weights, bias majorant, scale and final arrangement by the graph constructions above.
Form all chamber probes, in the original chamber-index and unperturbed-then-unit-perturbed order.
Evaluate the original input network with $\mathsf{Eval}_K$ at every scaled padded probe in parallel and compare with its coordinate maximum.
Select the least failing index by the displayed parallel-selection formula.
If there is none return empty.

Every loop length is bounded by the same polynomial number terms as the original construction, and the composition depth is fixed by $d,K$, so this is a chosen total $\mathsf{FTC}^0$ graph.
Every malformed, failed-degree and missing-pivot branch was specified before its invariant.

The preceding constructions make the finite cone argument available in $\mathbf{T}$.
They supply the rational ordered-field identities, fixed-dimensional algebra, bounded prefix identities and coordinate induction.

Suppose that all probes agree with the maximum.
Bias domination and the integer gap identify each homogeneous integer probe value with the scaled maximum.
The $d+1$ probes in each chamber then identify its gradient.

The product, refinement, locality and maximum-normalization identities give degree-$d$ value $a^d$, where $a=D^{k+1}$ and $D$ is the nonbias denominator product of \Cref{sec:quotients}, not the evaluator's product of all parameter denominators.
The direct prime-quotient identity and its integer twin express the same value as $qz$, where $z$ is the computed integer twin of \Cref{lem:fp-prime-quotients}.
The residue equations make $a^d$ nonzero modulo $q$ and $qz$ zero modulo $q$, a contradiction.

Thus at least one parallel test fails.
This is a disagreement for the original candidate, because the tests used its guarded evaluation and normalization preserves rational values.
The exact original probe-length calculation uses $d,K\le m$ only on successful candidates; it proves $|x|<2^{100}(m+2)^{40}\le p(m)$.
On the normalization branch $|N^\sharp|\le|N|$ and the nonnegative coefficients of $p$ preserve the same cutoff.
None of the larger auxiliary-array caps enters this coordinate calculation.

The deductions use bounded number induction with strings as parameters and the explicit chosen graphs, so they are deductions in the conservative extension $\mathbf{T}$.
Eliminate the finitely many auxiliary definitions, retaining the chosen refuter graph for its stated guarantee.
This gives the asserted $\mathbf{VTC}^0$ theorem.
It does not assert totality of the generic old $\FP$ scan schema or uniformity in growing $K$.
\end{proof}

\paragraph{\emph{An original-language formula retaining the chosen refuter.}}
Let $C^{\mathrm{thr}}_{q,K}(n,N)$ be the complete candidate test just used, including successful normalization, dimension and depth.
It and its negation have chosen bounded existential computation representations in $\mathbf{VTC}^0$.
Write $G_{\neg C}(n,N,u)$ for the representation of its negation, with auxiliary string tuple $u$.
Let $G_A(n,N,x,w)$ be the bounded existential computation graph of the \emph{chosen} refuter with auxiliary string tuple $w$, and let $G_E(n,N,x,v)$ represent $\mathsf{Err}_K$ with its actual evaluation histories and auxiliary string tuple $v$.

Choose respective polynomial bound terms $b,b',b''$ for $u,w,v$ from the proved all-string caps.
In the bounds for $w,v$, replace $|x|$ by $p(|N|)$, so all three bound terms depend only on the free parameters $n,|N|$.
Rename all tuple components apart.
Form
\[
\theta^{\mathrm{thr}}_{q,K}(n,N)\equiv
\exists x\le p(|N|)\,\exists u\le b\,\exists w\le b'\,\exists v\le b''\,
\left(G_{\neg C}(n,N,u) \vee
\left(G_A(n,N,x,w)\wedge G_E(n,N,x,v)\right)\right),
\]
where the displayed graph matrices have no string quantifiers after their witnesses are pulled into the displayed prefix; tuple coding can collapse each finite block.
This is an original-language $\Sigma^B_1$ formula.
Off the candidate guard choose $x$ and unused graph witnesses empty.
On the guard use the chosen refuter's total graph and the theorem's proved error computation.
Therefore $\mathbf{VTC}^0$ proves $\forall n,N\,\theta^{\mathrm{thr}}_{q,K}(n,N)$.

On a candidate, the first disjunct has no witnesses and the graph of $\mathcal A^{\mathsf{TC}}_{q,K}$ is unique.
Thus any satisfying tuple contains exactly the chosen refuter output, and that output is a genuine rational disagreement within the cutoff.
Conversely the chosen output supplies all these witnesses.
This variant retains the chosen algorithm; omitting $G_A$ would instead retain every error point, as in \Cref{cor:proofs} below.

The propositional consequence of this theorem is \Cref{tc:proofs} in \Cref{sec:proofs}.
It uses the polynomial-time translation theorem for $\Vone$, which applies because $\Vone$ proves every theorem of $\mathbf{VTC}^0$ (\Cref{lem:vtc-in-v1}).

\paragraph{\emph{Fixed-base power witnesses.}}
The fixed-base assertion of \Cref{cor:fixed-prime-depth-values} also has a threshold version.
For each separately fixed positive base $\beta$, the witness power $\beta^t$ is IMUL on $t$ copies of $\beta$, with $t$ a unary argument.
Its stage-$i$ value, for $i\le t$, is IMUL on $i$ copies.
The recurrence is multiplication by $\beta$ and the length cap is $2+(t+1)(|\beta|+2)$, as in the original graph.
At $\beta=1$ use the constant integer-one graph.
These equations prove idle clipping on valid power witnesses; no bound of $t$ by network length is needed.

The residue-product law gives nonzero residue of every such power when $q$ does not divide $\beta$.
For a stored nonbias fraction $u/v$, with $v>0$, let $a,t$ be its fixed-base representation witnesses, so $u\beta^t=va$.
If the denominator is divisible by $q$, the fixed-prime residue table applied to this equality forces divisibility of $u$.
Canceling the common $q$ in that integer equality preserves the same $t,a$ witness after a successful normalization step.
Bounded induction on its explicit stage history proves that the failure branch is impossible.
This also proves the original value-based fixed-base corollary without factoring or computing a general gcd.
\space

\section{Certified Constructions and a Matching Witnessing Correspondence}
\label{uc:section}

\Cref{tc:refuter} gives a refuter together with a proof that its output is a disagreement.
This section supplies the matching construction side: a network together with a certificate, and a proof of the checker's soundness in the same arithmetic theory.
It then gives the witnessing correspondence that pairs the two sides (\Cref{uc:duality}) and the resulting sharp threshold at prime two (\Cref{uc:sharp}).
It closes with the certified rational construction (\Cref{rc:body}).

The binary maximum tree and its depth bound are classical (cf.~\cite[Lemma~D.3]{ABMM18}).
The certificate for this tree labels each intermediate maximum by the set of input coordinates it represents.
These labels let us compute intermediate maxima directly from the original inputs and then verify every adjacent network equation in parallel, even when assembly depth grows.

Write $\bm{\mathcal T}$ for the conservative extension of $\mathbf{VTC}^0$ by the chosen $\mathsf{FTC}^0$ functions used below.
Its name differs from that of the theory $\mathbf{T}$ of \Cref{tc:parallel-arithmetic} because the chosen functions it adjoins are those of this section rather than those of the threshold refuter.
Like $\mathbf{T}$ in \Cref{tc:parallel-arithmetic}, it is conservative over $\mathbf{VTC}^0$ by \Cref{fact:tc-interface}.
Since each of $\mathbf{T}$ and $\bm{\mathcal T}$ also extends $\mathbf{VTC}^0$, the three theories $\mathbf{VTC}^0$, $\mathbf{T}$ and $\bm{\mathcal T}$ prove the same sentences of the original language.

\paragraph{\emph{Explicit bounds for the encodings.}}
We use the pairing $\langle v,b\rangle$ of \Cref{sec:foundations}.
For a unary size parameter $s$, define the fixed polynomial terms
\[
b_0(s)=100(s+2)^4,\qquad
b_{i+1}(s)=\langle b_i(s)+2,b_i(s)+2\rangle+2\quad(0\le i<8).
\]
This is a finite composition of terms, not recursion with an input-length iteration count.
An array with at most $b_i$ rows and row length at most $b_i$ has length below $b_{i+1}$ by (Arr).
The bounds below use $b_8$ only for the returned network, certificate and value history.
The auxiliary computation witnesses for the admitted arithmetic functions retain their own proved polynomial terms; composing a fixed number of those terms gives their bounds.
No bound on these witnesses is inferred merely from the lengths of final arithmetic values.

The balanced binary construction pads its leaves to the least power of two at least $n$.
We therefore first construct that leaf count and the intermediate powers that specify each block size, keeping their numerical values bounded by $2n$.
\begin{lemma}[Bounded dyadic indexing]
\label{uc:powers}
There are chosen $\mathsf{FTC}^0$ functions $L(n),m(n)$ and $\omega(n,i)$ whose equations $\bm{\mathcal T}$ proves as follows.
Let $U(v)$ be the canonical signed integer given by the admitted sum of $v$ copies of one, and let $P_2(i)$ be the chosen iterated multiplication IMUL applied to $i$ copies of two, the binary integer of value $2^i$.
For $n\ge1$, with $L=L(n)$ and $m=m(n)$,
\[
n\le m<2n,\quad \omega(n,0)=1,\quad \omega(n,L)=m,
\quad \omega(n,i+1)=2\omega(n,i)\quad(i<L),
\]
\[
\omega(n,i)\omega(n,L-i)=m\quad(i\le L),\qquad \omega(n,i)<n\quad(i<L),
\]
and $L$ is the least $i\le n$ with $P_2(i)\ge U(n)$.
All displayed powers are numbers at most $2n$.
In the standard model, $L=\lceil\log_2 n\rceil$ and $m=2^L$.
\end{lemma}
\begin{proof}
Construct the complete $U$ table through $4n+4$ and the complete $P_2$ table through $n+1$ by aggregate closure.
Their prefix equations are
\[
U(0)=0, U(v+1)=U(v)+1,\quad
P_2(0)=1, P_2(i+1)=2P_2(i).
\]
Since canonical one and two have lengths two and three, the integer sum and product length bounds give the valid component caps $2+2v$ for $U(v)$ and $2+4i$ for $P_2(i)$.
Then (Arr) bounds both tables by $b_1(n)$.
These are binary integer strings, even when the value of $P_2(i)$ is exponentially large.

Bounded number induction, with the tables as string parameters, proves $U(a+t)=U(a)+U(t)$ whenever the listed arguments are in range: start at $t=0$ and use the two successor equations at the step.
Consequently $U$ is strictly increasing, and $U(2n)=2U(n)$.
A second bounded number induction proves $P_2(i)\ge U(i+1)$: the base is $1\ge1$; at a step, positivity gives $2P_2(i)\ge P_2(i)+1\ge U(i+2)$.
Thus some $i\le n$ has $P_2(i)\ge U(n)$.
Let $L$ be the least such index, using bounded minimization.
If $L=0$, then $n=1$.
Otherwise minimality and the prefix equation give $U(n)\le P_2(L)=2P_2(L-1)<2U(n)=U(2n)$.

For every $i\le L$, choose the least $v\le2n$ with $P_2(i)\le U(v)$.
Such a $v$ exists since $P_2(i)\le P_2(L)\le U(2n)$; positivity proves the required monotonicity of $P_2$.
The chosen $v$ is positive.
Minimality gives $U(v-1)<P_2(i)\le U(v-1)+1$.
Discreteness of the signed integer order forces equality $P_2(i)=U(v)$.
Define $\omega(n,i)=v$ and $m=\omega(n,L)$.
Injectivity of $U$ now proves the stated numeric equalities, inequalities and doubling equations.
For $i<L$, minimality of $L$ gives $P_2(i)<U(n)$.
Hence $U(\omega(n,i))=P_2(i)<U(n)$, and strict monotonicity of $U$ gives $\omega(n,i)<n$.
This never identifies an unbounded binary integer value with a number-sort term.

All searches have explicit bounds $n$ or $2n$ and compare admitted arithmetic terms, so their bit graphs are $\mathsf{FTC}^0$.
For $n=0$, or an index $i>L$, return zero; the tables and their caps are defined before the successful guard is used.
For block indexing, use a direct number-sort induction on $i$ to prove $\omega(n,i)\omega(n,L-i)=m$, avoiding a multiplicative conversion of binary integer values.
At $i=0$ this reads $\omega(n,0)\omega(n,L)=m$, which holds since $\omega(n,0)=1$ and $\omega(n,L)=m$.
If $i<L$, the two doubling equations give
\[
\omega(n,i+1)\omega(n,L-i-1)
=2\omega(n,i)\omega(n,L-i-1)
=\omega(n,i)\omega(n,L-i)=m.
\]
Associativity and commutativity of number multiplication justify the middle equality; every entry is at most $2n$ and each product is a polynomial number term.
Thus $\omega(n,i)\omega(n,L-i)=m$ for all $i\le L$.
Use $\omega(n,L-i)$ for the number of blocks of length $\omega(n,i)$.
Every induction here has only bounded number quantifiers and fixed input strings in the conservative language.
\end{proof}

\paragraph{\emph{The certificate checker.}}
A maximum gadget on two inputs $a$ and $b$ has the three hidden outputs $\relu(b-a)$, $\relu(a)$ and $\relu(-a)$, in this order, and its output combines them as
\[
\relu(b-a)+\relu(a)-\relu(-a)=\max(a,b).
\]
This identity follows from $\relu(a)-\relu(-a)=a$ and the two cases $b\ge a$ and $b<a$; it includes ties and negative arguments.

The certified networks have the following restricted format.
Every parameter is a canonical denominator-one rational code.
All biases are zero and every matrix entry is $-1,0$ or $1$.
We make no completeness claim for arbitrary correct networks.

In a certified network every hidden layer consists of triples of neurons: for each $t$, the neurons $3t$, $3t+1$ and $3t+2$ of a layer, numbered from zero as in the history vector of \Cref{sec:foundations}, form its triple $t$.
A triple is read as one maximum gadget whose neuron $3t+r$, in position $r\in\{0,1,2\}$, carries the hidden output in position $r$ of the display above.
The readout of a triple is the combination of its three values with coefficients $(1,1,-1)$, which by the gadget identity is the maximum of its two inputs.

A certificate $c$ is an exact tuple consisting of a child-reference array, a label array and a selected final triple index.
A child reference at a layer after the first, and the selected final triple, name a triple by its within-layer index $t$.
The child-reference and label arrays list the triples layer by layer, by increasing $t$ within a layer.

A child-reference record is an exact pair of unary indices, naming the gadget's children $a$ and $b$ in that order.
A label is a string of length at most $n$ whose bit $i-1$ represents coordinate $i$, for $1\le i\le n$.
All label membership, singleton, union, maximum-selection and full-label checks below use this convention.

The checker $\mathsf{MCert}(n,N,c)$ performs all the following tests.
It requires $n\ge1$, complete validity of $N$ with input dimension $n$, and positive hidden widths divisible by three.

Every child index refers to an input coordinate at layer one, and to a triple of the immediately preceding layer afterwards.
Every label is nonempty, has no present bit at a position $i\ge n$, and equals the union of the two child singleton sets at layer one or the union of the two child labels at a later layer.

It checks every dense matrix entry: the three preactivation rows are exactly $b-a,a,-a$, interpreting a previous triple through its readout $(1,1,-1)$.
When child indices coincide, their contributions at the same column are added, giving a zero difference row; distinct columns are never canceled merely because their values happen to agree.
It checks every zero bias.
The output row selects exactly the recorded final triple's readout, whose label must be $[n]$.

The child and label arrays have exactly the required rows and zero unused bits; tuple shape and each unary record are checked by exact reconstruction.

There is one separate accepted branch: $n=1$, depth zero, $N$ the canonical affine identity, and $c$ the empty string.
Other depth-zero inputs are rejected.
Every failed shape, range or entry test returns false, before an invalid row is used.

Each test is a bounded number formula of the admitted arithmetic and array language.
Its characteristic function is consequently a chosen $\mathsf{FTC}^0$ function.
This is local checking of supplied labels, not an algorithm for finding reachable-input labels in an arbitrary unlabeled circuit.

Recall the local-history predicate $\mathsf{LEval}(N,n,x,Z)$ of the fixed network record in \Cref{sec:foundations}.
It requires complete network and input validity and an exact vector $Z$ of all hidden-neuron rational values in layer order followed by one scalar rational output, with no unused bits.
Each hidden-neuron entry of $Z$ equals the ReLU of that neuron's affine expression.
That affine expression reads $x$ at layer one and the preceding stored layer afterwards.
The final entry equals the affine output applied to the last stored hidden layer, or directly to $x$ at depth zero.

Every value equation uses cross-products for rational equality.
The checker computes each supplied adjacent-layer affine sum by the admitted common-denominator products and sums; all such checks are aggregated in parallel.
Its bounds depend on the total input length, including $|Z|$, so it is a total $\mathsf{FTC}^0$ predicate on arbitrary supplied histories, rejecting malformed shapes before their rows are used.
This defines a checker, not a total evaluator for arbitrary networks.

The certified histories of \Cref{uc:assemblies} and \Cref{rc:assembly-history} select maxima of input coordinates over labels and use one fact about these maxima, which we isolate.
\begin{lemma}[Maxima over labels]
\label{uc:label-max}
Let $n\ge1$, and let $V=(V_1,\ldots,V_n)$ be a supplied table of signed integers or of valid rationals ordered by cross-products.
For a nonempty $S\subseteq[n]$, let $M_S=V_i$ for the least $i\in S$ with $V_j\le V_i$ for every $j\in S$.
Then $\bm{\mathcal T}$ proves that this index exists for every nonempty $S$.
If $S=S_1\cup\cdots\cup S_k$ with every $S_t$ nonempty, then $\bm{\mathcal T}$ proves $M_{S_t}\le M_S$ for every $t\le k$, and $M_S=M_{S_t}$ for some $t\le k$.
For rationals, the order and the equality here are the cross-product order and rational-value equality.
\end{lemma}
\begin{proof}
Fix $S$ and $V$ as string parameters.
Bounded number induction on $c\le n$ proves that if some element of $S$ is at most $c$, then some $i\in S$ with $i\le c$ satisfies $V_j\le V_i$ for every $j\in S$ with $j\le c$.
At $c=0$ no element of $S\subseteq[n]$ is at most $c$.

At the step from $c$ to $c+1$, if $c+1\notin S$, the index retained at $c$ works.
If $c+1\in S$ and no element of $S$ is at most $c$, take $i=c+1$.
Otherwise take the retained index $i$ when $V_{c+1}\le V_i$, and $c+1$ when $V_i<V_{c+1}$; transitivity of the total order gives the bound at every $j\le c+1$ in $S$.

At $c=n$ the premise holds because $S$ is nonempty, and bounded minimization gives the least such index.
This is a proof of existence; the function computing $M_S$ uses all pairwise comparisons and bounded minimization in parallel, not sequential recursion.

For the union, let $i_t$ be the selected index of $S_t$ and $i$ that of $S$.
Since $i_t\in S_t\subseteq S$, the defining property of $i$ gives $M_{S_t}=V_{i_t}\le V_i=M_S$.
Since $i\in S$, some $t\le k$ has $i\in S_t$, and the defining property of $i_t$ gives $M_S=V_i\le V_{i_t}=M_{S_t}$.
Antisymmetry gives $M_S=M_{S_t}$ for this $t$.
\end{proof}

The checker is sound: for every rational input, the labels of a passing certificate determine a complete local history, computed directly from the input through \Cref{uc:label-max}, whose output is the maximum.
\begin{theorem}[Certified maximum assemblies; restatement of \Cref{ithm:upper-duality}]
\label{uc:assemblies}
There is a chosen total $\mathsf{FTC}^0$ function $H(n,N,c,x)$ such that $\bm{\mathcal T}$ proves:
if $\mathsf{MCert}(n,N,c)$ and $x$ is a valid rational $n$-vector, then $\mathsf{LEval}(N,n,x,H(n,N,c,x))$ holds and its output is $\MAX_n(x)$.
The history bound is $b_8(n+|N|+|c|+|x|)$.
The depth of a passing assembly may grow with its input length.
\end{theorem}
\begin{proof}
The history is built by clearing all input denominators to a common denominator and reading each triple's values off the maxima of its two child labels.
Reject a failed certificate or malformed point by returning empty.
For a valid point $x=((a_i,\epsilon_i))_{i=1}^n$, form
\[
E=\prod_i \epsilon_i>0,\qquad X_i=a_i\prod_{j\ne i}\epsilon_j.
\]
The excluded-entry product replaces the omitted factor by one; it never divides by a numerator.
The admitted product equations prove $X_i/E=x_i$.

For every supplied nonempty label $S$, let $M_S$ be its maximum in the table $(X_1,\ldots,X_n)$, as in \Cref{uc:label-max}.
Every checked label is the union of its two child labels, where a layer-one child $i$ has the singleton label $\{i\}$ and maximum $X_i$.
\Cref{uc:label-max} therefore gives $M_S=\max(A,B)$ whenever $A,B$ are the maxima of its two child labels.
For every triple with child maxima $A,B$, set its three history entries to
\[
\left(\max(0,B-A),E\right),\quad
\left(\max(0,A),E\right),\quad
\left(\max(0,-A),E\right).
\]
Their readout is $M_S/E$ by the gadget identity and the checked union equation.
In a later layer substitute the preceding triples' readouts into the explicitly checked three affine rows.
These rows give precisely the three preactivations just displayed; positive scaling by $E$ commutes with ReLU.
Thus every local equation holds, independently of the assembly depth.

The selected output label is $[n]$, so its value is $\max_i X_i/E=\MAX_n(x)$.
Set the scalar output entry to $(M_{[n]},E)$; the output row is the readout of the selected final triple, whose label is $[n]$, so this entry satisfies the final affine equation.
For the depth-zero branch, return the one output entry $(X_1,E)$.

For $s=n+|N|+|c|+|x|$, every numerator, denominator and difference above has component length at most $8(s+2)^2$.
Each product has at most $n$ factors of component length at most $|x|$, and subtraction gains at most two bits; maxima retain an existing component.

Store $H$ as the exact vector of all hidden-neuron values in layer order followed by the scalar output, with no recursively nested layer records.
There are at most $|N|+1$ rows.
A rational pair and this value vector add two array levels above the component bound, so $b_8(s)$ is larger than their explicit (Arr) terms.
The same bound covers the input-clearing and label-maximum tables: their row counts are at most $s^2$, and at most three additional array levels are used.

These are bounds on returned explicit data; arithmetic computation witnesses use their separately admitted terms.
All minima, comparisons, products and row constructors are aggregated with a fixed composition depth.
This defines a total chosen $\mathsf{FTC}^0$ graph and proves its guarantee in $\bm{\mathcal T}$.
\end{proof}

The certificate format has a uniform concrete instance.
The next construction supplies its wiring and labels directly from the input dimension.

\begin{theorem}[A uniform certified balanced construction; restatement of \Cref{ithm:upper-duality}]
\label{uc:balanced}
Chosen $\mathsf{FTC}^0$ functions $B(n),C(n)$ satisfy, provably in $\bm{\mathcal T}$, for every $n\ge1$:
$\mathsf{MCert}(n,B(n),C(n))$, $|B(n)|,|C(n)|\le b_8(n)$, and the hidden depth of $B(n)$ is exactly $L(n)$.
Thus the construction has weights in $\{-1,0,1\}$, zero biases and depth $\lceil\log_2 n\rceil$ in the standard model.
\end{theorem}
\begin{proof}
For $n=1$ use the separate affine identity branch.
For $n\ge2$, take $m,L,\omega$ from \Cref{uc:powers}.
Pad $m$ leaves by the map $\pi(i)=i+1$ for $0\le i<n$ and $\pi(i)=1$ for $n\le i<m$.
At layer $\ell$ there are $\omega(n,L-\ell)$ triples.
Their children are $\pi(2j),\pi(2j+1)$ at layer one and triples $2j,2j+1$ of the preceding layer afterwards.
The previous layer has twice as many triples by the proved doubling equations, so all references are valid.
No padding layer is introduced.
Define every dense entry directly by the gadget's three signed row formulas, and set the output to the sole final triple's readout.
Coincident input references are combined at their actual matrix column.

Label triple $(\ell,j)$ by
\[
S_{\ell,j}=\{\pi(i):j\omega(n,\ell)\le i<(j+1)\omega(n,\ell)\}.
\]
This is a bounded number definition of each bit, using a block contained in $[0,m)$: for $j<\omega(n,L-\ell)$, the product identity of \Cref{uc:powers} gives $(j+1)\omega(n,\ell)\le\omega(n,L-\ell)\omega(n,\ell)=m$.
The doubling equations split each parent block into its two child blocks; consequently all singleton and union checks hold.
Blocks are nonempty and the root block is all padded leaves, whose image is $[n]$.
The final layer has one triple.
Every remaining certificate test holds by the exact entry construction.
All bits of $B,C$ are therefore given by bounded formulas with the chosen arithmetic functions, not by an input-dependent iteration of network composition.

Bounded number induction on $j\le L$, using the doubling equations, proves
\[
\sum_{i<j}\omega(n,i)=\omega(n,j)-1.
\]
The sum is zero at $j=0$, and at a step $\left(\omega(n,j)-1\right)+\omega(n,j)=\omega(n,j+1)-1$ by $\omega(n,j+1)=2\omega(n,j)$.

Layer $\ell$ has $3\omega(n,L-\ell)$ hidden neurons, so there are $3\sum_{i<L}\omega(n,i)=3(m-1)<6n$ hidden neurons.
Put $d_0=n$, $d_\ell=3\omega(n,L-\ell)$ for $1\le\ell\le L$, and $d_{L+1}=1$; these are the layer widths from input to output.
Let $\mathcal W=\sum_{\ell\le L+1}d_\ell$, so that $\mathcal W=n+3(m-1)+1<7n+1$.

The matrix of layer $\ell$ has $d_\ell d_{\ell-1}$ entries, for $1\le\ell\le L+1$.
Bounded number induction on $j\le L+1$ proves $\sum_{1\le\ell\le j}d_\ell d_{\ell-1}\le\mathcal W\sum_{1\le\ell\le j}d_\ell$.
Both sides are zero at $j=0$, and the step adds $d_{j+1}d_j\le\mathcal Wd_{j+1}$, since $d_j$ is one of the nonnegative summands of $\mathcal W$.
At $j=L+1$ the right side is at most $\mathcal W^2<(7n+1)^2$, so there are fewer than $(7n+1)^2$ matrix entries.

There are $3(m-1)+1<6n+1$ biases.
Thus fewer than $50(n+1)^2$ constant-length rational parameters are stored, including every zero.
Unary metadata, layer dimensions and parameter counts are below $b_0(n)$.
The encoding uses a rational-pair level, a flat parameter-list level, a dimension-list level, a metadata level and the outer network record, fewer than eight array levels.
For $C$, child pairs, the flat child array, the flat label array and the outer tuple likewise use fewer than eight.

The (Arr) bound at each of these constructors proves $b_8(n)$ for both codes.
At $n=0$ return empty.
\end{proof}

The binary family $B(n)$ is the literal fixed record of \Cref{sec:foundations} formed from the dimensions, dense entries and biases just constructed.
Its certificate and history read those same stored entries, rather than an unspecified encoding of a semantically equivalent network.

The smallest dimension with a padded leaf shows the network and certificate of \Cref{uc:balanced} concretely.
\begin{example}[The balanced certificate for $n=3$]
\label{uc:example}
In the standard model, \Cref{uc:powers} gives $L=2$, $m=4$ and $\omega(3,i)=2^i$ for $i\le2$.
The padding map sends $0,1,2,3$ to $1,2,3,1$.
Layer one has $\omega(3,1)=2$ triples.

Triple $(1,0)$ has children $a=x_1$, $b=x_2$ and label $\{1,2\}$, so its preactivation rows are $x_2-x_1$, $x_1$ and $-x_1$.
Next, triple $(1,1)$ has children $a=x_3$, $b=x_1$ and label $\{1,3\}$, with rows $x_1-x_3$, $x_3$ and $-x_3$.
By the gadget identity their readouts are $\max(x_1,x_2)$ and $\max(x_1,x_3)$.

Layer two has the single triple $(2,0)$, whose children are these two triples and whose label is $\{1,2,3\}=[3]$; the output is its readout $\max(x_1,x_2,x_3)$.
The hidden depth is $2=\lceil\log_2 3\rceil$, the hidden widths are $6$ and $3$, and coordinate one lies in both layer-one labels because the fourth leaf repeats it.
\end{example}

\paragraph{\emph{Local histories and fixed-depth evaluation.}}
A supplied local history is checked by all adjacent rational affine/ReLU equations in parallel.
A rational affine sum of variable width is computed with a common denominator and iterated integer products and sums, so its chosen checker belongs to $\mathsf{FTC}^0$ even when the network depth varies.
Checking a supplied history is distinct from constructing one.

Any two such histories of the same network and point have the same rational values: fix both histories as string parameters and induct on the bounded number layer, comparing each coordinate by cross-products.
The previous-layer equalities and the local equations give the next equalities, and finally the same output.
This proves value uniqueness, not raw equality of unreduced codes.

For fixed $K$, the equations of the fixed-depth evaluator $\mathsf{Eval}_K$ of \Cref{tc:parallel-arithmetic} imply these local equations on its successful guard.
In particular its output on $B(n)$ with $L(n)\le K$ equals the output of $H$.
The parameter denominator product is one on this constructor; its cleared recurrence has denominator $E$, exactly the positive input denominator used above.
Thus this comparison does not import any generic growing-depth evaluator into $\bm{\mathcal T}$.

Let $\mathsf{Cert}(n,N,c)$ denote a chosen certificate checker, and let $a(n)$ be a fixed polynomial bound for the witnesses in the next statement.
\begin{proposition}[The two witnessing correspondences; restatement of \Cref{ithm:upper-duality}]
\label{uc:duality}
Fix polynomial bounds and chosen $\mathsf{FTC}^0$ checkers in the conservative language of $\mathbf{VTC}^0$.
In this proposition, a theorem is a theorem of the conservative extension of $\mathbf{VTC}^0$ by chosen $\mathsf{FTC}^0$ functions, and the theory is that extension.
\begin{enumerate}[label=\textup{(\alph*)},leftmargin=2em]
\item A theorem of bounded certified existence
$\forall n\ge1\,\exists N,c\le a(n)\,\mathsf{Cert}(n,N,c)$ is equivalent to a chosen $\mathsf{FTC}^0$ constructor of $(N,c)$ whose stated size and passing guarantee the theory proves.
To interpret this as an upper bound, the checker's soundness for the specified target must also be proved; \Cref{uc:assemblies} supplies it for $\mathsf{MCert}$.
\item A theorem that every guarded candidate has a bounded rational error is equivalent to a chosen $\mathsf{FTC}^0$ refuter whose size and error guarantee the theory proves.
Here the error relation must use an admitted chosen evaluator, such as $\mathsf{Eval}_K$ at fixed $K$.
\end{enumerate}
\end{proposition}
\begin{proof}
We apply the witnessing assertion of \Cref{fact:tc-interface} directly, keeping the output bounds in its matrix.
For part~\textup{(a)}, pack $N,c$ as the two entries of an exact tuple $Y$, with row zero empty as in (Arr).
If both entries have length at most $a(n)$, this tuple has length at most $\langle3,a(n)\rangle$.
The matrix consists of the exact two-entry reconstruction, the two component bounds, and $\mathsf{Cert}(n,N,c)$.
These conditions have bounded number quantifiers and no quantified strings in the chosen function language.
Extend the assertion to $n=0$ by permitting the empty tuple there.
The assumed certified-existence theorem therefore proves an assertion of the form required by \Cref{fact:tc-interface}, with number input $n$ and no free string input.

The witnessing function returns a chosen $\mathsf{FTC}^0$ tuple $Y(n)$ with a proved passing guarantee.
Its two row projections are chosen $\mathsf{FTC}^0$ functions, and for $n\ge1$ that guarantee gives the required $N(n),c(n)$, their original bounds $a(n)$, and acceptance by $\mathsf{Cert}$.
A separately proved soundness theorem for $\mathsf{Cert}$, as the statement requires, then gives the specified upper bound.

For part~\textup{(b)}, write $G(n,N)$ for the fixed candidate guard, $E(n,N,x)$ for the error check using its admitted evaluator, and $b(n,|N|)$ for the fixed polynomial error-length bound.
The bounded-refutation theorem is equivalent to
\[
\forall n\,\forall N\,\exists x\le b(n,|N|)\,
\left(\neg G(n,N) \vee E(n,N,x)\right).
\]
Off the guard the empty string is a witness; on the guard this is the assumed bounded error assertion.
Its matrix again has bounded number quantifiers and no quantified strings: $G$ and $E$ are the chosen checks stipulated in the statement.
Applying \Cref{fact:tc-interface} with number input $n$ and string input $N$ gives a chosen $\mathsf{FTC}^0$ refuter, with the same proved length bound and error guarantee on every guarded candidate.
For the refuters in this paper, $b(n,|N|)=p(|N|)$.

In the reverse direction, existentially quantify the chosen constructor or refuter outputs in their proved guarantees and eliminate their conservative definitions.
This recovers the respective bounded-existence assertions with the same bounds.
An error within the stated length bound contradicts the bounded universal correctness assertion that no such error exists, and negating that assertion yields precisely such an error on a valid candidate.
In neither part does the witnessing matrix contain a universally quantified string, so the argument extracts no constructor from a correctness assertion that carries an extra universal string quantifier.
\end{proof}

At prime two, the certified upper family and the modular refuter meet at the same known depth threshold.
The following statement records both witnesses in one arithmetic theory.
Write $\mathsf{IntNet}(N,n)$ for the conjunction of $\Net(N,n)$ with the condition that every nonbias weight of $N$ is a canonical denominator-one rational code; it permits arbitrary rational biases.

\begin{corollary}[A witnessed sharp integer-depth threshold; restatement of \Cref{icor:sharp}]
\label{uc:sharp}
For every separately fixed $K\ge1$, the conservative extension $\bm{\mathcal T}$ of $\mathbf{VTC}^0$ proves the following equivalence for $n\ge1$:
\[
n\le2^K\quad\leftrightarrow\quad
\exists N\le b_8(n)\,
\left(\mathsf{IntNet}(N,n)\wedge\Depth(N,K)
\wedge\forall x\le p(|N|)\,\neg\mathsf{Err}_K(N,n,x)\right).
\]
Replacing $\mathsf{IntNet}$, $\Depth$ and $\mathsf{Err}_K$ by their $\Delta^B_1$ definitions gives a theorem of $\mathbf{VTC}^0$.
On the left range, $B(n),C(n)$ give a certified constructor and its parallel history.
Above that range, the chosen $q=2$ refuter gives a rational disagreement for every such candidate.
\end{corollary}
\begin{proof}
Let $d=2^K$ be the external numeral and assume $n\le d$.
We first show $L(n)\le K$.
Suppose, for contradiction, that $K<L(n)$.
The doubling equations of \Cref{uc:powers} at the indices $0,\ldots,K-1$, all below $L(n)$, together with $\omega(n,0)=1$, give $\omega(n,K)=d$.
These are $K$ instances of proved equations for the fixed standard numeral $K$, not an induction on a variable.
Since $K<L(n)$, \Cref{uc:powers} also gives $\omega(n,K)<n$, so $d<n$, contrary to $n\le d$.
Hence $L(n)\le K$.
Every value used here is a value of $\omega$, at most $2n$, or the fixed numeral $d$; no additional table or numeric exponentiation is used.

\Cref{uc:balanced,uc:assemblies}, followed by the fixed-depth history comparison, give the forward implication with $N=B(n)$.
For the reverse implication assume $n>2^K$.
Every candidate on the right is eligible for \Cref{tc:refuter} at $q=2$.
Its nonbias denominators are one, which has nonzero residue modulo two, so the normalization of \Cref{tc:cancel} leaves every nonbias entry unchanged and sets no failure flag; together with $\Net(N,n)$ this gives $\mathsf{ValGood}_2(N,n)$.
Its dimension is at least $2^K+1$ and its depth is at most $K$.
The theorem's error within the same $p(|N|)$ contradicts the final conjunct.
Eliminate the auxiliary conservative definitions to obtain the original-language theorem.
\end{proof}

\paragraph{\emph{Finite proofs retaining the construction.}}
There is also a construction-side counterpart to \Cref{tc:proofs}.
Let $G_B(n,N,u)$ and $G_C(n,c,v)$ be the original-language bounded existential graphs of the chosen $B,C$, and let $G_M(n,N,c,w)$ be the corresponding computation matrix asserting acceptance by $\mathsf{MCert}$.
Choose polynomial bounds for their computation witnesses from their proved graph caps, retaining $|N|,|c|\le b_8(n)$.
After renaming and packing the witnesses, take the bounded existential formula
\[
\theta^{\mathrm{bal}}(n)\equiv\exists N,c\le b_8(n)\,\exists u,v,w\le e(n)\,
\left(n=0 \vee (G_B(n,N,u)\wedge G_C(n,c,v)\wedge G_M(n,N,c,w))\right).
\]
Here each $G$ denotes its quantifier-free-in-strings graph matrix after its bounded existential witnesses have been pulled into the displayed prefix; bounded number quantifiers may remain.
The fixed term $e$ is a polynomial envelope of the actual auxiliary graph bounds, independent of any depth cap $K$.

\Cref{uc:balanced} and the chosen graph equations prove $\forall n\,\theta^{\mathrm{bal}}(n)$ in $\mathbf{VTC}^0$ after conservative elimination; at $n=0$ use empty witnesses.
The propositional translation of $\theta^{\mathrm{bal}}$ is stated at the end of \Cref{sec:proofs}, after the proof system it uses has been defined.
Every satisfying existential assignment of that translation has $N=B(n)$ and $c=C(n)$ for $n\ge1$.

The upper constructor is uniform across $n$, whereas each lower refuter is defined for a separately fixed $K$, so combining the two parts does not give a polynomial refuter uniform in growing $K$.

Certified existence quantifies the network and its certificate in a bounded existential string prefix.
A growing-depth correctness assertion retaining evaluation histories instead has the form $\exists N\,\forall x\,\exists Z$: a history $Z$ must exist for each input $x$.
Quantifying universally over supplied histories alone would not establish this existence.
For fixed $K$, the admitted evaluator removes the history witness and gives the simpler form $\exists N\,\forall x$.
The certified construction correspondence does not extract a constructor from either correctness assertion merely by rewriting its matrix.


\subsection{Certified Rational Blocks of Growing Arity}
\label{rc:body}

The balanced construction of \Cref{uc:balanced}, with weights in $\{-1,0,1\}$, has depth $\lceil\log_2 n\rceil$ and serves as the uniform binary benchmark.
Rational weights permit a shallower family, of depth $O(\log n/\log\log n)$ by \Cref{rc:upper} and \Cref{ir:rate}, while preserving polynomial total encodings and complete ordinary histories.
Here an ordinary history of a network at an input is its history vector of \Cref{sec:foundations}, with one rational entry for every hidden neuron and one for the output, as checked by $\mathsf{LEval}$.
The semantic two-layer collapse is proved in~\cite{BHRSY26}; the direct block construction below adapts Zimmermann's integer-feature dictionary~\cite[Definitions~3--5 and Theorem~7]{Zim26}.
The contribution here is the explicit arithmetic version of this construction, verified in \(\mathbf{VTC}^0\).

We first describe the finite certificate used for a block.
Consider a homogeneous rational two-hidden-layer network \(Q(x)=c\relu(B\relu(Ax))\) on \(b\ge2\) inputs, that is, one with all biases zero, rational matrices \(A,B\), a rational row vector \(c\), and \(\relu\) applied coordinatewise.

For such a network, a cone certificate is a flat array of nodes.
A node carries a list \(G\) of rational rows, representing the closed cone on which their dot products are nonnegative.
The root list is empty.
An internal node carries a test row \(t\) and two child references whose lists are exactly \(G\) followed by \(t\) and \(G\) followed by \(-t\).
A leaf carries activation patterns and nonnegative rational multipliers for their sign inequalities.
For every coordinate \(a\), further multipliers certify both signs of its affine output minus \(x_a\) after adjoining all inequalities \(x_a\ge x_i\).
\Cref{rc:cones,rc:cc-sound} give the exact checker and soundness proof.

The global certificate stores either the binary benchmark certificate or one block and its cone certificate, together with powers, interval labels and the dense substitution equations.
In block mode the block and its cone certificate are stored only once, and the global certificate contains no certificate of a smaller assembly nested inside it (\Cref{rc:assembly-certificate}).
For \(n=1\) there is also the identity branch \(N=\mathsf{Id}_1,C=\varnothing\).
In block mode it records the block arity \(b\), the exponent \(h\) with \(b^{h-1}<n\le b^h<bn\), and the powers \(1,b,\ldots,b^h\).
The checker reconstructs every matrix entry and input-set label, as specified in \Cref{rc:assembly-certificate}.
It makes no completeness assertion for arbitrary growing-depth networks.

For a valid ordinary history \(Z\), write \(\operatorname{out}(Z)\) for its last scalar rational entry.
Let \(\ell\) be the total floor binary logarithm, with \(\ell(0)=\ell(1)=0\).
The following guarantee includes one uniform depth inequality inside the theory.

\begin{theorem}[A rational construction with sublogarithmic depth; restatement of \Cref{ithm:rational}]
\label{rc:upper}
There are a chosen total \(\mathsf{FTC}^0\) Boolean predicate \(S(n,N,C)\), chosen total \(\mathsf{FTC}^0\) functions \(F,C_F,k,H_F\), a fixed polynomial \(P\), and a fixed standard integer \(C_\ast\) with the following guarantees in \(\bm{\mathcal T}\).
For every unary \(n\ge1\),
\[
\Net(F(n),n),\qquad S(n,F(n),C_F(n)),\qquad
|F(n)|+|C_F(n)|\le P(n).
\]
Acceptance is sound: for every unary \(n\ge1\), all strings \(N,C,Z\) and every valid rational \(n\)-vector \(x\),
\[
S(n,N,C) \wedge \mathsf{LEval}(N,n,x,Z)
\rightarrow \operatorname{out}(Z)=_{\Q}\MAX_n(x).
\]
The literal hidden depth of \(F(n)\) is \(k(n)\).
The theory also proves, for every unary \(n\ge1\) and every valid rational \(n\)-vector \(x\),
\[
\mathsf{LEval}(F(n),n,x,H_F(n,x)),\qquad
\operatorname{out}(H_F(n,x))=_{\Q}\MAX_n(x),
\qquad |H_F(n,x)|\le P(n+|x|).
\]
It also proves the uniform numerical inequality
\[
\forall n\,\left(n\ge2\rightarrow(\ell(\ell(n))+1)k(n)\le C_\ast(\ell(n)+1)\right).
\]
In particular, in the standard model \(k(n)=O(\log n/\log\log n)\).
\end{theorem}

The complete proof is in \Cref{rc:rate}.
The checker \(S\) is the predicate \(\mathsf{SCert}\) in \Cref{rc:assembly-certificate}.
The constructor uses the fixed standard integers \(A=771\) and \(r_\ast=5\) of \Cref{ir:packets} (this constant \(A\) is not the matrix \(A\) of a block).
It selects the greatest arity \(b\) whose direct block budget \(2^{A(b+1)^{r_\ast}}\) is at most \(n\).
It uses the binary family when no such arity exists.
\Cref{ir:small} constructs the block and its actual cone certificate under that guard.
The construction takes a signed average of explicit two-hidden-layer root functions and writes its difference from the maximum as an expression whose terms have affine rank at most two less than the arity (\Cref{ir:span}).
\Cref{ir:direct} specifies each term of this expression directly, without recursion on earlier expressions, by a local key (a short code of local geometric data) and a short binary word recording the transformation used at each higher arity.
So all arrays exist before numerical induction verifies their identities.
\Cref{ir:rate} proves the displayed depth inequality.

The ordinary histories \(H_F(n,x)\) come from one function that works for every passing assembly, by \Cref{rc:assembly-history}.
It selects each block's inputs directly from the original input-coordinate sets, evaluates all depth-two blocks independently, and verifies the dense boundary equations.
Thus the construction needs no evaluator for arbitrary growing depth.

Cone case splitting, Farkas multipliers and rational checking have established roles in neural-network verification~\cite{IBZK22,Gok26,DISK25}.
The guarantee here includes chosen arithmetic graphs, literal dense matrices, polynomial total codes and every adjacent rational history equation for this particular growing-depth family.
It does not assert polynomial size at hidden depth two or an improvement in the semantic size-depth bounds alone.
\space
\space

\section{Witnessing and Quantified Propositional Proofs}
\label{sec:proofs}

The internal refutation theorem has two proof-complexity consequences.
Witnessing identifies a provable lower bound with a refuter whose guarantee is provable in the same theory.
Propositional translation gives finite quantified proofs that retain an existential disagreement witness.
We state these classical metatheorems separately from their application to the network-specific refuter.
All polynomials, defining graphs, and prime/depth parameters in this section are fixed as above.

\subsection{The Exact Refuter Interpretation}

The $\Delta^B_1$ definitions supplied by \Cref{fact:fp-interface} allow the classical witnessing theorem to be applied to our chosen predicates.
\begin{fact}[Polynomial-time witnessing]
\label{fact:fp-witnessing}
Suppose $\varphi(u,v,X,Y)$ is a $\Sigma^B_0$ formula and $\Vone$ proves $\forall u\,\forall X\,\exists v\,\exists Y\,\varphi(u,v,X,Y)$.
There are polynomial-time number and string functions $f,F$ whose chosen definitional extension proves $\forall u\,\forall X\,\varphi(u,f(u,X),X,F(u,X))$ \cite[Theorem~6.28]{CookNguyen}.
The corresponding conclusion holds for bounded existential assertions with a $\Delta^B_1$ matrix in the admitted conservative extensions, retaining any explicit length bound as part of the guarantee \cite[Corollary~6.29]{CookNguyen}.
Number inputs are measured in unary.
\end{fact}
For the second formulation, with a $\Delta^B_1$ matrix, replace the matrix by its proved $\Sigma^B_1$ representation, absorb its auxiliary bounded strings into the existential witness tuple, apply the cited witnessing theorem, and project the required components.
The representation equivalence then proves the original matrix and its retained bound for those components.

Let $C(n,N)$ be any polynomial-time candidate predicate whose implication to $\mathsf{Net}(N,n)$ is provable in $\Vone$.
It may include weight, depth, and size restrictions.
The equivalence below requires a proof of the refuter's guarantee inside $\Vone$; ordinary correctness of a polynomial-time algorithm is not sufficient.
Recall from \Cref{sec:foundations} that $\Err(N,n,x)$ says that $x$ is a valid rational $n$-vector at which the general-depth evaluator returns a nonempty valid rational output differing from $\MAX_n(x)$.
The output cutoff $p$ is the one fixed polynomial of \Cref{sec:foundations}, used for every prime and depth.
The correctness predicate $\Computes(N,n)$ is $\Net(N,n)$ together with $\neg\Err(N,n,x)$ for every string $x$ with $|x|\le p(|N|)$.
\begin{proposition}[Provable lower bounds and provable refuters]
\label{prop:witnessing}
For fixed $C$ and the fixed correctness predicate $\Computes$, the following are equivalent:
\begin{enumerate}[label=\textup{(\alph*)},leftmargin=2em]
\item $\Vone\vdash\forall n\ge1\,\forall N\,(C(n,N)\to\neg\Computes(N,n))$.
\item There is an $\FP$ function $A$ with a chosen definition such that $\Vone(A)$, the definitional extension of $\Vone$ by $A$, proves
\[
\forall n\ge1\,\forall N\,
\left(C(n,N)\to
\left(|A(1^n,N)|\le p(|N|)\wedge\Err(N,n,A(1^n,N))\right)\right).
\]
\end{enumerate}
\end{proposition}
\begin{proof}
On $C(n,N)$, network validity is already true.
Negating the definition of $\Computes$ therefore gives a string $x$ of length at most $p(|N|)$ satisfying $\Err(N,n,x)$.
Move $n\ge1$ and $C(n,N)$ into the matrix and take the empty string as the witness off this antecedent.
The resulting existence statement has a $\Delta^B_1$ matrix, including its explicit length bound.
The witnessing theorem (\Cref{fact:fp-witnessing}) gives (b) with exactly that bound in the proved guarantee.
Conversely, existentially quantify the output of $A$ in (b).
Its error within the cutoff contradicts $\Computes$.
The resulting sentence does not contain $A$, so conservativity of its definitional extension yields (a).
\end{proof}
The source of difficulty is proving either side for a particular class, not the abstract equivalence.
\Cref{thm:fixed-prime-depth,cor:fixed-prime-depth-values} do that work for the fixed-prime candidates by constructing and verifying the actual refuter.
No hardness assumption is used.

\subsection{Finite Proof Families Retaining the Disagreement Witness}

A quantified propositional formula extends ordinary Boolean formulas by quantifiers $\exists\mathsf p$ and $\forall\mathsf p$ over Boolean variables $\mathsf p$.
Define $\Sigma^q_1$ and $\Pi^q_1$ simultaneously as the smallest pair of classes containing the quantifier-free formulas and satisfying the following closure rules, as in \cite[Definition~7.30]{CookNguyen}.
Both classes are closed under conjunction and disjunction.
The class $\Sigma^q_1$ is closed under existential quantification, and $\Pi^q_1$ under universal quantification.
Negation interchanges the classes: if $A\in\Sigma^q_1$, then $\neg A\in\Pi^q_1$; if $A\in\Pi^q_1$, then $\neg A\in\Sigma^q_1$.

A sequent $\Gamma\Rightarrow\Delta$ is a pair of finite formula lists, its antecedent $\Gamma$ and succedent $\Delta$.
It asserts that if every formula of $\Gamma$ is true, then at least one formula of $\Delta$ is true.
An empty left list imposes no assumption, and an empty right list asserts falsity.
A proof is built from initial sequents by inference rules, each of which specifies premise sequents and a conclusion.
The cut rule combines $\Gamma\Rightarrow\Delta,A$ and $A,\Gamma\Rightarrow\Delta$ to infer $\Gamma\Rightarrow\Delta$; $A$ is its cut formula.
A tree-like proof uses each premise occurrence only once, so reusing a derivation requires a separate copy.

Weakening adds a formula on either side, exchange interchanges adjacent formulas on one side, and contraction replaces two adjacent copies by one.
A proof is cut-free when it has no cut inference.
In a connective- or quantifier-introduction rule, the formula introduced in the conclusion is principal; the premise formulas used to form it are auxiliary.

We spell out the propositional rules used below \cite[Section~2A.1]{CookNguyen}.
Write $\top$ and $\bot$ for the always-true and always-false formulas.
The initial sequents are $A\Rightarrow A$ for every formula $A$, $\Rightarrow\top$, and $\bot\Rightarrow$.
Negation-left infers $\neg A,\Gamma\Rightarrow\Delta$ from $\Gamma\Rightarrow\Delta,A$; negation-right infers $\Gamma\Rightarrow\Delta,\neg A$ from $A,\Gamma\Rightarrow\Delta$.

Conjunction-left replaces $A,B$ in the antecedent by $A\wedge B$.
Conjunction-right infers $\Gamma\Rightarrow\Delta,A\wedge B$ from $\Gamma\Rightarrow\Delta,A$ and $\Gamma\Rightarrow\Delta,B$.

Disjunction-left infers $A\vee B,\Gamma\Rightarrow\Delta$ from $A,\Gamma\Rightarrow\Delta$ and $B,\Gamma\Rightarrow\Delta$.
Disjunction-right replaces $A,B$ in the succedent by $A\vee B$.

The quantified sequent calculus $\mathbf G$ has these propositional rules and four quantifier-introduction rules.
For a formula $A(z)$, a quantified variable $z$, a substituting expression $s$ and a fresh variable $a$, the four patterns are
\[
\frac{\Gamma\Rightarrow\Delta,A(a)}
{\Gamma\Rightarrow\Delta,\forall z\,A(z)}
\qquad
\frac{A(s),\Gamma\Rightarrow\Delta}
{\forall z\,A(z),\Gamma\Rightarrow\Delta}
\qquad
\frac{\Gamma\Rightarrow\Delta,A(s)}
{\Gamma\Rightarrow\Delta,\exists z\,A(z)}
\qquad
\frac{A(a),\Gamma\Rightarrow\Delta}
{\exists z\,A(z),\Gamma\Rightarrow\Delta}.
\]
Here $A(s)$ means capture-avoiding substitution for the free occurrences of $z$.
In $\mathbf G$, $z,a$ are Boolean variables.
In the rules introducing an existential quantifier on the right and a universal quantifier on the left, the substituted Boolean formula is quantifier-free.
In the rules introducing a universal quantifier on the right and an existential quantifier on the left, the quantified variable is replaced by a fresh Boolean variable, the eigenvariable, which does not occur in the conclusion.

The tree-like system $\mathbf G_1^\star$ uses each premise at most once and restricts cut formulas to $\Sigma^q_1\cup\Pi^q_1$.
We use the standard calculus of \cite[Sections~7C--7D]{CookNguyen}.

We next recall how bounded formulas become propositional ones.
For a bounded arithmetic formula $\theta(n,N)$, its translation $\theta[n;\nu]$ substitutes the number $n$ and the string length $\nu$.
The free bits of $N$ become free Boolean variables; bounded number quantifiers become finite conjunctions or disjunctions; bounded existential string quantifiers become existential Boolean variables together with their length choices.

At a fixed string length, a string is a finite set whose largest present bit determines its length: for length $\nu>0$, its top bit is true and the lower $\nu-1$ bits are free; length zero has no bits.
These are the conventions of \cite[Sections~7B.1 and 7E]{CookNguyen}, with the one departure for string equalities fixed in \Cref{sec:translation-cases}.
With these conventions the classical translation theorem has the following precise form.
\begin{fact}[Polynomial-time propositional translation]
\label{fact:fp-translation}
Let $\varphi(u,X)$ be a bounded formula in the original two-sorted language whose universal closure is provable in $\Vone$.
A polynomial-time function of the unary numerical values $u$ and string lengths $\nu$ produces a $\mathbf G_1^\star$ proof of $\varphi[u;\nu]$.
The proof size is polynomial in these parameters, with the polynomial depending on the fixed arithmetic proof.
In particular, a $\Sigma^B_1$ formula has a $\Sigma^q_1$ translation \cite[Proposition~7.56 and Theorem~7.57]{CookNguyen}.
\end{fact}
The source proves this fact by translating, rule by rule, an anchored sequent proof of $\varphi$ \cite[Definition~6.40 and Corollary~6.43]{CookNguyen}.
In an anchored proof, every cut formula occurs in a non-logical axiom or as $A(0)$ or $A(t)$ in an induction instance.
The source leaves the number quantifier cases, a propositional lemma for the comprehension axioms, and the string universal quantifier cases as exercises \cite[Exercises~7.27, 7.58 and~7.59]{CookNguyen}, and \Cref{sec:translation-cases} writes them out.
We use the anchoring step as stated in the source, although the source leaves the construction of a countermodel in its proof as an exercise \cite[Exercise~6.44]{CookNguyen}.
We apply this fact to the bounded refutation assertion below.

For fixed $q,K$, let
\[
C_{q,K}(n,N)\equiv
n\ge q^K+1 \wedge \mathsf{ValGood}_q(N,n) \wedge \Depth(N,K).
\]
It has the chosen polynomial-time definition from the normalization and residue scans.
We specify a bounded formula retaining the actual error point, not just some formula equivalent to a theorem.
\begin{samepage}
\begin{corollary}[Polynomial-size quantified propositional proofs]
\label{cor:proofs}
For every separately fixed $q,K$ there is a $\Sigma^B_1$ formula $\theta_{q,K}(n,N)$ in the original language of $\Vone$, with designated existential string $x$ bounded by $p(|N|)$, such that:
\begin{enumerate}[label=\textup{(\arabic*)},leftmargin=2em]
\item $\Vone\vdash\forall n\,\forall N\,\theta_{q,K}(n,N)$.
\item A polynomial-time function of $1^n,1^\nu$ produces a $\mathbf G_1^\star$ proof of $\theta_{q,K}[n;\nu]$, of size polynomial in $n+\nu$.
\item For a candidate $N$ with $C_{q,K}(n,N)$, a string $x$ can satisfy the matrix of $\theta_{q,K}$ with its auxiliary existential strings if and only if it is a rational $n$-vector of length at most $p(|N|)$ at which the evaluated output of $N$ differs from $\MAX_n(x)$.
Accordingly the existential assignments in the translated formula decode to precisely such error points, with auxiliary computation witnesses.
\end{enumerate}
\end{corollary}
\end{samepage}
\begin{proof}
The predicate $C_{q,K}$ has a $\Delta^B_1$ definition, as does rational evaluation and comparison in $\Err$.
Choose $\Sigma^B_1$ computation representations of $\neg C_{q,K}(n,N)$ and $\Err(N,n,x)$ by expanding their actual chosen defining graphs.
Their polynomial auxiliary length bounds are fixed terms in $n,|N|,|x|$.
The graph existence, uniqueness and value equations ensure both directions of equivalence, in $\Vone$ and in the standard model; these are not arbitrary predicates assumed to compute the same functions.

Start with
\[
\exists x\,\left(|x|\le p(|N|) \wedge
\left(\neg C_{q,K}(n,N) \vee \Err(N,n,x)\right)\right).
\]
Insert the two computation representations, rename their auxiliary strings apart, and pull all their bounded existential strings in front of the disjunction.
Their idle values can be the empty strings in the branch where they are unused.
Replace $|x|$ in auxiliary bounds by $p(|N|)$ to get bounds depending only on the free parameters.
Take the resulting prenex $\Sigma^B_1$ formula as $\theta_{q,K}$.
The displayed length bound for $x$ remains a conjunct of its matrix.
The defining-graph equivalences prove the same equivalence in the original language of $\Vone$ after eliminating the abbreviations.

On $C_{q,K}$, \Cref{cor:fixed-prime-depth-values} supplies the required $x$.
Off $C_{q,K}$, the empty string suffices and the computation representation of $\neg C_{q,K}$ has its proved witnesses.
Thus $\Vone$ proves (1).
Apply \Cref{fact:fp-translation} to this fixed original-language theorem to get (2).
The polynomial bounding the proof size may depend on $q,K$ and on the proof, as allowed by the statement.

For (3), on a candidate the $\neg C_{q,K}$ branch has no computation witnesses, by the established representation equivalence.
The other branch has witnesses exactly when $\Err(N,n,x)$ holds, and $x$'s length bound is explicit.
By the definition of $\Err$, these are exactly the stated rational disagreements.

To check the translation assertion, fix the bits of $N$ and then the length choice and bits for each existential string.
Induction on the bounded formula shows that each translated numerical atom has its arithmetic truth value, each string atom has the specified bit value, Boolean connectives preserve their truth tables, and number quantifiers enumerate their actual finite domains.
A bounded string quantifier enumerates precisely the strings of each allowed length, with the top bit fixed as described above.
Hence a satisfying assignment of the translated matrix is exactly a satisfying arithmetic tuple, and its designated $x$ has the property just proved.
Conversely each such arithmetic tuple supplies that Boolean assignment.
\end{proof}
This corollary gives finite proofs of the fixed-parameter refutation assertions.
It uses a quantified proof system because the counterexample is existentially quantified.
Neither the translation theorem nor its application here establishes a lower bound on proof length or a uniform polynomial in growing $K$.

The same translation applies to the threshold refuter of \Cref{tc:refuter}, through the original-language formula $\theta^{\mathrm{thr}}_{q,K}$ of \Cref{tc:parallel-arithmetic}, which retains the chosen refuter's output.
\begin{samepage}
\begin{corollary}[Quantified propositional proofs retaining the threshold refuter; restatement of \Cref{icor:proofs}]
\label{tc:proofs}
Fix $q,K$.
A polynomial-time function of $1^n$ and $1^\nu$ produces a $\mathbf G_1^\star$ proof of $\theta^{\mathrm{thr}}_{q,K}[n;\nu]$, of size polynomial in $n+\nu$, where the polynomial may depend on $q$, $K$ and the fixed arithmetic proof.
Let an assignment to the free Boolean variables of this translation encode a string $N$ of length $\nu$ that passes the threshold candidate test $C^{\mathrm{thr}}_{q,K}(n,N)$.
Then, whenever an assignment to the existential Boolean variables, together with a choice of length disjuncts, satisfies the translated matrix, its designated string $x$ is the chosen refuter's output, which is a rational disagreement of length at most $p(|N|)$.
Conversely, the chosen output together with its computation witnesses gives such a satisfying assignment.
\end{corollary}
\end{samepage}
\begin{proof}
The formula $\theta^{\mathrm{thr}}_{q,K}$ is a bounded $\Sigma^B_1$ formula of the original two-sorted language, and $\mathbf{VTC}^0$ proves $\forall n\,\forall N\,\theta^{\mathrm{thr}}_{q,K}(n,N)$, as shown where it is defined in \Cref{tc:parallel-arithmetic}.
By \Cref{lem:vtc-in-v1}, $\Vone$ proves the same universal closure.
\Cref{fact:fp-translation} applies with the numerical input $n$ and the free-string length $\nu$ of $N$.
It gives a polynomial-time function of $1^n$ and $1^\nu$ whose value is a $\mathbf G_1^\star$ proof of $\theta^{\mathrm{thr}}_{q,K}[n;\nu]$ of size polynomial in $n+\nu$.
Its polynomial depends on the fixed arithmetic proof, which is fixed once $q$ and $K$ are fixed.

For the decoding assertion, fix an assignment to the free Boolean variables encoding $N$ with $C^{\mathrm{thr}}_{q,K}(n,N)$.
The truth-preservation argument in the last paragraph of the proof of \Cref{cor:proofs} uses only the form of a bounded formula and its translation, so it applies to $\theta^{\mathrm{thr}}_{q,K}$.
It shows that an existential assignment, together with its length choices, satisfies the translated matrix if and only if the decoded tuple $(x,u,w,v)$ satisfies the arithmetic matrix of $\theta^{\mathrm{thr}}_{q,K}$ within the displayed bounds.

The standard two-sorted model satisfies $\mathbf{VTC}^0$, so the representation equivalences and graph uniqueness proved there, and the guarantee of \Cref{tc:refuter}, hold in it.
Since $C^{\mathrm{thr}}_{q,K}(n,N)$ holds, the representation equivalence for $G_{\neg C}$ shows that no $u$ within its bound satisfies $G_{\neg C}(n,N,u)$.
Hence a satisfying tuple satisfies $G_A(n,N,x,w)\wedge G_E(n,N,x,v)$, and the uniqueness of the graph of the chosen refuter, noted after the definition of $\theta^{\mathrm{thr}}_{q,K}$, makes $x$ its output.
By \Cref{tc:refuter} this output is a rational disagreement of length at most $p(|N|)$.
Conversely, the chosen output and its computation witnesses satisfy the arithmetic matrix, as noted after the definition of $\theta^{\mathrm{thr}}_{q,K}$, and therefore supply a satisfying existential assignment of the translation.
\end{proof}
The proved formula $\theta^{\mathrm{thr}}_{q,K}$ retains its existential quantifiers, and replacing its witnesses by chosen functions would give a different formula.
The corollary asserts no bounded-depth threshold proofs.
Cook and Nguyen state a translation of bounded $\mathbf{VTC}^0$ theorems into tree-like quantified threshold proofs whose cuts are quantifier-free of bounded depth \cite[Corollary~10.83]{CookNguyen}.
The theorem on which that corollary rests is given in their draft only in outline, with the details left as an exercise \cite[Theorem~10.82 and Exercise~10.84]{CookNguyen}, and we do not use it.

The construction side of \Cref{uc:section} has the same propositional form.
The formula $\theta^{\mathrm{bal}}(n)$ defined there is a bounded $\Sigma^B_1$ formula of the original two-sorted language with no free string variable, and $\mathbf{VTC}^0$ proves $\forall n\,\theta^{\mathrm{bal}}(n)$.
Applied with the numerical input $n$ and no string input, the proof of \Cref{tc:proofs} gives a polynomial-time function of $1^n$ whose value is a $\mathbf G_1^\star$ proof of the translation $\theta^{\mathrm{bal}}[n]$, of size polynomial in $n$.

The proof of \Cref{tc:proofs} also shows, for $n\ge1$, that every satisfying existential assignment of the translated matrix decodes to a tuple $(N,c,u,v,w)$ satisfying the arithmetic matrix of $\theta^{\mathrm{bal}}(n)$.
The disjunct $n=0$ of $\theta^{\mathrm{bal}}(n)$ is then false.
The chosen graphs of the network constructor $B$ and the certificate constructor $C$ of \Cref{uc:balanced} are unique in the standard model of $\mathbf{VTC}^0$, and this uniqueness makes $N=B(n)$ and $c=C(n)$.
Thus the two proof families retain their respective constructive witnesses: a certified network on the upper side and a disagreement on the lower side.
\space

\phantomsection\addcontentsline{toc}{section}{References}
\begingroup
\makeatletter
\let\addcontentsline\@gobblethree
\makeatother
\newcommand{\etalchar}[1]{$^{#1}$}

\endgroup

\appendix
\crefalias{section}{appendix}\crefalias{subsection}{subappendix}


\section{Short Rational Disagreements}
\label{app:witness}

This appendix connects a bounded predicate on finite strings to agreement on all real inputs.
It is ordinary mathematics about rational networks, not an extra real-field axiom in the arithmetic theory.

We use the rational polyhedral representation theorem in a quantitative form.
For a finite set $X$, its convex hull $\operatorname{conv}(X)$ consists of sums $\sum_{x\in X}\lambda_xx$ with $\lambda_x\ge0$ and $\sum_{x\in X}\lambda_x=1$; these are its convex combinations.
For a finite set $Y$, its cone $\operatorname{cone}(Y)$ consists of nonnegative linear combinations of its vectors.
The sum of two sets, written with $+$, consists of all sums of one point from each.
Points and coefficients here are real, while the vectors of $X$ and $Y$ in the following fact are rational.
\begin{fact}[Quantitative rational polyhedral representation]
\label{fact:rational-polyhedra}
A polyhedron is a solution set of finitely many weak affine linear inequalities.
Let $A$ be a rational matrix and $b$ a rational vector of matching row dimension.
The inequality $Az\le b$ is interpreted coordinatewise.
There are finite rational sets $X,Y$ such that
\[
\{z:Az\le b\}=\operatorname{conv}(X)+\operatorname{cone}(Y),
\]
and every coordinate of every vector in $X\cup Y$ has binary length bounded by a fixed polynomial in the total binary description length of $(A,b)$.
If the polyhedron is nonempty, $X$ is nonempty \cite[Theorems~3, 8 and~9]{Kaibel11}.
\end{fact}
For the coordinate bound, Theorem 8(i) of \cite{Kaibel11} takes every coordinate of $X\cup Y$ to be a ratio of two subdeterminants of $(A,b)$, and Theorem 3 there bounds the binary length of such a ratio by a constant times the square of the number of columns times the largest entry length of $(A,b)$, by the Leibniz formula.
We use this imported theorem only in ordinary mathematics, outside $\mathbf V^1$.

We now apply this fact to the network encoding of \Cref{sec:foundations}.
\begin{lemma}[Short rational disagreement]
\label{lem:short}
There is a polynomial $p_0$ with nonnegative integer coefficients such that, for every network code $N$ with $\Net(N,n)$, if $N\ne\MAX_n$ on $\R^n$, then some valid rational $n$-vector $x$ has code length $|x|$ at most $p_0(|N|)$ and satisfies $N(x)\ne\MAX_n(x)$.
\end{lemma}
\begin{proof}
Let $s=|N|$ and fix a real disagreement point $z$.
Choose one activation pattern, that is, a selection of the linear or the zero branch of every hidden ReLU, consistent with the signs of all preactivations at $z$.
Positive preactivations select the linear branch and negative ones the zero branch, and either branch may be chosen at a zero preactivation.
Replace each ReLU by its selected linear or zero branch.
This gives rational affine expressions for every neuron and output, and weak affine sign inequalities defining a polyhedron $P$ containing $z$.
At every point of $P$ the selected expressions agree with actual evaluation, by induction through the layers; at a zero preactivation both branches have value zero.

The resulting expressions have polynomial coefficient length in $s$.
To verify this without enumerating patterns, take the positive product $D$ of all parameter denominators.
There are at most $s$ factors of at most $s$ bits, so $D$ has $O(s^2)$ bits.
At layer $j$, multiply each affine expression computed there by $D^j$.
Its coefficients are integers: a weight denominator divides $D$, and a bias denominator also divides $D$.

We bound the scaled integer coefficients by induction on the layer.
A coefficient at layer $j$ is formed from at most $s$ products of previous coefficients and integers of $O(s^2)$ bits, together with a scaled bias term.
For a stored bias $a/b$, that term is $a(D/b)D^{j-1}$, an integer because $b$ is a factor of $D$, and has $O(js^2)$ bits.

A sum of at most $s$ integers increases their maximum bit length by at most $O(s)$.
There are at most $s$ layers, so induction bounds every coefficient by a polynomial number of bits in $s$.
The same applies to the affine inequalities describing $P$.
There are at most $s$ hidden sign inequalities and at most $s$ variables.

Choose an index $i$ such that $z_i=\MAX_n(z)$, and let $P'$ be the polyhedron obtained from $P$ by adding the weak inequalities $x_i\ge x_j$ for all $j$.
It is described by at most $2s$ weak rational inequalities in at most $s$ variables, with coefficients of polynomial bit length in $s$, and it contains $z$.
Write $h(x)$ for the selected output affine form, put $g(x)=h(x)-x_i$, and write $g(x)=\ell(x)+g(0)$ with $\ell$ linear.
At every point of $P'$ the coordinate $x_i$ is a maximum coordinate and the selected expressions agree with actual evaluation.
So a point of $P'$ is a disagreement exactly when $g$ is nonzero there; in particular $g(z)\ne0$.

\Cref{fact:rational-polyhedra} gives finite rational sets $X,Y$, with $X$ nonempty, such that $P'=\operatorname{conv}(X)+\operatorname{cone}(Y)$ and every coordinate of their vectors has binary length polynomial in $s$.
Write $z=\sum_{v\in X}\lambda_vv+\sum_{y\in Y}\mu_yy$ with $\lambda_v,\mu_y\ge0$ and $\sum_{v\in X}\lambda_v=1$.
Since the $\lambda_v$ sum to one,
\[
0\ne g(z)=\sum_{v\in X}\lambda_vg(v)+\sum_{y\in Y}\mu_y\ell(y).
\]

If some $v\in X$ has $g(v)\ne0$, then $v$ is a rational disagreement.
Otherwise $g(v)=0$ for every $v\in X$, so some $y\in Y$ has $\ell(y)\ne0$.
Fix any $v_0\in X$.
Then $v_0+y\in P'$ and $g(v_0+y)=g(v_0)+\ell(y)=\ell(y)\ne0$, so $v_0+y$ is a rational disagreement.
In the second case, a coordinate of $v_0+y$ is a sum of two rationals whose numerators and denominators have length at most some $L$ polynomial in $s$, and the common-denominator sum has numerator and denominator of length at most $2L+3$.
In either case the disagreement has ordinary numerator and denominator lengths polynomial in $s$.

Converting the component numerators and positive denominators of the disagreement to the canonical signed codes, then to rational pairs and the $n$-vector array, uses only the fixed quadratic pairing a fixed number of times.
Because $n\le s$, the total length of the resulting string remains polynomial in $s$.
Choose a polynomial upper bound with nonnegative integer coefficients and call it $p_0$.
This polynomial is chosen independently of the prime, depth cap, probe algorithms and the cutoff $p$.
Since the cutoff $p$ is at least $p_0$, the bounded tests with cutoff $p$ characterize agreement on all real inputs for every network.
This holds whether or not $\mathsf{GoodDen}_q$ or $\mathsf{ValGood}_q$ holds, for any prime $q$.
\end{proof}
\space

\section{Rational Block Certificates and Dense Assembly}
\label{rc:appendix}

This appendix proves \Cref{rc:upper}.
It first gives a finite checker for rational two-layer blocks.
It then defines guarded direct arity selection, literal network matrices, and the ordinary histories used by the uniform family.
An ordinary history of a network at an input is its history vector of \Cref{sec:foundations}, with one rational entry for each hidden neuron and one for the output, in the order fixed by the network record.
The predicate \(\mathsf{LEval}\) checks the equations that link each layer of such a history to the one before it.

The arithmetic theory is the conservative extension \(\bm{\mathcal T}\) of \(\mathbf{VTC}^0\) used in \Cref{uc:section}.
The arithmetic operations and their exact prefix equations are those of \Cref{tc:arithmetic} and the shifted signed and rational codes of \Cref{sec:foundations}.
All new functions below are defined by bounded-number comprehension, minimization, aggregate closure, and a fixed number of compositions of those chosen arithmetic functions.
Induction is bounded-number induction with constructed strings as parameters; the limited-recursion principle for arbitrary polynomial-time functions is not used.

\subsection{Closed Cones and Their Finite Certificates}
\label{rc:cones}

We use zero-based indices for neurons, matrix entries, original coordinates and nodes.
An exact tuple still places its entries in rows \(1,\ldots,r\), with empty row zero.
Below, a row of a matrix, of a cone list or of a coefficient table is a row vector, while a row of an exact tuple or array is one of its entry positions.
Because indices are zero-based while exact tuples start at row one, a coordinate \(i<n\) is addressed in two ways.
In a label, which is a string of length at most \(n\) coding a set of coordinates as in \Cref{uc:section}, it is bit \(i\); in an input tuple, its rational value is tuple entry \(i+1\).
This is the coordinate numbered \(i+1\) in the main text; \(\MAX_n\) is unchanged by this local reindexing.

Write \(e_i\) for the coordinate row with entry one at index \(i\) and zero at every other index; its length is the dimension \(b\) of the rows it is combined with, and \(i<b\).
A rational vector of length \(b\) is an exact tuple of \(b\) valid rational pairs.
An \(a\)-by-\(b\) rational matrix uses entry-row index \(\langle i,j\rangle\) for \(i<a\) and \(j<b\), with zero unused bits.
If either matrix dimension is zero, its exact code is the empty string.
All equality of coefficients below means rational-value equality unless exact bit reconstruction is explicitly required.

For a finite list \(G=(g_0,\ldots,g_{r-1})\) of rational rows in \(\Q^b\), its closed cone is
\[
\{x\in\R^b: g_j\cdot x\ge0\text{ for every }j<r\}.
\]
The empty list represents \(\R^b\).
A block \(Q\) has exactly \(b\ge2\) inputs and positive hidden widths \(\mathfrak{w}_1,\mathfrak{w}_2\).
Its matrices are \(A\in\Q^{\mathfrak{w}_1\times b}\) and \(B\in\Q^{\mathfrak{w}_2\times \mathfrak{w}_1}\), and its output row is \(c\in\Q^{\mathfrak{w}_2}\).
All its biases have zero rational value, and
\[
Q(x)=c\relu(B\relu(Ax)).
\]
The complete network record is the common record fixed in the paragraph ``A Fixed Network Record'' of \Cref{sec:foundations}; in particular all dense zero entries and all bias entries are stored.

A cone certificate \(d\) is the exact two-entry tuple \((\operatorname{Un}(t),\mathsf{Nodes})\), where the number of nodes \(t\) satisfies \(1\le t\le|d|\).
The second entry is an exact array with node rows \(0,\ldots,t-1\).
A node is the exact triple
\[
(\operatorname{Un}(\mathit{tag}),\mathsf{Cone},\mathsf{Payload}),
\qquad \mathsf{Cone}=(\operatorname{Un}(r),G).
\]
Here \(G\) is an exact \(r\)-by-\(b\) rational matrix with \(r\le|d|\).
The only tags are zero for an internal node and one for a leaf.
Node zero has \(r=0\).
An internal payload is the exact triple \((w,\operatorname{Un}(p),\operatorname{Un}(q))\).
Its row \(w\) has length \(b\), and the child indices satisfy \(p,q<t\) and \(p\ne q\).
The cone list of the first child \(p\) is exactly \(G\) followed by \(w\), and that of the second child \(q\) is exactly \(G\) followed by \(-w\).
The code for \(-w\) negates each canonical signed numerator and retains its denominator.
These child identities are bitwise identities of the specified reconstructed lists.
Zero test rows are allowed.

A leaf payload is the exact six-entry tuple
\[
(\chi,\psi,\Lambda_1,\Lambda_2,\Lambda_+,\Lambda_-).
\]
The entries $\chi$ and $\psi$ are bit patterns that satisfy \(|\chi|\le \mathfrak{w}_1,|\psi|\le \mathfrak{w}_2\); missing trailing bits are zero.
Their lengths need not equal their dimensions, so patterns whose final bits are zero, including the all-zero pattern, are allowed.

A bit $\chi_i=1$ chooses the nonnegative branch of the preactivation $(Ax)_i$ of first-layer neuron $i$, while zero chooses its nonpositive branch; $\psi_j$ makes the same choice for the preactivation $(B\relu(Ax))_j$ of second-layer neuron $j$.
Either branch is allowed at value zero.
The multiplier identities below certify that these chosen signs hold throughout the node's cone.
The multiplier matrices \(\Lambda_1\) and \(\Lambda_2\) have dimensions \(\mathfrak{w}_1\)-by-\(r\) and \(\mathfrak{w}_2\)-by-\(r\).
Each of \(\Lambda_+\) and \(\Lambda_-\) has dimension \(b\)-by-\((r+b)\).
Every multiplier is a valid nonnegative rational.

Write \(A_i\) for row \(i\) of \(A\).
For \(i<\mathfrak{w}_1\) and \(j<\mathfrak{w}_2\), the checker computes, rather than trusts, the following coefficient rows in \(\Q^b\):
\[
\alpha_i=A_i,\quad \beta_i=\chi_i\alpha_i,\quad
\gamma_j=\sum_{i<\mathfrak{w}_1}B_{ji}\beta_i,\quad
\delta_j=\psi_j\gamma_j,\quad f=\sum_{j<\mathfrak{w}_2}c_j\delta_j.
\]
Multiplication by a bit selects either the same row or the zero row.
The signs \(2\chi_i-1\) and \(2\psi_j-1\) mean the chosen signed constants \(1\) or \(-1\), according to the bit; no negative number-sort subtraction is used.
Every sum is the chosen parallel rational sum in its displayed index order.
The multiplier matrices \(\Lambda_1\) and \(\Lambda_2\) must satisfy
\begin{equation}
\label{rc:sign-identities}
\begin{aligned}
(2\chi_i-1)\alpha_i&=\sum_{k<r}(\Lambda_1)_{ik}g_k,\\
(2\psi_j-1)\gamma_j&=\sum_{k<r}(\Lambda_2)_{jk}g_k.
\end{aligned}
\end{equation}
For each candidate winner \(a<b\), let \(G^{(a)}\) be \(G\) followed, in order, by \(e_a-e_0,\ldots,e_a-e_{b-1}\).
Write \(G^{(a)}_k\) for its row \(k\), so that \(G^{(a)}_k=g_k\) for \(k<r\) and \(G^{(a)}_{r+i}=e_a-e_i\) for \(i<b\).
The remaining identities are
\begin{equation}
\label{rc:winner-identities}
\begin{aligned}
f-e_a&=\sum_{k<r+b}(\Lambda_+)_{ak}G^{(a)}_k,\\
e_a-f&=\sum_{k<r+b}(\Lambda_-)_{ak}G^{(a)}_k.
\end{aligned}
\end{equation}
Empty sums are zero; a zero cone row or a zero left-hand side in these identities causes no exception.
All \(b\) winner indices are checked.

Define \(\mathsf{CC}(b,Q,d)\) to require the valid block, the exact certificate schema, and every internal and leaf identity just specified.
It rejects an unknown tag, nonunary index, malformed rational, extra row or nonpair bit, incorrect dimension, or out-of-range child before using that field.
It need not check reachability from node zero.
The strict increase of row count along every edge rules out cycles, but the soundness proof does not require an algorithm for graph traversal.

The arithmetic needed for this definition, and for the consistency tests of a rational \(b\)-vector \(x\) against the node cones in the proof of \Cref{rc:cc-lemma}, has fixed composition depth.
Put \(s=b+|Q|+|d|+|x|+2\), with \(|x|=0\) for \(\mathsf{CC}\) itself.
Every guarded count and input component is at most \(s\).
Every flattened coefficient table, and the membership table recording which node cones contain \(x\), has at most \(s^4\) entries.
Each scalar computation uses fewer than eight successive rational-product or rational-sum stages.

A product of \(v\) rationals of component length \(L\) has component length at most \(v(L+2)+2\).
A sum of \(v\) already formed rationals of component length \(L\), using a product denominator and excluded-entry products, has component length at most \(v(L+2)+2v+2\).
These follow by adding the bit lengths in the integer products and using the signed-sum bound.
With \(v\le s\), eight such stages are bounded by \(100(s+2)^{20}\).

For returned strings put
\[
\mathfrak{Q}_0(s)=100(s+2)^{20},\qquad
\mathfrak{Q}_{i+1}(s)=\langle \mathfrak{Q}_i(s)+2,\mathfrak{Q}_i(s)+2\rangle+2\quad(0\le i<16).
\]
These are sixteen fixed term compositions.
They dominate rational pairs, coefficient rows, cone and payload tuples, the flat node array, the outer certificate tuple, and flattened auxiliary tables.
Matrix coordinate pairing changes the polynomial index bound, rather than adding a variable number of tuple levels.
The arithmetic graphs retain their own polynomial computation-witness bounds; compose those bounds a fixed number of times, separately from \(\mathfrak{Q}_{16}\).
All guards are tested before allocation.
Thus \(\mathsf{CC}\) is a chosen total \(\mathsf{FTC}^0\) predicate with all-string polynomial graph bounds.

\subsection{Soundness and Fixed-Depth Ordinary Histories}
\label{rc:cc-sound}

The checker obtains its soundness from ordered rational arithmetic, rather than from an internal linear-programming algorithm.
The proof selects a leaf by comparing all supplied node cones independently, without traversing the certificate.

\begin{lemma}[Soundness of a Cone Certificate]
\label{rc:cc-lemma}
If \(\mathsf{CC}(b,Q,d)\) holds, then \(Q(x)=\MAX_b(x)\) for all real \(x\).
There is a chosen ordinary two-hidden-layer history function \(\mathsf{Hist}_2(Q,x)\) with the following property.
The theory \(\bm{\mathcal T}\) proves that, for every valid rational network \(Q\) of hidden depth two and input dimension \(b\), every valid rational \(b\)-vector \(x\) and every string \(d\), \(\mathsf{LEval}(Q,b,x,\mathsf{Hist}_2(Q,x))\) holds and, when \(\mathsf{CC}(b,Q,d)\) also holds, the output of this history equals \(\MAX_b(x)\).
The history is defined for every valid rational network of hidden depth two, independently of \(\mathsf{CC}\).
\end{lemma}
\begin{proof}
Call a node consistent at \(x\) when all its cone rows have nonnegative dot product with \(x\).
For rational inputs all these tests are computed in parallel.
The selection and the identities below use only ordered-field arithmetic on the finitely many supplied rows, so for real \(x\) the same argument proves the first sentence of the lemma in the standard model; inside \(\bm{\mathcal T}\) it is carried out for valid rational \(x\).
The root is consistent because its list is empty.
Among consistent nodes choose one with greatest row count, breaking ties by least node index.
This is a bounded numerical maximum and minimum over the supplied finite array.

The selected node cannot be internal.
If its test row \(w\) satisfies \(w\cdot x\ge0\), its first child is consistent and has one more cone row.
If \(w\cdot x<0\), its second child is consistent because \((-w)\cdot x>0\), again with one more row.
Both cases contradict maximality.
At equality the first case applies, even if both children are consistent.
Consequently the selected node is a leaf.
This proof uses neither parent pointers nor recursive traversal.

Fix that leaf, a first-layer neuron \(i<\mathfrak{w}_1\) and a second-layer neuron \(j<\mathfrak{w}_2\).
Dotting the leaf's first identity in \Cref{rc:sign-identities} with \(x\) gives \((2\chi_i-1)\,\alpha_i\cdot x=\sum_{k<r}(\Lambda_1)_{ik}\,(g_k\cdot x)\ge0\), since every multiplier is nonnegative and every cone row \(g_k\) of the consistent leaf satisfies \(g_k\cdot x\ge0\).
Hence \(\alpha_i\cdot x\ge0\) when \(\chi_i=1\) and \(\alpha_i\cdot x\le0\) when \(\chi_i=0\), which is the sign chosen by \(\chi_i\).
When \(\chi_i=1\), the rectified value of first-layer neuron \(i\) is \(\alpha_i\cdot x=\beta_i\cdot x\); when \(\chi_i=0\), it is \(0=\beta_i\cdot x\).
Both statements hold at zero.
As \(i\) was arbitrary, distributivity identifies the preactivation \((B\relu(Ax))_j\) of second-layer neuron \(j\) with \(\gamma_j\cdot x\).
The second identity in \Cref{rc:sign-identities}, dotted with \(x\) in the same way, gives \((2\psi_j-1)\,\gamma_j\cdot x\ge0\), the sign chosen by \(\psi_j\), so the rectified value of second-layer neuron \(j\) is \(\delta_j\cdot x\).
As \(j\) was arbitrary, the output \(c\relu(B\relu(Ax))\) is \(\sum_{j<\mathfrak{w}_2}c_j\delta_j\cdot x=f\cdot x\).

Choose a maximizing coordinate \(a\) of \(x\).
It exists in the finite nonempty vector, and all rows of \(G^{(a)}\) have nonnegative dot product with \(x\).
The two identities in \Cref{rc:winner-identities} give \(f\cdot x-x_a\ge0\) and \(x_a-f\cdot x\ge0\).
Antisymmetry yields \(f\cdot x=x_a=\MAX_b(x)\).
Every winner is checked, so ties require no additional choice convention.

We now construct the ordinary history for rational inputs.
Let \(D\) be the product of all parameter denominators, including those of the biases, and let \(E\) be the product of input denominators.
In the rest of this proof \(U_\lambda\), \(A_{ji}\) and \(\bar A_{ji}\) are the cleared integers of \(\mathsf{Eval}_2\) in \Cref{tc:parallel-arithmetic}, with its one-based neuron indices shifted to the zero-based indices of this section.
They are not the function \(U\) of \Cref{rc:selection} or the block matrix \(A\).
For parameter numerator \(\nu_\lambda\) and denominator \(\kappa_\lambda\), form its cleared integer \(U_\lambda=\nu_\lambda\prod_{\mu\ne\lambda}\kappa_\mu\).
For input \(x_i=p_i/\epsilon_i\), form \(A_{0i}=p_i\prod_{h\ne i}\epsilon_h\).

For the three affine stages \(j=1,2,3\), use the exact integer equations
\[
\bar A_{ji}=\sum_h U_{jih}A_{j-1,h}
+U_{ji,\mathrm{bias}}D^{j-1}E,\qquad
A_{ji}=
\begin{cases}
\max(0,\bar A_{ji})&j=1,2,\\
\bar A_{3i}&j=3.
\end{cases}
\]
The last stage has a single output.
Return the exact flat vector of the pairs \(A_{1i}/(DE)\) over the first-layer neurons \(i\), then \(A_{2i}/(D^2E)\) over the second-layer neurons \(i\), then \(\bar A_{30}/(D^3E)\).

Positive denominators let rectification commute with this scaling.
Dividing the displayed equation by \(D^jE\) proves the corresponding adjacent affine equation.
There are only three stages, so this is a fixed composition of the admitted arithmetic and array constructors.
Its last pair is exactly the scalar convention of \(\mathsf{Eval}_2\); the history lists all three stages and not only this scalar.
There is no coprimality or denominator-freeness restriction.

For explicit bounds, let \(m=|Q|\) and \(z=|x|\), and put
\[
B_D=2+m(m+3),\quad B_E=2+z(z+3),\quad L_0=m+z+B_D+B_E+10,
\]
\[
\bar L=5(B_D+B_E+m+z+10)+2(m+1),\qquad L_j=L_0+j\bar L\quad(j\le3).
\]
The products \(D,E\) have component lengths at most \(B_D,B_E\); cleared parameters have length at most \(m+B_D+1\), and cleared inputs at most \(z+B_E+1\).
A weighted predecessor adds at most \(m+B_D+2\) bits, and a sum has at most \(m\) terms.
The bias bound is \(m+B_D+1+(j-1)(B_D+1)+B_E+2\).
The displayed \(\bar L\) dominates these increases, proving the \(L_j\) bounds at all three stages.
Exact flat reconstruction then bounds the return by a fixed polynomial; invalid networks, wrong depth, malformed inputs or an input dimension different from the block's declared dimension return empty before allocation.

For the internal proof, all consistency tests, coefficient sums and selections use these chosen terms.
Existence of a greatest finite entry follows by bounded-number induction with the already formed table as parameter.
Cross-products reduce every equality or order comparison to the signed ordered ring.
Multiplication of nonnegative quantities and distributivity justify the dotted multiplier identities.
No quantified-string induction or internal Farkas existence is needed.
The same finite argument proves the claimed rational statement in \(\bm{\mathcal T}\).
\end{proof}

This soundness argument is compatible with cone certificates having unreachable nodes, zero test rows, flat activation regions and lower-dimensional intersections with winning-coordinate cones.
All nodes are checked, and the maximum-row-count selector remains valid in each case.
For the selected blocks, \Cref{ir:cones} constructs these certificates inside \(\bm{\mathcal T}\).
The soundness proof here uses only the stored rational identities.

\subsection{The Finite Integer-Feature Dictionary}
\label{rc:effective}

The direct construction uses a raw generating list for Zimmermann's finite dictionary.
Each generator carries a tag: either its generating data \((\pi,S,\tau)\) below, or the name of a coordinate hinge or a pair-difference hinge (the functions \(\relu(\pm x_i)\) and \(\relu(x_i-x_j)\) appended below).
Through the formulas below, a tag determines the rows of its generator, so every generator is given explicitly by its tag, and generators with different tags are kept apart even when they define the same function.
For $b\ge3$, choose a permutation $\pi$ of the coordinates $0,\ldots,b-1$ and a nonempty set $S\subseteq\{1,\ldots,b-2\}$.
Put $q=\max S$ and $p(j)=\max(\{0\}\cup(S\cap\{1,\ldots,j-1\}))$.
The head row and edge rows are
\[
h=e_{\pi(q+1)}-e_{\pi(0)},\qquad
l_j=e_{\pi(p(j)+1)}-e_{\pi(j+1)}\quad(1\le j\le b-2).
\]
Position $j$ carries coordinate $\pi(j+1)$, and position zero together with the positions in $S$ are called marked.
The edge $l_j$ joins coordinate $\pi(j+1)$ to the coordinate $\pi(p(j)+1)$ of its nearest preceding marked position $p(j)$.
So the coordinates at marked positions form a spine, the other non-apex coordinates attach to it as pendant vertices, and $\pi(0)$ is the separate apex.
These are the caterpillar descriptions of \cite{Zim26}.
A ternary state is a map $\tau$ from the edge indices $1\le j\le b-2$ to $\{\mathrm{omit},\mathrm{linear},\mathrm{hinge}\}$; its value $\tau_j$ omits edge $l_j$, keeps $l_j\cdot x$ linearly, or keeps $\relu(l_j\cdot x)$, respectively.

For generating data \((\pi,S,\tau)\) as above, the feature \(g_{\pi,S,\tau}\) uses the vector \(a\) that collects the head row \(h\) and the edges \(l_j\) that \(\tau\) keeps linearly:
\[
g_{\pi,S,\tau}(x)=\relu\left(a\cdot x+
\sum_{j:\tau_j=\mathrm{hinge}}\relu(l_j\cdot x)\right),
\qquad a=h+\sum_{j:\tau_j=\mathrm{linear}}l_j.
\]
Append the coordinate hinges $\relu(\pm x_i)$ for $i<b$ and the pair-difference hinges $\relu(x_i-x_j)$ for $i<j<b$.
Retain every raw $(\pi,S,\tau)$ and auxiliary tag, including aliases that define the same function.

\begin{fact}[Finite dictionary for the maximum, {\cite[Definitions~3--5 and Theorem~7]{Zim26}}]
\label{rc:dictionary}
For every \(b\ge3\), there is an enumerable tagged list \(g_0,\ldots,g_{K_b-1}\) whose real span contains \(\MAX_b\), with
\[
K_b\le b!(2^{b-2}-1)3^{b-2}+2b+\binom b2
=2^{O(b\log(b+1))}.
\]
Each feature is a coordinate hinge, a pair-difference hinge, or
\[
g(x)=\relu\left(a\cdot x+\sum_{l\in M}\relu(l\cdot x)\right),
\qquad |M|\le b-2,
\]
where \(M\) is a set of signed coordinate differences and \(a\in\Z^b\) has \(|a_i|\le b\).
\end{fact}

The span statement is Zimmermann's Theorem~7 for the function dictionary of Definitions~3--5~\cite{Zim26}.
Our raw list maps onto that function set; its tags are finer than Zimmermann's collected adjusted-head and retained-edge-set tags.
Each contribution is assigned to its own raw generating tag, so no division among aliases is needed.
There are at most $b-2$ linear edges and one head difference, so every entry of $a$ has magnitude at most $b$.
\Cref{ir:appendix} adapts the deletion and correction strategy of Zimmermann's Step~7.1 to construct rational coefficients directly from finite face identities and integer subset-chain sums.
No rational precision bound is inferred from Zimmermann's real coefficient prescription.

\subsection{Guarded Conversion and Arity Selection}
\label{rc:selection}

Write \(C_B(n)\) for the certificate function that \Cref{uc:balanced} calls \(C(n)\), keep the name \(B(n)\) of its binary network function, and write \(H_B(n,N,c,x)\) for the ordinary-history function in \Cref{uc:assemblies}; the network \(B(n)\) is unrelated to the block matrix \(B\).
The binary family and the direct construction below use the common explicit network record.
The binary family supplies a fallback whenever the direct arity guard admits no block.

We spell out the numerical guards, since a large binary power is not itself a number-sort allocation bound.
Recall from \Cref{uc:powers} that \(U(v)\) is the canonical signed integer given by the admitted sum of \(v\) copies of one.
Its prefix equation proves
\[
U(0)=0,\quad U(v+1)=U(v)+1,\quad
U(u+w)=U(u)+U(w),\quad U(uw)=U(u)U(w)
\]
whenever the arguments lie in the polynomially bounded tables being used.
For addition, induct on \(w\) using the successor equation.
For multiplication, distribute the sum of \(w\) copies of \(U(u)\), or induct on that bounded factor.
In particular \(U\) is strictly increasing and injective.

For a number \(M\), the cap, there is a guarded conversion from an integer string \(X\) satisfying \(0\le X\le U(M)\) to a number at most \(M\).
Choose the least \(v\le M\) with \(X\le U(v)\).
If \(v=0\), then \(X=0\).
Otherwise minimality and the successor equation give \(U(v-1)<X\le U(v-1)+1\), so discreteness forces \(X=U(v)\).
This proves existence and uniqueness.
The function uses bounded minimization, returns zero on a failed guard, and never converts an integer larger than its stated cap.

Aggregate the \(U\) table through \((n+2)^4\).
For \(n\ge1\), form, as in \Cref{uc:powers}, the complete binary-string power table
\[
P_2(j)=\operatorname{IMUL}(\underbrace{2,\ldots,2}_{j\text{ entries}})
\quad(0\le j\le n+1).
\]
Its component lengths are \(O(n)\), including powers whose values exceed \(n\).
The recurrence \(P_2(0)=1,P_2(j+1)=2P_2(j)\) proves \(P_2(j)\ge U(j+1)\) by bounded-number induction.

Let \(L\) be the largest \(j\le n\) with \(P_2(j)\le U(n)\).
It exists, and \(P_2(n+1)>U(n)\), so
\[
P_2(L)\le U(n)<P_2(L+1).
\]
Bounded-number induction on \(j\), using the recurrence and \(P_2(j)\ge0\), proves \(P_2(j)\le P_2(j+1)\), so \(0\le P_2(j)\le P_2(L)\le U(n)\) for every \(j\le L\).
For \(j\le L\), convert \(P_2(j)\) with cap \(n\), obtaining \(r_j\le n\).
Injectivity and the recurrence give \(r_0=1,r_{j+1}=2r_j\) while both indices are in range.
No power with \(j>L\) is converted.

This \(L\) is the internal value \(\ell(n)\) used in \Cref{rc:upper}, distinct from the ceiling exponent \(L(n)\) of \Cref{uc:powers}; externally \(L=\lfloor\log_2 n\rfloor\) and \(r_j=2^j\).
At \(n=0\) put \(\ell(0)=0\), as in \Cref{ir:appendix}.
For \(n\ge2\) the definition gives \(\ell(n)\ge1\), since \(P_2(1)=2=U(2)\le U(n)\).
Hence \Cref{ir:rate}, which evaluates \(\ell\) only at \(n\ge2\) and at \(\ell(n)\), never uses the value \(\ell(0)\).

Take the fixed standard integers \(A=771\) and \(r_\ast=5\), whose allocation bounds are proved in \Cref{ir:packets}; the integer \(A\) is unrelated to the matrix \(A\) of a block.
For each \(2\le b\le n\), let \(\Gamma_b=P_2(A(b+1)^{r_\ast})\) be the binary comparison string, an entry of the same IMUL power table extended through the index bound \(A(n+1)^{r_\ast}\).
The following statement supplies the chosen block functions before we use them in the arity selector.
\begin{lemma}[Uniform direct supply of certified blocks]
\label{ir:small}
For the fixed values \(A=771,r_\ast=5\), there are chosen total \(\mathsf{FTC}^0\) functions \(Q_{\mathrm{dir}}(n,b),D_{\mathrm{dir}}(n,b)\) such that \(\bm{\mathcal T}\) proves the following.
If \(2\le b\le n\) and the guarded comparison \(\Gamma_b\le U(n)\) holds, they return a homogeneous rational two-hidden-layer block \(Q\) and a certificate \(d_b\) with
\[
\mathsf{CC}(b,Q,d_b),\qquad U(b+2|Q|+2|d_b|+4)\le\Gamma_b\le U(n),\qquad b+2|Q|+2|d_b|+4\le n.
\]
Off the guard they return empty strings.
\end{lemma}
The proof follows the direct construction and allocation bounds in \Cref{ir:packets}.
The arity selector tests \(\Gamma_b\le U(n)\) for each candidate arity \(b\).
The comparison string \(\Gamma_b\) has length polynomial in unary \(n\), including on a failing candidate.

Select the greatest passing \(b\), using a separate found flag.
Only under a passing guard call \(Q_{\mathrm{dir}}(n,b),D_{\mathrm{dir}}(n,b)\).
By \Cref{ir:small}, they supply a block \(Q\) and a certificate \(d\) with \(\mathsf{CC}(b,Q,d)\) and \(b+2|Q|+2|d|+4\le n\).
Thus their widths and scalar lengths have the required numerical bounds.
All comparisons and choices are bounded numerical tests and a fixed aggregate of the chosen functions.

At \(n=0\), return the declared empty defaults.
At \(n=1\), return \(\mathsf{Id}_1\), its empty certificate and one-output identity history.
If \(n\ge2\) has no passing arity, return \(B(n),C_B(n)\) with the fallback wrapper below.
Otherwise use the selected block in the dense compiler.
\subsection{Assembly Powers, Labels and Dense Matrices}
\label{rc:compiler}

Suppose the direct arity guard admits a block, with \(2\le b\le n\), block \(Q\), certificate \(d\), and positive widths \(\mathfrak{w}_1,\mathfrak{w}_2\).
The code length gives \(\mathfrak{w}_1+\mathfrak{w}_2\le|Q|\le n\).
Form, in parallel, the binary-string powers \(P_b(j)=\operatorname{IMUL}(j\text{ copies of }U(b))\), \(0\le j\le n\).
Their component length is at most \(2+j(2b+3)\), a polynomial in unary \(n\).
The recurrence \(P_b(0)=1\), \(P_b(j+1)=U(b)P_b(j)\) and \(b\ge2\) give \(P_b(j)\ge U(j+1)\), by bounded-number induction on \(j\).
The base case is \(P_b(0)=1=U(1)\); at a successor, \(U(b)\ge U(2)=2\) and \(P_b(j)\ge U(j+1)\ge1\) give \(P_b(j+1)=U(b)P_b(j)\ge2P_b(j)\ge P_b(j)+1\ge U(j+2)\).

Among these powers, let \(h\le n\) be the least exponent with \(P_b(h)\ge U(n)\).
It exists, is positive, and
\[
U(n)\le P_b(h)=U(b)P_b(h-1)<U(bn).
\]
Only the powers with index at most \(h\) are converted to numbers, using cap \(bn\) and the \(U\) table through that polynomial bound.
The guard of each conversion holds, since \(P_b(j+1)=U(b)P_b(j)\ge P_b(j)\ge0\) gives \(P_b(j)\le P_b(h)<U(bn)\) for \(j\le h\) by bounded-number induction.
Denote them by \(\omega_j\).
Then \(\bm{\mathcal T}\) proves
\[
\omega_0=1,\quad \omega_{j+1}=b\omega_j\text{ for }j<h,\qquad
\omega_{h-1}<n\le m:=\omega_h<bn\le n^2.
\]
The bound \(\omega_{h-1}<n\) holds because minimality of \(h\) gives \(U(\omega_{h-1})=P_b(h-1)<U(n)\), and \(U\) is strictly increasing.
On a request \(j>h\) the numerical-power function returns zero.
The unconverted power table still has its polynomial string cap on every input.

Set \(\theta_r=\omega_{h-r}\) for \(0\le r\le h\).
Level zero consists of the \(m\) leaves \((0,j)\) for \(j<m\), of which those with \(n\le j<m\) are padding leaves; level \(r\) has the nodes \((r,j)\) for \(j<\theta_r\).
A node \((r,j)\) with \(1\le r\le h\) is an internal node of the assembly tree, and its children are \((r-1,bj+t)\) for \(t<b\).
The power equations give \(\theta_{r-1}=b\theta_r\).
Bounded induction proves \(\omega_r\theta_r=m\).
At a successor step with \(r<h\), \(\omega_{r+1}\theta_{r+1}=b\omega_r\theta_{r+1}=\omega_r\theta_r\), because \(\omega_{r+1}=b\omega_r\) and \(\theta_r=b\theta_{r+1}\).
All factors are numbers bounded by \(n^2\).

Define \(\pi(i)=i\) for \(i<n\) and \(\pi(i)=0\) for \(n\le i<m\).
The exact label of node \((r,j)\) is
\[
S_{r,j}=\{\pi(i): j\omega_r\le i<(j+1)\omega_r\}.
\]
Its bit \(a<n\) is the bounded assertion that such an \(i<m\) has \(\pi(i)=a\).
These intervals are nonempty and lie within the padded leaf range.
Division of \(i-j\omega_r\) by the positive \(\omega_{r-1}\) proves that the \(b\) child intervals partition the parent.
The quotient is less than \(b\), and the remainder is less than \(\omega_{r-1}\); both follow from the interval length \(b\omega_{r-1}\).
Thus
\[
S_{0,j}=\{\pi(j)\},\qquad
S_{r,j}=\bigcup_{t<b}S_{r-1,bj+t},\qquad S_{h,0}=\{0,\ldots,n-1\}.
\]
A wholly padded subtree has label \(\{0\}\), never the empty set.

For a numerical count bound, form \(J_q=\sum_{s<q}\omega_s\) by the admitted prefix sums and guarded conversion with cap \(qbn\).
Its recurrence proves \((b-1)J_q+1=\omega_q\) by bounded induction.
At \(q=0\), both sides are one.
At a successor, \((b-1)(J_q+\omega_q)+1=\omega_q+(b-1)\omega_q=b\omega_q=\omega_{q+1}\).
The number of internal nodes is \(I=\sum_{r=1}^h \theta_r=J_h\), so \((b-1)I+1=m<bn\).
If \(I\ge2n\), then \(2n(b-1)+1\ge bn+1\), a contradiction.
Therefore \(I<2n\).

Write the selected block entries as \(A_{\alpha t}\), \(B_{\beta\alpha}\), \(c_\beta\).
Let \(\operatorname{Sum}_{\Q}\) and multiplication denote the actual chosen unreduced rational operations, in increasing listed index order.
The global hidden depth is \(2h\), and the layer dimensions are
\[
d_0=n,\quad d_{2r-1}=\theta_r\mathfrak{w}_1,\quad d_{2r}=\theta_r\mathfrak{w}_2\text{ for }1\le r\le h,
\quad d_{2h+1}=1.
\]
All dimensions are positive.
The dense matrices are specified by the following exact code equations.
At the bottom, for \(j<\theta_1\), \(\alpha<\mathfrak{w}_1\) and \(i<n\),
\begin{equation}
\label{rc:bottom}
W_1[j\mathfrak{w}_1+\alpha,i]
=\operatorname{Sum}_{\Q}\left(V_t:t<b\right),\qquad
V_t=\begin{cases}A_{\alpha t}&\pi(bj+t)=i,\\0_{\Q}&\text{otherwise}.\end{cases}
\end{equation}
This adds every collision at the same actual input column; keeping just one matching padded slot would be incorrect.
An all-zero sum has exactly the unreduced code given by the chosen sum function.
For \(1\le r\le h\), \(j,k<\theta_r\), \(\alpha<\mathfrak{w}_1\) and \(\beta<\mathfrak{w}_2\),
\begin{equation}
\label{rc:within}
W_{2r}[j\mathfrak{w}_2+\beta,k\mathfrak{w}_1+\alpha]
=\begin{cases}B_{\beta\alpha}&k=j,\\0_{\Q}&k\ne j.\end{cases}
\end{equation}
For \(2\le r\le h\), \(j<\theta_r\), \(k<\theta_{r-1}\), \(\alpha<\mathfrak{w}_1\) and \(\beta<\mathfrak{w}_2\),
\begin{equation}
\label{rc:boundary}
W_{2r-1}[j\mathfrak{w}_1+\alpha,k\mathfrak{w}_2+\beta]
=\begin{cases}
A_{\alpha t}c_\beta&k=bj+t\text{ for some }t<b,\\
0_{\Q}&\text{otherwise}.
\end{cases}
\end{equation}
The slot \(t\) in this expression is unique.
The final affine row is \(W_{2h+1}[0,\beta]=c_\beta\), since \(\theta_h=1\).
Every global bias is exactly \(0_{\Q}\).
Zero biases stored in the block need only have zero rational value; their denominators remain in its local histories.
When \(h=1\), \Cref{rc:boundary} has no instances.

Construct the network with the common metadata and parameter offsets from these equations.
Every dense matrix entry and every bias has one valid prescribed rational code; all unused bits are zero.
The boundary entry contains a product of two block coefficients, not a growing product along a path.
Consequently the compiler is a bounded bit formula with a fixed number of arithmetic/array stages, and exact reconstruction proves \(\Net\).
It constructs a literal layered network, including its zeros; it does not merely describe an equivalent sparse graph.

For later all-string bounds, in the block branch
\[
q_{\mathrm{hid}}=(\mathfrak{w}_1+\mathfrak{w}_2)I<2n^2,\qquad 2h\le2n,\qquad
M\le100(n+2)^6,
\]
for the hidden-neuron count \(q_{\mathrm{hid}}\), the hidden depth \(2h\) and the parameter count \(M\) of the record.
Indeed every hidden width is at most \(2n^2\), and summing the \(2h+1\) dense sizes \(d_jd_{j-1}+d_j\) gives this coarse bound.
A boundary product has component length at most \(2n+1\).
A bottom sum has at most \(n\) entries of length at most \(n\), so its common-denominator numerator and denominator lengths are bounded by \(10(n+2)^3\).
The other entries are original block entries or fixed zeros.
This bounds every returned parameter without cancellation or normalization.

Define the following fixed polynomial terms:
\[
\mathfrak{D}_0(s)=2^{20}(s+10)^{16},\qquad
\mathfrak{D}_{i+1}(s)=\langle \mathfrak{D}_i(s)+2,\mathfrak{D}_i(s)+2\rangle+2\quad(0\le i<12).
\]
The scalar and numerical bounds lie below \(\mathfrak{D}_0\).
Recall the array bound (Arr) of \Cref{sec:foundations}: an array whose rows \(0,\ldots,v-1\) have length at most \(\lambda\), and whose later rows are empty, has length at most \(\langle v,\lambda\rangle\).
The network code is built from its scalars through the rational pairs, the flat parameter tuple, the dimension tuple, the metadata tuple and the outer network tuple, fewer than eight array levels in all.
Each level raises the bound from \(\mathfrak{D}_i\) to at most \(\mathfrak{D}_{i+1}\) by (Arr), so the constructed network has length at most \(\mathfrak{D}_8(n)\).
The terms \(\mathfrak{D}_1,\ldots,\mathfrak{D}_{12}\) are twelve fixed term compositions, not a recursion whose iteration count depends on \(n\).

\subsection{Assembly Certificates and Their Ordinary Histories}
\label{rc:assembly-certificate}

In block mode return the exact seven-entry certificate
\[
C_F(n)=(1,Q,d,\operatorname{Un}(b),\operatorname{Un}(h),\Omega,\mathsf{Lab}).
\]
The tag \(1\) is the one-bit word, not the shifted signed integer code for one.
The tuple \(\Omega\) consists of the unary fields for \(\omega_0,\ldots,\omega_h\).
The label array \(\mathsf{Lab}\) has row \(\langle r,j\rangle+1\) equal to \(S_{r,j}\) for \(0\le r\le h\) and \(j<\theta_r\), and every other row empty.
The block and its certificate occur once; no child certificate is recursively nested.
The largest possible label row index is at most \(\langle n,n^2\rangle+1\), and every label has length at most \(n\).
Exact reconstruction, including nonpair and unused bits, gives \(|C_F(n)|\le \mathfrak{D}_8(n)\).

The fallback certificate has the same seven-entry tuple shape, empty tag, entry two \(C_B(n)\), and entries three through seven empty.
For an exact tuple \(C\), write \(C^{[2]}\) for its second entry; the checker below reads it only after the checker's exact-shape guard, which tests that \(C\) has the seven-entry tuple shape above.
At \(n=1\) the construction outputs the identity network \(\mathsf{Id}_1\) with the empty certificate.

The checker \(\mathsf{SCert}(n,N,C)\) first rejects \(n=0\).
For \(n=1\) it accepts precisely \(N=\mathsf{Id}_1,C=\varnothing\).
For \(n\ge2\) it requires the exact seven-entry tuple shape above, rejecting a failed shape before reading its fields.
Its two modes in this range are as follows.
\begin{enumerate}[label=\textup{(\alph*)},leftmargin=2em]
\item With empty tag, require the exact fallback tuple, empty unused fields, and
\(\mathsf{MCert}(n,N,C^{[2]})\).
\item With the one-bit tag, write \(C=(1,Q,d,\operatorname{Un}(b),\operatorname{Un}(h),\Omega,\mathsf{Lab})\) and let \(\omega_0,\ldots,\omega_h\) be the numbers coded by the unary entries of \(\Omega\).
Require \(\mathsf{CC}(b,Q,d)\), \(2\le b\le n\), \(1\le h\le n\), and exact unary powers bounded by \(bn\).
Require
\[
\omega_0=1,\quad \omega_{j+1}=b\omega_j\text{ for }j<h,\quad
\omega_{h-1}<n\le\omega_h<bn.
\]
Require \(\mathsf{Lab}\) to equal its exact interval-label reconstruction and \(N\) to equal the network given by the literal dense compiler of \Cref{rc:bottom,rc:within,rc:boundary}, with the common metadata and zero biases.
\end{enumerate}
It need not check that the chosen block maximizes the admitted arity; that is a property of the constructor, not a hypothesis needed for assembly soundness.

On arbitrary supplied strings, let \(s=n+|N|+|C|+2\).
Read a count or index only after its unary and range guards.
The block dimensions and coefficient lengths are at most \(|Q|\le|C|\); powers are at most \(bn\le n^2\).
Every compiler count and arithmetic cap above therefore has a polynomial bound after replacing \(n\) and \(|Q|\) by \(s\).
Unknown tags, extra bits, invalid denominators, nonpositive divisors, incorrect offsets or failed reconstructive equalities reject before an unsafe read.
All acceptance conditions are bounded numerical formulas using the chosen total functions.
Thus \(\mathsf{SCert}\) is a chosen total \(\mathsf{FTC}^0\) predicate.
It extends the restricted construction interface through a new checker; it does not assert that the original \(\mathsf{MCert}\) accepts arbitrary rational blocks.

For the constructed block branch, \(\mathsf{CC}(b,Q,d)\) holds by \Cref{ir:small}, and all other checks hold by the power, compiler and label equations.
The fallback and identity branches satisfy their tests by \Cref{uc:assemblies,uc:balanced} and the exact identity record.
The same \(\mathfrak{D}_8\) bound dominates the binary returned network and certificate bounds, since their seed \(100(n+2)^4\) is below \(\mathfrak{D}_0(n)\).
The fallback wrapper adds one tuple level above \(C_B(n)\), so by (Arr) its length is at most \(\langle8,\mathfrak{D}_8(n)\rangle<\mathfrak{D}_9(n)\).
Hence \(2\mathfrak{D}_{12}(n)\) bounds the returned network and assembly certificate together on every branch.

It remains to construct the histories in the theorem.
The following statement also establishes soundness for arbitrary passing certificates.

\begin{lemma}[Histories for a Passing Assembly]
\label{rc:assembly-history}
There is a chosen total \(\mathsf{FTC}^0\) function \(\mathsf{AH}(n,N,C,x)\), polynomially bounded in its total argument length, such that \(\bm{\mathcal T}\) proves
\[
\begin{aligned}
&\mathsf{SCert}(n,N,C) \wedge x\text{ a valid rational }n\text{-vector}\\
&\quad\rightarrow
\mathsf{LEval}(N,n,x,\mathsf{AH}(n,N,C,x))\wedge
\operatorname{out}(\mathsf{AH}(n,N,C,x))=_{\Q}\MAX_n(x).
\end{aligned}
\]
\end{lemma}
\begin{proof}
Return empty on either failed guard.
The identity and fallback modes use the single-output identity history and \(H_B(n,N,C^{[2]},x)\), respectively.
Their rational validity, adjacent equations and maximum output are already proved, using the common record.
Assume the block mode passes.
The history is assembled from independent local histories of the blocks, each evaluated on the maxima of its children's labels.

For each nonempty label \(S_{r,j}\), let \(M_{r,j}\) be its maximum in the input table \(x\), as in \Cref{uc:label-max} under the zero-based indexing of \Cref{rc:cones}; it is the selected original rational coordinate code.
This selection uses pairwise cross-product comparisons and bounded minimization, in parallel.
Its numerator and denominator lengths are at most \(|x|\); no new reduced fraction is formed.

For every internal node \((r,j)\) of the assembly tree, with \(1\le r\le h\) and \(j<\theta_r\), independently form the exact rational input tuple
\[
y_{r,j}(t)=M_{r-1,bj+t}\quad(t<b)
\]
and compute \(\mathsf{Hist}_2(Q,y_{r,j})\) in parallel.
Denote its first hidden values by \(a_{r,j,\alpha}\), its second hidden values by \(z_{r,j,\beta}\), and its output by \(o_{r,j}\).
The local equations proved in \Cref{rc:cc-lemma} give
\begin{equation}
\label{rc:local-history}
\begin{split}
a_{r,j,\alpha}&=_{\Q}\relu\left(\sum_{t<b}A_{\alpha t}M_{r-1,bj+t}\right),\\
z_{r,j,\beta}&=_{\Q}\relu\left(\sum_{\alpha<\mathfrak{w}_1}B_{\beta\alpha}a_{r,j,\alpha}\right),\\
o_{r,j}&=_{\Q}\sum_{\beta<\mathfrak{w}_2}c_\beta z_{r,j,\beta}.
\end{split}
\end{equation}
Biases are suppressed in this display only as rational values equal to zero; their stored denominators remain in the actual local history.

The maximum of the child maxima is the parent maximum: by the exact union identity and \Cref{uc:label-max}, \(M_{r-1,bj+t}\le M_{r,j}\) for every \(t<b\), with equality for some \(t\).
Apply the soundness of the accepted block to each input tuple \(y_{r,j}\); its output is the maximum of the entries \(M_{r-1,bj+t}\), which is \(M_{r,j}\) by the previous paragraph.
Hence
\begin{equation}
\label{rc:local-max}
o_{r,j}=_{\Q}M_{r,j}.
\end{equation}
There is no computational dependency from a child history's output to its parent's input.
Every \(M_{r,j}\) was selected directly from \(x\).
The local histories can therefore all be computed in one aggregate fixed-depth call.

Form \(Z\) as the exact flat history vector of the record, with the hidden neurons in layer order followed by the output.
The entry of neuron \(j\mathfrak{w}_1+\alpha\) of layer \(2r-1\) is \(a_{r,j,\alpha}\).
The entry of neuron \(j\mathfrak{w}_2+\beta\) of layer \(2r\) is \(z_{r,j,\beta}\), and the last entry is \(o_{h,0}\).
Its number of entries is exactly \(q_{\mathrm{hid}}+1\), with offsets supplied by the numerical prefix sums of global layer widths.
Every denominator is positive by the local-history construction.
Unused rows and nonpair bits are zero.
A zero numerator does not cause its denominator to be replaced by one.

We check every global adjacent equation explicitly.
For a first-layer row, \Cref{rc:bottom} gives
\begin{equation}
\label{rc:first-row}
\sum_{i<n}W_1[j\mathfrak{w}_1+\alpha,i]x_i
=_{\Q}\sum_{t<b}A_{\alpha t}x_{\pi(bj+t)}
=_{\Q}\sum_{t<b}A_{\alpha t}M_{0,bj+t}.
\end{equation}
Distribute the literal collision sum and interchange the two finite sums.
For each slot \(t\), exactly one column is \(\pi(bj+t)\); every other term is zero.
This retains all coincident padded slots.
Since the leaf label is a singleton, the second equality follows.
The first local equation in \Cref{rc:local-history}, zero global bias and congruence of ReLU prove this row's stored value.

For every even layer \(2r\), \Cref{rc:within} reads only the same node's values \(a_{r,j,\alpha}\) in layer \(2r-1\) and has coefficients \(B_{\beta\alpha}\).
Its affine value is the second dot product in \Cref{rc:local-history}; its zero bias and rectifier give \(z_{r,j,\beta}\).

For every later odd layer \(2r-1\), \(r\ge2\), \Cref{rc:boundary} gives
\begin{equation}
\label{rc:boundary-row}
\begin{split}
&\sum_{\substack{k<\theta_{r-1}\\\beta<\mathfrak{w}_2}}
W_{2r-1}[j\mathfrak{w}_1+\alpha,k\mathfrak{w}_2+\beta]z_{r-1,k,\beta}
=_{\Q}\sum_{\substack{t<b\\\beta<\mathfrak{w}_2}}
(A_{\alpha t}c_\beta)z_{r-1,bj+t,\beta}\\
&\quad=_{\Q}\sum_{t<b}A_{\alpha t}
\left(\sum_{\beta<\mathfrak{w}_2}c_\beta z_{r-1,bj+t,\beta}\right)\\
&\quad=_{\Q}\sum_{t<b}A_{\alpha t}o_{r-1,bj+t}
=_{\Q}\sum_{t<b}A_{\alpha t}M_{r-1,bj+t}.
\end{split}
\end{equation}
The first equality discards exactly the zero columns and lists each valid child slot once.
The next equality is distributivity.
The third uses the child's local affine readout, and the last uses \Cref{rc:local-max}.
This is the parent's independently constructed first preactivation in \Cref{rc:local-history}.
Its rectification therefore equals the stored \(a_{r,j,\alpha}\).
These identities cover the case \(h=1\), in which there is no later odd layer, as well as all-negative inputs, repeated coordinates, equal-valued distinct inputs and zero preactivations.

At the output there is exactly one root block, and the literal final row is \(c\).
Its value is \(\sum_{\beta<\mathfrak{w}_2}c_\beta z_{h,0,\beta}=_{\Q}o_{h,0}\), the stored scalar.
By \Cref{rc:local-max} and the full root label this is \(\MAX_n(x)\).

All finite rational manipulations used here have internal proofs with the supplied arrays as fixed string parameters.
Multiply a compared equality by the positive product of its denominators, reducing it to the signed-ring laws.
Finite distributivity follows by bounded-number induction using the admitted sum-prefix recurrence.
Rectangular sum reshaping follows by induction on its row and column indices; masking zeros and grouping the \(b\) contiguous child blocks use that same recurrence.
Equality substitution is cross-product equality, and ReLU respects the total rational order.
Thus no assertion of raw equality of differently unreduced sums is required.

For completeness, the returned histories have a fixed polynomial cap independent of the global depth.
Let \(s=n+|N|+|C|+|x|+2\).
A selected coordinate has component lengths at most \(|x|\), rational-pair length at most \(\langle3,|x|\rangle\), and node-input length at most
\[
\langle b+1,\langle3,|x|\rangle\rangle\le(s+10)^4.
\]
The block length is at most \(s\).
Substitute these values into the explicit \(B_D,B_E,L_0,\bar L,L_j\) bounds in \Cref{rc:cc-lemma}.
They are quadratic in block length plus node-input length, so \(1000(s+10)^{10}\) dominates every local numerator, denominator and named cleared integer.
There are polynomially many nodes and hidden values, with the count bound \(q_{\mathrm{hid}}<2n^2\) for the actual constructed family.
For a general passing certificate, replacing \(n\) and block length by \(s\) gives the corresponding polynomial count bound.
The label maxima, node inputs, aggregate local histories and final flat vector use fewer than twelve fixed array levels above \(\mathfrak{D}_0(s)\), which dominates both component and count bounds.
Hence \(\mathfrak{D}_{12}(s)\) bounds these returned data.

Each local history and arithmetic operation retains its own admitted computation-witness polynomial.
Only a fixed number of aggregate/composition stages occurs: selection from the original input, formation of node tuples, one aggregate of \(\mathsf{Hist}_2\), and flat copying.
Substituting the proved argument-length bounds into those witness polynomials supplies a fixed polynomial for the complete graph.
This establishes chosen total \(\mathsf{FTC}^0\) functions, not merely small output strings.
The identity and binary fallback use their own already proved polynomial witness and return bounds.
Combining the finitely many branches proves the lemma.
\end{proof}

The proof checks every ordinary \(\mathsf{LEval}\) row.
It does not require its fractions to have the denominators \(D^jE\) that the construction in the proof of \Cref{rc:cc-lemma} would give for the global network \(N\), where \(D\) is the product of all parameter denominators of \(N\) and \(E\) is the product of the input denominators.
The predicate $\mathsf{LEval}$ requires only rational-value equations, and these are precisely what the construction supplies.

\subsection{The Uniform Certified Construction}
\label{rc:rate}

We finish by assembling the chosen functions, using the block supplied by \Cref{ir:small} and the depth inequality of \Cref{ir:rate}.

\begin{restatedtheorem}[\Cref{rc:upper}]
There are a chosen total \(\mathsf{FTC}^0\) Boolean predicate \(S(n,N,C)\), chosen total \(\mathsf{FTC}^0\) functions \(F,C_F,k,H_F\), a fixed polynomial \(P\), and a fixed standard integer \(C_\ast\) with the following guarantees in \(\bm{\mathcal T}\).
For every unary \(n\ge1\),
\[
\Net(F(n),n),\qquad S(n,F(n),C_F(n)),\qquad
|F(n)|+|C_F(n)|\le P(n).
\]
Acceptance is sound: for every unary \(n\ge1\), all strings \(N,C,Z\) and every valid rational \(n\)-vector \(x\),
\[
S(n,N,C) \wedge \mathsf{LEval}(N,n,x,Z)
\rightarrow \operatorname{out}(Z)=_{\Q}\MAX_n(x).
\]
The literal hidden depth of \(F(n)\) is \(k(n)\).
The theory also proves, for every unary \(n\ge1\) and every valid rational \(n\)-vector \(x\),
\[
\mathsf{LEval}(F(n),n,x,H_F(n,x)),\qquad
\operatorname{out}(H_F(n,x))=_{\Q}\MAX_n(x),
\qquad |H_F(n,x)|\le P(n+|x|).
\]
It also proves the uniform numerical inequality
\[
\forall n\,\left(n\ge2\rightarrow(\ell(\ell(n))+1)k(n)\le C_\ast(\ell(n)+1)\right).
\]
In particular, in the standard model \(k(n)=O(\log n/\log\log n)\).
\end{restatedtheorem}

\begin{proof}[Proof of \Cref{rc:upper}]
Define \(F,C_F\) by the direct arity selector and its identity, fallback and block branches.
The successful branch uses the literal compiler and certificate above.
Define \(k(0)=0\) and, for \(n\ge1\), define \(k(n)\) by reading the common metadata of this actual \(F(n)\): it is zero at \(n=1\), the binary benchmark's depth on fallback, and exactly \(2h\) in block mode.
Thus this is an operational hidden depth; no removal of zero units or semantic minimization changes it.
All definitions are total chosen \(\mathsf{FTC}^0\) functions.
The preceding reconstruction proofs give \(\Net\), \(\mathsf{SCert}\) and the returned-code bounds in \(\bm{\mathcal T}\).

Take \(S=\mathsf{SCert}\).
If \(S(n,N,C)\) and \(\mathsf{LEval}(N,n,x,Z)\) hold for a valid rational input, \Cref{rc:assembly-history} supplies \(\mathsf{AH}(n,N,C,x)\) for the same network and input, with output \(\MAX_n(x)\).
Both $Z$ and $\mathsf{AH}(n,N,C,x)$ satisfy $\mathsf{LEval}$ for this same valid network and input.
The value-uniqueness proof in the paragraph ``Local histories and fixed-depth evaluation'' of \Cref{uc:section} therefore applies: it requires only the adjacent rational equations, and does not require denominator-one weights or reduced fractions.
It gives
\[
\operatorname{out}(Z)=_{\Q}\operatorname{out}(\mathsf{AH}(n,N,C,x))
=_{\Q}\MAX_n(x).
\]
This proves the stated soundness clause in $\bm{\mathcal T}$.

Set
\[
H_F(n,x)=\mathsf{AH}(n,F(n),C_F(n),x).
\]
Its fixed composition and \Cref{rc:assembly-history} prove all adjacent equations and the maximum output for every valid rational input.
On invalid input it has the already specified empty default.
For the block branch the bounds above give \(2\mathfrak{D}_{12}(n)\) for network plus certificate, and a fixed polynomial in \(n+|x|\) for history.
The binary history bound is the fixed polynomial \(b_8(n+|B(n)|+|C_B(n)|+|x|)\).
Substitute the proved \(b_8(n)\) bounds for its network and certificate, and choose a single polynomial \(P\) that dominates this bound, \(2\mathfrak{D}_{12}\) and the identity case.
These are finitely many fixed polynomial compositions.
This proves the operational and ordinary-history clauses.
\Cref{ir:rate} proves the uniform depth inequality for these chosen functions.

\end{proof}
\space

\section{Direct Rational Blocks and the Internal Depth Bound}
\label{ir:appendix}

This appendix constructs the blocks used by the rational family directly.
Each term is specified by local geometric data and a short binary word.
The theory forms all arrays in parallel and verifies their meaning by numerical induction with those arrays fixed.
The block and its cone certificate are each of size at most one exponential of a fixed polynomial in the arity.

Write \(\ell(t)=\lfloor\log_2t\rfloor\) for \(t\ge2\), and set \(\ell(0)=\ell(1)=0\).
The guarded binary power tables of \Cref{rc:selection} define this function internally.
Set \(L_1=\ell(n)\) and \(L_2=\ell(L_1)\).
Throughout, \(b\) is a candidate arity \(2\le b\le n\) of a direct maximum block, as tested by the arity guard of \Cref{rc:selection}, and \(P_2\) is the binary power table defined there.
Exponential expressions are binary integer strings, converted to numbers only after comparison with a stated numerical bound.

\begin{theorem}[An internally verified depth bound; restatement of \Cref{ithm:rational}]
\label{ir:rate}
There is a fixed standard integer \(C_\ast\) such that \(\bm{\mathcal T}\) proves, for the direct family \(F(n)\) of \Cref{rc:upper}, whose literal hidden depth is the chosen function \(k(n)\) of that theorem,
\[
\forall n\,\left(n\ge2\rightarrow(\ell(\ell(n))+1)k(n)\le C_\ast(\ell(n)+1)\right).
\]
In the standard model, \(k(n)=O(\log n/\log\log n)\).
\end{theorem}

The geometric identities are adapted from Zimmermann's finite dictionary proof~\cite[Theorem~7]{Zim26}.
The additional ingredients here are the direct term and vertex functions, their explicit code bounds, and their verification in the conservative arithmetic language.
No semantic span principle or arbitrary polynomial-time recursion is added to \(\bm{\mathcal T}\).

The appendix proceeds in seven steps.
\Cref{ir:arithmetic} fixes the rational sums, linear solves, integer kernel bases and matrix words used throughout, with explicit component caps.
\Cref{ir:span} defines the root two-layer functions \(C_{\pi,S}\) and their signed average \(F_m\).
It then takes the error \(f_m=F_m-M_m\) of this average against the maximum \(M_m\), and, by induction on the arity \(m\), writes \(f_m\) as an expression whose terms have affine rank at most \(m-2\).
\Cref{ir:direct} specifies every term of these low expressions directly, by a local key and a binary word recording the transformation used at each higher arity, following the recurrence in \Cref{ir:paired}, and evaluates every vertex from its primitive root by one matrix word.

\Cref{ir:readout} removes the low-rank terms by face corrections.
This leaves the relations of \Cref{ir:relation}, each of which expresses a simplex support through linear rows, hinge features and smaller simplices.
Solving these relations gives the readout coefficients and the block \(Q\).

\Cref{ir:tiers} proves, by numerical induction on arrays of rational points formed in advance, that these arrays compute the intended functions, so that \(Q(x)=\MAX_b(x)\) at every rational point \(x\).
\Cref{ir:cones} constructs the cone certificate checked by \(\mathsf{CC}\), and \Cref{ir:packets} fixes the size allocation, proves the block-supply result \Cref{ir:small} and derives \Cref{ir:rate}.
The arrays of \Cref{ir:direct,ir:readout} are formed from \(b\), local data and words before \Cref{ir:tiers} proves anything about their values, and the verification takes the completed arrays as string parameters.
\Cref{tab:ir-guide} locates the direct definitions, bounds and verification steps after the constructed objects have been introduced.
The final precision bounds are collected in \Cref{tab:ir-precision}.

\subsection{Controlled Arithmetic}
\label{ir:arithmetic}

We fix the rational graphs used by the static construction.
For a list \(p_i/q_i\), with positive \(q_i\), form
\[
D=\prod_iq_i,\qquad \tilde p_i=p_i\prod_{j\ne i}q_j,
\qquad \sum_i p_i/q_i=_{\Q}(\sum_i\tilde p_i)/D.
\]
The final sum is an integer sum.
For \(\mathfrak{n}\) input components of length at most \(L\), all cleared components have length at most \(4(\mathfrak{n}+1)(L+2)\).
This follows by adding product lengths; an integer sum of \(\mathfrak{n}\) terms increases the maximum summand length by at most \(\lceil\log_2(\mathfrak{n}+1)\rceil+2\).
In particular, the cleared length grows linearly in \(\mathfrak{n}\): the denominators are not multiplied again for each summand.

For an \(r\)-square rational system, \(r\le b\), clear all matrix and right-hand-side denominators together.
If the resulting system is \(Uz=Y\), let $U_i$ be $U$ with column $i$ replaced by $Y$ and print \(z_i=\det(U_i)/\det(U)\), changing both signs when necessary to make the denominator positive.
Singular systems return a declared default; the order-zero system returns the empty coordinate tuple.
Determinants use integer Leibniz products and an integer sum.
The adjugate equation gives $U\operatorname{adj}(U)Y=(\det U)Y$, and expansion in the replaced column gives $(\operatorname{adj}(U)Y)_i=\det(U_i)$.
Thus these coordinates satisfy every equation, and multiplying a homogeneous equation by the adjugate proves uniqueness.

Here the order varies up to $b$, so we justify the determinant identities for these actual arrays rather than invoke only the fixed-order \Cref{fact:dt-det}.
A row interchange reindexes each permutation by the corresponding transposition, reversing its sign.
Indeed, let the transposition exchange positions \(a<c\).
An index pair avoiding both \(a\) and \(c\) keeps its inversion status.
For \(x<a\) the statuses of the pairs \((x,a)\) and \((x,c)\) are exchanged, and likewise for \(x>c\), so these pairs keep their total.

For each \(a<x<c\), the pairs \((a,x)\) and \((x,c)\) together change the inversion count by \(-2\), \(0\) or \(2\).
The pair \((a,c)\) changes its status, so the count changes by an odd number; bounded counting over the preformed pair arrays proves this.
Equal rows therefore cancel in pairs.

Grouping permutations by the selected column gives each cofactor expansion.
The diagonal adjugate entries are those expansions, and the off-diagonal ones are determinants with a repeated row; the transpose argument gives the product in the other order.
These are bounded reindexings and regroupings of preformed permutation arrays.
Bounded-number induction through their sum prefixes proves the equalities with those arrays fixed, without a new polynomial or vector string in the induction formula.
The component bound for a solve, justified below, is
\[
\mathsf{Solve}(r,L)=100(r+1)^4(L+2).
\]
For $r\ge1$, the Leibniz products are formed by enumerating all $r^r$ tuples of $r$ base-$r$ digits and rejecting those that are not permutations; the empty determinant is one.
Thus the bound $r^r$ on the number of enumerated tuples covers rejected candidates as well as retained ones, and it is this bound that the count below uses.

There are at most $r(r+1)$ denominators to clear for a solve, so each cleared entry is a product of at most $r(r+1)+1$ factors of length at most $L$, and the product length bound of \Cref{tc:arithmetic} gives it length at most $2+(r(r+1)+1)(L+1)\le3(r+1)^2(L+2)$.
Each determinant product has $r$ such factors, hence length at most $2+r\left(3(r+1)^2(L+2)+1\right)\le4(r+1)^3(L+2)$.

A canonical integer of length $\lambda$ has magnitude below $P_2(\lambda)$.
Number induction on $t$ proves that a sum of $t$ integers of magnitude below $P_2(\lambda)$ has magnitude below $tP_2(\lambda)$.
Since $r^r\le P_2(r^2)$, summing the at most $r^r$ products adds at most $r^2+1$ bits, so every determinant has length at most $5(r+1)^3(L+2)$.
These bounds are dominated by the displayed $\mathsf{Solve}(r,L)$, including sign normalization.

For a matrix of at most $b$ columns, first clear each row's denominators separately, preserving its kernel.
Choose a nonzero minor of the largest order of the resulting integer matrix; its rows are the pivot rows, its columns the pivot coordinates, and the other columns the free coordinates.
This minor supplies integer kernel vectors: give one free coordinate the pivot determinant, and the pivot coordinates the signed replaced-column determinants.
Each row clearing uses at most $b$ components, and each minor has order at most $b$, so the same larger bound $\mathsf{Solve}(b,L)$ covers these components.

These vectors form a basis of the kernel over \(\Q\), as follows.
For the cleared integer matrix \(A\), let \(I,J\) be the pivot row and column sets, put \(M=A_{I,J}\), and write \(\delta=\det M\ne0\).
For each free column \(f\), the vector \(v^f\) has free coordinate \(v^f_f=\delta\), all its other free coordinates zero, and pivot coordinates \(v^f_J=-\operatorname{adj}(M)A_{I,f}\).

On a pivot row, the adjugate identity gives $Mv^f_J+\delta A_{I,f}=0$.
For a nonpivot row $i$, put $a=A_{I,f}$, $c=A_{i,J}$ and $t=A_{if}$.
The bordered minor is zero by maximality of the pivot rank, and
\[
0=\det\begin{pmatrix}M&a\\c&t\end{pmatrix}
=\delta t-c\operatorname{adj}(M)a=(Av^f)_i.
\]
To verify the middle equality, subtract $cM^{-1}$ times the top rows from the last row; row multilinearity together with the vanishing of determinants with a repeated row preserves the determinant and leaves the last entry $t-c\operatorname{adj}(M)a/\delta$.
Expansion in that last row gives the displayed identity.
Thus every row equation holds, not only the pivot equations.

The free coordinates of these vectors prove their independence.
For any \(z\) in the kernel, subtracting \(\sum_f(z_f/\delta)v^f\) leaves a kernel vector with all free coordinates zero; invertibility of \(M\) makes its pivot coordinates zero as well.

Moreover, if a row \(w\) annihilates every \(v^f\), subtract the combination of the pivot rows whose entries on \(J\) equal \(w_J\).
The residual row \(w'\) has \(w'_J=0\) and \(w'\cdot v^f=\delta w'_f=0\), so \(w'=0\) and \(w\) lies in the row span.
Consequently a row outside that span is nonzero on some kernel-basis vector.
For rank zero use the coordinate basis, and for full column rank the kernel is zero and the pivot rows span the whole row space.
All these equations use the same minor graphs and bounded coordinate sums.

A matrix word means an ordered product $M_h\cdots M_1$, with the rightmost factor acting first.
For its $(i,j)$ entry, a coordinate walk is a sequence $i_0=j,i_1,\ldots,i_h=i$; its contribution is $\prod_{t=1}^h M_t[i_t,i_{t-1}]$.
Summing these contributions over all intermediate indices is exactly matrix multiplication, by repeated distribution.
A local frame is the ordered list of coordinate indices currently in use; a map's source and target frames specify its column and row coordinates.
Only maps with matching intermediate frames are composed; coordinate walks use the corresponding index ranges, each bounded by the declared dimension.

A word of \(h\) matrices of dimension \(r\) is evaluated from its original factors.
Clear the \(hr^2\) entry denominators once, writing each factor \(M_j\) as \(\tilde M_j/D\).
Sum the integer products over all coordinate walks and print the matrix product over \(D^h\).
This gives the component cap
\[
\mathsf{Word}_{\Q}(r,h,L)=
20(h+1)(hr^2+1)(L+2)+20(h+1)\lceil\log_2(r+1)\rceil+20.
\]
The identity for extending a word follows by grouping walks according to their last coordinate.
For integral maps acting on a rational root vector, clear only that vector's denominators, once.
For a word of \(h\) integral matrices of dimension \(r\) whose entries have component length at most \(L_M\), applied to a rational vector whose components have length at most \(L_V\), a valid cap for the components of its image is
\[
\mathsf{Word}_{\Z}(r,h,L_M,L_V)=
10\left((h+1)(L_M+2)+r(L_V+2)
+(h+1)\lceil\log_2(r+1)\rceil+2\right).
\]
The empty word is the identity.
For an affine map $y\mapsto Ay+t$, homogeneous coordinates append one coordinate:
\[
\begin{pmatrix}y\\1\end{pmatrix}\mapsto
\begin{pmatrix}A&t\\0&1\end{pmatrix}\begin{pmatrix}y\\1\end{pmatrix}.
\]
More generally $(Y,d)$ with $d>0$ represents the rational point $Y/d$; when a displayed homogeneous matrix changes $d$, its output represents the new numerator divided by its new positive last coordinate.
A base vector can be placed in the first column of one extra matrix, so the same direct word bound applies to its image.

For magnitude bounds, put $\|y\|_\infty=\max_i|y_i|$ and $\|A\|_\infty=\max_i\sum_j|A_{ij}|$.
The triangle inequality gives $\|Ay\|_\infty\le\|A\|_\infty\|y\|_\infty$; an affine offset adds at most $\|t\|_\infty$.

All these graphs have a fixed number of aggregate product, integer-sum and reconstruction stages.
Their \(\mathsf{FTC}^0\) definitions take as an extra unary parameter the ambient arity \(n\), or \(n+|x|\) for the arrays of rational points \(x\) in \Cref{ir:tiers}.
They include explicit polynomial count guards in that parameter, and they are not polynomial-output functions of unary \(b\) alone.

\subsection{Finite Support Identities and a Coefficient Construction}
\label{ir:span}

This subsection defines the root functions and their signed average, and proves by induction on the arity that the average differs from the maximum by an expression whose terms have affine rank at most the arity minus two.
For root arity \(3\le m\le b\), all calculations take place in the hyperplane \(\{y\in\Q^m:y_0=0\}\), with local coordinates \(y_1,\ldots,y_{m-1}\).

A polytope record \(P\) is a nonempty rational vertex list; its support is \(h_P(y)=\max_{p\in P}p\cdot y\), the maximum of the listed dot products.
We identify a linear form with its coefficient vector, so a point vector $a$ has support $y\mapsto a\cdot y$.
We call such a linear form a linear row, and the support of a one-point list a point row.
An expression is a flat list of rational coefficient--polytope pairs \((a_k,P_k)\); its value is the function \(y\mapsto\sum_ka_kh_{P_k}(y)\), and an operation defined on single supports acts on an expression term by term, keeping the coefficients.

For a nonempty finite point list, its affine hull consists of combinations of its points whose coefficients sum to one.
A nonempty list $p_1,\ldots,p_t$ is affinely independent when $p_2-p_1,\ldots,p_t-p_1$ are linearly independent; for $t=1$ this difference list is empty.
We use rational coefficients for these affine computations; their real geometric interpretation uses the same formulas over $\R$.
The direction space of a polytope is the span of its vertex differences.
Its affine rank is the dimension of this space, computed by minors of order at most \(b\).
A term whose vertex list would be empty is omitted, so every listed polytope record is nonempty.
For a direction \(u\), let \(P^u\) be the sublist of vertices attaining the maximum at \(u\), called the exposed face in direction \(u\).
A face is proper when its convex hull is a strict subset of the original polytope, and a facet is a face whose dimension is one less than that of the polytope.
Put
\[
P_u h_P=h_{P^u},\qquad D_u=P_u-P_{-u}.
\]
These symbols denote finite expression operations.
Their transport of functional identities will be justified in \Cref{ir:tiers}.

We now define the root functions and name their parts.
In the reduced local frame $y_0=0$, use $e_0=0$ and let $e_i$, $1\le i<m$, be the usual coordinate unit vectors.
Thus $y_i=e_i\cdot y$, including label zero.
For a permutation \(\pi\) of \(\{0,\ldots,m-1\}\) and a nonempty cut set \(S\subseteq\{1,\ldots,m-2\}\), whose members we call cuts, put
\[
q=\max S,\quad p(j)=\max(\{0\}\cup(S\cap\{1,\ldots,j-1\})),
\]
\[
h=y_{\pi(q+1)}-y_{\pi(0)},\quad
l_j=y_{\pi(p(j)+1)}-y_{\pi(j+1)},
\]
\[
U_{\pi,S}=\relu\left(h+\sum_{j=1}^{m-2}\relu(l_j)\right),\quad
C_{\pi,S}=y_{\pi(0)}+U_{\pi,S}.
\]
The rows \(l_j\) form a tree on the non-apex labels \(\pi(1),\ldots,\pi(m-1)\), rooted at \(\pi(1)\).
Its spine is the sequence of \(\pi(1)\) followed by the labels \(\pi(j+1)\) for \(j\in S\) in increasing order of \(j\).
The edge \(l_j\) joins \(\pi(p(j)+1)\) to \(\pi(j+1)\): it is a spine edge for \(j\in S\), and otherwise its latter endpoint is a leaf attached to the preceding spine label.

Deleting one edge from this non-apex tree is a spine cut or a leaf cut, according to the kind of edge: a spine cut deletes \(l_k\) for a cut \(k\in S\), and a leaf cut deletes \(l_j\) for some \(j\notin S\).
The head \(h\) joins the last spine label \(\pi(q+1)\) to the apex \(\pi(0)\); it is excluded from these cuts and non-apex degree counts.

The head and edge rows are independent.
To check this, first retain all $m$ label coordinates in each difference row, before imposing $y_0=0$.
In a vanishing combination of these lifted rows, the apex coordinate first forces the head coefficient to zero; eliminating children in decreasing position then forces each edge coefficient to zero, since every parent precedes its child.
Every lifted row has coordinate sum zero, so deleting coordinate zero is injective on their span and preserves independence in the local frame.

The vertex list for \(U\) is the origin together with all \(h+\sum_{j\in A}l_j\) for \(A\subseteq\{1,\ldots,m-2\}\); translating it by the apex unit vector, that is, adding \(e_{\pi(0)}\) in this reduced frame, gives the list for \(C\).
The origin of the list for \(U\), and its translate in the list for \(C\), is the apex vertex, and the other listed points are base vertices.
A base face is a face with only base vertices, and an apex-containing face is a face containing the apex vertex.
Define
\[
F_m=\frac1{m!}\sum_{\pi,S\ne\varnothing}(-1)^{m+|S|}C_{\pi,S},
\qquad M_m=\max(0,y_1,\ldots,y_{m-1}),\qquad f_m=F_m-M_m.
\]
We retain every permutation and cut in this sum, including terms that define the same function.
Each face selected below from a root likewise keeps its face-state tag, a ternary state of \Cref{rc:effective} that assigns each edge one of the uses omitted, linear or hinged.
The total translation weight of \(F_m\) is \(W_m=\sum_{S\ne\varnothing}(-1)^{m+|S|}=(-1)^{m+1}\).

\paragraph{\emph{Cuts and simplex reconstruction.}}
An extreme point of a convex set cannot be written as $(1-t)u+tv$ for distinct points $u,v$ of that set and $0<t<1$.
Write $[v,w]=\{(1-t)v+tw:0\le t\le1\}$ for the segment joining $v$ and $w$, with real coefficients in geometric statements and rational coefficients in the internal point calculations.
A full simplex is the convex hull of \(m\) affinely independent points of the hyperplane \(y_0=0\), so that it has dimension \(m-1\).
This paragraph writes each root support as a signed sum of full-simplex supports and lower-rank terms, and then rebuilds the support of a full simplex from its facets that are not parallel to the first coordinate direction.

Let a hyperplane cut a polytope record \(P\).
Write \(P_+,P_-\) for the vertex lists of the two closed halves, and \(Q\) for the vertex list of the section, the intersection of \(P\) with the hyperplane.
The finite support identity is
\[
h_P+h_Q=h_{P_+}+h_{P_-}.
\]
To verify it at an arbitrary rational vector $y$, choose maximizers on both halves.
If their values differ, take a maximizer on the half with the smaller value.
If it lay strictly off the section, consider the segment from it to a maximizer of the other half.
A short enough step along this segment, taken before the segment crosses the cut, stays in the smaller half and increases the value at $y$, because the other endpoint has the larger value; this contradicts maximality.
So the smaller-side maximum is attained on the section; since the section lies in both halves, it equals \(h_Q(y)\), while the larger maximum is \(h_P(y)\).
If both values agree, the segment joining the two maximizers meets the cutting plane at a point of the section, and since the value at $y$ is affine along the segment, that point has the same value.
A cut is forced when the list lies on one closed side of it, and all-tied when the list lies in the cutting plane; in both cases the same equality holds directly.

The parameter pyramid $\mathfrak{P}=\{0\le s_i\le\lambda\le1\}$, with $m-2$ coordinates $s_i$, maps to each root by $(\lambda,s)\mapsto\lambda h+\sum_i s_i l_i$, followed by its apex translation.
The head and edge independence makes this an invertible map of the $(m-1)$-dimensional parameter space.
Fix an order of the $N=\binom{m-2}{2}$ unordered comparisons $s_i\ge s_j$ and cut successively by their two closed sides.
For a binary word $\omega$ of length $t$, let $P_\omega$ be the root image after its $t$ choices, and let $Q_\omega$ be its section at the next comparison.
The cut identity gives the prefix invariant
\[
h_{\mathrm{root}}=\sum_{|\omega|=t}h_{P_\omega}-\sum_{|\omega|<t}h_{Q_\omega}.
\]
The base $t=0$ is immediate; replacing each depth-$t$ term by its two sides minus its section proves the successor.
Pieces that coincide as polytopes but carry different tags are kept as separate terms.
If a cut is forced, its section cancels the smaller closed side; if its parent lies in the plane, the equation is $h_P=h_P+h_P-h_P$.
Multiplying by the root's signed weight preserves this invariant.

At any prefix, the vertices of a piece are exactly the origin and the points $(1,1_B)$, where $1_B\in\{0,1\}^{m-2}$ is the indicator vector of $B$.
Here $B\subseteq\{1,\ldots,m-2\}$ ranges over the binary masks that satisfy the comparisons chosen so far, which we call feasible masks.
To see completeness, sort the distinct members of $\{0,\lambda,s_1,\ldots,s_{m-2}\}$ as $0=t_0<t_1<\cdots<t_r=\lambda$ and put $B_j=\{i:s_i\ge t_j\}$.
Then
\[
(\lambda,s)=(1-\lambda)0+\sum_{j=1}^r(t_j-t_{j-1})(1,1_{B_j}).
\]
An empty threshold mask is allowed, and the coefficients sum to one.
Every threshold mask respects every comparison imposed on $s$.
Conversely each feasible mask is an extreme point of the original pyramid and remains an extreme point of its intersection with these halfspaces.
If $\lambda=0$, the point is simply the origin.

Every comparison section satisfies a nonzero equation $s_i=s_j$ in the original parameter space and therefore has ambient rank at most $m-2$, even if its parent was already flat.
At a final leaf the comparisons orient every pair.
We classify a final leaf as full when its orientation is transitive, that is, when for all distinct \(i,j,k\) the chosen comparisons \(s_i\ge s_j\) and \(s_j\ge s_k\) imply the chosen comparison \(s_i\ge s_k\), and as flat otherwise.
A flat leaf therefore contains a directed triangle \(s_i\ge s_j\), \(s_j\ge s_k\), \(s_k\ge s_i\), since the pair \(i,k\) is oriented one way or the other, and this forces equality around the triangle.

Without a cycle, the comparisons give one total coordinate order; its feasible masks are the empty mask and the nested initial segments.
Together with the origin they form a full simplex: successive differences supply all $s_i$ directions and the $\lambda$ direction.
The prefix invariant thus gives the full-simplex sum and the rank-at-most-$m-2$ remainder.
An internal section has the negative root weight and a final flat leaf the positive root weight; the separate target simplex $\operatorname{conv}(e_0,\ldots,e_{m-1})$, whose support is $M_m$, has weight $-1$.
There are at most $2^N$ leaves and fewer than $2^{N+1}$ nodes per root, including forced branches.

The next operation evaluates an expression at the projection $y-y_1e_1$ whenever $y_1<0$; it is what allows a full simplex to be rebuilt from its facets.
In these local coordinates define the linear operation
\[
(T_m E)(y)=\begin{cases}E(y)&y_1\ge0,\\
E(y-y_1e_1)&y_1<0.\end{cases}
\]
For a polytope \(P\), choose an integer \(c\ge0\) making all first coordinates of \(P+ce_1\) nonnegative.
Define \(\operatorname{pr}_m z=z-z_1e_1\) for points and vertices in the same local frame, and write \(\operatorname{pr}=\operatorname{pr}_m\) while the arity is fixed.
This projection sets coordinate one to zero and retains the ambient frame.
A literal support representation is
\[
T_mh_P=h_{\operatorname{conv}((P+ce_1)\cup\operatorname{pr}(P+ce_1))}
-c h_{[0,e_1]}.
\]
For \(y_1\ge0\) each translated point dominates its projection; for \(y_1<0\) its projection dominates it.
This proves the formula and linearity of \(T_m\).
The operation \(T_m\) increases the affine dimension of a support term by at most one.

For a full local simplex $P=\operatorname{conv}(v_0,\ldots,v_{m-1})$, let $\lambda_i$ be the affine functions that are one at $v_i$ and zero at the other vertices, its barycentric functions.
Write $a_i$ for the coefficient of $y_1$ in $\lambda_i$.
The identities $\sum_i\lambda_i=1$ and $y=\sum_i\lambda_i(y)v_i$ give
\[
\sum_i a_i=0,\qquad \sum_i a_i v_i=e_1.
\]
Thus both nonzero signs occur among the $a_i$.
Write $\lambda_i=c_i-n_i\cdot y$, so $n_i\cdot y\le c_i$ is the outward inequality for facet $i$, opposite $v_i$.
Write $n_{i1}$ for the first coordinate of $n_i$, and $n_F,n_{F1},c_F$ for $n_i,n_{i1},c_i$ when $F$ is facet $i$.
A facet is vertical when $n_{i1}=0$, meaning its hyperplane is parallel to the first coordinate direction.

For a pair $a_i,a_j$ of the same nonzero sign, their opposite ridge is the common face on all vertices other than $v_i,v_j$.
The cut $a_j\lambda_i-a_i\lambda_j=0$ contains that ridge and meets $[v_i,v_j]$ at
\[
z=\frac{a_iv_i+a_jv_j}{a_i+a_j}.
\]
Its two weights are positive.
The children are $\operatorname{conv}(z,\{v_r:r\ne i\})$ and $\operatorname{conv}(z,\{v_r:r\ne j\})$.
Each remains full, since replacing an endpoint by an interior point of that edge preserves affine independence.
The cut is vertical, since its $y_1$ coefficient is $a_ja_i-a_ia_j=0$.
One child retains original facet $i$ whole and replaces $j$ by the cut facet; the other retains $j$ and replaces $i$.
Every other original facet is divided into two closed pieces.

All subsequent nonvertical facets are inherited originals, possibly with positive barycentric rescalings.
If $\lambda_i^{\mathrm{cur}}=\kappa_i\lambda_i$ and $\lambda_j^{\mathrm{cur}}=\kappa_j\lambda_j$, with $\kappa_i,\kappa_j>0$, their current cut function equals $\kappa_i\kappa_j(a_j\lambda_i-a_i\lambda_j)$.
Its plane and signs therefore remain those of the fixed original pool
\[
(n_i/n_{i1}-n_j/n_{j1})\cdot y=c_i/n_{i1}-c_j/n_{j1}.
\]
Every split removes one of a same-sign pair and adds only a vertical facet, preserving at least one nonvertical facet of each sign.
After at most $m-2$ splits precisely two remain.

Put $S(P)=\sum_F\operatorname{sgn}(n_{F1})h_F$, summed over the facets $F$ of $P$, each taken as its vertex list, and $R(P)=h_P-T_mS(P)$.
At a terminal simplex let $F_+,F_-$ be its positive and negative nonvertical facets.
Their opposite vertices lie on the other $m-2$ vertical facets, whose common direction is the line spanned by $e_1$.
The two vertices therefore have the same projection, and both facets share all remaining vertices, giving $\operatorname{pr}(F_+)=\operatorname{pr}(F_-)$.

Over a projected point the simplex fiber is an interval with endpoints on $F_+$ and $F_-$.
Thus $F_+$ supplies $h_P$ for $y_1\ge0$, while $F_-$ supplies it for $y_1\le0$.
When $y_1<0$, $T_m(h_{F_+}-h_{F_-})$ is zero because the projected supports agree; when $y_1\ge0$ it is their ordinary difference.
Consequently $R(P)=h_{F_-}$, with the lower-facet sign retained.

At a split with section $Q$, the two selected endpoint facets each survive whole on one child.
Every other facet $F$ satisfies the cut identity, and the new vertical facets have signed coefficient zero.
Therefore
\[
S(P_+)+S(P_-)=S(P)+\sum_{F\text{ other}}\operatorname{sgn}(n_{F1})h_{F\cap Q}.
\]
Combining this with $h_{P_+}+h_{P_-}=h_P+h_Q$ gives
\[
R(P)=R(P_+)+R(P_-)-h_Q
+T_m\sum_{F\text{ other}}\operatorname{sgn}(n_{F1})h_{F\cap Q}.
\]
The section $Q$ has rank $m-2$, and each other-facet section has rank at most $m-3$, so every residual after $T_m$ has rank at most $m-2$.
Unfolding the split tree writes $R(P)$ through the terminal lower facets with positive parent weight, sections with negative parent weight, and transported other-facet sections with their outward sign.

\paragraph{\emph{The top-error induction.}}
Recall that $f_m=F_m-M_m$ is the error of the averaged root formula against the maximum.
By induction on $m$, we construct an explicit expression for $f_m$ whose terms have affine rank at most $m-2$.
At $m=3$, each $C_{\pi,\{1\}}$ is $\max(y_0,y_1,y_2)$ by applying $v+\relu(w-v)=\max(v,w)$ twice, so $f_3=0$.
Assume $m\ge4$.
For a root and a direction $u$, put
\[
a=h(u)+\sum_j\relu(l_j(u)),\qquad
H_u=h+\sum_{l_j(u)>0}l_j+\sum_{l_j(u)=0}\relu(l_j).
\]
A maximizing base vertex includes exactly the positive edges, excludes the negative ones, and permits either endpoint of every zero edge.
Comparing its value $a$ with the origin gives selected supports $0,H_u,\relu(H_u)$ for $a<0,a>0,a=0$, respectively.
The apex translation adds its linear row, which cancels under $D_u$.
A base face is a point row plus separate segment supports.
Call a non-apex edge of value zero a free hinge edge, since $H_u$ keeps its segment $\relu(l_j)$.
If at least two non-apex edges have nonzero values, an apex-containing face has at most $m-4$ free hinge edges and rank at most $m-3$.
The head $h$ is excluded from this edge count.

For $u=0$, $D_u=0$.
Call the coordinate $u_i$ the level of the label $i$ at $u$.
If no non-apex edge has nonzero value and $u\ne0$, all non-apex labels share a level and the apex is the single exception.

If exactly one edge $l_k$ is nonzero and the selected face contains both apex and base, then $a=0$ and deleting that edge leaves two constant-level components.
For a spine cut $k\in S$, let their levels be $\mathfrak{a}$ in positions $1,\ldots,k$ and $\mathfrak{b}$ in positions $k+1,\ldots,m-1$.
The tie equation is $u_{\pi(0)}=\max(\mathfrak{a},\mathfrak{b})$.
If the component at the lower level is a singleton, its label is the one exceptional label; otherwise both levels repeat, since the apex shares the higher level.

For a leaf cut, let its leaf level be $d$ and every other non-apex level be $c$.
The tie equation is $u_{\pi(0)}=c+\max(c-d,0)$.
If $d>c$, only that leaf is exceptional; if $d<c$, the levels are $c+t,c-t,c,\ldots,c$ with $t=c-d>0$.
Applying the same analysis at $-u$ covers negative exposed faces.
These cases exhaust the directions $u$ at which a selected apex-containing face may have rank greater than $m-3$.

We treat the remaining configurations in turn: three levels, two repeated levels, and a single exceptional label.
For three levels, let $a_+,a_-$ have values $\gamma+t,\gamma-t$, with all other labels at $\gamma$ and $t\ne0$.
Let $\tau$ exchange these labels and put $(By)_i=y_{\tau(i)}-y_{\tau(0)}$.
If $\delta_i=[i=a_+]-[i=a_-]$, where the brackets are zero-one indicators, then $u_i=t(\delta_i-\delta_0)$ and
\[
(By)_i-y_i=-(\delta_i-\delta_0)(y_{a_+}-y_{a_-}).
\]
Hence $B=I-2u\beta$, where $\beta(y)=(y_{a_+}-y_{a_-})/(2t)$ and $\beta(u)=1$, so $Bu=-u$.
Permutation and recentering reindex the tagged roots and give $F_m\circ B=F_m+\lambda$, with $\lambda=-W_my_{\tau(0)}$.
For each polytope, $(B^TP)^u=B^T(P^{-u})$.
On the negative face $p\cdot u=-h_P(-u)$, so $B^Tp=p+2h_P(-u)\beta$ and
\[
P_u(h_P\circ B)=P_{-u}h_P+2h_P(-u)\beta.
\]
Apply this to each tagged term and compare with $P_u(F_m+\lambda)=P_uF_m+\lambda$.
It follows that
\[
D_uF_m=2F_m(-u)\beta-\lambda.
\]
The target simplex has a unique highest and lowest label at these levels, so its two selected supports are point rows.
Thus $D_uf_m$ is linear, and its term $-\lambda$ keeps the sign displayed above.

For two repeated levels, $k$ is the only spine cut whose edge $l_k$ joins labels at different levels; we call $l_k$ the crossing edge, and the apex is at the higher level.
First suppose the latter component is higher, $\mathfrak{b}>\mathfrak{a}$.
Since the level $\mathfrak{a}$ of the former component is repeated and the apex is at level $\mathfrak{b}$, that component has at least two labels, so $k\ge2$.

Toggle $k-1$ in $S$ while retaining $k$; this changes only the crossing edge's parent, and that edge has negative value $\mathfrak{a}-\mathfrak{b}$ and is absent from $H_u$.
The head, zero edges and apex row are unchanged, so the apex-containing faces coincide.
The map is an involution and reverses the parity of $|S|$.

If the former component is higher, $\mathfrak{a}>\mathfrak{b}$, then the level $\mathfrak{b}$ of the latter component is repeated, so position $k+2$ exists.
Exchange $r=\pi(k+1)$ and $s=\pi(k+2)$ and toggle $k+1$ in $S$, retaining $k$.
Starting with $k+1\notin S$, put $t=y_r-y_s$ and write $l_k=y_c-y_r$ for the preceding spine label $c$.
The two relevant rows change from $(l_k,t)$ to $(l_k+t,-t)$.
All later parent labels and the last spine label agree, so the other rows and head agree.
The crossing row is active and the other is tied; the identity
\[
l_k+\relu(t)=(l_k+t)+\relu(-t)
\]
therefore proves equality of the two apex-containing face supports.
The labels \(r\) and \(s\) occupy positions \(k+1\) and \(k+2\) of the latter component, so both have level \(\mathfrak{b}\), and exchanging them leaves every level unchanged; repeating the swap and toggle restores the original description.
Each map keeps $S$ nonempty because cut $k$ remains, and the paired roots have opposite weights, so their apex-containing faces cancel.
At $-u$ the same pairing handles negative apex faces with their common $D_u$ sign.
The other selected base faces need not coincide, so each is kept as its point row and its individual segment supports, all of rank at most $1\le m-3$.

For one exceptional label, we use first the direction \(u_*=(0,1,\ldots,1)\).
If \(\pi(0)=0\), then \(l_j(u_*)=0\) and \(h(u_*)=1\).
The positive exposed face is the base and the negative exposed face is the apex, so their support difference is \(h+\sum_j\relu(l_j)\).
This is a linear row plus segment supports, already included in the base-face terms.

For the remaining occurrences, degree means degree in the non-apex tree defined above.
Only degree-one occurrences of label zero remain after the preceding rank tests.
Let \(\hat p=\pi^{-1}(0)\in\{1,\ldots,m-1\}\) be its position in the upper permutation.
Delete that label to obtain the lower order \(\bar\pi:\{0,\ldots,m-2\}\to\{1,\ldots,m-1\}\), retaining the other labels in their original order.

For a lower cut set \(T\subseteq\{1,\ldots,m-3\}\), its upper preimages are listed below.
Only preimages with \(S\ne\varnothing\) are admitted, so a pendant preimage occurs only when \(T\ne\varnothing\).
\[
\begin{array}{ll}
\text{first:}&\hat p=1,\quad S=\{1\}\cup\{t+1:t\in T\},\\
\text{pendant }i:&\hat p=i+2,\quad S=\{\iota_i(t):t\in T\},\quad0\le i\le m-3,\\
\text{last:}&\hat p=m-1,\quad S=T\cup\{m-2\},
\end{array}
\]
where \(\iota_i(t)=t\) for \(t<i+1\) and \(t+1\) otherwise.
Their inverses delete the displayed added cut or undo \(\iota_i\).

The corresponding active tests are \(\hat p=1,1\in S\); \(\hat p=i+2,i+1\notin S\); and \(\hat p=m-1,m-2\in S\), respectively.
They partition the degree-one occurrences.

At the first non-apex position, degree one requires cut $1$, since otherwise the root has more than one child.
A non-spine position has only its incoming edge, giving precisely the pendant test.
A spine position has both an incoming and an outgoing edge unless it is last; a last spine position still has later attached leaves unless it is position $m-1$.
These observations give the last-endpoint test and exhaust the possibilities, including the two distinct tests at position $m-1$.

\begin{samepage}
To specify the residuals, put \(q_T=\max(\{0\}\cup T)\), \(p_T(j)=\max(\{0\}\cup(T\cap\{1,\ldots,j-1\}))\), and
\[
y_{\mathrm{last}}=y_{\bar\pi(q_T+1)},\qquad
E_{\mathrm{low}}=\sum_{j=1}^{m-3}\relu(y_{\bar\pi(p_T(j)+1)}-y_{\bar\pi(j+1)}),\qquad
H=y_{\mathrm{last}}-y_{\bar\pi(0)}+E_{\mathrm{low}}.
\]
\end{samepage}

Thus $y_{\mathrm{last}}$ is the last spine coordinate, and $C_{\mathrm{low}}=y_{\bar\pi(0)}+\relu(H)$.
Write $C_{\mathrm{up}}=C_{\pi,S}$.
For a pendant let $l=y_{\mathrm{parent}}-y_0$, with its preceding spine parent; for the first endpoint let $l=y_0-y_{\bar\pi(1)}$.
The pendant's nonzero edge is positive at $u_*$, while its head and other edges are zero, so its positive and negative untranslated supports are $H+l$ and $\relu(H)$.
For the first endpoint these two supports are $\relu(H)$ and $H+l$.
At the last endpoint the positive translated support is $C_{\mathrm{low}}$ and the negative one is $y_0+E_{\mathrm{low}}$.
The apex row cancels in each difference, and $y_0=0$, giving
\[
\begin{array}{ll}
\text{pendant:}&D_{u_*}C_{\mathrm{up}}+C_{\mathrm{low}}=y_{\mathrm{last}}+l+E_{\mathrm{low}},\\
\text{first:}&D_{u_*}C_{\mathrm{up}}-C_{\mathrm{low}}=-y_{\mathrm{last}}-l-E_{\mathrm{low}},\\
\text{last:}&D_{u_*}C_{\mathrm{up}}-C_{\mathrm{low}}=-y_0-E_{\mathrm{low}}.
\end{array}
\]
Each upper root here arises from its lower root by inserting label zero, so we call these three residuals the insertion residuals and the count of preimages of a lower root, computed next, the insertion count.
Each residual is a point row plus signed individual segments.

For every lower cut set $T$, including $T=\varnothing$, put $w_T=(-1)^{m-1+|T|}/(m-1)!$.
A pendant has upper weight $-w_T/m$ and coefficient $-1$ on $C_{\mathrm{low}}$, while either endpoint has weight $w_T/m$ and coefficient $+1$.
For nonempty $T$ there are $m-2$ pendant slots and two endpoints, so the lower-root coefficient is exactly $m(w_T/m)=w_T$.
For empty $T$ no pendant is admitted; only the two endpoints contribute, with coefficient $(2/m)w_\varnothing$.

The empty-cut lower root is an auxiliary expression, not an admitted empty upper cut.
For a lower order $\bar\pi$ on positions $0,\ldots,m-2$, its apex is at position zero and root at position one:
\[
C_{\bar\pi,\varnothing}=y_{\bar\pi(0)}+\relu\left(y_{\bar\pi(1)}-y_{\bar\pi(0)}+
\sum_{j=2}^{m-2}\relu(y_{\bar\pi(1)}-y_{\bar\pi(j)})\right).
\]
Put $J=\{0,2,\ldots,m-2\}$, so that $|J|=m-2$, put $d_a=y_{\bar\pi(1)}-y_{\bar\pi(a)}$, and
\[
V_A=\relu\left(\sum_{a\in A}d_a+\sum_{a\notin A}\relu(d_a)\right).
\]
The Boolean sum $\sum_A(-1)^{|A|}V_A$ vanishes.
If some $d_a\ge0$, toggling that $a$ leaves the argument unchanged and reverses its sign in the sum.
If all $d_a<0$, a nonempty $A$ gives a negative argument and the empty $A$ gives zero, so every term vanishes.
Consequently
\[
\sum_{a\in J}V_{\{a\}}=V_\varnothing+\sum_{|A|\ge2}(-1)^{|A|}V_A.
\]
Here $V_\varnothing=\sum_a\relu(d_a)$ is a sum of separate segments.
For $|A|\ge2$, its support polytope is $\operatorname{conv}(0,\sum_{a\in A}d_a+\sum_{a\notin A}[0,d_a])$.
Its rank is at most $1+(m-2)-|A|\le m-3$.

For each $a\in J$, exchange apex position zero with position $a$, fixing root position one.
Direct substitution gives $C_{\bar\pi^{(a)},\varnothing}=y_{\bar\pi(a)}+V_{\{a\}}$.
Every fixed positional swap is a bijection on lower orders, including the identity when $a=0$.
Thus
\[
(m-2)\sum_{\bar\pi}C_{\bar\pi,\varnothing}
=\sum_{\bar\pi}\left(\sum_{a\in J}y_{\bar\pi(a)}
+V_\varnothing+\sum_{|A|\ge2}(-1)^{|A|}V_A\right).
\]
Division by $m-2>0$ gives the empty-star expression, so named because the tree of an empty-cut lower root is a star centered at position one.
Its terms are the star components, the vertex lists of the \(V_A\) are the star lists, the subsets \(A\subseteq J\) are the star masks, and the prefactor below is the star factor.
Together with the two endpoint weights its prefactor is $\frac{2}{m(m-2)}\frac{(-1)^{m-1}}{(m-1)!}$, multiplying every bracketed term with its displayed Boolean sign.

At $u_*$, the difference of the two selected supports of the target simplex is the maximum over the retained labels.
Subtracting it from the nonempty lower-root sum leaves one lower error.
Translate those labels by $-y_1$ to rebase them.
Every root gains $y_1$ before rebasing, so $F_{m-1}$ gains $W_{m-1}y_1$ and the maximum gains $y_1$.
The raw lower error is therefore
\[
f_{m-1}(0,y_2-y_1,\ldots,y_{m-1}-y_1)+(W_{m-1}-1)y_1.
\]
Retain this final linear row; the coefficient of the single lower-error occurrence remains one.

A nonstandard exceptional direction is a direction \(u\ne u_*\) whose coordinates take exactly two values, one of them at a single label.
For such a direction choose a permutation \(\tau\) of the labels whose integral permutation and recentering map \((B_\tau y)_i=y_{\tau(i)}-y_{\tau(0)}\) satisfies \(B_\tau u=tu_*\) for a rational \(t\ne0\).
Every \(\tau\) with \(\tau(0)\) equal to that single label has this property.

The top-error identity for the standard exceptional direction $u_*$ is established first, from the identity already established at arity $m-1$, and the nonstandard directions are then reduced to it.
The top-error induction reaches the case of a nonstandard exceptional direction only for \(m\ge4\).
The argument uses the following symmetry identities, each stated explicitly:
\[
F_m\circ B_\tau=F_m-W_my_{\tau(0)},\qquad
M_m\circ B_\tau=M_m-y_{\tau(0)},\qquad
f_m\circ B_\tau=f_m+(1-W_m)y_{\tau(0)}.
\]
The vertex identity $(B_\tau^Tp)\cdot u=p\cdot B_\tau u$ identifies positive faces of a pullback with positive faces at $u_*$ when $t>0$, and negative faces when $t<0$.
Thus $D_u(f_m\circ B_\tau)=\operatorname{sgn}(t)(D_{u_*}f_m)\circ B_\tau$.
On the other hand, the third displayed identity, for $f_m\circ B_\tau$, gives $D_u(f_m\circ B_\tau)=D_uf_m$, since a linear row has the same selected point support at $u$ and $-u$.
Combining these equations gives
\[
D_uf_m=\operatorname{sgn}(t)(D_{u_*}f_m)\circ B_\tau.
\]
The identity $D_uf_m=\operatorname{sgn}(t)(D_{u_*}f_m)\circ B_\tau$ therefore rests only on the symmetry identities for $F_m$ and $M_m$, the vertex identity, and the two expressions for $D_u(f_m\circ B_\tau)$ stated above.

To finish the top-error induction, write the completed subdivision of $f_m$ as $f_m=\sum_P\alpha_Ph_P+\rho$, where every term of $\rho$ has rank at most $m-2$.
Group normalized nonvertical facet rows $u=n_F/n_{F1}$ and define
\[
B_u(P)=\sum_{F:n_F/n_{F1}=u}\operatorname{sgn}(n_{F1})h_F.
\]
A matching facet with $n_{F1}>0$ occurs in $P_u h_P$, and one with $n_{F1}<0$ occurs in $-P_{-u}h_P$.
Subtracting its signed occurrence from $D_uh_P$ removes that facet literally and leaves the opposite vertex.
If neither selected face is a facet, both have rank at most $m-3$.
A full simplex of dimension at least two has no pair of parallel opposite facets, so these cases exhaust the possibilities.
For a term of $\rho$, a direction constant on its vertex list selects the whole list on both sides and cancels; otherwise both selected faces are proper and have rank at most $m-3$.

The preceding face cases give $D_uf_m$ a rank-at-most-$m-3$ expression, with its single lower-error term supplied by the arity induction hypothesis.
Hence the left-hand expression in
\[
G_u=D_uf_m-D_u\rho-\sum_P\alpha_P(D_uh_P-B_u(P))
=\sum_P\alpha_PB_u(P)
\]
has that rank bound.
The right side follows by applying the local face identities to the completed subdivision and collecting equal normalized rows.
It is a facet regrouping identity, not a claim that each facet on its right has low rank.

Every nonvertical facet occurs in exactly one such group, with its outward sign.
Using $h_P=R(P)+T_m\sum_u B_u(P)$ therefore gives
\[
f_m=\rho+\sum_P\alpha_PR(P)+T_m\sum_uG_u.
\]
The reconstruction terms and $\rho$ have rank at most $m-2$, and each $T_mG_u$ starts at rank at most $m-3$ and increases it by at most one.
These identities specify the local terms and lower-arity occurrences used below.
All regroupings use fixed arrays and numerical induction.

\subsection{Direct Arrays for the Low Expressions}
\label{ir:direct}

We now define every low expression without using a previously constructed expression as a computational input.
The local identities above supply the terminal terms.
A short binary word specifies how a terminal is carried to higher arities.

A primitive root is a vertex of a local term as computed before any local recentering, pullback or \(T_m\) translation.
The primitive roots are the vertices of root, face, star, segment and point lists, the Cramer-quotient vertices of split, section and other-facet keys, and the three-level point vertex, that is, cases (i)--(iii) of the length bound in \Cref{ir:tiers}.
A primitive vertex is a primitive root or its image under the bounded local recentering or pullback word of its key, case (iv) there.
All vertices are evaluated from their primitive roots by one matrix-word function.

\paragraph{\emph{Original facets and split paths.}}
For an original full simplex with vertices \(v_0,\ldots,v_{m-1}\), choose its facet rows by the systems
\[
n_i\cdot v_j-c_i=-[i=j]\qquad(0\le j<m).
\]
The system has order \(m\le b\); affine independence makes it nonsingular.
Its solution has \(n_i\) outward, and \(c_i-n_i\cdot y\) is the barycentric coordinate of vertex \(i\).
Normalize each nonvertical occurrence to \(u=n_i/n_{i1}\), with zeroth coordinate zero.
Keep the least occurrence of each rational-value normal, testing equality by cross multiplication in every coordinate.
The resulting list \(\mathcal U_m\) includes the facets of the target simplex, which occurs with coefficient \(-1\) in the subdivision of \(f_m\).
It depends only on original simplices.

A split-path key contains \(t\le m\), masks \(\mathcal{M}_0,\ldots,\mathcal{M}_t\) of original facets, and deletion labels \(d_0,\ldots,d_{t-1}\).
Require \(\mathcal{M}_0=\{0,\ldots,m-1\}\).
At each \(a<t\), let \((i_a,j_a)\) be the least pair in \(\mathcal{M}_a\) with \(i_a<j_a\) and \(n_{i_a1}n_{j_a1}>0\).
Require \(d_a\in\{i_a,j_a\}\) and \(\mathcal{M}_{a+1}=\mathcal{M}_a\setminus\{d_a\}\).
All these conditions are checked simultaneously on the given masks.
All unused fields are zero.
In particular the test does not construct a child vertex from a parent vertex.

For the step at \(a\), put
\[
g_a(y)=\left(\frac{n_{i_a}}{n_{i_a1}}-\frac{n_{j_a}}{n_{j_a1}}\right)\cdot y
-\left(\frac{c_{i_a}}{n_{i_a1}}-\frac{c_{j_a}}{n_{j_a1}}\right).
\]
Its first-coordinate coefficient is zero.
The new supporting inequality is \(s_ag_a(y)\le0\), where \(s_a=\operatorname{sgn}(n_{i_a1})\) if \(d_a=i_a\), and \(s_a=-\operatorname{sgn}(n_{i_a1})\) otherwise.
Indeed \(g_a\) has sign \(-\operatorname{sgn}(n_{i_a1})\) at the vertex opposite \(i_a\) and the opposite sign at the vertex opposite \(j_a\).
Moreover the barycentric cut used above is \(n_{i_a1}n_{j_a1}g_a=0\), with a positive multiplier.
Thus this convention selects exactly the two children of that cut.

At depth \(t\), the supporting system consists of the original rows in \(\mathcal{M}_t\) and the \(t\) new vertical rows.
There are exactly \(m\) constraints, indexed by slots \(r<m\): slot \(r\) holds the original row \(n_r\cdot y-c_r\) when \(r\in \mathcal{M}_t\), and otherwise the row \(s_ag_a\) of the step \(a<t\) with \(d_a=r\).
Each cut replaces one endpoint of an edge by an interior point, so each valid child remains a full simplex.

A split-vertex choice omits one slot and solves the remaining \(m-1\) equalities, rejecting singular candidates and candidates at which some row of the current system is positive.
The facet at slot \(r\) is the stored vertex list with its slot-\(r\) entry omitted.
A section key adds the next cut equality and omits two current constraints; a section of another facet additionally imposes that facet equality and omits three current constraints, including the already imposed facet.
There are polynomially many such omitted-set choices.

These omitted-set choices cover all vertices.
At the new edge point $z$, the $m-2$ old facets other than the selected pair, together with the cut plane, give $m-1$ independent equations.
At a retained old section vertex, discard one of the two old equalities whose combination is the cut plane and keep that plane instead.
The resulting equations remain independent because both coefficients of the selected pair are nonzero.

For intersection with another facet, impose its equality separately and apply the same argument, omitting that facet from the remaining old-equality choices to avoid repetition.
Thus omitting one, two or three constraints as specified enumerates every child, section and other-facet-section vertex, with singular or infeasible extra choices rejected.
Every vertex coordinate is therefore a single Cramer quotient of a small system whose equations come from the fixed pool of original planes.

Both signs of the nonvertical facet coefficients occur initially, since the barycentric derivatives sum to zero and are not all zero.
Deleting one member of a same-sign pair preserves both signs.
Call a valid path terminal when its last mask \(\mathcal{M}_t\) contains no pair \(i<j\) with \(n_{i1}n_{j1}>0\).
Consequently a terminal path has exactly two nonvertical facets, one of each sign.
Its lower facet is the one with negative outward first component.
This also proves, by numerical induction on the supplied mask paths, that every requested reconstruction child and terminal key exists.

\paragraph{\emph{The local expression.}}
For each \(m\ge4\), let \(\mathcal L_m\) contain the following unmerged terms.
Their coefficients and vertex selectors are the formulas in the preceding local identities.
\begin{enumerate}[label=\textup{(\alph*)},leftmargin=2em]
\item Include every section term of the order-prefix subdivision with coefficient \(-w_{\pi,S}\), and every final flat leaf with coefficient \(w_{\pi,S}\), where \(w_{\pi,S}=(-1)^{m+|S|}/m!\).
These constitute \(\rho_m\).
\item For every original full simplex of weight \(\alpha_P\), unfold its reconstruction along the valid split paths: include the lower facet of each terminal path with coefficient \(\alpha_P\), every internal section with coefficient \(-\alpha_P\), and every other-facet section after \(T_m\) with coefficient \(\alpha_P\operatorname{sgn}(n_{F1})\).
Vertical facets have coefficient zero.
Each \(T_m\) image is represented by its union-image term and its negative segment term.
\item For each \(u\in\mathcal U_m\), include \(T_m\) applied to the local part of
\[
D_uf_m-D_u\rho_m-\sum_P\alpha_P(D_uh_P-B_u(P)).
\]
For \(D_u\rho_m\), constant-direction terms cancel, and otherwise the two selected faces are proper.
For each simplex, remove the matching signed facet literally; its remaining selected faces have rank at most \(m-3\).
The local part of \(D_uf_m\) is supplied by the point and segment reductions, the three-level linear row, the repeated-level involution residuals, or the exceptional insertion and star residuals above.
Base faces are always split into their point and individual segment terms.

The exceptional case retains the row \((W_{m-1}-1)y_1\), but excludes its one occurrence of the lower \(f_{m-1}\).
For an empty lower cut, its star components have coefficient \(\frac{2}{m(m-2)}\frac{(-1)^{m-1}}{(m-1)!}\).
Nonstandard exceptional residuals are pulled back by their integral recentering map before the outer \(T_m\).
\end{enumerate}

A key of \(\mathcal L_m\) consists of finite selectors: the original root, cut, order-prefix, split-path, facet, sign, source-case and residual-component fields.
The key needs no field for the coordinate \(y_{\mathrm{last}}=y_{\bar\pi(q_T+1)}\) entering an insertion residual: it is the last spine coordinate of the lower root of that residual, and this coordinate is defined before the insertion residuals, so the root and cut fields of the key determine it.

We code a local key as a number with at most sixteen digits in base \(P_2(2(b+1)^2)\), one digit for each field.
A permutation of at most \(b\) labels is a base-\(b\) number with at most \(b\) digits, below \(P_2(b^2)\) because \(b\le P_2(b)\).
A cut set, a lower cut set or a star mask is a string of at most \(b\) bits.
An order-prefix word with its length, where \(N=\binom{m-2}{2}\) is the number of comparisons, is below \(\langle N,P_2(N)\rangle\le P_2(2N+4)\le P_2(b^2+4)\), since \(2N=(m-2)(m-3)\le b^2\).
A split-path key is a string of at most \((m+1)m+m^2+m\le2(b+1)^2\) bits.
Slot, position, sign, case, category and component fields are below \(4b\le P_2(b+1)\).

A key of kind (a) uses at most five fields, a key of kind (b) at most eight, and a key of kind (c) at most fourteen.
The fields of a key of kind (c) are a category tag, four for the direction \(u\) (its root permutation, cut set, order-prefix word of its final leaf, and slot), the sign, the source category, at most four source fields, the component index, and the image and pullback bits.
Hence every local key is below \(P_2(32(b+1)^2)\).
An invalid key has coefficient zero and the zero point as its polytope, and it is dropped when the valid keys are compacted.
We set \(\mathcal L_3=0\).

Every local coefficient is a fixed product of signs, factorial factors, the displayed star factor and, for a local segment term, its translation factor.
The three-level row \(2F_m(-u)\beta-\lambda\) is a single point vertex, since \(F_m(-u)\) is a scalar.

Write $u=\widetilde u/d_u$ by clearing its coordinate denominators once, with $d_u>0$.
Every root $C_{\pi,S}$ has integer coefficients and is positively homogeneous, so its value at $-u$ is its integer value at $-\widetilde u$ divided by $d_u$.
Consequently $F_m(-u)$ is evaluated by one integer sum over the single denominator $m!d_u$.
This $d_u$ is the common denominator of $u$, independent of the earlier cut index and of the star factor.
It is not a recursively obtained coefficient.

All masks are selected on original local lists.
Thus the local library is a fixed number of aggregate small-solve, product, integer-sum, comparison and reconstruction stages.
Its construction uses no lower-arity expression.

Each local term has rank at most \(m-2\).
For the terms inside the final \(T_m\), the bound before that operation is \(m-3\): it follows from the proper-face tests, the explicit insertion and star components, and the literal facet cancellation.
The remaining lower expression will be supplied separately by the recurrence below.

\paragraph{\emph{The two exceptional directions.}}
Every normalized direction has \(u_0=0,u_1=1\).
If its exceptional label were different from both zero and one, these two coordinates would have to be equal.
The only one-exception directions are therefore
\[
u_+=u_*=(0,1,\ldots,1),\qquad u_-=e_1.
\]
Both occur in \(\mathcal U_m\): the target simplex has the opposite-zero facet with normal \((1,\ldots,1)\), and the facet \(y_1=0\) with normal \(-e_1\).
Normalizing the latter gives \(e_1\).
The standard exceptional identity has sign \(+1\) and lower map
\[
\Phi_+(y)=(0,y_2-y_1,\ldots,y_{m-1}-y_1).
\]
For \(u_-\), swap labels zero and one and recenter.
The resulting coordinates are $z_1=-y_1$ and $z_i=y_i-y_1$ for $i\ge2$.
This recentering sends $u_-$ to $-u_+$, so the general covariance formula has $t=-1$ and contributes the negative sign.
Composing its coordinates with the standard lower map $\Phi_+$ gives
\[
\Phi_-(y)=(0,y_2,\ldots,y_{m-1}).
\]
The insertion count in the local proof gives coefficient exactly one on the lower \(f_{m-1}\): the \(m-2\) pendant cases and two endpoint cases each contribute \(w_T/m\), for total \(w_T\).
The empty-cut contribution is already in \(\mathcal L_m\).
Thus the only recursive part of the top-error identity is
\[
T_m(f_{m-1}\circ\Phi_+)-T_m(f_{m-1}\circ\Phi_-).
\]
In the construction, \(f_{m-1}\) enters only through the lower array \(E_{m-1}\) specified directly by the recurrence \Cref{ir:paired} below, and \(E_{m-1}\) is not an argument to the function defining \(E_m\).

\paragraph{\emph{Uniform translations and primitive roots.}}
Before choosing the translations used by \(T_m\), we bound the components of the primitive library.
For binary rational operation depth \(k\), put \(C_k(L)=2^k(L+3)-3\).
A binary product or difference has cap \(2L+3\), and guarded inversion interchanges components and signs.
Put \(\mathfrak{i}=10(b+1)^2\), which bounds the component lengths of the original integer vertices, whose magnitudes are at most \(b+1\), and hence the entries of their facet systems.
Put \(\mathfrak{f}=\mathsf{Solve}(b,\mathfrak{i})\), \(B_d^0=C_1(\mathfrak{f})\) and \(\mathfrak{j}=C_2(\mathfrak{f})\).
These quantities bound the components of the normalized original facet rows and of the fixed pair planes.

For \(m\le b\), choose
\[
B_F=4(m+1)(B_d^0+2)+\lceil\log_2(m(m+1)+1)\rceil+2\left(m\lceil\log_2(m+1)\rceil+m+2\right)+10.
\]
Clearing \(u\) once and summing the at most \(m^m2^m\) original root values over \(m!d_u\) proves this cap for \(F_m(-u)\).
The row \(2F_m(-u)\beta-\lambda\) has cap \(B_{\mathrm{lin}}=C_3(\max(B_F,C_1(B_d^0),3))\).

Take monotone maxima over \(m\le b\), and put
\[
B_{\mathrm{root}}=\mathsf{Solve}(b,\mathfrak{j})+B_{\mathrm{lin}}+10,
\qquad J_g=10(b+1)^2,\qquad R_g=4(b+1),
\]
\[
B_t^0=B_{\mathrm{root}}+4+(2J_g+3)\lceil\log_2R_g\rceil,
\qquad B_0=\mathsf{Word}_{\Z}(b+1,J_g,B_t^0,B_{\mathrm{root}}).
\]
Here \(J_g\) bounds the number of factors of the local matrix words and of the lineage matrix words displayed below.
The quantity \(B_t^0\) bounds the component lengths of the local recentering and pullback matrices, and \(R_g\) is the per-arity growth base of the translations chosen next.
The primitive roots and their bounded local integral recenterings have this component cap; the generous word bound also covers the point, segment and star lists.
It is a fixed polynomial of degree at most twelve.
This bound concerns the library before any new \(T_m\) translation.

A single translation at each arity serves both signs and every local term:
\[
\mathfrak{c}=2^{B_0+4},\qquad c_m=\mathfrak{c}R_g^{8m+4},\qquad \mathcal B_m=\mathfrak{c}R_g^{8m+8}.
\]
These are binary integer strings; they are never used as numbers that size an array.
In these vertex bounds, the norm means \(\|v\|_\infty=\max_i|v_i|\).
Every terminal in \(\mathcal L_m\) is a list of primitive vertices, each of norm at most \(\mathfrak{c}\), or contains just one new local \(T_m\) image of such a list.
The nonstandard recentering precedes that operation.
Thus every terminal has norm at most \(2c_m<\mathcal B_m\).

If a lower vertex has norm at most \(\mathcal B_{m-1}=\mathfrak{c}R_g^{8m}\), its image under \(\Phi_\pm^T\) has norm at most \(b^2\mathfrak{c}R_g^{8m}\le \mathfrak{c}R_g^{8m+2}<c_m\).
Translation makes its first coordinate nonnegative, and either translated or projected image has norm at most \(2c_m<\mathcal B_m\).
This proves that the same \(c_m\) is valid for every term and both signs.

Put
\[
O_{m,\pm}(P)=\operatorname{conv}\left((\Phi_\pm^TP+c_me_1)
\cup\operatorname{pr}_m(\Phi_\pm^TP+c_me_1)\right).
\]
For each signed coefficient \(a\), the two segment contributions cancel separately:
\[
aT_mh_{\Phi_+^TP}-aT_mh_{\Phi_-^TP}
=ah_{O_{m,+}(P)}-ah_{O_{m,-}(P)}.
\]
This identity uses neither a sum of earlier coefficients nor a test that \(a\ne0\).
It specifies the recurrence
\begin{equation}
E_3=0,\qquad
E_m=\mathcal L_m+\sum_{(a,P)\in E_{m-1}}
\left(ah_{O_{m,+}(P)}-ah_{O_{m,-}(P)}\right).
\label{ir:paired}
\end{equation}
The term and vertex keys of the paragraph \emph{Closed term and vertex keys} below specify all its terms directly.

The base case can be seen directly from the roots.
For \(m=3\), the only nonempty cut set is \(S=\{1\}\), and
\[
\begin{aligned}
C_{\pi,\{1\}}(y)
&=y_{\pi(0)}+\relu\left(y_{\pi(2)}-y_{\pi(0)}
+\relu(y_{\pi(1)}-y_{\pi(2)})\right)\\
&=\max(y_{\pi(0)},y_{\pi(1)},y_{\pi(2)})=M_3(y).
\end{aligned}
\]
For rational scalars \(u\) and \(v\), the identity \(v+\relu(u-v)=\max(u,v)\) holds.
Apply it first to the two non-apex coordinates and then to their maximum and the apex coordinate.
Each of the six permutations has weight \((-1)^4/3!=1/6\), so \(F_3=M_3\) and \(f_3=0\), explaining the empty low array \(E_3\).
Thus the root average already gives a two-hidden-layer expression for the reduced maximum at arity three.
For an original rational input \(x=(x_0,x_1,x_2)\), restoring coordinates gives \(x_0+F_3(0,x_1-x_0,x_2-x_0)=\MAX_3(x)\).
The fixed-array verification of this base case is given in the arity induction of \Cref{ir:tiers}.

\paragraph{\emph{Closed term and vertex keys.}}
A term key for \(E_m\) is \((m,s,\mathfrak{l},\varepsilon_{s+1},\ldots,\varepsilon_m)\), where \(3\le s\le m\), \(\mathfrak{l}\) is a valid local key of \(\mathcal L_s\), and every \(\varepsilon_i\) is a plus or minus sign.
Unused slots are zero.
Its coefficient is
\[
a_{s,\mathfrak{l}}(-1)^{\#\{i:\varepsilon_i=-\}}.
\]
Here \(a_{s,\mathfrak{l}}\) is the coefficient of the local term with key \(\mathfrak{l}\) in \(\mathcal L_s\).
Its polytope starts at the local terminal and applies the indicated \(O_{i,\varepsilon_i}\) at each higher arity.
A vertex key adds three fields to a term key: a primitive-root selector; the bounded local wrapper choices, namely the bounded local recentering or pullback word of the primitive root's key that turns that primitive root into a primitive vertex (case (iv) of the length bound in \Cref{ir:tiers}); and one translated/projected image bit per higher arity.
At each higher arity \(i\), with its translation \(c_i\) and projection \(\operatorname{pr}_i\), use the corresponding homogeneous lineage matrices
\[
\begin{pmatrix}\Phi_\pm^T&c_ie_1\\0&1\end{pmatrix},
\qquad
\begin{pmatrix}\operatorname{pr}_i\Phi_\pm^T&0\\0&1\end{pmatrix}.
\]
Concatenate the local wrapper matrices and all lineage matrices, and evaluate this entire integral word once from its pre-translation primitive root.
The word always starts at a pre-translation primitive root, never at a local vertex already translated by \(T_m\), so its starting vertex obeys the caps of \emph{Uniform translations and primitive roots}, which concern the library before any \(T_m\) translation.
The matrices carry their source and target frames; padded coordinates are zero and the homogeneous coordinate is one.

The following quantities bound every matrix factor and every vertex component, respectively:
\[
B_{\mathrm{tr}}=B_0+8+(8b+4)\lceil\log_2R_g\rceil,
\qquad B_v=\mathsf{Word}_{\Z}(b+1,J_g,B_{\mathrm{tr}},B_0+4).
\]
There are at most a fixed number of local factors plus \(b\) lineage factors, below \(J_g\).
Thus this cap applies to the complete concatenated word and has degree at most fourteen.
Grouping coordinate walks gives every local and lineage prefix equation.

Enumerate the complete padded numerical domains of these keys and compute all arity arrays in parallel.
Validity uses the local tables and the supplied sign and image words only.
Count valid keys and use their ranks for compact indices, with least inverse lookup.
All scalar summands and output headers use these actual counts.
Distinct term keys remain distinct even when their polytopes coincide; duplicate vertices remain in their maxima.
Invalid keys do not enter the compact arrays.

Partition the keys at arity \(m\) into \(s=m\) and \(s<m\).
The first class is exactly \(\mathcal L_m\).
Deleting the last sign is a bijection from the second class to an arity-\(m-1\) key and one sign.
Coefficient parity and vertex-word grouping therefore prove \Cref{ir:paired} as an identity of the directly specified flat arrays.
No induction asserting the existence of a new string is used.

The rank and norm proofs are numerical inductions on these fixed arrays.
The rank of \(O_{m,\pm}(P)\) is at most \(\dim P+1\), since differences of its vertices lie in the space spanned by the vertex differences of \(\Phi_\pm^TP\) and the one vector \(e_1\).
The local ranks and the base point-zero expression therefore give rank at most \(m-2\) for every term of \(E_m\).
The preceding inequalities give its norm bound \(\mathcal B_m\).

A term key is a tuple \((m,s,\mathfrak{l},\varepsilon)\) with \(m,s\le b\), a local key \(\mathfrak{l}\) and at most \(b\) signs, so it is below \(P_2(32(b+1)^2+3(b+1))\le P_2(33(b+1)^2)\), using \(3(b+1)\le(b+1)^2\).
In particular every \(E_m\) has fewer than \(P_2(33(b+1)^2)\) terms.
Within a term, a vertex selector is below \(P_2(5(b+1))\).
A primitive list has at most \(2^b+1\) entries or is indexed by at most three omitted slots, below \(b^3\le P_2(3(b+1))\); the selector adds two wrapper bits and at most \(b\) image bits.
So vertex keys are below \(P_2(35(b+1)^2)\), using \(5(b+1)\le2(b+1)^2\).

An auxiliary key adds to a vertex key either a minor permutation tuple, below \((b+1)^{b+1}\le P_2((b+1)^2)\), or a word position, at most \(J_g\le P_2(2(b+1)+4)\), and in the second case also a coordinate walk of at most \(J_g+1\) indices below \(b+2\le P_2(b+1)\).
The walk alone is below \(P_2((J_g+1)(b+1))=P_2(10(b+1)^3+(b+1))\).
Multiplying this bound by the vertex-key bound \(P_2(35(b+1)^2)\) and by both alternative bounds \(P_2((b+1)^2)\) and \(P_2(2(b+1)+4)\), and using \(P_2(a)P_2(a')=P_2(a+a')\), shows that every auxiliary key is below \(P_2(10(b+1)^3+36(b+1)^2+3(b+1)+4)\).
Since \(36(b+1)^2+3(b+1)+4\le14(b+1)^3\) for \(b\ge2\), this exponent is at most \(24(b+1)^3\), which also exceeds the exponents of the local, term and vertex keys, so every padded domain of this subsection is below \(P_2(w_D(b))\), with \(w_D(b)=24(b+1)^3\).
These comparisons are instances of those collected in \Cref{ir:packets}.
All computations are a fixed number of aggregate arithmetic and selection stages on the original keys.

\subsection{Direct Face Corrections and Integer Readouts}
\label{ir:readout}

This subsection turns the low-rank expressions into the final network.
Repeated face operations remove the low-rank terms and leave an exact relation that expresses each simplex support through linear rows, hinge features and smaller simplices; solving these relations gives the dictionary coefficients and the block \(Q\).

The completed low arrays determine the following closure and directions before any input-dependent proof.
Their definitions do not assume the functional equality verified in \Cref{ir:tiers}.
Close every low-rank list just constructed under masks consisting of its vertices in the affine hull of \(J\), where \(J\) is an affinely independent sublist of at most \(b+1\) original vertices.
For \(M\) lists of length at most \(v\), at most \(2M(v+1)^{b+1}\) records suffice.

This family is face-closed.
For a list with vertex set \(\mathcal V\) and the affine hull \(A\) of such a \(J\), the mask is \(\mathcal V\cap A\).
Suppose a face of \(\operatorname{conv}(\mathcal V\cap A)\) has affine hull \(A'\subseteq A\).
Every original vertex in \(A'\) satisfies its exposing equality, hence the mask of the face is exactly \(\mathcal V\cap A'\).
An independent spanning sublist of that face occurs among the enumerated \(J\).

For each rank \(k\), choose one representative of every direction space in this closure.
The orthogonal complement $V^\perp$ consists of the vectors whose dot product with every vector of $V$ is zero.
A simplex-subset space is the direction space of the convex hull of a nonempty subset of the vertices \(0,e_1,\ldots,e_{b-1}\).
For a representative space \(V\), choose an integer direction \(u_V\in V^\perp\) outside the orthogonal complement of every other rank-\(k\) space and of every simplex-subset space of dimension greater than \(k\).

Such a direction exists, and the following finite search finds one.
Compute an integer kernel basis \(z_0,\ldots,z_{c-1}\) by small minors and test \(u(t)=\sum_{j<c}t^jz_j\).
For each forbidden space choose a spanning row not in \(V\); its scalar product with \(u(t)\) is a nonzero polynomial of degree at most \(c-1\).
If there are \(s\) forbidden spaces, one of \(1,\ldots,s(c-1)+1\) works.

For completeness, we prove the fact used here, that a nonzero polynomial of degree at most \(c-1\) cannot vanish at \(c\) distinct integers, by a finite determinant argument.
Evaluate at \(c\) distinct trial integers \(t_a\) and replace monomial columns by the Newton columns \(\prod_{a<j}(X-t_a)\).
Their subset-expanded coefficients give a triangular change matrix with diagonal one; the evaluation determinant is \(\prod_{a<j}(t_j-t_a)\ne0\).
Thus a nonzero coefficient row cannot vanish at \(c\) trials.
All coefficient arrays and permutation expansions are formed before the numerical counting argument.
This uses no induction quantifying polynomial strings.

To lower the rank of every record one level at a time, we use the face operation \(P_u\) of \Cref{ir:span} and define the fixed operator
\[
\mathfrak{A}_k=I-\sum_{V:\dim V=k}P_{u_V}.
\]
For $k\ge1$, consider a rank-$k$ record with direction space $W$.
Its own direction $u_W$ annihilates every vertex difference, so all vertex dot products tie and $P_{u_W}$ selects the entire list.

For any other rank-$k$ representative $V$, the choice of $u_V$ excludes $W^\perp$.
Hence its dot products are not all equal on this list, and its maximizing face is proper.
A proper exposed face has dimension at most $k-1$: it lies in a level hyperplane of a nonconstant linear functional on the affine hull.
The whole-record contributions from $I$ and $P_{u_W}$ cancel, leaving only these lower-dimensional faces.
Selecting a face of a record already below rank $k$ cannot increase its rank.
Thus $\mathfrak{A}_k$ lowers rank $k$ and preserves the required bound on all lower-rank records.

The directions do not depend on any coefficients, so the same operators serve every later target.
Expanding at most \(b\) such operators uses at most \((R+1)^b\) face words, for the fixed closure size \(R\).
Face words select faces in lexicographic order, defined as follows.
For distinct equal-length rational tuples \(a,a'\), lexicographic comparison uses their first differing entry: we write \(a<_{\mathrm{lex}}a'\), or \(a'>_{\mathrm{lex}}a\), when \(a\) has the smaller entry there.
A tuple is lexicographically positive or negative according to its first nonzero entry, and is zero when every entry is zero.
Each word selects a mask of its original vertex list by successive exposure in this order.

We now apply these operators to the support of a single simplex.
Let $\Delta_T$ be the simplex on a nonempty subset $T$ of the vertices of $\Delta_b=\operatorname{conv}(0,e_1,\ldots,e_{b-1})$.
A singleton has a coordinate linear support, so consider $s=|T|\ge2$.
Order the $d=b-s$ outside vertices as $v_1,\ldots,v_d$ and write $y_v=v\cdot y$ for a vertex's coordinate linear support.
For a nonzero vertex $e_i$ use deletion direction $\hat u_{e_i}=e_i$, and for vertex zero use $\hat u_0=-\sum_{i=1}^{b-1}e_i$.
The dot product with $\hat u_v$ is uniquely largest at the deleted vertex $v$ and equal at all the other vertices of $\Delta_b$.
For a subset $A$ of the vertices of $\Delta_b$ with $|A|\ge2$ and $v\in A$, this gives
\[
D_{\hat u_v}h_{\Delta_A}=y_v-h_{\Delta_{A\setminus\{v\}}},
\qquad L_T=D_{\hat u_{v_d}}\cdots D_{\hat u_{v_1}},\quad \varepsilon_T=(-1)^d.
\]

Every $D_u$ kills linear rows.
For $d\ge1$, induction through the deletions therefore gives
\[
L_Th_{\Delta_b}=\varepsilon_T h_{\Delta_T}+(-1)^{d-1}y_{v_d};
\]
for $d=0$ the operator is the identity and the final point row is absent.
The surviving row belongs to the last deleted outside vertex, and will be kept as a linear contribution.
Each deletion \(D_{\hat u_v}=P_{\hat u_v}-P_{-\hat u_v}\) acts termwise.
On a constant-direction support, that is, a term whose vertex differences are all annihilated by \(\hat u_v\), both exposures select the whole list and the two contributions cancel; on every other term both selected faces are proper, so each has rank at most one less than the term.
The terms of the low expression for $f_b$ have rank at most $b-2$, so after the $d=b-s$ deletions every remaining term has rank at most $b-2-d=s-2$.

Put $C_T=\mathfrak{A}_1\mathfrak{A}_2\cdots \mathfrak{A}_{s-2}$, with the rightmost operator acting first, and $O_T=C_TL_T$.
The empty correction product is the identity.
Expanding the correction means choosing at each rank either its identity term or one selected face with coefficient $-1$.
For a choice word $w$, let $\lambda_w=(-1)^{\text{number of face choices}}$ and let $P_w$ be its ordered face operation.
The all-identity choice is denoted $0$, with $\lambda_0=1$.
Every nonidentity choice first selects a proper face of $\Delta_T$: its chosen direction avoids the orthogonal complement of the $(s-1)$-dimensional direction space of $\Delta_T$.
All later faces stay within that proper face.
Writing $T(w)\subsetneq T$ for the retained nonempty vertex set gives
\[
C_T=\sum_w\lambda_wP_w,\qquad
C_Th_{\Delta_T}=h_{\Delta_T}+\sum_{w\ne0}\lambda_wh_{\Delta_{T(w)}}.
\]
The rank cancellations already proved make $r_T=O_Tf_b$ homogeneous linear.
Recover it from the original expressions, rather than from the low-array coefficients:
\[
r_{Ti}=(O_TF_b)(e_i)-(O_Th_{\Delta_b})(e_i),
\qquad r_T(y)=\sum_{i=1}^{b-1}r_{Ti}y_i.
\]
A homogeneous linear function is determined by these basis values.
Both expressions are evaluated by masks of their original vertex lists.

We now specify exactly how such a face word becomes dictionary data.
For its chronological directions $u_1,\ldots,u_t$, put $\mathbf v(a)=(a\cdot u_1,\ldots,a\cdot u_t)$.
Partition the root edges into $I_+,I_-,I_0$ according as $\mathbf v(l_j)$ is lexicographically positive, negative or zero, and put
\[
H_w=h+\sum_{j\in I_+}l_j+\sum_{j\in I_0}\relu(l_j),
\qquad a_w=\mathbf v(h)+\sum_{j\in I_+}\mathbf v(l_j).
\]
Maximization on the base chooses each edge endpoint independently: positive tuples contribute the full edge, negative tuples contribute zero and zero tuples retain the segment.
Comparison with the apex tuple zero gives the selected translated-root support
\[
\begin{cases}
y_{\pi(0)},&a_w<_{\mathrm{lex}}0,\\
y_{\pi(0)}+H_w,&a_w>_{\mathrm{lex}}0,\\
y_{\pi(0)}+\relu(H_w),&a_w=0.
\end{cases}
\]
The last case is exactly the raw ternary tag with linear, omitted and hinged states on $I_+,I_-,I_0$.
The middle case is a linear row plus pair hinges.
The pair hinges of the middle case, those of the edges in \(I_0\), stand outside any outer feature, and each is stored as an oriented pair hinge, with the smaller coordinate index first.
For such a hinge in the reverse orientation use $\relu(y_i-y_j)=\relu(y_j-y_i)+(y_i-y_j)$ when $i>j$.
The outer features keep their original oriented inner hinges; they need no such reversal.
Every apex row and hinge-reversal row is retained.

Combining the deletion and correction equations yields the literal identity
\[
h_{\Delta_T}=\varepsilon_TO_TF_b-\varepsilon_Tr_T
+\begin{cases}C_Ty_{v_d},&d\ge1,\\0,&d=0,\end{cases}
-\sum_{w\ne0}\lambda_wh_{\Delta_{T(w)}}.
\]
Thus no last-vertex term is used when the deletion word is empty.
The positive sign of the outside point row follows from $-\varepsilon_T(-1)^{d-1}=1$.
In \Cref{ir:relation} below, \(g_j\) ranges over the tagged features produced by the face rule above, namely the coordinate linear rows \(y_1,\ldots,y_{b-1}\), the oriented pair hinges and the raw outer features \(\relu(H_w)\); the final substitution below rewrites them through the dictionary features of \Cref{rc:dictionary}.
Expanding the first three terms by the face rule above and the coordinate linear rows defines $a_{Tj}$.
Explicitly, the proper-face coefficient is $M_{TA}=-\sum_{w\ne0:T(w)=A}\lambda_w$.
Thus
\begin{equation}
h_{\Delta_T}=\sum_j a_{Tj}g_j+
\sum_{\varnothing\ne A\subsetneq T}M_{TA}h_{\Delta_A}.
\label{ir:relation}
\end{equation}
There is no remaining deletion-sign factor in $M_{TA}$.

At $s=2$ the correction is empty.
For a singleton target $T=\{0\}$, set every $a_{Tj}$ to zero.
For $T=\{e_i\}$, set $a_{Tj}$ to one at the face-rule tag for the coordinate linear row $y_i$, and to zero at every other tag.
All these entries are formed from original masks and fixed words before any smaller-target substitution.

All coefficients in \Cref{ir:relation} have the chosen common denominator $D_0=b!$.
Expand $L_T$ into its signed face choices and $C_T$ into the words $w$ before merging any terms.
Each original root has numerator $\pm1$ over $D_0$, and each original simplex term has integer coefficient represented over $D_0$.

Exposure only chooses a mask of the original integer vertices, whose component magnitudes are at most $b+1$.
Consequently every basis value used in $r_{Ti}$ is an integer root or simplex maximum multiplied by one of those original coefficients.
The generic directions determine the masks, but none of their denominators becomes a readout coefficient.
The low-array star factors, smaller-arity coefficients and rational translations prove rank bounds only and never multiply these readout atoms.

For a selected base face, the head and hinge reversals give coordinate coefficients of magnitude at most $2b+2$.
An apex-containing face gives one raw outer feature with coefficient one plus its apex row.
After the final substitution $y_i=x_i-x_0$, a linear row contributes the negative sum of at most $b-1$ such coefficients to coordinate zero; outer features and pair hinges are translation invariant.
Thus $\mathfrak{m}=4(b+1)^2$ bounds every individual integer conversion multiplier.
Coordinate rows use $x_i=\relu(x_i)-\relu(-x_i)$.
For a recovered basis atom, its value has magnitude at most $b+1$, and its coordinate-row conversion has only signs $0,1,-1$.

Let $N_T$ count the actual elementary signed-word, component and basis-value contributions after these conversions and before merging by a dictionary or proper-face tag.
When $d\ge1$, retain $C_Ty_{v_d}=\sum_w\lambda_wy_{v_d}$ as separate word contributions; its potentially large summed coefficient is not an individual multiplier bounded by $\mathfrak{m}$.
When $d=0$, this last-vertex contribution is absent.
Each such unmerged numerator over $D_0$ has magnitude at most $\mathfrak{m}(b+1)D_0$.
An integer sum therefore gives numerator magnitude at most $N_T\mathfrak{m}(b+1)D_0$, and the component cap
\[
B_T=\lceil\log_2(b+2)\rceil+3+\lceil\log_2(N_T+1)\rceil+
\lceil\log_2(\mathfrak{m}+1)\rceil+\lceil\log_2(D_0+1)\rceil+10
\]
dominates its signed binary length.
Divide once by $D_0$; no product denominator over all summands is formed.

For each face-rule feature tag \(j\), let \(c_{Tj}\) be the sum, over every chain \(T=A_0\supsetneq A_1\supsetneq\cdots\supsetneq A_t\) of nonempty sets, of the product \(M_{A_0A_1}\cdots M_{A_{t-1}A_t}a_{A_tj}\).
There are at most \(b\) factors.
If their integer numerators are \(z_1,\ldots,z_k\), print the chain numerator \(z_1\cdots z_kD_0^{b-k}\), and sum these integers over the single denominator \(D_0^b\).
Partition chains by their first step.
Integer distributivity gives \(c_{Tj}=a_{Tj}+\sum_{\varnothing\ne A\subsetneq T}M_{TA}c_{Aj}\).
Numerical induction on \(|T|\), with these arrays fixed, substitutes \Cref{ir:relation} and proves the final dictionary expansion.

Let $T_{\mathrm{full}}$ be the full vertex set of $\Delta_b$, whose labels are $\{0,\ldots,b-1\}$, and put $c_j=c_{T_{\mathrm{full}},j}$.
These coefficients still index the face-rule features.
Finally precompose their sum by \(x\mapsto(0,x_1-x_0,\ldots,x_{b-1}-x_0)\) and add \(x_0\).
All linear rows use coordinate-sign hinges.
The first hidden layer stores those hinges and all oriented pair hinges, of width \(\mathfrak{w}_1=b^2+b\); the second stores the dictionary features and copies of needed first-layer hinges.
This gives the homogeneous block \(Q\), stored in the common network record of \Cref{sec:foundations}.
All dense zeros and positive denominators are explicitly stored.

\Cref{tab:ir-guide} separates the direct definitions from the identities that verify them.
In particular, the arity induction verifies the completed low arrays; the subset-chain induction then verifies the readout of the completed block.
All count and precision bounds are consolidated into the guarded allocation of \Cref{ir:packets}.

\begin{table}[t]
\caption{Construction and verification of the direct block. The paragraph names identify the relevant interfaces within the cited subsections.}
\label{tab:ir-guide}
\centering
\begin{tabular}{@{}p{.22\linewidth}p{.36\linewidth}p{.35\linewidth}@{}}
\toprule
Object & Direct data and bounds & Verifying identity or invariant\\
\midrule
Local terms (a): subdivision &
\Cref{ir:direct}, \emph{The local expression}: section coefficient \(-w_{\pi,S}\), flat-leaf coefficient \(w_{\pi,S}\); local-key count there, and primitive cap \(B_0\) in \emph{Uniform translations and primitive roots}. &
\Cref{ir:tiers}, \emph{Local identities at a point} and \emph{The arity induction}: the terms give \(\rho_m\), with rank at most \(m-2\).\\[.6em]
Local terms (b): reconstruction &
\Cref{ir:direct}, \emph{The local expression}: coefficients from \(\alpha_P\) and facet signs, with each \(T_m\) image split into its union and segment terms. &
\Cref{ir:tiers}, the same two paragraphs: split-path reconstruction gives the stored contribution; \emph{Ranks} verifies its rank bound.\\[.6em]
Local terms (c): face residuals &
\Cref{ir:direct}, \emph{The local expression}: signed local residuals, including the exceptional star factor; rank at most \(m-3\) before the outer \(T_m\). &
\Cref{ir:tiers}, \emph{Local identities at a point}: (I7)--(I9) and the exceptional identities; \emph{Ranks}: rank at most \(m-2\) after transport.\\[.6em]
Low arrays \(E_m\) and vertices &
\Cref{ir:paired} and \Cref{ir:direct}, \emph{Closed term and vertex keys}: local coefficient times the lineage sign; direct matrix words, key counts and component caps. &
\Cref{ir:tiers}, \emph{The arity induction}: equality of the root error and low expression at point keys satisfying the frame and depth guard; \emph{Ranks}: the rank bound persists.\\[.6em]
Face masks, corrections and readout &
\Cref{ir:readout}: closure counts, fixed directions, common denominator \(b!\), coefficient caps and subset-chain coefficients. &
\Cref{ir:tiers}: lexicographic mask identity; \emph{Deletion and correction} proves \Cref{ir:relation}; \emph{Subset chains and the block} proves \(Q(x)=_{\Q}\MAX_b(x)\).\\[.6em]
Rational point arrays &
\Cref{ir:tiers}: fixed point-word arrays, count guard, perturbation and component bounds, formed before numerical induction. &
The same subsection: transport of the local identities at the guarded point keys; the completed point and value arrays are string parameters of the arity induction.\\[.6em]
Cone certificate and supplied block &
\Cref{ir:cones}: projected solves and compressed cover; \Cref{ir:packets}: total length and count allocation. &
\Cref{ir:cones}: certificate equations; \Cref{ir:small}: admitted block and certificate; \Cref{ir:rate}: the final depth inequality.\\
\bottomrule
\end{tabular}
\end{table}

\subsection{Verification on Fixed Arrays of Rational Points}
\label{ir:tiers}

This subsection proves that the arrays constructed above compute the intended functions, and hence that \(Q(x)=\MAX_b(x)\) at every rational point \(x\).
Every expression is evaluated on a finite array of rational points formed in advance, and the proof is a numerical induction on that array.

The local, lineage, face-closure and coefficient arrays are chosen before their functional correctness is proved.
The closure inspects the completed arity-\(b\) low array of \Cref{ir:paired}; it does not feed an evolving expression back into the arity construction.
The complete deletion and correction words are also formed in advance.

For a face word with directions \(u_1,\ldots,u_t\), applied in this order, and an original vertex list \(p_0,\ldots,p_{v-1}\), write \(\mathbf v_j(p)=(p\cdot u_1,\ldots,p\cdot u_j)\) for \(j\le t\), so that \(u_1\), the direction applied first, is compared first.
For rational tuples of equal length, \(a\le_{\mathrm{lex}}a'\) means that \(a=a'\) or that \(a_k<a'_k\) at the least index \(k\) with \(a_k\ne a'_k\).
Comprehension forms, for all \(j\le t\) at once, the masks
\[
\Lambda_j=\{i<v:\forall i'<v\,\mathbf v_j(p_{i'})\le_{\mathrm{lex}}\mathbf v_j(p_i)\},
\]
whose defining formula has only bounded number quantifiers over vertex, coordinate and word positions.
We verify in \(\bm{\mathcal T}\) that \(\Lambda_0\) is the whole list and that, for every \(j<t\),
\[
\Lambda_{j+1}=\{i\in\Lambda_j:\forall i'\in\Lambda_j\,p_{i'}\cdot u_{j+1}\le p_i\cdot u_{j+1}\},
\]
so that \(\Lambda_{j+1}\) is the face of the list \(\Lambda_j\) exposed by \(u_{j+1}\), with all maximizers retained.
No induction on \(j\) is needed for this, because every \(\Lambda_j\) is formed directly; the verification is a separate argument for each \(j\), with \(j\) as a parameter.

First, \(\le_{\mathrm{lex}}\) is a total order on tuples of length \(j\).
Totality and antisymmetry follow from the trichotomy of the rational order at the least differing index, which bounded minimization supplies.
For transitivity, suppose \(a\le_{\mathrm{lex}}a'\le_{\mathrm{lex}}a''\).
If \(a=a'\) or \(a'=a''\), then \(a\le_{\mathrm{lex}}a''\) is one of the two hypotheses.
Otherwise let \(k\) and \(k'\) be the least differing indices of the pairs \(a,a'\) and \(a',a''\).
Then \(a\) and \(a''\) agree below \(\min(k,k')\) and satisfy \(a_{\min(k,k')}<a''_{\min(k,k')}\), as the cases \(k<k'\), \(k>k'\) and \(k=k'\) show.

Second, \(\Lambda_j\) is nonempty.
For \(c\ge1\), bounded number induction on \(c\le v\) proves the formula stating that some \(i<c\) has \(\mathbf v_j(p_{i'})\le_{\mathrm{lex}}\mathbf v_j(p_i)\) for every \(i'<c\); this is the argument of \Cref{uc:label-max} for the order \(\le_{\mathrm{lex}}\).

Third, two members of \(\Lambda_j\) have equal tuples by antisymmetry, and a vertex whose tuple equals that of a member of \(\Lambda_j\) is a member, by transitivity.

Finally, directly from the definition, \(\mathbf v_{j+1}(p_{i'})\le_{\mathrm{lex}}\mathbf v_{j+1}(p_i)\) holds if and only if either \(\mathbf v_j(p_{i'})<_{\mathrm{lex}}\mathbf v_j(p_i)\), or \(\mathbf v_j(p_{i'})=\mathbf v_j(p_i)\) and \(p_{i'}\cdot u_{j+1}\le p_i\cdot u_{j+1}\).
Hence \(i\in\Lambda_{j+1}\) if and only if \(i\in\Lambda_j\) and \(p_{i'}\cdot u_{j+1}\le p_i\cdot u_{j+1}\) for every \(i'\) whose \(j\)-tuple equals that of \(i\).
By the third fact these \(i'\) are exactly the members of \(\Lambda_j\).
At \(j=0\) all tuples are empty, so \(\Lambda_0\) is the whole list.
The argument uses only bounded comparisons of the supplied vertices, directions and word positions, and the ordered-field laws of \Cref{lem:code-field}, whose proof uses only the integer laws of \Cref{fact:binary-arithmetic} and is therefore valid in \(\bm{\mathcal T}\), as recorded in \Cref{tc:parallel-arithmetic}.

The padded count bounds of \Cref{ir:readout} and of this subsection are explicit, with \(\bar b=b+1\) and the comparisons of \Cref{ir:packets}.
A closure record is given by a term key of \(E_b\), below \(P_2(33\bar b^2)\), and at most \(b+1\) vertex selectors of that term, each below \(P_2(5\bar b)\), so there are fewer than \(R_b=P_2(39\bar b^2)\) records, which also bounds the representatives.
The forbidden spaces of one representative number fewer than \(R_b+P_2(b)\le P_2(40\bar b^2)\), and the trial integers are below \(bP_2(40\bar b^2)\le P_2(41\bar b^2)\).

A rank or kernel computation on one record chooses at most \(b\) of its fewer than \(P_2(10\bar b)\) difference rows, at most \(b\) columns and a permutation below \(P_2(\bar b^2)\), so all such computations have fewer than \(P_2(51\bar b^2)\) keys.
A correction word has at most \(b\) letters, each the identity or a representative, so with its deletion sign word and its target there are fewer than \(P_2(b(39\bar b^2+1)+2b)\le P_2(41\bar b^3)\) words.
Their masks on all closure lists have fewer than \(P_2(41\bar b^3+39\bar b^2+5\bar b)\le P_2(56\bar b^3)\) keys.

The contributions counted by \(N_T\) are indexed by a word, an original root or simplex term of arity \(b\), below \(P_2(2\bar b^2)\), and a component and basis index, below \(P_2(2\bar b)\), so \(N_T<P_2(42\bar b^3)\).
The subset chains, with their target and dictionary tag, number fewer than \(P_2(5\bar b^2)\).
Every rejected tuple is included in these padded domains.
So every count of \Cref{ir:readout} and of this subsection that does not involve points is below \(P_2(w_R(b))\), with \(w_R(b)=56(b+1)^3\).

To bound the precision of the directions, put
\[
B_z=\mathsf{Solve}(b,10(B_v+2))+10,
\qquad \mathfrak{t}_b=c_R(b+1)^4+\lceil\log_2(b+1)\rceil+4,
\]
\[
B_d=B_z+b(\mathfrak{t}_b+2)+\lceil\log_2(b+1)\rceil+B_{\mathrm{tr}}+\mathfrak{j}+20,
\]
where \(c_R=41\), so that \(c_R(b+1)^4\) bounds the logarithm of the closure and forbidden-space counts and of the trial integers, all below \(P_2(41(b+1)^2)\).
The small-minor kernels and the tested moment-curve directions satisfy this cap, of degree at most eighteen.
All allocation logarithms are replaced by their fixed polynomial majorants.
The readout coefficients have polynomial component length by their integer chain formula; auxiliary low-expression coefficients never enter them.

Put \(\mathfrak{H}=6(b+1)\) as the maximum length of a point word, that is, a word in the point maps listed next.
It exceeds \(\tau_b+b+2\), where \(\tau_b=4(b-2)\) is the depth allowance of the arity induction below, and this is all that the local transformations and the later deletion, correction, basis and recentering operations use.
The alphabet of point maps is typed by source and target frame.
It contains the local projections, \(\Phi_\pm\), permutation and recentering maps, the perturbations \(y\mapsto u+\epsilon y\) for a rational \(0<\epsilon<1\) fixed below, and constant maps to each required \(u\) or basis vector.
Start words at the arbitrary input and at the basis of every local frame.

With the operator norm induced by the vector norm \(\|\cdot\|_\infty\), every nonperturbative linear map has norm at most \(2(b+1)\).
Perturbation maps contract by \(\epsilon<1\) and add an offset of component cap \(B_d\).
For \(s=n+|x|\), the logarithm of each point magnitude at depth at most \(\mathfrak{H}\) is therefore at most
\[
L_x=|x|+B_d+\mathfrak{H}(\lceil\log_2(2b+4)\rceil+2)+10.
\]

We next choose the perturbation size \(\epsilon\) so that a perturbation selects its maximizer only among the vertices already maximal at the direction \(u\).
For two vertices and a direction $u$, clear all coordinate denominators in their dot-product difference.
There are at most $b$ coordinates; each denominator uses at most two vertex components of cap $B_v$ and one direction component of cap $B_d$.
The common denominator is therefore at most $2^{b(2B_v+B_d)}$.
A nonzero integer numerator has absolute value at least one, so a nonzero exposed-vertex gap is at least $2^{-b(2B_v+B_d)}$.

A perturbation changes a difference of vertex values by at most $\epsilon b2^{B_v+L_x+1}$, using the component bounds on the two vertices and $y$.
Thus use
\[
\mathfrak{p}(s)=10(s+2)^2\left(B_v(s+2)+B_d(s+2)+\mathfrak{H}(s+2)+s+20\right),
\qquad \epsilon=2^{-\mathfrak{p}(s)}.
\]
Here $B_v(s+2)$, $B_d(s+2)$ and $\mathfrak{H}(s+2)$ denote $B_v$, $B_d$ and $\mathfrak{H}$ with every logarithm replaced by its polynomial majorant from \Cref{ir:packets} and with $b$ replaced by $s+2$.
These polynomials have nonnegative coefficients in $b+1$, hence are nondecreasing in $b$, so under the ambient count guard they dominate the corresponding quantities in this computation.
For $s=n+|x|$ on that guard, $b\le s+2$, and the displayed $\mathfrak{p}(s)$ is greater than
\[
b(2B_v+B_d)+\lceil\log_2(b+1)\rceil+B_v+L_x+2.
\]
Indeed $\lceil\log_2(2b+4)\rceil\le2(s+2)$, and substituting the bound for $L_x$ leaves only linear multiples of $B_v,B_d,\mathfrak{H},s+20$, each dominated by $10(s+2)^2$ times their sum.
Thus $\epsilon b2^{B_v+L_x+1}<\tfrac12 2^{-b(2B_v+B_d)}$.
The exponent \(\mathfrak{p}(s)\) is a fixed polynomial in \(s\) of degree at most twenty, since \(B_v\) has degree at most fourteen, \(B_d\) at most eighteen and \(\mathfrak{H}\) one, and the factor \(10(s+2)^2\) adds two.
The denominator of \(\epsilon\), a positive signed integer, has length \(\mathfrak{p}(s)+2\); its magnitude is never converted to a number.

We now count the typed point words explicitly.
The alphabet consists of the projections, the maps \(\Phi_\pm\), the input recentering, and a recentering map for each \(u\in\mathcal U_m\).
It also has the perturbation and constant letters of the directions \(\pm u\) for \(u\in\mathcal U_m\), of the \(2b\) deletion directions and of the correction directions, together with the \(b^2\) basis constants.

The normalized directions of all arities number fewer than \(P_2(2\bar b^2+3\bar b)\), with \(\bar b=b+1\), since an original facet is given by a root, the order-prefix word of a final leaf and a slot.
The correction directions number fewer than \(R_b=P_2(39\bar b^2)\).
So there are fewer than \(P_2(40\bar b^2)\) letters.
Hence there are fewer than \((P_2(40\bar b^2)+1)^{\mathfrak{H}+1}\le P_2(41\bar b^2(6\bar b+1))\le P_2(287\bar b^3)\) typed point words of length at most \(\mathfrak{H}\).
Adjoining a term and a vertex selector of \(E_b\), below \(P_2(35\bar b^2)\), shows that every point and value domain is below \(P_2(w_{\mathrm{pt}}(b))\), with \(w_{\mathrm{pt}}(b)=300(b+1)^3\).

Evaluate the point of each word independently by the complete rational matrix-word graph, including its base-vector factor.
Evaluate every required expression by one flat rational sum of finite maxima, using its actual term and vertex ranges.
Under a sufficiently large fixed exponential count guard, all these point, value and computation-witness arrays have polynomial bounds in \(n+|x|\).
They are fixed string parameters in the following numerical proofs.

We fix notation for these arrays.
Write \(N_{\mathrm{pt}}\) for the number of point keys, and for a key \(\kappa<N_{\mathrm{pt}}\) write \(\operatorname{fr}(\kappa)\) for its target frame, \(\operatorname{dep}(\kappa)\) for its word length and \(y_\kappa\) for its stored rational point.
For a stored vertex list \(P\), \(h_P[\kappa]\) is the stored maximum of the dot products \(p\cdot y_\kappa\) over the listed vertices, selected as in \Cref{uc:label-max}.
For an expression \(E\), a flat array of coefficient--list pairs, \(E[\kappa]\) is the stored flat rational sum of the coefficients times these maxima.
Thus \(F_m[\kappa]\), \(M_m[\kappa]\), \(E_m[\kappa]\) and \(f_m[\kappa]=F_m[\kappa]-M_m[\kappa]\) are entries of preformed arrays, and \(h_P(z)\) denotes the same maximum at any stored rational vector \(z\).

For a key \(\kappa\) and an alphabet letter \(g\) whose source frame is \(\operatorname{fr}(\kappa)\), write \(g\kappa\) for the key obtained by appending \(g\), of depth \(\operatorname{dep}(\kappa)+1\).
For a direction \(u\), \(\mathsf{per}_u\) is the perturbation letter \(z\mapsto u+\epsilon z\) and \(\mathsf{con}_u\) the constant letter \(z\mapsto u\).
The prefix equations of the matrix-word graph, obtained by grouping coordinate walks as in \Cref{ir:arithmetic}, give \(y_{g\kappa}=_{\Q}g(y_\kappa)\) coordinatewise whenever \(\operatorname{dep}(\kappa)<\mathfrak{H}\).

We call a formula admissible if its only quantifiers are number quantifiers bounded by number terms, its string parameters are the preformed arrays, and its function symbols are chosen \(\mathsf{FTC}^0\) functions.
By \Cref{fact:tc-interface}, \(\bm{\mathcal T}\) proves comprehension, minimization and number induction for admissible formulas, and every induction below is of this kind.
The number terms bounding the quantifiers of the admissible formulas used below are at most a fixed polynomial in \(n+|x|\), by the count guard.

The facts below are proved in order and used by every later argument: termwise comparison of sums, congruence of values at equal points, a bound on point magnitudes, a lower bound on exposed gaps, the perturbation identity that reads an exposed face off a perturbed and an unperturbed value, and its iterated form.

\emph{Termwise comparison.}
Let \(a_j\) and \(a'_j\), for \(j<r\), be rationals given by the preformed arrays and chosen functions of \(j\), with \(a_j\le a'_j\) for every \(j<r\).
Then \(\bm{\mathcal T}\) proves \(\sum_{j<r}a_j\le\sum_{j<r}a'_j\), with \(\sum_{j<r}a_j=_{\Q}\sum_{j<r}a'_j\) if and only if \(a_j=_{\Q}a'_j\) for every \(j<r\).

Bounded number induction on \(t\le r\) proves the same statement for the sums over \(j<t\), which is an admissible formula.
The base is the pair of empty sums.
For the step, let \(S\) and \(S'\) be the sums of \(a_j\) and \(a'_j\) over \(j<t\).
The prefix equations of \Cref{tc:arithmetic} add \(a_t\) and \(a'_t\) to the two sums.
The compatibility of the order with addition in \Cref{lem:code-field} then gives \(S+a_t\le S'+a'_t\) whenever \(S\le S'\), with equality if and only if \(S=_{\Q}S'\) and \(a_t=_{\Q}a'_t\).

The same induction applies to rational vectors of one length, added coordinatewise and ordered lexicographically, since that total order is also compatible with addition.

Two consequences of termwise comparison are used repeatedly.
Termwise equality gives equality of the sums.
The termwise bounds \(-|a_j|\le a_j\le|a_j|\) and linearity of bounded sums give the triangle inequality \(\left|\sum_{j<r}a_j\right|\le\sum_{j<r}|a_j|\).

\emph{Congruence.}
If two stored rational vectors \(z,z'\) of one frame satisfy \(z_c=_{\Q}z'_c\) at every coordinate \(c\), then \(h_P(z)=_{\Q}h_P(z')\) for every list \(P\), and hence every expression has equal values at \(z\) and \(z'\).
For each vertex \(p\), the congruence of products in \Cref{lem:code-field} gives \(p_cz_c=_{\Q}p_cz'_c\) for every \(c<b\), so termwise comparison gives \(p\cdot z=_{\Q}p\cdot z'\).
Each of the two selected maxima is at least every entry of the other list of dot products, since corresponding entries are equal, and each is one of its own entries; antisymmetry of the rational order gives their equality.

\emph{Magnitudes.}
Let \(\lceil\log_2(2b+4)\rceil\) denote the least \(j\) with \(U(2b+4)\le P_2(j)\), with \(U\) and the binary power table \(P_2\) as in \Cref{rc:selection}, the table being formed by the same IMUL definition through a fixed polynomial index bound in \(n+|x|\).
Put
\[
\mu(d)=|x|+B_d+d\left(\lceil\log_2(2b+4)\rceil+2\right)+10,
\]
so that \(\mu(\mathfrak{H})=L_x\).
Number induction on \(d\le \mathfrak{H}\) proves the admissible formula
\[
\forall\kappa<N_{\mathrm{pt}}\,\left(\operatorname{dep}(\kappa)\le d\rightarrow\forall c<b\,\left(|(y_\kappa)_c|\le P_2(\mu(d))\right)\right).
\]
At \(d=0\) the key is the input point, whose components have magnitude at most \(P_2(|x|)\), or a basis point, whose components have magnitude at most \(P_2(B_d)\).
For the step, let \(\kappa\) have depth at most \(d\) and put \(z=y_\kappa\), so that \(|z_c|\le P_2(\mu(d))\) for every \(c<b\) by the induction hypothesis.

A nonperturbative letter is a linear map \(z\mapsto Az\) whose rows have absolute sums at most \(2(b+1)\).
The components of its image are \(\sum_cA_{ic}z_c\), and \(|A_{ic}z_c|\le|A_{ic}|P_2(\mu(d))\) at every \(c\), so the triangle inequality, termwise comparison and linearity of bounded sums bound them by \(\sum_c|A_{ic}|P_2(\mu(d))\le2(b+1)P_2(\mu(d))\le P_2(\mu(d+1))\).
A perturbation letter gives \(|u_c+\epsilon z_c|\le P_2(B_d)+P_2(\mu(d))\le P_2(\mu(d+1))\), since \(0<\epsilon<1\), and a constant letter gives at most \(P_2(B_d)\).
The comparisons used are \(2(b+1)\le P_2(\lceil\log_2(2b+4)\rceil)\), \(P_2(a)P_2(a')=P_2(a+a')\) and \(P_2(a)+P_2(a')\le P_2(\max(a,a')+1)\); the last two are among the power-table comparisons of \Cref{ir:packets}.

\emph{The gap.}
We now prove that a nonzero difference of two vertex values at a direction exceeds the change that a perturbation of size \(\epsilon\) can produce.

Let \(p,p'\) be vertices of one stored list and \(u\) a direction used with it.
The rational sum \((p-p')\cdot u\), computed with the full-denominator convention of \Cref{tc:arithmetic}, has a positive integer denominator \(D\) that is the product of at most \(2b\) vertex denominators of component length at most \(B_v\) and at most \(b\) direction denominators of length at most \(B_d\).
Each factor is less than \(P_2\) of its component length, so monotonicity of multiplication and the power identity \(P_2(a)P_2(a')=P_2(a+a')\) give \(D<P_2(b(2B_v+B_d))\), where \(B_v,B_d\) are the caps at \(b\), which \(B_v(s+2)\) and \(B_d(s+2)\) dominate on the count guard.

If the value is nonzero, its integer numerator has magnitude at least one, since the least positive integer is one (\Cref{sec:foundations}), so \(|(p-p')\cdot u|\ge1/D>1/P_2(b(2B_v+B_d))\).

For a key \(\kappa\) with \(\operatorname{dep}(\kappa)\le \mathfrak{H}\), the triangle inequality and the magnitude bound give \(|(p-p')\cdot y_\kappa|\le2b P_2(B_v)P_2(L_x)\).
Before comparing this bound with \(\epsilon\), we bound the two logarithms \(\lceil\log_2(2b+4)\rceil\) and \(\lceil\log_2(b+1)\rceil\), which occur in \(L_x\) and in the lower bound for \(\mathfrak{p}(s)\).
Here \(\lceil\log_2(b+1)\rceil\) is the least \(j\) with \(U(b+1)\le P_2(j)\).
The comparisons \(U(2j)\le P_2(j)\) for \(j\ge1\) and \(U(j+1)\le P_2(j)\) of \Cref{ir:packets} give \(U(2b+4)\le P_2(b+2)\) and \(U(b+1)\le P_2(b)\).
So minimality and \(b\le s+2\) give \(\lceil\log_2(2b+4)\rceil\le b+2\le s+4\le2(s+2)\) and \(\lceil\log_2(b+1)\rceil\le b\le s+2\).

After these substitutions the displayed inequality for \(\mathfrak{p}(s)\) is an inequality between number terms with nonnegative coefficients; \(\bm{\mathcal T}\) proves it from the semiring and order laws of the number sort by comparing monomials, using \(b\le s+2\) on the guard.
Together with the bound on \(|(p-p')\cdot y_\kappa|\) above, \(2b\le P_2(\lceil\log_2(b+1)\rceil+1)\) and the power identity, this inequality gives
\[
\epsilon|(p-p')\cdot y_\kappa|<\frac{1}{2P_2(b(2B_v+B_d))}.
\]

\emph{The perturbation identity.}
Let \(P\) be a stored list, \(u\) a direction used with it, \(\kappa\) a key with \(\operatorname{dep}(\kappa)<\mathfrak{H}\), and \(y=y_\kappa\).
Write \(P^u\) for the face exposed by \(u\), formed as the masks \(\Lambda_j\) at the start of this subsection.
We show
\[
h_P[\mathsf{per}_u\kappa]=_{\Q}h_P[\mathsf{con}_u\kappa]+\epsilon h_{P^u}[\kappa].
\]
By the prefix equations and congruence, the two values \(h_P[\mathsf{per}_u\kappa]\) and \(h_P[\mathsf{con}_u\kappa]\) are \(h_P(u+\epsilon y)\) and \(h_P(u)\).
Let \(i^\ast\) be the index selected for \(h_P(u+\epsilon y)\), and let \(k\) be an index in \(P^u\), so that \(p_k\cdot u=h_P(u)\).
If \(i^\ast\notin P^u\), then \((p_k-p_{i^\ast})\cdot u>1/P_2(b(2B_v+B_d))\), and the bound on \(\epsilon|(p_k-p_{i^\ast})\cdot y|\) gives \(p_k\cdot(u+\epsilon y)>p_{i^\ast}\cdot(u+\epsilon y)\), contradicting the selection.
Hence \(i^\ast\in P^u\) and \(p_{i^\ast}\cdot u=h_P(u)\).

For every \(i\in P^u\), \(p_i\cdot(u+\epsilon y)=h_P(u)+\epsilon(p_i\cdot y)\), so the maximality of \(i^\ast\) and \(\epsilon>0\) give \(p_{i^\ast}\cdot y\ge p_i\cdot y\).
Since \(i^\ast\in P^u\), antisymmetry gives \(p_{i^\ast}\cdot y=h_{P^u}(y)\), and the identity follows.
This argument uses no induction: it consists of the selections, the gap inequality and the ordered-field laws.
Linearity of flat sums, from the prefix equations, extends it to every expression \(E\):
\[
E[\mathsf{per}_u\kappa]-E[\mathsf{con}_u\kappa]=_{\Q}\epsilon(P_uE)[\kappa],
\]
where \(P_uE\) is the expression whose lists are the faces exposed by \(u\), with the same coefficients.

\emph{Iterated faces.}
Let \(w=(u_1,\ldots,u_t)\) be a direction word, applied in this order, and let \(P_wE\) be the expression whose lists are the masks \(\Lambda_t\) defined at the start of this subsection, formed termwise.
For \(\sigma<2^t\), write \(\sigma_1,\ldots,\sigma_t\) for its bits, where \(\sigma_j\) is the bit at place \(j-1\), and \(|\sigma|\) for the number of one bits.
For a key \(\kappa\), let \(\kappa_\sigma\) be obtained from \(\kappa\) by appending, for \(j=t,t-1,\ldots,1\) in this order, the letter \(\mathsf{per}_{u_j}\) when \(\sigma_j=1\) and \(\mathsf{con}_{u_j}\) when \(\sigma_j=0\).
If \(\operatorname{dep}(\kappa)+t\le \mathfrak{H}\), then
\[
\epsilon^t(P_wE)[\kappa]=_{\Q}\sum_{\sigma<2^t}(-1)^{t-|\sigma|}E[\kappa_\sigma].
\]

By linearity of flat sums it suffices to prove the displayed identity for each list \(P\) of \(E\).
For \(j\le t\) and \(\sigma''<2^{t-j}\), encoding bits \(\sigma_{j+1},\ldots,\sigma_t\), let \(\kappa^{(j)}_{\sigma''}\) be \(\kappa\) followed by the letters for \(u_t,\ldots,u_{j+1}\) only.
For \(\sigma'<2^j\), the number \(\sigma'+2^j\sigma''\) has the bits of \(\sigma'\) followed by those of \(\sigma''\).
Number induction on \(j\le t\) proves the admissible formula
\[
\forall\sigma''<2^{t-j}\,\left(
\epsilon^jh_{\Lambda_j}[\kappa^{(j)}_{\sigma''}]=_{\Q}\sum_{\sigma'<2^j}(-1)^{j-|\sigma'|}h_P[\kappa_{\sigma'+2^j\sigma''}]\right),
\]
where the powers of \(\epsilon\) come from the interval-product table of \Cref{fact:tc-interface} and the sums for all \(j,\sigma''\) are formed in parallel beforehand.
At \(j=0\), \(\Lambda_0\) is the whole list and the sum has the single term \(\sigma'=0\) of length zero.

For the step from \(j\) to \(j+1\), fix \(\sigma'''\) encoding \(\sigma_{j+2},\ldots,\sigma_t\), and put \(\zeta=\kappa^{(j+1)}_{\sigma'''}\).
The perturbation identity, applied to the list \(\Lambda_j\) with direction \(u_{j+1}\), and the exposure property \(\Lambda_{j+1}=(\Lambda_j)^{u_{j+1}}\) give \(\epsilon h_{\Lambda_{j+1}}[\zeta]=h_{\Lambda_j}[\mathsf{per}_{u_{j+1}}\zeta]-h_{\Lambda_j}[\mathsf{con}_{u_{j+1}}\zeta]\).
These two keys are \(\kappa^{(j)}_{\sigma''}\) for the two values of the bit \(\sigma_{j+1}\).
Multiplying by \(\epsilon^j\), applying the induction hypothesis to each, and splitting the sum over \(\sigma<2^{j+1}\) by its bit \(\sigma_{j+1}\) through the injective reindexing equation of \Cref{tc:parallel-arithmetic} proves the formula at \(j+1\).
In this splitting, a one bit \(\sigma_{j+1}\) leaves the sign \((-1)^{j-|\sigma'|}\) unchanged, and a zero bit reverses it.
At \(j=t\) the formula is the claim.
Every key \(\kappa_\sigma\) has depth \(\operatorname{dep}(\kappa)+t\).

\paragraph{\emph{Pairing and supports of masks.}}
Two further facts are used repeatedly.
Let \(a_k,v_k\), for \(k<\mathfrak{k}\), be entries of preformed rational arrays, let \(Z\) be an admissible predicate, and let \(\iota\) be a chosen function.
Suppose that every \(k<\mathfrak{k}\) with \(Z(k)\) satisfies \(\iota(k)<\mathfrak{k}\), \(Z(\iota(k))\), \(\iota(\iota(k))=k\), \(\iota(k)\ne k\) and \(a_{\iota(k)}v_{\iota(k)}=_{\Q}-a_kv_k\).
Then \(\bm{\mathcal T}\) proves \(\sum_{k<\mathfrak{k}}[Z(k)]a_kv_k=_{\Q}0\).
Indeed, the indicator identity \([Z(k)]=[Z(k)\wedge k<\iota(k)]+[Z(k)\wedge\iota(k)<k]\) and additivity of prefix sums (\Cref{tc:arithmetic}) split the sum into two sums.
The function \(\iota\) is injective on the first class and maps it onto the second, so the injective reindexing equation of \Cref{tc:parallel-arithmetic} rewrites the second sum as \(\sum_{k<\mathfrak{k}}[Z(k)\wedge k<\iota(k)]a_{\iota(k)}v_{\iota(k)}\), which is the negative of the first.
We call this the pairing principle.

For rationals \(c,c_0,\ldots,c_{r-1}\), consider the values \(c+\sum_{j\in A}c_j\) over all masks \(A\subseteq\{0,\ldots,r-1\}\).
Then \(\bm{\mathcal T}\) proves that every such value is at most \(c+\sum_{j<r}\relu(c_j)\), with equality exactly for the masks \(A\) such that \(c_j>0\) implies \(j\in A\) and \(c_j<0\) implies \(j\notin A\), for every \(j<r\).

For a fixed mask and every \(j<r\), \([j\in A]c_j\le\relu(c_j)\), with equality if and only if \(c_j>0\) implies \(j\in A\) and \(c_j<0\) implies \(j\notin A\).
Indeed, for \(j\in A\) the two sides are \(c_j\) and \(\relu(c_j)\), equal if and only if \(c_j\ge0\), and for \(j\notin A\) they are \(0\) and \(\relu(c_j)\), equal if and only if \(c_j\le0\).
Termwise comparison, and adding \(c\) to both sides, give the claim.
The mask \(\{j:c_j>0\}\), formed by comprehension, attains the bound.
The same proof applies verbatim to vectors \(c_j\in\Q^t\) ordered lexicographically, by the vector form of termwise comparison, with \(\relu(c_j)\) meaning \(c_j\) when \(c_j>_{\mathrm{lex}}0\) and zero otherwise.
We call this the mask principle.

For a root \((\pi,S)\), its stored list is the apex vertex together with the apex-translated points \(h+\sum_{j\in A}l_j\); the mask principle with \(c=h\cdot y\) and \(c_j=l_j\cdot y\), and comparison with the apex value, give \(h_{C_{\pi,S}}(y)=_{\Q}C_{\pi,S}(y)\), the value of the displayed \(\relu\) formula, at every stored point \(y\).
We call this the root mask principle.

\paragraph{\emph{Local identities at a point.}}
Fix \(m\ge4\) and a key \(\zeta\) of frame \(m\) with \(\operatorname{dep}(\zeta)+3\le \mathfrak{H}\).
Each identity below is a rational value equality between stored entries at \(\zeta\) or at keys obtained from \(\zeta\) by at most three letters; (I1)--(I3) append no letters and hold at every key.
For an expression \(E\) of frame \(m\) and a direction \(u\), write \(D_uE=P_uE-P_{-u}E\); by the perturbation identity,
\[
\epsilon(D_uE)[\zeta]=_{\Q}E[\mathsf{per}_u\zeta]-E[\mathsf{con}_u\zeta]-E[\mathsf{per}_{-u}\zeta]+E[\mathsf{con}_{-u}\zeta].
\]
Write \((T_mE)[\zeta]\) for \(E[\zeta]\) when \((y_\zeta)_1\ge0\), and for \(E[\operatorname{pr}_m\zeta]\) otherwise.

\emph{(I1) Order-prefix cuts.}
Fix a root \((\pi,S)\) and the order of its \(N\) comparisons.
For a binary word \(\omega\) of length \(t\le N\), the stored list of \(P_\omega\) consists of the apex-translated images, under \((\lambda,s)\mapsto\lambda h+\sum_is_il_i\), of the origin and of the points \((1,1_B)\) for the masks \(B\) feasible for \(\omega\).
Here a chosen comparison \(s_i\ge s_j\) makes \(B\) feasible when \(j\in B\) implies \(i\in B\).

For \(t<N\), with next comparison between \(i\) and \(j\), the stored section \(Q_\omega\) consists of the images of the origin and of the feasible masks with \(i\in B\) if and only if \(j\in B\); by the threshold decomposition in \Cref{ir:span}, these are exactly its vertices.
Write \(\operatorname{val}(B)=(y_\zeta)_{\pi(0)}+h\cdot y_\zeta+\sum_{k\in B}l_k\cdot y_\zeta\) for the value at \(y_\zeta\) of the vertex of mask \(B\); the origin has value \((y_\zeta)_{\pi(0)}\).
We prove
\[
h_{P_{\omega0}}[\zeta]+h_{P_{\omega1}}[\zeta]=_{\Q}h_{P_\omega}[\zeta]+h_{Q_\omega}[\zeta].
\]
Put \(a=h_{P_{\omega0}}[\zeta]\), \(a'=h_{P_{\omega1}}[\zeta]\), \(r=h_{Q_\omega}[\zeta]\) and \(p=h_{P_\omega}[\zeta]\).
Every feasible mask satisfies at least one of the two orientations of the new comparison, so the list of \(P_\omega\) is the union of the two child lists, and \Cref{uc:label-max} gives \(p=\max(a,a')\).
The list of \(Q_\omega\) is contained in both child lists, so \(r\le\min(a,a')\).

For the reverse inequality, let \(X\) and \(Y\) be selected maximizers for \(a\) and \(a'\).
If either lies in \(Q_\omega\), then \(r\) is at least its value, which is at least \(\min(a,a')\).
Otherwise \(X\) is a mask with \(i\in X\) and \(j\notin X\), and \(Y\) is a mask with \(j\in Y\) and \(i\notin Y\).

Every comparison is an implication between memberships, so feasibility for \(\omega\) is preserved by unions and intersections of masks.
Hence \(X\cup Y\) and \(X\cap Y\) are feasible masks containing both or neither of \(i,j\), that is, entries of \(Q_\omega\).
The termwise identity \([k\in X\cup Y]+[k\in X\cap Y]=[k\in X]+[k\in Y]\) and linearity of bounded sums give \(\operatorname{val}(X\cup Y)+\operatorname{val}(X\cap Y)=a+a'\), so one of these two values is at least \((a+a')/2\ge\min(a,a')\).
Thus \(r=\min(a,a')\), and \(p+r=\max(a,a')+\min(a,a')=a+a'\).

\emph{(I2) The prefix invariant.}
Number induction on \(t\le N\) proves the admissible formula
\[
\forall(\pi,S)\,h_{C_{\pi,S}}[\zeta]=_{\Q}\sum_{|\omega|=t}h_{P_\omega}[\zeta]-\sum_{|\omega|<t}h_{Q_\omega}[\zeta],
\]
in which root keys range below their padded count, words are numbers below \(2^N\) with a length field, and the sums for every \(t\) are formed in parallel beforehand.
At \(t=0\) the only word is empty, \(P_\varnothing\) is the root list, and the second sum is empty.
For the step, the injective reindexing equation splits the sum over words of length \(t+1\) according to their last letter.
Then (I1) replaces each pair \(h_{P_{\omega0}}+h_{P_{\omega1}}\) by \(h_{P_\omega}+h_{Q_\omega}\).
Finally, the sum over \(|\omega|<t+1\) is the sum over \(|\omega|<t\) plus the sum over \(|\omega|=t\).

\emph{(I3) The subdivision.}
Multiply (I2) at \(t=N\) by the root weights \(w_{\pi,S}\), sum over all roots, and subtract \(M_m[\zeta]=h_\Delta[\zeta]\), where \(\Delta\) is the stored list of the target simplex.
Regrouping the resulting flat sum by the classification of each final leaf as an original full simplex or a flat leaf, through the injective reindexing equation, gives
\[
f_m[\zeta]=_{\Q}\sum_P\alpha_Ph_P[\zeta]+\rho_m[\zeta],
\]
where \(P\) ranges over the original full simplices with their weights \(\alpha_P\), the target simplex included with weight \(-1\), and \(\rho_m\) is the expression of \Cref{ir:direct}(a).

\emph{(I4) Face transport.}
Applying (I3) at the four keys \(\mathsf{per}_{\pm u}\zeta\) and \(\mathsf{con}_{\pm u}\zeta\), which have frame \(m\), and combining them by the formula for \(D_u\) gives, for every \(u\in\mathcal U_m\),
\[
(D_uf_m)[\zeta]=_{\Q}(D_u\rho_m)[\zeta]+\sum_P\alpha_P(D_uh_P)[\zeta],
\]
where \(D_uf_m=D_uF_m-D_uM_m\) is formed termwise from the literal root and target lists.
For each \(u\), let \(G_u\) be the termwise expression \(D_uf_m-D_u\rho_m-\sum_P\alpha_P(D_uh_P-B_u(P))\); the displayed identity is equivalent to \(G_u[\zeta]=_{\Q}\sum_P\alpha_PB_u(P)[\zeta]\).

Let \(P\) be an original simplex with a facet \(F\).
The solved facet row satisfies \(n_F\cdot v-c_F=0\) at the vertices \(v\) of \(F\) and \(n_F\cdot v-c_F=-1\) at the opposite vertex.
This follows from the nonsingularity of the facet systems and the correctness of the Cramer solution in \Cref{ir:arithmetic}.
The nonsingularity is proved for the original simplices and the target simplex as (F1) in (I5) below, without use of (I4).

Hence, for \(u=n_F/n_{F1}\), the face \(P^u\) is exactly the list of \(F\) when \(n_{F1}>0\), and the face \(P^{-u}\) is exactly the list of \(F\) when \(n_{F1}<0\).
So the literal removal of matching facets in \Cref{ir:direct}(c) does not change any stored value.

Every nonvertical facet of \(P\) has its normalized row in \(\mathcal U_m\), and distinct facets give distinct pairs of a row index and a facet, so the injective reindexing equation gives \(S(P)[\zeta']=_{\Q}\sum_{u\in\mathcal U_m}B_u(P)[\zeta']\) at every key \(\zeta'\), where \(S(P)=\sum_F\operatorname{sgn}(n_{F1})h_F\).

\emph{(I5) Reconstruction.}
Fix an original full simplex \(P\).
For a valid split-path key \(\theta\) of \(P\), write \(P_\theta\) for its stored simplex and \(S(P_\theta)=\sum_F\operatorname{sgn}(n_{F1})h_F\) for the sum over its current facets.
Write \(R_\theta\) for the unfolded expression of the subtree below \(\theta\).
It consists of three kinds of terms: the lower facets of the terminal descendants of \(\theta\), with coefficient one; the sections of its nonterminal descendants, \(\theta\) included, with coefficient \(-1\); and the transported other-facet sections of these nonterminal descendants, with coefficient \(\operatorname{sgn}(n_{F1})\).
Thus \(R(P)=R_\varnothing\) is the expression of \Cref{ir:direct}(b) before the factor \(\alpha_P\).

We now derive the structure of the stored split arrays inside \(\bm{\mathcal T}\).
For a valid key \(\theta\) of length \(t\), write \(\gamma^\theta_r\) for the row in slot \(r\), as fixed in \Cref{ir:direct}, \(\ell^\theta_r\) for its linear part, and \(v^\theta_r\) for the stored solution of the slot-\(r\) vertex system \(\gamma^\theta_{r'}(y)=0\), \(r'\ne r\).
Write \(\gamma_r(y)=n_r\cdot y-c_r\) for the original rows.

Three facts are used for square rational systems of order at most \(b\).
(K) A system is nonsingular if and only if its homogeneous system has only the zero solution, and then its Cramer quotient is its unique solution.
If the determinant of the system matrix \(U\) vanishes, the maximal-minor kernel basis of \Cref{ir:arithmetic}, or a coordinate vector at rank zero, is a nonzero homogeneous solution; if it does not vanish, the adjugate equation gives \((\det U)w=0\) for every homogeneous solution \(w\), and uniqueness is the adjugate argument of \Cref{ir:arithmetic}.

(Dep) A square matrix one of whose rows is a rational combination of the others has determinant zero: the Leibniz sum is linear in that row, by distributivity and the exchange of bounded double sums, and each resulting determinant has a repeated row, so it vanishes as in \Cref{ir:arithmetic}.

(Aff) For an affine row \(\gamma\), points \(a,a'\) and a rational \(\vartheta\), \(\gamma(\vartheta a+(1-\vartheta)a')=_{\Q}\vartheta\gamma(a)+(1-\vartheta)\gamma(a')\), by the ring laws at each coordinate and the linearity of bounded sums.

\emph{Original simplices.}
Let \(P\) be an original full simplex, with root \((\pi,S)\).
For a comparison coordinate \(i\), let \(\operatorname{pos}(i)\) be the number of coordinates \(i'\ne i\) whose chosen comparison is \(s_{i'}\ge s_i\).
By transitivity and bounded counting, the comparison \(s_i\ge s_{i'}\) is chosen if and only if \(\operatorname{pos}(i)<\operatorname{pos}(i')\); so \(\operatorname{pos}\) is injective and, by counting, a bijection onto \(\{0,\ldots,m-3\}\).

A feasible mask whose member of largest \(\operatorname{pos}\) is \(i\) contains every \(i'\) with \(\operatorname{pos}(i')<\operatorname{pos}(i)\) and nothing else, so the feasible masks are exactly the sets \(B_k=\{i:\operatorname{pos}(i)<k\}\), \(0\le k\le m-2\).
Relabeling the stored vertices and their facet rows together, which changes none of the statements below, we index the vertices of \(P\) as \(v_0\), the image of the origin, and \(v_{k+1}\), the image of \((1,1_{B_k})\).
With \(\sigma(k)\) the coordinate whose \(\operatorname{pos}\) is \(k-1\), this gives \(v_1-v_0=h\) and \(v_{k+1}-v_k=l_{\sigma(k)}\) for \(1\le k\le m-2\).

The facts (F0)--(F3) are proved in the order (F0), (F2), (F1), (F3), because (F1) uses (F2).

(F0) If \(\mu h+\sum_j\nu_jl_j=0\) in the local frame, then \(\mu=0\) and every \(\nu_j=0\).
Lift each row to all \(m\) label coordinates; the lifted combination vanishes at labels \(1,\ldots,m-1\) and has coordinate sum zero, so it also vanishes at label zero.
At label \(\pi(0)\) only \(h\) has a nonzero coefficient, so \(\mu=0\).

Number induction on \(c\le m-2\) proves that \(\nu_j=0\) for every \(j>m-2-c\).
The base is vacuous.
For the step, put \(j=m-2-c\).
At label \(\pi(j+1)\) the edge \(l_j\) has entry \(-1\), and every other edge with a nonzero entry there has parent position \(j+1\), hence index at least \(j+1\) and a vanishing multiplier \(\nu\) by the hypothesis.
So the combination has entry \(-\nu_j\) at that label, and \(\nu_j=0\).

(F2) Every vector \(w\) of the local frame is a rational combination of \(v_1-v_0,\ldots,v_{m-1}-v_0\).
If \(\sum_k\beta_k(v_k-v_0)=0\), the telescoping differences give \((\sum_k\beta_k)h+\sum_{k'=1}^{m-2}(\sum_{k>k'}\beta_k)l_{\sigma(k')}=0\), so (F0) makes all these suffix sums zero, and hence every \(\beta_k\) is zero.
By (K) the system whose columns are the \(v_k-v_0\) is nonsingular, and its Cramer solution at \(w\) gives the combination.

(F1) The facet systems of \Cref{ir:direct} are nonsingular, \(\gamma_i(v_j)=_{\Q}-[i=j]\), and \(\sum_in_i=0\).
A homogeneous solution \((n,c)\) has \(n\cdot(v_k-v_0)=0\) for every \(k\), so \(n=0\) by (F2), and then \(c=n\cdot v_0=0\); (K) gives nonsingularity, and the Cramer correctness of \Cref{ir:arithmetic} gives the displayed values.
Then \(\sum_i\gamma_i(v_k)=-1\) for every \(k\), so \(\sum_in_i\) annihilates every \(v_k-v_0\), and (F2) gives \(\sum_in_i=0\).

(F3) Both signs occur among the \(n_{i1}\).
Their sum is zero by (F1).
If all the \(n_{i1}\) vanished, writing \(e_1=\sum_{k\ge1}\beta_k(v_k-v_0)\) by (F2) would give \(0=n_k\cdot e_1=-\beta_k\) for every \(k\ge1\), since \(n_k\cdot(v_{k'}-v_0)=-[k=k']\) for \(k\ge1\), and hence \(e_1=0\).
The target simplex, with \(v_j=e_j\), satisfies (F1), (F2) and (F3) by the same arguments, (F2) being immediate since \(e_0=0\).

\emph{The vertex invariant.}
Number induction on \(t\le m\) proves the admissible formula \(\mathrm V(t)\) stating that, for every valid key \(\theta\) of length \(t\) and every slot \(r<m\), the slot-\(r\) vertex system is nonsingular, \(\gamma^\theta_{r'}(v^\theta_r)=_{\Q}0\) for every \(r'\ne r\), and \(\gamma^\theta_r(v^\theta_r)<0\), and that \(\mathcal{M}_t\) contains a row with \(n_{r1}>0\) and a row with \(n_{r1}<0\).
Its quantifiers range over the padded key, slot and coordinate domains, and the rows, determinants and solutions are preformed arrays.

At \(t=0\), the slot-\(r\) system has the rows \(n_{r'}\), \(r'\ne r\).
If \(w\) solves its homogeneous system, write \(w=\sum_{k\ge1}\beta_k(v_k-v_0)\) by (F2); then \(n_{r'}\cdot w=-\beta_{r'}\) for \(r'\ge1\) and \(n_0\cdot w=\sum_k\beta_k\), so \(\beta_{r'}=0\) for \(r'\notin\{0,r\}\), then \(\beta_r=0\) from the row \(r'=0\) when \(r\ne0\), and \(w=0\).
By (K) and (F1), \(v^\varnothing_r=v_r\), with \(\gamma_r(v_r)=-1\), and (F3) gives both signs.

For the step, let \(\theta'\) extend \(\theta\), of length \(t\), by the label \(d\) of its selected pair \((i,j)\), let \(e\) be the other member of the pair, and write \(v_r=v^\theta_r\), \(\gamma_r=\gamma^\theta_r\) and \(\ell_r=\ell^\theta_r\).
By \(\mathrm V(t)\), the cut row \(g=\gamma_i/n_{i1}-\gamma_j/n_{j1}\) satisfies \(g(v_r)=0\) for \(r\ne i,j\), while \(g(v_i)=\gamma_i(v_i)/n_{i1}\) and \(g(v_j)=-\gamma_j(v_j)/n_{j1}\) are nonzero of opposite signs, since \(n_{i1}n_{j1}>0\).
Put \(\vartheta=g(v_j)/(g(v_j)-g(v_i))\), so that \(0<\vartheta<1\), and \(z=\vartheta v_i+(1-\vartheta)v_j\).
By (Aff), \(g(z)=0\), \(\gamma_r(z)=0\) for \(r\ne i,j\), \(\gamma_i(z)=\vartheta\gamma_i(v_i)<0\) and \(\gamma_j(z)=(1-\vartheta)\gamma_j(v_j)<0\).

The rows of \(\theta'\) are those of \(\theta\) except at slot \(d\), which holds \(sg\) with the sign \(s\) of \Cref{ir:direct}, and \(sg(v_d)=\gamma_d(v_d)/|n_{d1}|<0\).
At slot \(d\), the system of \(\theta'\) is that of \(\theta\), so its stored solution is \(v_d\) by (K).
At a slot \(r\notin\{i,j\}\), a homogeneous solution \(w\) is annihilated by \(\ell_e\) and by the linear part of \(g\), hence by \(\ell_d\), hence by every row of the slot-\(r\) system of \(\theta\), so \(w=0\) by \(\mathrm V(t)\) and (K); and \(v_r\) solves the system, so it is the stored solution.

At slot \(e\), let \(w\) be a homogeneous solution and put \(\mu=(n_i\cdot w)/\gamma_i(v_i)\).
Since \(n_i\cdot(v_i-v_j)=\gamma_i(v_i)\) and \(\ell_r\cdot(v_i-v_j)=0\) for \(r\ne i,j\), the vector \(w-\mu(v_i-v_j)\) is annihilated by every row of the slot-\(j\) system of \(\theta\), so it is zero; then the linear part of \(g\) gives \(0=\mu(g(v_i)-g(v_j))\), so \(\mu=0\) and \(w=0\).
The point \(z\) solves this system, so it is the stored solution, and \(\gamma_e(z)<0\).
Finally \(\mathcal{M}_{t+1}=\mathcal{M}_t\setminus\{d\}\) still contains \(e\), whose first coefficient has the sign of \(n_{d1}\), so both signs remain.
This proves \(\mathrm V(t+1)\); each child list is that of \(\theta\) with \(v_e\) replaced by \(z\), and no entry is rejected, since each solution makes every row nonpositive.

\emph{Sections.}
A section system of \(\theta\) consists of the rows of \(\theta\) outside two omitted slots \(r_1,r_2\), together with \(g\).
If \(\{r_1,r_2\}=\{i,j\}\), it is the slot-\(e\) system of a child up to the sign of \(g\), with solution \(z\).

If exactly one omitted slot lies in \(\{i,j\}\) and the other is \(r\), the system contains the other member of the pair and \(g\), and the argument for slot \(r\) gives the solution \(v_r\).
If neither lies in \(\{i,j\}\), the system contains \(\ell_i\), \(\ell_j\) and the linear part of \(g\), which is their combination, so (Dep) makes it singular and it is rejected.
The points \(z\) and \(v_r\) make every row of \(\theta\) nonpositive, so neither is rejected.

For a section of another facet, at a slot \(f\notin\{i,j\}\), the three omitted slots are \(f\) and two slots \(r_1,r_2\ne f\), and imposing the slot-\(f\) equality separately gives the section system with omitted slots \(r_1,r_2\); so the same cases apply with \(r\ne f\).

These computations give the two facts used below.
(S1) At a nonterminal key with selected pair \(i,j\), the stored list of one child is that of \(P_\theta\) with \(v_i\) replaced by \(z=\vartheta v_i+(1-\vartheta)v_j\), \(0<\vartheta<1\), and that of the other child has \(v_j\) replaced by the same \(z\).
The entries of the stored section are \(z\) and the \(v_r\) with \(r\ne i,j\), each of which occurs.
For every other facet \(F\) of \(P_\theta\), at a slot \(f\), its stored pieces and section are the same three lists with \(v_f\) omitted.

(S2) At a terminal key, \(\mathcal{M}_t\) contains no pair of rows whose first coefficients have equal nonzero signs, so by \(\mathrm V(t)\) it contains exactly one row \(F_+\) with \(n_{F_+1}>0\) and one row \(F_-\) with \(n_{F_-1}<0\).
Every other slot holds a vertical row, either an original row with zero first coefficient or a cut row, whose first coefficient is \(1-1=0\).
The stored list consists of the \(m\) points \(v_F\), one for each slot \(F\), each satisfying every other row with equality and its own row strictly.

\emph{Terminal identity.}
At a terminal key, put \(v_\pm=v_{F_\pm}\), so that the list of \(F_+\) consists of \(v_-\) and the other \(m-2\) points, and the list of \(F_-\) of \(v_+\) and the same points.
Both \(v_+\) and \(v_-\) satisfy the \(m-2\) vertical equalities, so their difference is annihilated by the vertical linear parts, and so is \(e_1\).

Put \(w=v_+-v_-\) and \(t=(n_{F_+}\cdot w)/n_{F_+1}\).
The vector \(w-te_1\) is annihilated by \(n_{F_+}\) and by every vertical row, that is, by every row of the slot system of \(v_-\), which is nonsingular by \(\mathrm V\); so (K) gives \(v_+-v_-=te_1\).
Since \(n_{F_+}\cdot v_-=c_{F_+}>n_{F_+}\cdot v_+\) and \(n_{F_+1}>0\), we get \(t<0\).

If \((y_\zeta)_1\ge0\), then \(v_+\cdot y_\zeta\le v_-\cdot y_\zeta\), so \(h_{P_\theta}[\zeta]=h_{F_+}[\zeta]\), and \((T_mS(P_\theta))[\zeta]=h_{F_+}[\zeta]-h_{F_-}[\zeta]\), the vertical facets having coefficient zero.
If \((y_\zeta)_1<0\), then \(h_{P_\theta}[\zeta]=h_{F_-}[\zeta]\); moreover \(\operatorname{pr}_mv_+=\operatorname{pr}_mv_-\) and \(p\cdot\operatorname{pr}_m(y)=\operatorname{pr}_m(p)\cdot y\), so \(h_{F_+}[\operatorname{pr}_m\zeta]=h_{F_-}[\operatorname{pr}_m\zeta]\) and \((T_mS(P_\theta))[\zeta]=0\).
In both cases \(h_{P_\theta}[\zeta]-(T_mS(P_\theta))[\zeta]=_{\Q}h_{F_-}[\zeta]=R_\theta[\zeta]\).

\emph{Split identity.}
At a nonterminal key, let \(\mathfrak{v}_r=v_r\cdot y\) at a stored point \(y\), and \(\mathfrak{v}_z=\vartheta\mathfrak{v}_i+(1-\vartheta)\mathfrak{v}_j\), which lies between \(\mathfrak{v}_i\) and \(\mathfrak{v}_j\).
For a nonempty list \(\mathcal{J}\) of further values, the identity
\[
\max(\mathcal{J},\mathfrak{v}_i,\mathfrak{v}_j)+\max(\mathcal{J},\mathfrak{v}_z)=\max(\mathcal{J},\mathfrak{v}_z,\mathfrak{v}_j)+\max(\mathcal{J},\mathfrak{v}_z,\mathfrak{v}_i)
\]
holds by the two cases \(\mathfrak{v}_i\le\mathfrak{v}_z\le\mathfrak{v}_j\) and \(\mathfrak{v}_j\le\mathfrak{v}_z\le\mathfrak{v}_i\); in the first both sides equal \(\max(\mathcal{J},\mathfrak{v}_j)+\max(\mathcal{J},\mathfrak{v}_z)\), and the second is symmetric.
With \(\mathcal{J}\) the values of the \(v_r\), \(r\ne i,j\), (S1) gives \(h_{P_+}+h_{P_-}=h_{P_\theta}+h_Q\) at every stored point, where \(P_\pm\) are the children and \(Q\) the section.

With \(\mathcal{J}\) the values of the \(v_r\), \(r\ne i,j,f\), which is nonempty since \(m\ge4\), it gives \(h_{F\cap P_+}+h_{F\cap P_-}=h_F+h_{F\cap Q}\) for every other facet \(F\), where \(F\cap P_\pm\) and \(F\cap Q\) denote the stored pieces and section of \(F\) given by (S1).
One child keeps the facet opposite \(v_i\) whole and the other keeps the facet opposite \(v_j\) whole, while the new cut facets are vertical and have coefficient zero.
Hence \(S(P_+)+S(P_-)=S(P_\theta)+\sum_{F\text{ other}}\operatorname{sgn}(n_{F1})h_{F\cap Q}\) at \(\zeta\) and at \(\operatorname{pr}_m\zeta\), and therefore
\[
h_{P_\theta}-T_mS(P_\theta)=\left(h_{P_+}-T_mS(P_+)\right)+\left(h_{P_-}-T_mS(P_-)\right)-h_Q+T_m\sum_{F\text{ other}}\operatorname{sgn}(n_{F1})h_{F\cap Q}
\]
at \(\zeta\).

\emph{The induction.}
Every valid key has length at most \(m-1\), since each step deletes one member of a pair in the current mask and \(\mathcal{M}_0\) has \(m\) members.
Number induction on \(k\le m\) proves the admissible formula stating that every valid key \(\theta\) of length at least \(m-k\) satisfies \(h_{P_\theta}[\zeta]-(T_mS(P_\theta))[\zeta]=_{\Q}R_\theta[\zeta]\).
At \(k=0\) no valid key has length at least \(m\).

For the step, let \(\theta\) have length \(m-k-1\).
If it is terminal, the terminal identity applies.
Otherwise its two children are valid keys of length \(m-k\), covered by the hypothesis, and the split identity, together with the decomposition of \(R_\theta\) by the first step of each descendant through the injective reindexing equation, gives the formula at \(\theta\).
At \(k=m\) and \(\theta=\varnothing\), it gives \(h_P[\zeta]=_{\Q}R(P)[\zeta]+(T_mS(P))[\zeta]\).

\emph{(I6) Literal transport.}
Let \(P\) be a stored list of frame \(m\) and \(c\ge0\) a stored integer with \(p_1+c\ge0\) for every vertex \(p\), and let \(O\) be the stored list of the points \(p+ce_1\) and \(\operatorname{pr}_m(p+ce_1)\).
Then \(h_O[\zeta]-c\cdot h_{[0,e_1]}[\zeta]=_{\Q}(T_mh_P)[\zeta]\).
If \((y_\zeta)_1\ge0\), a translated vertex has value \(p\cdot y_\zeta+c(y_\zeta)_1\) and its projection \(p\cdot y_\zeta-p_1(y_\zeta)_1\), which is not larger because \((p_1+c)(y_\zeta)_1\ge0\); the maximum is \(h_P[\zeta]+c(y_\zeta)_1\), and \(c\cdot h_{[0,e_1]}[\zeta]=c(y_\zeta)_1\).
If \((y_\zeta)_1<0\), each projection dominates its translate, its value \(p\cdot y_\zeta-p_1(y_\zeta)_1\) equals \(p\cdot\operatorname{pr}_m(y_\zeta)\), and \(h_{[0,e_1]}[\zeta]=0\); since \(y_{\operatorname{pr}_m\zeta}=_{\Q}\operatorname{pr}_m(y_\zeta)\), congruence gives the value \(h_P[\operatorname{pr}_m\zeta]\).

The hypothesis \(p_1+c_m\ge0\) holds for every list to which \Cref{ir:direct} applies \(T_m\).
For the local lists it is the primitive bound \(\|p\|_\infty<\mathfrak{c}\le c_m\), which we prove as follows.
A canonical integer of length \(\lambda\) has magnitude below \(P_2(\lambda)\), and a rational code with positive denominator has magnitude at most that of its numerator, so a vertex whose components have length at most \(B_0\) has \(\|p\|_\infty<P_2(B_0)<P_2(B_0+4)=\mathfrak{c}\).
Moreover \(\mathfrak{c}\le c_m\), since \(R_g^a\ge1\) by the power-table comparisons of \Cref{ir:packets}.

It remains to show that every component of every primitive vertex has length at most \(B_0\); this is a statement about the stored arrays, proved by cases on the category of the local key, using the length bounds of \Cref{tc:arithmetic} and the determinant bound of \Cref{ir:arithmetic}.
\begin{enumerate}[label=(\roman*)]
\item Root, face, star, segment and point vertices are integer vectors whose components are sums of at most \(2m\) terms in \(\{0,1,-1\}\), so they have length at most \(\lceil\log_2(2b+1)\rceil+2\le B_0\).
\item Each original facet system has order \(m\le b\) and entries of length at most \(\mathfrak{i}\), so by the determinant bound its solution components have length at most \(\mathfrak{f}=\mathsf{Solve}(b,\mathfrak{i})\).
Normalization by \(n_{i1}\) and the difference of two normalized rows are at most two binary operations of cap \(2L+3\), so the normalized rows and pair planes have components of length at most \(\mathfrak{j}=C_2(\mathfrak{f})\).
Split, section and other-facet vertices are Cramer quotients of systems of order at most \(m-1\le b\) with entries from these rows, so their components have length at most \(\mathsf{Solve}(b,\mathfrak{j})\le B_{\mathrm{root}}\).
\item The three-level point vertex has components of length at most \(B_{\mathrm{lin}}\le B_{\mathrm{root}}\), by the computation of its single integer sum in \Cref{ir:direct}.
\item A recentered or pulled-back vertex is the image of a vertex of cases (i)--(iii) under a local integral word of at most \(J_g\) factors with entries of length at most \(B_t^0\), so the word bound \(\mathsf{Word}_{\Z}\) of \Cref{ir:arithmetic} gives length at most \(\mathsf{Word}_{\Z}(b+1,J_g,B_t^0,B_{\mathrm{root}})=B_0\).
\end{enumerate}

Each case is an inequality between the actual length of a stored entry and a number term, proved with the arrays fixed and no induction beyond the prefix inductions inside those length bounds.
For the lineage lists, number induction on \(m\le b\) proves the admissible formula stating that every component of every vertex of every term of \(E_m\) has magnitude at most \(\mathcal B_m\), where \(\mathfrak{c}\), \(c_m\) and \(\mathcal B_m\) are binary integers formed by the interval-product table of \Cref{fact:tc-interface}.
At \(m=3\) there are no terms.

For the step, a local terminal has components of magnitude at most \(2c_m\).
By the category of its local key, each of its vertices is a primitive vertex, of norm below \(\mathfrak{c}\le c_m\) by the bound just proved, an endpoint of the segment \([0,e_1]\), of norm at most one, or a vertex \(p+c_me_1\) or \(\operatorname{pr}_m(p+c_me_1)\) of a \(T_m\) image of a primitive vertex \(p\), recentered beforehand when required, whose components are \(p_c\), \(p_1+c_m\) or zero, of magnitude at most \(\mathfrak{c}+c_m\le2c_m\).

A lineage vertex is a translated or projected image of \(\Phi_\pm^Tp\) for a vertex \(p\) of a term of \(E_{m-1}\), by the prefix equations of its matrix word, and the rows of \(\Phi_\pm^T\) have absolute sums at most \(b\), so its components are at most \(b\mathcal B_{m-1}+c_m\le2c_m\).

The comparisons \(b\mathcal B_{m-1}\le \mathfrak{c}R_g^{8m+1}\le c_m\) and \(2c_m<\mathcal B_m\) follow from \(2\le b\le R_g=4(b+1)\) and the power-table identities \(R_g^aR_g^{a'}=R_g^{a+a'}\) of \Cref{ir:packets}.
In particular every vertex of \(\Phi_\pm^TP\) has first component at least \(-c_m\).

\emph{(I7) Literal faces.}
For a root list and a direction \(u\), the mask principle with \(c_j=l_j\cdot u\) identifies the maximizing base vertices as those whose masks contain every edge with \(l_j\cdot u>0\) and no edge with \(l_j\cdot u<0\), and comparison of their common value with the apex value selects the base face, the apex, or both.
For a base face, the mask principle at \(y_\zeta\), applied to the edges with \(l_j\cdot u=0\), gives
\[
h_{\mathrm{face}}[\zeta]=_{\Q}\left(e_{\pi(0)}+h+\sum_{l_j\cdot u>0}l_j\right)\cdot y_\zeta+\sum_{l_j\cdot u=0}\relu(l_j\cdot y_\zeta),
\]
so replacing a base face by its point row and its individual segment terms, as \Cref{ir:direct}(c) does, preserves the stored value.
Every other selected face in \Cref{ir:direct}(c), of a root, of the target simplex, of a term of \(\rho_m\) or of an original simplex, is stored as a literal mask and needs no identity.

\emph{(I8) Three-level directions.}
Let \(u\in\mathcal U_m\) have the three-level form of \Cref{ir:span}, with exchange \(\tau\), map \(B\) and functional \(\beta\), and put \(\lambda(w)=-W_mw_{\tau(0)}\).
The identities \(Bw=w-2u\beta(w)\), \(Bu=-u\) and \(\beta(u)=1\) hold for every rational \(w\), by the coordinate formula there and the ring laws, checked at each of the \(m\) coordinates under a bounded quantifier.

For every root \((\pi,S)\) and every stored point \(w\), the root formula gives \(C_{\pi,S}(Bw)=_{\Q}C_{\tau\pi,S}(w)-w_{\tau(0)}\), since \((Bw)_{\pi(k)}=w_{\tau\pi(k)}-w_{\tau(0)}\) at every position \(k\) and the head and edge rows are differences.
Composition with \(\tau\) is a bijection of root keys preserving \(S\), hence the weights, so the injective reindexing equation gives \(F_m(Bw)=_{\Q}F_m(w)+\lambda(w)\).
Apply this with \(w\) the stored points of \(\mathsf{per}_u\zeta\) and \(\mathsf{con}_u\zeta\), so that \(Bw\) is stored at the keys \(B\mathsf{per}_u\zeta\) and \(B\mathsf{con}_u\zeta\).
Since \(B(u+\epsilon y)=-u+\epsilon By\), congruence identifies the values of \(F_m\) there with \(F_m[\mathsf{per}_{-u}B\zeta]\) and \(F_m[\mathsf{con}_{-u}B\zeta]\), and the perturbation identity and linearity of \(\lambda\) give
\[
(P_uF_m)[\zeta]=_{\Q}(P_{-u}F_m)[B\zeta]-\lambda(y_\zeta).
\]

To compare the values of \(P_{-u}F_m\) at \(B\zeta\) and at \(\zeta\), fix a root \((\pi,S)\) and a vertex \(p\) of the face of its root list exposed by \(-u\).
Then \(p\cdot u=-h_{C_{\pi,S}}(-u)\), where \(h_{C_{\pi,S}}(-u)\) is the maximum of that list at \(-u\), so the identity \(Bw=w-2u\beta(w)\) gives \(p\cdot By_\zeta=p\cdot y_\zeta+2h_{C_{\pi,S}}(-u)\beta(y_\zeta)\).
All vertices of that face are shifted by the same amount, so its maximum is shifted by it; summing with the weights gives \((P_{-u}F_m)[B\zeta]=_{\Q}(P_{-u}F_m)[\zeta]+2F_m(-u)\beta(y_\zeta)\), where \(F_m(-u)\) is the stored value at the constant point \(-u\).
Therefore
\[
(D_uF_m)[\zeta]=_{\Q}2F_m(-u)\beta(y_\zeta)-\lambda(y_\zeta),
\]
which is the stored three-level point vertex at \(\zeta\); its scalar \(F_m(-u)\) equals the integer root sum of \Cref{ir:direct} by positive homogeneity of \(\relu\).
The faces of the target simplex are stored literally.

\emph{(I9) Repeated-level and exceptional directions.}
For a direction \(\xi\) and a root \((\pi,S)\), let \(\operatorname{nz}_\xi(\pi,S)\) be the number of edges \(j\) with \(l_j\cdot\xi\ne0\), and put \(a_\xi(\pi,S)=h\cdot\xi+\sum_j\relu(l_j\cdot\xi)\).

By (I7), the face exposed by \(\xi\) is the base face when \(a_\xi>0\), the apex when \(a_\xi<0\), and contains the apex and the base face when \(a_\xi=0\); in the last case its value at \(y=y_\zeta\) is \(y_{\pi(0)}+\relu\left(h\cdot y+\sum_{l_j\cdot\xi>0}l_j\cdot y+\sum_{l_j\cdot\xi=0}\relu(l_j\cdot y)\right)\), by the mask principle.
These quantities are bounded sums of stored rationals, so the classes defined from them below are admissible predicates.

If \(\operatorname{nz}_\xi=0\) and \(a_\xi=0\), then \(\xi\) annihilates \(h\) and every \(l_j\), hence every difference of the root list, so both exposures select the whole list and the two faces cancel in \(D_u\); the stored local part omits them.

(Lv) Let \(\operatorname{nz}_\xi(\pi,S)=1\), with nonzero edge \(l_k\), and write \(\bar\xi_a=\xi_{\pi(a)}\) for positions \(a\).
If \(k\in S\), then \(\bar\xi_a=\bar\xi_1\) for \(1\le a\le k\) and \(\bar\xi_a=\bar\xi_{k+1}\) for \(k<a\le m-1\).
If \(k\notin S\), then \(\bar\xi_a=\bar\xi_1\) for every \(a\ne k+1\) with \(1\le a\le m-1\).
In both cases \(a_\xi=\bar\xi_{q+1}+\relu(l_k\cdot\xi)-\bar\xi_0\), where \(q=\max S\).

Number induction on \(a\le m-1\) proves the stated equalities for all positions up to \(a\).
At position \(j+1\) with \(j\ne k\), the edge \(l_j\) has value zero, so \(\bar\xi_{j+1}=\bar\xi_{p(j)+1}\) with \(p(j)<j\).
If \(k\in S\), then \(p(j)+1\le j\le k-1\) for \(j<k\), and \(p(j)\ge k\) for \(j>k\), because \(k\in S\cap\{1,\ldots,j-1\}\), so parent and child lie in the same block.
If \(k\notin S\), then \(p(j)\ne k\) because \(p(j)\in S\cup\{0\}\), so the parent position is not \(k+1\).
For the head, \(q\ge k\) when \(k\in S\), so position \(q+1\) lies in the second block, and \(q\ne k\) when \(k\notin S\), so position \(q+1\) is not \(k+1\); the formula for \(a_\xi\) follows because every other edge has value zero.

\emph{Two repeated levels.}
Say that \(u\) has two repeated levels when its coordinates \(u_0,\ldots,u_{m-1}\), with \(u_0=0\), take exactly two values, each at least twice.
For such \(u\) and \(\varsigma\in\{1,-1\}\), let \(Z_{\varsigma u}\) be the class of roots with \(\operatorname{nz}_{\varsigma u}=1\) and \(a_{\varsigma u}=0\), and put \(\xi=\varsigma u\).

For a root in \(Z_{\varsigma u}\) with nonzero edge \(l_k\), we show that \(k\in S\).
Suppose instead that \(k\notin S\).
Then by (Lv) every non-apex label other than \(\pi(k+1)\) has level \(c=\bar\xi_1\), the label \(\pi(k+1)\) has a level \(d\ne c\), and \(a_\xi=0\) puts the apex at \(c+\relu(c-d)\), which differs from \(d\).
So the value \(d\) of \(\xi\), and hence the corresponding value of \(u\), is attained once, which is impossible for two repeated levels.

So \(k\in S\), and by (Lv) the positions \(1,\ldots,k\) have a level \(\mathfrak{a}\), the positions \(k+1,\ldots,m-1\) a level \(\mathfrak{b}\ne \mathfrak{a}\), and \(a_\xi=0\) puts the apex at \(\max(\mathfrak{a},\mathfrak{b})\).
The lower of the two levels is attained only on its block, so that block has at least two positions.

Define \(\iota\) on \(Z_{\varsigma u}\) as follows.
If \(\mathfrak{b}>\mathfrak{a}\), then \(k\ge2\), and \(\iota(\pi,S)=(\pi,S\mathbin{\triangle}\{k-1\})\).

Toggling \(k-1\) changes only the parent of \(l_k\), from position \(p(k)+1\) to another position at most \(k\).
Indeed, \(p(j)\) for \(j\ne k\) depends only on \(S\cap\{1,\ldots,j-1\}\), which is unchanged for \(j<k\) and has maximum at least \(k\) for \(j>k\).
The head is unchanged as well, because \(k\in S\) keeps \(q\ge k>k-1\).
The new \(l_k\) still has value \(\mathfrak{a}-\mathfrak{b}<0\), so the image lies in \(Z_{\varsigma u}\) with the same \(k\), every other row is unchanged, and the face formula above excludes \(l_k\) on both sides; so the two face values agree at every stored point.

If \(\mathfrak{a}>\mathfrak{b}\), then \(k+2\le m-1\); let \(\iota\) exchange \(\pi(k+1)\) and \(\pi(k+2)\) and toggle \(k+1\) in \(S\).
Suppose \(k+1\notin S\), and write \(r=\pi(k+1)\), \(s=\pi(k+2)\) and \(t=y_r-y_s\).

Before the map, \(l_k=y_{\pi(p(k)+1)}-y_r\) and \(l_{k+1}=y_r-y_s=t\), since \(p(k+1)=k\); after it, the edge with index \(k\) is \(y_{\pi(p(k)+1)}-y_s\), which equals the old \(l_k+t\), and the edge with index \(k+1\) is \(y_s-y_r=-t\), since \(p(k+1)=k\) and \(p(k)\) is unchanged.
Edges with \(j<k\) are unchanged; for \(j\ge k+2\), the parent label is \(r\) on both sides when \(S\cap\{k+2,\ldots,j-1\}=\varnothing\), at position \(k+1\) before and \(k+2\) after, and is unchanged otherwise; the head label is \(r\) on both sides when \(q=k\), and is unchanged otherwise.

At \(\xi\) the edge \(l_k\) has value \(\mathfrak{a}-\mathfrak{b}>0\) and \(t\) has value zero on both sides, so the image lies in \(Z_{\varsigma u}\) with the same \(k\), and the face values agree at every stored point by the ordered-field identity \(l_k+\relu(t)=(l_k+t)+\relu(-t)\).

Since the exchange and the toggle are involutions, so is \(\iota\); when \(k+1\in S\), the computation above applies to \(\iota(\pi,S)\), which has \(k+1\notin S\), and gives the same equality of face values.
In both cases \(\iota\) has no fixed points, and \(|S|\) changes by one, so the weights are opposite.

The pairing principle, with the indicator of \(Z_{\varsigma u}\), the weights \(w_{\pi,S}\) and the stored face values, gives \(\sum_{(\pi,S)\in Z_{\varsigma u}}w_{\pi,S}h_{P_{\varsigma u}C_{\pi,S}}[\zeta]=_{\Q}0\).
For such \(u\), the stored local part \(\mathcal K_u\) consists of the faces \(P_{\varsigma u}C_{\pi,S}\) with coefficient \(\varsigma w_{\pi,S}\) for the roots outside \(Z_{\varsigma u}\) whose faces at \(\pm u\) are not both the whole list, base faces split as in (I7), and the two target faces with coefficients \(-1\) and \(+1\).
So (I7) and the two pairing identities give \((D_uf_m)[\zeta]=_{\Q}\mathcal K_u[\zeta]\).

\emph{The direction \(u_+\).}
Write \(u_\ast=u_+\), so label zero has level zero and every other label level one, and \(l_j\cdot u_\ast\) is \(1\), \(-1\) or \(0\) according as the child label, the parent label or neither of \(l_j\) is zero.
Split the roots into the class \(Z^0\) with \(\pi(0)=0\), the class \(Z^1\) with \(\pi(0)\ne0\) and exactly one edge incident to label zero, and the class \(Z^2\) of the others.
On \(Z^0\) every edge has value zero and \(h\cdot u_\ast=1\), so the faces at \(u_\ast\) and \(-u_\ast\) are the base face and the apex; on \(Z^2\) at least two edges are nonzero.
Both classes keep their stored faces, with values given by (I7).

A lower key is a triple \((\bar\pi,T,\mathsf c)\) of a bijection \(\bar\pi\) from \(\{0,\ldots,m-2\}\) onto \(\{1,\ldots,m-1\}\), a set \(T\subseteq\{1,\ldots,m-3\}\), and a case \(\mathsf c\) that is \(\mathrm{first}\), \(\mathrm{last}\), or \(\mathrm{pend}_i\) with \(0\le i\le m-3\); it is valid unless \(\mathsf c\) is a pendant and \(T=\varnothing\).
Put \(\hat p=1\), \(\hat p=m-1\) or \(\hat p=i+2\) accordingly.
Let \(\pi\) insert label zero at position \(\hat p\), that is, \(\pi(a)=\bar\pi(a)\) for \(a<\hat p\), \(\pi(\hat p)=0\) and \(\pi(a)=\bar\pi(a-1)\) for \(a>\hat p\), and let \(S\) be given by the table of \Cref{ir:span}.
Write \(\Psi(\bar\pi,T,\mathsf c)=(\pi,S)\).
The map \(\Psi\) is injective on valid keys and its image is \(Z^1\).

The label at position \(\hat p\) has the incoming edge \(l_{\hat p-1}\) when \(\hat p\ge2\), and the outgoing edges \(l_j\), \(j\ge\hat p\), with \(p(j)=\hat p-1\); the edge \(l_{\hat p}\) is outgoing when \(\hat p\le m-2\) and \(\hat p-1\in S\cup\{0\}\), and so is \(l_2\) when \(\hat p=1\) and \(1\notin S\).

Conversely, every outgoing edge \(l_j\) has \(p(j)=\hat p-1\in S\cup\{0\}\) and \(\hat p\le j\le m-2\); so for \(\hat p\ge2\) there is none when \(\hat p-1\notin S\) or \(\hat p=m-1\), and for \(\hat p=1\) and \(1\in S\) only \(l_1\) is outgoing, since then \(p(j)\ge1\) for \(j\ge2\).

Hence label zero, at a non-apex position \(\hat p\), has exactly one incident edge if and only if \(\hat p=1\) and \(1\in S\), or \(\hat p\ge2\) and \(\hat p-1\notin S\), or \(\hat p=m-1\) and \(m-2\in S\); these are the first, pendant and last cases, distinguished at \(\hat p=m-1\) by whether \(m-2\in S\).
Deleting position \(\hat p\) recovers \(\bar\pi\), and inverting the displayed map from \(T\) to \(S\) recovers \(T\), which gives injectivity and the image.

For a valid key with \(\Psi(\bar\pi,T,\mathsf c)=(\pi,S)\), write \(\bar h\) and \(\bar l_{j'}\) for the head and edges of the lower root of \((\bar\pi,T)\), in the notation of \Cref{ir:span}.
The upper rows are as follows.
In the pendant case, \(h=\bar h\), the edge \(l_{\iota_i(j')}\) equals \(\bar l_{j'}\) for every \(j'\), and the remaining edge is \(l_{i+1}=y_{\bar\pi(p_T(i+1)+1)}-y_0\).
In the first case, \(h=\bar h\), \(l_{j'+1}=\bar l_{j'}\) for every \(j'\), and \(l_1=y_0-y_{\bar\pi(1)}\).
In the last case, \(l_{j'}=\bar l_{j'}\) for every \(j'\le m-3\), \(h=y_0-y_{\bar\pi(0)}\) and \(l_{m-2}=y_{\bar\pi(q_T+1)}-y_0\), so that \(h+l_{m-2}=\bar h\).

Each of these identities is an equality of label pairs checked at every index, using \(p_S(j)=\max(\{0\}\cup(S\cap\{1,\ldots,j-1\}))\).
In the pendant case, \(S\cap\{1,\ldots,j-1\}\) is the image under \(\iota_i\) of \(T\cap\{1,\ldots,j'-1\}\) for \(j=\iota_i(j')\), and \(\iota_i\) is increasing and skips \(i+1=\hat p-1\), so the parent and child positions correspond under the insertion; the head uses \(q_S=\iota_i(q_T)\), with \(q_T\ge1\) since \(T\ne\varnothing\).
In the first case, \(p_S(j'+1)=p_T(j')+1\) and \(q_S=q_T+1\), and in the last case \(p_S(j')=p_T(j')\) for \(j'\le m-3\), \(q_S=m-2\) and \(p_S(m-2)=q_T\).

Now put \(y=y_\zeta\), \(E_{\mathrm{low}}=\sum_{j'}\relu(\bar l_{j'}\cdot y)\), \(H_{\mathrm{low}}=\bar h\cdot y+E_{\mathrm{low}}\), and \(C_{\mathrm{low}}=y_{\bar\pi(0)}+\relu(H_{\mathrm{low}})\), the value of the lower root formula, and let \(l\) be the edge incident to label zero.
In the pendant case \(l\cdot u_\ast=1\) and \(h\cdot u_\ast=0\), so \(a_{u_\ast}=1\) and \(a_{-u_\ast}=0\); in the first case \(l\cdot u_\ast=-1\) and \(h\cdot u_\ast=0\), so \(a_{u_\ast}=0\) and \(a_{-u_\ast}=1\); in the last case \(l\cdot u_\ast=1\) and \(h\cdot u_\ast=-1\), so \(a_{u_\ast}=0\) and \(a_{-u_\ast}=1\).
The face formulas of (I7) therefore give the three residual identities of \Cref{ir:span},
\[
(D_{u_\ast}C_{\pi,S})[\zeta]=_{\Q}
\begin{cases}
-C_{\mathrm{low}}+y_{\mathrm{last}}+l\cdot y+E_{\mathrm{low}},&\text{pendant},\\
C_{\mathrm{low}}-y_{\mathrm{last}}-l\cdot y-E_{\mathrm{low}},&\text{first},\\
C_{\mathrm{low}}-y_0-E_{\mathrm{low}},&\text{last},
\end{cases}
\]
with \(y_{\mathrm{last}}=y_{\bar\pi(q_T+1)}\), by the ordered-field laws.
The upper weight \((-1)^{m+|S|}/m!\) equals \(-w_T/m\) in the pendant case, where \(|S|=|T|\), and \(w_T/m\) in the two endpoint cases, where \(|S|=|T|+1\), with \(w_T=(-1)^{m-1+|T|}/(m-1)!\).
By the injective reindexing equation along \(\Psi\), and the count of \(m-2\) pendant slots and two endpoints,
\[
\sum_{(\pi,S)\in Z^1}w_{\pi,S}(D_{u_\ast}C_{\pi,S})[\zeta]
=_{\Q}\sum_{\bar\pi}\left(\sum_{T\ne\varnothing}w_TC_{\mathrm{low}}(\bar\pi,T)+\frac2mw_\varnothing C_{\mathrm{low}}(\bar\pi,\varnothing)\right)+\mathrm{Res}[\zeta],
\]
where \(\mathrm{Res}\) is the stored sum of the residual point rows and segments with their upper weights.

\emph{Rebasing.}
Put \(\pi'(a)=\bar\pi(a)-1\), a bijection from lower orders onto the orders of arity \(m-1\).
Every lower row is a difference of two coordinates, and \((\Phi_+y)_c=y_{c+1}-y_1\), so the ring laws give \(C_{\mathrm{low}}(\bar\pi,T)=y_1+C^{(m-1)}_{\pi',T}(\Phi_+y)\), where \(C^{(m-1)}_{\pi',T}\) is the root formula of arity \(m-1\).

By the root mask principle and congruence at the key \(\Phi_+\zeta\), the injective reindexing equation along \(\bar\pi\mapsto\pi'\) gives \(\sum_{\bar\pi,T\ne\varnothing}w_TC_{\mathrm{low}}(\bar\pi,T)=_{\Q}W_{m-1}y_1+F_{m-1}[\Phi_+\zeta]\), where \(W_{m-1}\) is the sum of the weights of \(F_{m-1}\).
Number induction on \(r\le b\) shows that there are \(r!\) bijections of an \(r\)-element label set, by the injective reindexing equation for the map recording the position of the largest label and the order of the others.
The pairing principle, toggling the element one, gives \(\sum_{T\subseteq\{1,\ldots,m-3\}}(-1)^{|T|}=0\), since \(m\ge4\).
Hence \(W_{m-1}=\sum_{T\ne\varnothing}(-1)^{m-1+|T|}=(-1)^m\).
The target faces at \(u_\ast\) and \(-u_\ast\) are the lists of \(e_1,\ldots,e_{m-1}\) and of \(e_0\), with values \(\max_{a\ge1}y_a\) and zero, and \Cref{uc:label-max}, applied to the translated coordinates, gives \(\max_{a\ge1}y_a=_{\Q}y_1+M_{m-1}[\Phi_+\zeta]\).

\emph{The empty star.}
Fix \(\bar\pi\), and let \(J\), \(d_a\) and \(V_A\) be as in \Cref{ir:span}, evaluated at \(y\).
By the mask principle, the stored list of \(V_A\) has value \(V_A(y)\).

First, \(\sum_{A\subseteq J}(-1)^{|A|}V_A(y)=_{\Q}0\).
If some \(d_a(y)\ge0\), let \(a_0\) be the least such \(a\in J\); toggling \(a_0\) leaves the argument of \(V_A\) unchanged, since \(d_{a_0}=\relu(d_{a_0})\) there, and reverses the sign, so the pairing principle applies.
Otherwise every \(d_a(y)<0\), every nonempty \(A\) gives a negative argument, and \(A=\varnothing\) gives argument zero, so every term is zero.

Second, for each \(a\in J\), the exchange of positions zero and \(a\) is an involution \(\bar\pi\mapsto\bar\pi^{(a)}\) of lower orders, and substituting into the root formula gives \(C_{\mathrm{low}}(\bar\pi^{(a)},\varnothing)=_{\Q}y_{\bar\pi(a)}+V_{\{a\}}(y)\).
Summing over \(\bar\pi\) through the injective reindexing equation, then over the \(m-2\) members of \(J\), exchanging the bounded sums, and using the first identity, gives
\[
(m-2)\sum_{\bar\pi}C_{\mathrm{low}}(\bar\pi,\varnothing)=_{\Q}\sum_{\bar\pi}\left(\sum_{a\in J}y_{\bar\pi(a)}+V_\varnothing(y)+\sum_{|A|\ge2}(-1)^{|A|}V_A(y)\right),
\]
and \(V_\varnothing(y)=\sum_a\relu(d_a(y))\), since that argument is nonnegative.

Collecting these identities, the stored local part \(\mathcal K_+\) for \(u_+\) consists of the stored faces of \(Z^0\) and \(Z^2\) with their signed weights, the residuals \(\mathrm{Res}\), the star components with coefficient \(\frac{2}{m(m-2)}\frac{(-1)^{m-1}}{(m-1)!}\) times their displayed Boolean signs, and the row \((W_{m-1}-1)y_1\).
The flat sums then give \((D_{u_+}f_m)[\zeta]=_{\Q}\mathcal K_+[\zeta]+f_{m-1}[\Phi_+\zeta]\), with \(f_{m-1}[\Phi_+\zeta]=F_{m-1}[\Phi_+\zeta]-M_{m-1}[\Phi_+\zeta]\).

For \(u_-\), let \(B\) be its integral recentering, which exchanges labels zero and one, so that \(Bu_-=-u_+\) and \(\Phi_+B=\Phi_-\) coordinatewise.
The root formula and the reindexing \(\pi\mapsto\tau\pi\), where \(\tau\) now exchanges labels zero and one, give \(F_m(Bw)=_{\Q}F_m(w)-W_mw_1\), and \Cref{uc:label-max}, applied to the permuted and translated coordinates, gives \(M_m(Bw)=_{\Q}M_m(w)-w_1\).
Hence \(f_m(Bw)=_{\Q}f_m(w)+\lambda'(w)\) with the linear function \(\lambda'(w)=(1-W_m)w_1\).
The argument of (I8) then gives \((P_{\pm u_-}f_m)[\zeta]=_{\Q}(P_{\mp u_+}f_m)[B\zeta]-\lambda'(y_\zeta)\), so \((D_{u_-}f_m)[\zeta]=_{\Q}-(D_{u_+}f_m)[B\zeta]\).
Combining the identity for \(u_+\) at \(B\zeta\) with the transposition \(h_{B^TP}[\zeta]=h_P[B\zeta]\), which follows from \((B^Tp)\cdot y=p\cdot By\) by exchanging bounded sums, and with congruence for \(\Phi_+B=\Phi_-\), gives
\[
(D_{u_-}f_m)[\zeta]=_{\Q}-\mathcal K_-[\zeta]-f_{m-1}[\Phi_-\zeta],
\]
where \(\mathcal K_-\) is the pullback by \(B\) of \(\mathcal K_+\), so that the stored local part for \(u_-\) is \(-\mathcal K_-\).
For every other \(u\in\mathcal U_m\), (I7), (I8) and the pairing identity for two repeated levels give \((D_uf_m)[\zeta]=_{\Q}\mathcal K_u[\zeta]\) for the stored local part \(\mathcal K_u\).

\paragraph{\emph{The arity induction.}}
Put \(\tau_m=4(m-2)\), so that \(\tau_3=4\) and \(\tau_m-\tau_{m-1}=4\).
Number induction on \(m\) from \(3\) to \(b\) proves the admissible formula
\[
\Theta(m):\quad\forall\kappa<N_{\mathrm{pt}}\,\left(\operatorname{fr}(\kappa)=m\wedge\operatorname{dep}(\kappa)\le \mathfrak{H}-\tau_m\rightarrow f_m[\kappa]=_{\Q}E_m[\kappa]\right).
\]
Its single quantifier is bounded by \(N_{\mathrm{pt}}\); the point, list and value arrays are string parameters; and no string is quantified.

At \(m=3\), \(E_3\) is empty, so \(E_3[\kappa]=0\).
For each root, \(S=\{1\}\), and the root formula gives \(C_{\pi,\{1\}}(y)=y_{\pi(0)}+\relu(\max(y_{\pi(1)},y_{\pi(2)})-y_{\pi(0)})=\max(y_0,y_1,y_2)\), by applying \(v+\relu(w-v)=\max(v,w)\) twice, with \(y_0=0\).
The six valid roots have weight \((-1)^{4}/3!=1/6\) each and the padded invalid keys have coefficient zero, so the injective reindexing equation reduces \(F_3[\kappa]\) to a sum of six equal terms, and \(F_3[\kappa]=_{\Q}\max(0,y_1,y_2)=M_3[\kappa]\) by \Cref{uc:label-max}.

For the step, assume \(\Theta(m-1)\) with \(4\le m\le b\), and let \(\kappa\) have frame \(m\) and depth at most \(\mathfrak{H}-\tau_m\).
Put \(\zeta=\kappa\) if \((y_\kappa)_1\ge0\) and \(\zeta=\operatorname{pr}_m\kappa\) otherwise, so \(\operatorname{dep}(\zeta)\le\operatorname{dep}(\kappa)+1\).
The local identities are used at \(\kappa\), at \(\zeta\) and, for \(u_-\), at \(B\zeta\), whose depth is at most \(\operatorname{dep}(\kappa)+2\).
Their hypothesis that the depth plus three is at most \(\mathfrak{H}\) holds at all three keys, since \(\operatorname{dep}(\kappa)+5\le \mathfrak{H}\) by \(\tau_m=4(m-2)\ge8\) for \(m\ge4\).

By (I3) at \(\kappa\), (I5) at \(\kappa\) for each original simplex, the regrouping of (I4) at \(\zeta\) and linearity of \(T_m\), we obtain
\[
f_m[\kappa]=_{\Q}\rho_m[\kappa]+\sum_P\alpha_PR(P)[\kappa]+\sum_{u\in\mathcal U_m}\sum_P\alpha_PB_u(P)[\zeta].
\]
By (I4) at \(\zeta\), the inner sum is \(G_u[\zeta]\).
By (I7)--(I9) at \(\zeta\), \(G_u[\zeta]\) is the value at \(\zeta\) of the stored local part of \Cref{ir:direct}(c) for \(u\), plus \(f_{m-1}[\Phi_+\zeta]\) for \(u=u_+\) and minus \(f_{m-1}[\Phi_-\zeta]\) for \(u=u_-\).
The stored terms of \(\mathcal L_m\) are those of \(\rho_m\), of the unfolded \(R(P)\) with the factors \(\alpha_P\), and the transported local parts, represented literally; so (I6) and the injective reindexing equation give
\[
f_m[\kappa]=_{\Q}\mathcal L_m[\kappa]+f_{m-1}[\Phi_+\zeta]-f_{m-1}[\Phi_-\zeta].
\]

On the other side, the key partition of \Cref{ir:paired}, read as an identity of flat sums through the injective reindexing equation, gives
\[
E_m[\kappa]=_{\Q}\mathcal L_m[\kappa]+\sum_{(a,P)\in E_{m-1}}\left(a\cdot h_{O_{m,+}(P)}[\kappa]-a\cdot h_{O_{m,-}(P)}[\kappa]\right).
\]
By (I6) with \(c=c_m\) and the cancellation of the two segment terms, each bracket is \(a\cdot(T_mh_{\Phi_+^TP})[\kappa]-a\cdot(T_mh_{\Phi_-^TP})[\kappa]\).
The transposition \((\Phi^Tp)\cdot y=p\cdot\Phi y\) and congruence for the stored points of \(\Phi_\pm\zeta\) turn this into \(a\cdot h_P[\Phi_+\zeta]-a\cdot h_P[\Phi_-\zeta]\), and summing over the lower terms gives \(E_{m-1}[\Phi_+\zeta]-E_{m-1}[\Phi_-\zeta]\).
The keys \(\Phi_\pm\zeta\) have frame \(m-1\) and depth at most \(\operatorname{dep}(\kappa)+2\le \mathfrak{H}-\tau_{m-1}\), since \(\tau_m-\tau_{m-1}=4\).
The induction hypothesis at these two keys replaces \(E_{m-1}\) by \(f_{m-1}\) there, so the two displayed expressions agree, that is, \(E_m[\kappa]=_{\Q}f_m[\kappa]\), which is \(\Theta(m)\).

\paragraph{\emph{Ranks.}}
The rank of a stored list is the largest order of a nonzero minor of the matrix of its pairwise vertex differences, computed by the minor graphs of \Cref{ir:arithmetic}.
Five facts hold for these actual arrays of variable order at most \(b\).
(R1) A sublist has rank at most that of the list, since its difference rows are rows of the larger matrix.

(R2) If every pairwise difference of a list is a rational combination of \(k\) fixed vectors, the list has rank at most \(k\).
Indeed, a minor of order \(k+1\) expands by multilinearity in its rows, by bounded number induction on the number of rows already expanded, into a sum over maps from its \(k+1\) rows to the \(k\) vectors.
No such map is injective: by the bracketed form of the injective reindexing equation with all entries one, an injective map would give \(k+1\le k\).
So each term is a determinant with a repeated row and vanishes, as in \Cref{ir:arithmetic}.

(R3) If a list has rank \(k\), then every difference is a rational combination of any \(k\) differences forming a nonzero minor of order \(k\).
This is the bordered-minor computation of \Cref{ir:arithmetic}: every bordered minor of order \(k+1\) vanishes by maximality, so the difference annihilates every kernel-basis vector of those \(k\) rows and hence lies in their row span.

(R4) If a direction \(u\) does not annihilate every difference of a list of rank \(k\ge1\), then the face exposed by \(u\) has rank at most \(k-1\).
Suppose instead that the face has rank \(k\), the largest possible by (R1).
Then it has \(k\) differences forming a nonzero minor, all annihilated by \(u\), and (R3) writes every difference of the list as their combination, so \(u\) annihilates them all, contrary to the hypothesis.

(R5) A mask \(\Lambda\) obtained from a list \(V\) by successive exposure is a record of the face closure of \Cref{ir:readout}.
Let \(J\subseteq\Lambda\) consist of a vertex and \(\operatorname{rank}(\Lambda)\) further vertices whose differences from it form a nonzero minor; these exist, since if the differences from the first vertex had a smaller rank \(k'\), (R3) for them and (R2) would give \(\operatorname{rank}(\Lambda)\le k'\).

Each vertex of \(\Lambda\) is an affine combination of \(J\) by (R3); conversely, bounded number induction on the exposure stage shows that a vertex of \(V\) that is an affine combination of \(J\) attains every stage maximum, since all of \(J\) does, and hence lies in \(\Lambda\).
So \(\Lambda\) is the closure record of \(J\), and its direction space has a closure representative \(W\); by (R3), applied to the differences of both lists together, a direction annihilates the differences of \(\Lambda\) if and only if it annihilates those of \(W\).

(Rk) Number induction on \(m\) proves the admissible formula stating that every term of \(E_m\) has rank at most \(m-2\).
At \(m=3\) there are no terms.
For the step, the differences of a lineage term \(O_{m,\pm}(P)\) are combinations of the images under \(\Phi_\pm^T\) of \(\operatorname{rank}(P)\) differences of \(P\), by (R3), and of \(e_1\), so (R2) gives rank at most \(\operatorname{rank}(P)+1\le m-2\).
It remains to show that the local terms of \(\mathcal L_m\) have rank at most \(m-2\), and that those inside the final \(T_m\) have rank at most \(m-3\) before that operation.

Two consequences of (R2) and (R3) are used.
(R6) A list whose entries lie among \(k+1\) points \(p_0,\ldots,p_k\) has rank at most \(k\), since each difference \(p_a-p_{a'}\) is \((p_a-p_0)-(p_{a'}-p_0)\).

(R7) If every difference of a list \(P'\) is a combination of the images, under one linear map, of differences of a list \(P\), together with \(c\) further vectors, then \(\operatorname{rank}(P')\le\operatorname{rank}(P)+c\), by (R3) and (R2).
In particular a \(T_m\) union-image term has rank at most one more than its source, since the differences of translated and projected points are differences of the source plus multiples of \(e_1\), and the segment term has rank one.
Each case below is a bound for every key of the corresponding category, proved with the arrays fixed.

(a) A section \(Q_\omega\) at the comparison of \(i\) and \(j\) has entries that are images of parameter points with \(s_i=s_j\), so each difference is a combination of the \(m-2\) vectors \(h\), \(l_i+l_j\) and \(l_k\), \(k\ne i,j\), and (R2) gives rank at most \(m-2\).
A final flat leaf contains a directed triangle \(s_i\ge s_j\), \(s_j\ge s_k\), \(s_k\ge s_i\), so a feasible mask containing \(i\) contains \(k\) and hence \(j\), and one containing \(j\) contains \(i\); its entries therefore also have \(s_i=s_j\), with the same bound.

(b) By (S1), a section of the reconstruction has entries among the \(m-1\) points \(z\) and \(v_r\), \(r\ne i,j\), and by (S2) a terminal lower facet has \(m-1\) entries, so (R6) gives rank at most \(m-2\).
An other-facet section has entries among \(m-2\) points, hence rank at most \(m-3\), and its \(T_m\) image has rank at most \(m-2\) by (R7).

(c) For \(u\in\mathcal U_m\), the stored local part of \(G_u\) consists of the following terms, each of rank at most \(m-3\).
First, the faces at \(\pm u\) of a term of \(\rho_m\), which has rank at most \(m-2\) by (a): they are omitted when \(u\) annihilates every difference of its list, and otherwise have rank at most \(m-3\) by (R4).

Second, the faces of an original simplex \(P\), or of the target simplex, after the literal removal of matching facets.
By (F2), \(u\) does not annihilate every difference of \(P\), since \(u_1=1\).

If the face \(P^u\) omits exactly one vertex \(v_i\), choose \(k\ne i\) and put \(\varkappa=-u\cdot(v_i-v_k)\), which is positive; by (F1) the row \(u-\varkappa n_i\) annihilates every \(v_{k'}-v_k\), hence by (F2) it is zero, so \(\varkappa n_{i1}=1\), \(n_{i1}>0\) and \(u=n_i/n_{i1}\).
So that facet occurs in \(B_u(P)\) and is removed, and by (F1) the face \(P^{-u}\) is \(v_i\) alone.
The case where \(P^{-u}\) omits exactly one vertex is symmetric, with \(n_{i1}<0\).

The two faces cannot both omit exactly one vertex, since for \(m\ge3\) they would share a vertex at which \(u\) attains both its maximum and its minimum on \(P\), making \(u\) annihilate every difference.
Moreover, by (I4) a facet in \(B_u(P)\) is one of these faces.
Every remaining face has at most \(m-2\) entries, hence rank at most \(m-3\) by (R6).

Third, the stored local part of \(D_uf_m\).
A base face is split into its point row, of rank zero, and its segments, of rank at most one, and \(1\le m-3\); the apex alone has rank zero.
An apex-containing face of a root at \(\xi=\varsigma u\), for a sign \(\varsigma\in\{1,-1\}\), with \(\operatorname{nz}_\xi\ge2\) has differences that are combinations of \(h+\sum_{l_j\cdot\xi>0}l_j\) and the at most \(m-4\) edges of value zero, so its rank is at most \(m-3\) by (R2); one with \(\operatorname{nz}_\xi=0\) is omitted, as shown in (I9).

Let \(\operatorname{nz}_\xi=1\) and \(a_\xi=0\), with nonzero edge \(l_k\).
By (Lv), if \(k\in S\) the coordinates of \(\xi\) take exactly two values, and one of them is attained once exactly when the lower block has one position.

If \(k\notin S\), the leaf has a level \(d\), every other non-apex label a level \(c\ne d\), and the apex the level \(c+\relu(c-d)\).
So for \(d>c\) the value \(d\) is attained once, and for \(d<c\) the coordinates take the three values \(c+t\), \(c-t\) and \(c\), with \(t=c-d\), each of the first two once.

The same holds for \(u\), since \(\xi=\varsigma u\) has the same coordinate pattern up to sign, so \(u\) has three levels, two repeated levels, or exactly two values with one attained once.
In the last case let \(e\) be the label attaining its value once; since \(u_0=0\ne1=u_1\), the label \(e\) is zero or one, so \(u=u_+\) or \(u=e_1=u_-\).

For a three-level \(u\), the stored local part of \(D_uf_m\) is the point vertex of (I8) and the target faces, each a single label because the extreme values are attained once.
For \(u\) with two repeated levels, the faces with \(\operatorname{nz}_\xi=1\) and \(a_\xi=0\) are those of \(Z_{\varsigma u}\), which are omitted.
For \(u\) of none of these kinds, no such face occurs.

For two repeated levels and for \(u\) of none of these kinds, a target face is the list of labels at an extreme value of \(u\), which has at most \(m-2\) labels, since otherwise the remaining label would attain its value once and \(u\) would be \(u_\pm\); so it has rank at most \(m-3\).

For \(u_+\), the faces of \(Z^2\) have \(\operatorname{nz}\ge2\), those of \(Z^0\) are a base face and the apex, the residuals are point rows and segments, the star components are point rows, segments and the lists of \(V_A\), \(|A|\ge2\), whose differences are combinations of \(\sum_{a\in A}d_a\) and the \(d_a\), \(a\notin A\), at most \(1+(m-2)-|A|\le m-3\) vectors, and the rebasing row has rank zero.
For \(u_-\), every term is the pullback \(B^TP\) of a term \(P\) for \(u_+\), and (R7) with \(c=0\) gives \(\operatorname{rank}(B^TP)\le\operatorname{rank}(P)\).
Every term in (c) therefore has rank at most \(m-3\) before \(T_m\), and its \(T_m\) image terms have rank at most \(m-2\) by (R7).

This completes the step of (Rk): every local term of \(\mathcal L_m\) has rank at most \(m-2\).

\paragraph{\emph{Deletion and correction.}}
Fix a nonempty target \(T\) with \(s=|T|\ge2\), its \(d=b-s\) outside vertices \(v_1,\ldots,v_d\), and write \(\hat u_k=\hat u_{v_k}\).
Expand \(L_T=D_{\hat u_d}\cdots D_{\hat u_1}\) into sign words \(\vartheta\in\{+,-\}^d\) and \(C_T=\mathfrak{A}_1\cdots \mathfrak{A}_{s-2}\) into choice words \(w\), as in \Cref{ir:readout}.
This writes \(O_T=\sum_{\vartheta,w}(-1)^{\#\{k:\vartheta_k=-\}}\lambda_wP_{\vartheta,w}\), where \(P_{\vartheta,w}\) is exposure along the direction word that lists \(\vartheta_1\hat u_1,\ldots,\vartheta_d\hat u_d\) and then the face choices of \(w\), at most \(b-2\) letters in all.
Let \(\kappa\) be a key of frame \(b\) with \(\operatorname{dep}(\kappa)\le \mathfrak{H}-\tau_b-(b-2)\); the recentered input and the basis keys qualify, because \(\mathfrak{H}=6(b+1)\) exceeds \(\tau_b+b\).

\emph{Transport.}
For every pair \((\vartheta,w)\), the iterated-face identity, applied to \(F_b\), \(h_{\Delta_b}\) and \(E_b\), writes \(\epsilon^t(P_{\vartheta,w}E)[\kappa]\) as the same signed sum of the values \(E[\kappa_\sigma]\), where \(t\) is the word length.
The keys \(\kappa_\sigma\) have frame \(b\) and depth at most \(\mathfrak{H}-\tau_b\), so \(\Theta(b)\) gives \(f_b[\kappa_\sigma]=_{\Q}E_b[\kappa_\sigma]\).
Summing over \(\sigma\) and then over \((\vartheta,w)\), with the exchange of bounded double sums, gives \((O_TF_b)[\kappa]-(O_Th_{\Delta_b})[\kappa]=_{\Q}(O_TE_b)[\kappa]\).

\emph{Cancellation.}
Consider the triples \((\mathfrak{e},\vartheta,w)\) of a term \(\mathfrak{e}\) of \(E_b\) and the two words, and the masks of \(\mathfrak{e}\) after each of the \(d+s-2\) stages, taken in the order of application: the deletions, then \(\mathfrak{A}_{s-2},\ldots,\mathfrak{A}_1\), the stage for \(\mathfrak{A}_k\) having level \(k\).

A deletion stage \(k\) is a toggle position when \(\hat u_k\) annihilates every difference of the current mask; a correction stage of level \(k\) is a toggle position when the current mask has rank \(k\) and the choice is either the identity or the face \(P_{u_W}\) for the representative \(W\) of (R5).

Let \(\iota\) change the choice at the least toggle position, between \(\pm\hat u_k\) or between the identity and \(P_{u_W}\).
At that position both choices select the whole current mask, since the direction annihilates it, so all later masks and the least toggle position are unchanged; thus \(\iota\) is an involution that reverses the sign and preserves the final mask, and the pairing principle cancels all triples with a toggle position.

For a triple without a toggle position, number induction on the stage index proves that the rank before a deletion stage \(k\) is at most \(b-1-k\) and before a correction stage of level \(k\) at most \(k\).
The base is (Rk) at \(m=b\).

At a deletion stage the direction does not annihilate the mask, so (R4) lowers the rank.
At a correction stage of level \(k\), a mask of rank below \(k\) keeps rank at most \(k-1\) by (R1); a mask of rank \(k\) receives a face \(P_{u_V}\) with \(V\ne W\), whose direction does not annihilate it by the choice of \(u_V\) and (R5), so (R4) lowers its rank.
Hence every surviving final mask has rank zero, so all its vertices are equal and its value at \(y_\kappa\) is \(p\cdot y_\kappa\) for its first vertex \(p\).

Exchanging bounded sums gives \((O_TE_b)[\kappa]=_{\Q}\hat r\cdot y_\kappa\) for one rational vector \(\hat r\) independent of \(\kappa\); at the basis keys, \((O_TE_b)[e_i]=_{\Q}\hat r_i\).
With the transport at \(\kappa\) and at every \(e_i\), this proves \((O_Tf_b)[\kappa]=_{\Q}\sum_i(O_Tf_b)[e_i](y_\kappa)_i=r_T(y_\kappa)\).

\emph{The relation.}
For an outside vertex \(v\) and a subset \(A\ni v\) containing at least two vertices, the faces of \(\Delta_A\) exposed by \(\hat u_v\) and \(-\hat u_v\) are the masks \(\{v\}\) and \(A\setminus\{v\}\), by direct comparison of the coordinates.
In the expansion of \(L_Th_{\Delta_b}\), a sign word whose first plus sign is at position \(k<d\) has a singleton mask from then on, and its partner with the sign at position \(d\) reversed has the same mask; the pairing principle cancels these.
The remaining words give \((L_Th_{\Delta_b})[\kappa]=_{\Q}\varepsilon_Th_{\Delta_T}[\kappa]+(-1)^{d-1}(y_\kappa)_{v_d}\) for \(d\ge1\).

For a choice word \(w\ne0\), the first face choice uses a direction outside the orthogonal complement of the direction space of \(\Delta_T\).
That direction space has dimension \(s-1\), greater than the level of that choice, so by the exposure property the choice selects a proper subset \(T(w)\), and later faces stay inside it.
Faces of a singleton are the singleton.
Combining these with \(O_Th_{\Delta_b}=O_TF_b-O_Tf_b\) and the linearity just proved gives the literal identity of \Cref{ir:readout} for \(h_{\Delta_T}[\kappa]\).

For the readout, the lexicographic mask principle, applied to the tuples \(\mathbf v(l_j)\) and \(\mathbf v(h)\), identifies each selected root face as stated there, and the ordinary mask principle evaluates it at \(y_\kappa\).
The hinge reversal \(\relu(y_i-y_j)=\relu(y_j-y_i)+(y_i-y_j)\) and the identity \(x_i=\relu(x_i)-\relu(-x_i)\) are ordered-field identities.
Merging the resulting contributions by the injective reindexing equation gives \Cref{ir:relation} at \(\kappa\).

\paragraph{\emph{Subset chains and the block.}}
The chain recurrence \(c_{Tj}=a_{Tj}+\sum_{\varnothing\ne A\subsetneq T}M_{TA}c_{Aj}\) holds as a rational identity of the preformed integer arrays, by partitioning the chains through \(T\) according to their first step, the injective reindexing equation and integer distributivity.
Number induction on \(k\le b\) proves the admissible formula
\[
\forall T<2^b\,\left(1\le|T|\le k\rightarrow h_{\Delta_T}[\kappa]=_{\Q}\sum_jc_{Tj}g_j[\kappa]\right),
\]
where subsets are coded by numbers below \(2^b\), \(|T|\) is computed by counting, and \(g_j[\kappa]\) is the value of face-rule feature \(j\) at \(y_\kappa\).
At \(k=0\) there is nothing to prove.
For a singleton $T=\{v\}$, we have $h_{\Delta_T}[\kappa]=v\cdot y_\kappa$.
The explicit singleton assignment in \Cref{ir:readout} gives $v\cdot y_\kappa=\sum_j a_{Tj}g_j[\kappa]$: every coefficient is zero for $v=0$, and the coefficient of the coordinate linear row $y_i$ is one for $v=e_i$.
Since $T$ has no proper nonempty subset, the chain recurrence gives $c_{Tj}=a_{Tj}$ for every $j$.
For \(|T|\ge2\), \Cref{ir:relation} at \(\kappa\), the hypothesis for every proper nonempty \(A\subsetneq T\), the chain recurrence and the exchange of bounded double sums give the formula at \(T\).
At \(k=b\) and \(T=T_{\mathrm{full}}\), with the abbreviation $c_j=c_{T_{\mathrm{full}},j}$ fixed in \Cref{ir:readout}, it gives \(M_b(y)=_{\Q}\sum_jc_jg_j(y)\) at the recentered point \(y=(0,x_1-x_0,\ldots,x_{b-1}-x_0)\).

Let \(Q(x)\) be the output entry of the ordinary history \(\mathsf{Hist}_2(Q,x)\) of \Cref{rc:cc-lemma}, which equals \(c\relu(B\relu(Ax))\) by its adjacent affine equations, where \(A\), \(B\) and \(c\) are the two weight matrices and the output row of \(Q\).
Each second-layer preactivation of \(Q\) is the inner argument of its dictionary feature at \(y\), or the copied first-layer hinge, by the termwise identities \(y_i=x_i-x_0=\relu(x_i)-\relu(-x_i)-\relu(x_0)+\relu(-x_0)\) and \(y_i-y_j=x_i-x_j\), and by regrouping bounded sums; a copied hinge satisfies \(\relu(\relu(z))=\relu(z)\).

Regrouping the output sum gives \(Q(x)=_{\Q}x_0+\sum_jc_jg_j(y)=x_0+M_b(y)\).
By \Cref{uc:label-max}, \(\MAX_b(x)=x_a\) for a coordinate with \(x_a\ge x_i\) for every \(i\), so \(\MAX_b(x)-x_0=\max(0,y_1,\ldots,y_{b-1})=M_b(y)\), and hence \(Q(x)=_{\Q}\MAX_b(x)\).

For fixed admissible \(n\) and \(b\), the network \(Q\) is chosen independently of \(x\) and \(\epsilon\); no equality of its returned strings for different ambient \(n\) is needed here.
The proof treats \(x\) as an arbitrary valid rational \(b\)-vector, so generalization proves correctness for every such input in \(\bm{\mathcal T}\) after these fixed-array proofs.
It uses only the conservative \(\mathsf{FTC}^0\) choice, aggregation and induction properties of \(\bm{\mathcal T}\) recorded in \Cref{fact:tc-interface}, not a general polynomial-time evaluator or an elementary-arithmetic reflection principle.

\subsection{A Direct Bounded-Dimensional Cone Certificate}
\label{ir:cones}

The remaining output is an actual \(\mathsf{CC}\) certificate, not just a proof of rational-input equality.
We construct it from small-dimensional projections and a compressed hyperplane cover.
The certificate predicate \(\mathsf{CC}\) is the one defined in \Cref{rc:cones}.
All enumerations range over counts determined by \(b\) and \(Q\).

We first prove a conic alternative: either a target vector is a nonnegative combination of given rows, or a rational point separates it from them.
Let \(g_1,\ldots,g_m,w\in\Q^b\).
In matrix products we identify each vector, including a normal stored as a row, with its coordinate column; its transpose is the corresponding row.
Thus an outer product $uv^T$ has entries $u_iv_j$.
In this subsection squared norms are Euclidean: \(\|z\|^2=\sum_{i<b}z_i^2\).
Two vectors are orthogonal when their dot product is zero.

For an independent subset \(I\) of at most \(b\) rows, its Gram matrix is \(G_{ij}=g_i\cdot g_j\), for \(i,j\in I\).
Solve
\[
\sum_{j\in I}G_{ij}\beta_j=g_i\cdot w\quad(i\in I),
\qquad p_I=\sum_{j\in I}\beta_jg_j.
\]
These equations say that \(w-p_I\) is orthogonal to every row in \(I\).
Thus $p_I$ is the orthogonal projection of $w$ onto their span.
Independence makes \(G\) positive definite: for any nonzero coefficient vector \(v\), \(v^TGv=\|\sum_{i\in I}v_ig_i\|^2>0\).
A nonzero kernel vector would contradict this inequality, so the maximal-minor kernel calculation proves nonsingularity and justifies the Cramer solve.
Clearing the selected rows and \(w\) once gives integer Gram determinants of order at most \(b\), with polynomial component bounds in \(b\) and the input precision.

There are at most $(m+1)^b$ padded index tuples for these subsets; reject repetitions and dependent tuples.
Enumerate them and retain the projections whose coefficient vectors are nonnegative, including \(p_\varnothing=0\).
A point is feasible when it is a nonnegative combination of an independent set of at most \(b\) rows.
Choose a retained \(p\) of minimum squared distance from \(w\), with the fixed least-index tie rule, and discard its zero coefficients.
Its support is the set of rows with positive coefficients.
The residual \(r=w-p\) is orthogonal to every row of that support.

Suppose \(a=r\cdot g_j>0\).
Since \(r\) is orthogonal to the current support, \(g_j\) is outside its span.
Adjoining it therefore gives an independent set of at most \(b\) rows.
With \(v=\|g_j\|^2\) and \(\theta=a/(1+v)>0\), the feasible point \(y=p+\theta g_j\) satisfies
\[
\|w-y\|^2-\|w-p\|^2
=-2\theta a+\theta^2v
=-\frac{a^2(2+v)}{(1+v)^2}<0.
\]
Thus it is strictly closer to \(w\).

Next we show that for every feasible point there is a retained projection that is no farther from \(w\); applied to the point \(y\) above, this contradicts the minimality of \(p\).
For any independent support \(J\) and feasible point \(y=\sum_{i\in J}\alpha_i g_i\), with \(\alpha\ge0\), let \(q\) be the unconstrained Gram projection with coordinates \(\beta\).
If \(\beta\ge0\), orthogonality makes \(q\) no farther than \(y\).
Otherwise choose the least minimizing negative-coordinate ratio
\[
t=\min_{\beta_i<0}\frac{\alpha_i}{\alpha_i-\beta_i}\in[0,1).
\]
Every coordinate of \(y'=(1-t)y+tq\) is nonnegative in this basis: for \(\beta_i<0\) the chosen bound on \(t\) gives this directly, and for \(\beta_i\ge0\) it follows by convex combination.
A minimizing coordinate vanishes.
Since \(w-q\) is orthogonal to \(y-q\),
\[
\|w-y'\|^2=\|w-q\|^2+(1-t)^2\|y-q\|^2\le\|w-y\|^2.
\]
Delete a vanished coordinate, retaining an independent support, and repeat on that smaller support.
This includes \(t=0\), when the point stays fixed but the support still shrinks.
After at most \(b\) deletions a retained projection is reached; it is no farther than the original \(y\), contradicting minimality of \(p\).

It follows that \(r\cdot g_j\le0\) for every row.
Since \(r\cdot p=0\), either \(r=0\), giving a nonnegative representation of \(w\) on at most \(b\) independent rows, or
\[
x=-r,\qquad g_j\cdot x\ge0\quad(j\le m),\qquad w\cdot x=-\|r\|^2<0.
\]
This is a finite rational conic alternative, including nonpointed cones and zero rows.

We now show how \(\bm{\mathcal T}\) carries out the deletion argument above, using only matrix words formed in advance.
For each independent support \(S\), form its Gram matrix \(G_{ij}=g_i\cdot g_j\) and its projection \(q_S\), with coordinates \(\beta_i\).
Put \(c_i=\sum_{j\in S}(G^{-1})_{ij}g_j\).
The inverse equation gives \(c_i\cdot g_j=\delta_{ij}\), where \(\delta_{ij}\) is one for \(i=j\) and zero otherwise.
Thus \(c_i\cdot y=\alpha_i\) for \(y=\sum_{j\in S}\alpha_jg_j\), and \(c_i\cdot q_S=\beta_i\).
These are the coordinate-dual vectors used as the rows \(c_i^T\) below.
Deletion of \(i\) is the homogeneous matrix action
\[
\begin{pmatrix}q_Sc_i^T-\beta_iI&0\\c_i^T&-\beta_i\end{pmatrix}
\begin{pmatrix}Y\\d\end{pmatrix},\qquad y=Y/d.
\]
Its new denominator is \(d(c_i\cdot y-\beta_i)>0\) on a valid step.
At a deletion prefix, that is, after a sequence of such deletions, the remaining original support determines its projection and update matrices; the current point is obtained from the preformed matrix word.
Index the initial support by its at most \(b\) local positions, and use these positions as deletion labels.

Enumerate every deletion word of length at most \(b\) and evaluate its complete matrix word directly.
The prefix equations follow by grouping coordinate walks.
A valid prefix tests the minimum ratios, signs and nonnegative coordinates at every prefix.
Bounded-number induction appends a valid deletion label unless a feasible projection has already been reached.
Numerical word codes have at most \(O(b\log(b+1))\) bits; their derived coefficient arrays stay strings.
This proves existence of a terminal word without an induction quantifying new rational vectors.

We now construct the hyperplane cover.
Consider every first-layer row $A_i$ and every expanded second-layer row $\gamma_j(\sigma)=\sum_i B_{ji}\sigma_i A_i$, for every first-layer Boolean pattern $\sigma$.
List only the nonzero rows among these two kinds, together with the coordinate rows $e_i$.
Retain a separate lookup for discarded zero rows, including those from the all-zero pattern, and let $\mathfrak{q}$ be the length of the nonzero list.
Write \(h_1,\ldots,h_{\mathfrak{q}}\) for the listed rows, which we call the listed normals.
For \(\mathfrak{w}_1\le b^2+b\) first-layer rows and \(\mathfrak{w}_2\) second-layer rows,
\[
\mathfrak{q}\le \mathfrak{w}_1+\mathfrak{w}_2 2^{\mathfrak{w}_1}+b=2^{\operatorname{poly}(b)}.
\]
No enumeration of \(2^{\mathfrak{q}}\) signs is used.
Enumerate both signs of every cofactor kernel ray of \(b-1\) independent normal rows; there are at most \(R_0=2(\mathfrak{q}+b)^{b-1}\) rays.
At \(b=1\) use the two coordinate rays.
Enumerate sums of \(b\) independent listed rays and retain those avoiding all listed hyperplanes.
Their number is at most \(R_0^b\).

The sign word of a rational point \(x\) records, for each listed normal \(h_j\), whether \(h_j\cdot x\) is positive, negative or zero; it is strict when no entry is zero, and \(x\) is then a strict point.

For a strict sign word, its signed normals are the rows \(a_j\) equal to \(h_j\) or \(-h_j\) according as its entry at \(j\) is positive or negative.
Its chamber is the set of points \(y\) with \(a_j\cdot y>0\) for every \(j\), and its closed chamber is the set of points \(y\) with \(a_j\cdot y\ge0\) for every \(j\).
We prove that the enumerated sums represent every strict sign word realized by a rational point.

Fix such a word, a rational strict point realizing it, and the signed normals \(a_j\) of the word.
Because the coordinate rows are listed, the word fixes the coordinate signs \(\varepsilon_i\in\{-1,1\}\), and \(\varepsilon_ie_i\) is among the signed normals.
On the closed chamber of this word, put \(c=\sum_i\varepsilon_ie_i\).
Then \(c\cdot y=\sum_i|y_i|>0\) for \(y\ne0\), so every nonzero point can be normalized to the section \(c\cdot y=1\).

We walk from the strict point, normalized to the section \(c\cdot y=1\), to a listed ray of the closed chamber, making one more normal tight at each step.
After \(t\) steps of this walk, retain \(c\) and \(t\) normals.
Together these \(t+1\) rows are independent; the \(t\) normals are tight at the current point \(y\), while \(c\cdot y=1\).
The invariant is
\[
c\cdot y=1,\qquad a_j\cdot y\ge0\text{ for every }j,
\qquad a_{j_s}\cdot y=0\quad(1\le s\le t).
\]
It does not require every tight normal to be retained.

While \(t<b-1\), the \(t+1\) retained rows have rank below \(b\), so their integer kernel basis, computed from a nonzero minor of largest order as in \Cref{ir:arithmetic}, contains a nonzero direction \(d\) annihilating them; we choose \(d\) from this small-minor kernel basis.
Since \(c\cdot d=0\), the coordinate normals give both positive and negative values of \(a_j\cdot d\).
Define
\[
t_+=\min_{a_j\cdot d<0}\frac{a_j\cdot y}{-a_j\cdot d},
\qquad
t_-=\min_{a_j\cdot d>0}\frac{a_j\cdot y}{a_j\cdot d}.
\]
Both sets are nonempty and both times are nonnegative.
The endpoints \(y_+=y+t_+d\) and \(y_-=y-t_-d\) satisfy every inequality: rows whose value decreases are protected by the corresponding minimum, and the other values do not decrease.
They preserve the retained equalities and \(c\cdot y=1\).

For the endpoint \(y_+\), respectively \(y_-\), the hit row is the signed normal \(a_j\) of least index \(j\) attaining the minimum \(t_+\), respectively \(t_-\).
Its product with \(d\) is nonzero, so it is independent of the retained rows, even if the endpoint time is zero.
For either sign the update is the direct linear map
\[
y\mapsto\left(I-\frac{d a_j^T}{a_j\cdot d}\right)y.
\]
Thus the invariant persists and the retained rank increases by one.
After \(b-1\) updates, the point is nonzero and annihilates \(b-1\) independent original normals.
It is therefore a positive multiple of one of their two listed cofactor rays, and that ray lies in the closed chamber.

Fix the starting point and enumerate all hit-row/sign words of length at most \(b-1\).
The retained rows at a prefix, and hence its chosen kernel direction and update matrix, depend only on that prefix.
Compute all complete matrix products directly and evaluate them at the fixed starting point before induction.
A valid prefix checks the displayed invariant, retained independence and the minimizing-hit conditions at every earlier step.
All these checks use bounded number indices in the already formed arrays.

The next induction shows that the listed rays in the closed chamber span the section points reached by the walk, in the following sense.
Suppose a rational row annihilates every listed ray in the closed chamber.
Backward numerical induction on valid prefix length proves that it annihilates every corresponding section point.
The terminal case follows from the listed-ray conclusion.
At a shorter prefix both valid extensions are available.
If \(t_++t_->0\), use
\[
y=\frac{t_-y_++t_+y_-}{t_++t_-}.
\]
If both times are zero, both endpoints equal \(y\), but each extension has a strictly longer retained basis, so the same induction still applies.
No new rational-vector string is quantified in this induction.
The word count is bounded by \((2\mathfrak{q}+2)^b\), with the same single-exponential arity bounds as the direct arrays.

For a strict rational point \(x\), choose a positive rational \(\delta\) smaller than every \(a_j\cdot x/(2(1+\sum_i|a_{ji}|))\).
Then the \(b\) points \(x+\delta\varepsilon_i e_i\) remain strict.
They are independent: after changing coordinate signs their columns are \(z+\delta e_i\), with \(z_i>0\).
A vanishing linear combination with coefficients \(\alpha_i\) and \(A=\sum_i\alpha_i\) gives \(Az_i+\delta\alpha_i=0\).
Summing yields \(A(\sum_i z_i+\delta)=0\), so \(A=0\) and then every \(\alpha_i=0\).

If the listed rays in the closed chamber had rank less than \(b\), the small-minor kernel basis of the matrix whose rows are these rays would contain a nonzero row annihilating every one of them, and by the backward induction above it would annihilate these independent points after normalization, a contradiction.
Hence the closed chamber contains \(b\) independent listed rays.
Their sum is strict: every signed normal is nonnegative on each ray, and a nonzero normal cannot vanish on their basis.
This sum is one of the retained representatives.

The direction search of \Cref{ir:readout} tests \(u(t)=\sum_{j<c}t^jz_j\) for an integer kernel basis \(z_0,\ldots,z_{c-1}\).
When the search is run with the standard basis of \(\Q^b\) in place of the kernel basis, so that \(c=b\) and \(z_j=e_j\), and with the listed nonzero normals as the rows to avoid, it gives a rational direction \(v\) with \(h_j\cdot v\ne0\) for every listed nonzero normal \(h_j\).
Indeed each \(h_j\cdot u(t)\) is then a nonzero polynomial of degree at most \(b-1\), so by the root count proved there one of the trials \(1,\ldots,\mathfrak{q}(b-1)+1\) is a root of none of them.
For any rational \(x\), choose a positive rational \(\delta'\) smaller than \( |h_j\cdot x|/(2(1+|h_j\cdot v|))\) whenever \(h_j\cdot x\ne0\); if there are no such rows, use any positive \(\delta'\).
Then \(x+\delta' v\) preserves every nonzero sign of \(x\), while a zero value becomes \(\delta' h_j\cdot v\ne0\).
Its strict sign word has a representative by the preceding argument.
Since \(x\) satisfies the weak inequalities of that word, every rational point lies in the closure of a represented chamber.
These choices require only rational row products, comparisons and a minimum over the listed rows.

A prefix of a bit word is its initial segment; the longest common prefix of two words ends just before their first differing bit.
Sort the representatives' \(\mathfrak{q}\)-bit sign words by lexicographic order, comparing at the first differing bit, and remove duplicates by all-pairs comparison and rank counting.

A prefix is retained when it is a prefix of one of these words, and branching when both of its one-bit extensions are retained.
For each retained prefix, its descendant interval is the consecutive group of words beginning with that prefix.

At a branching prefix $P$, test the next bit, at position $|P|$.
Partition its descendant interval into the two nonempty groups extending $P0$ and $P1$.
Each child stores its group's longest common prefix; a singleton group gives the complete word.
The distinct longest common prefixes of adjacent words give exactly these internal nodes.
This is the compressed binary trie: only branching prefixes and complete words are stored.

The root prefix is empty, since a strict word and its opposite both occur.
Number nodes by increasing prefix length and then lexicographic order, so the root is zero and every child has a larger index.
A node's cone list consists exactly of its proper internal ancestors' signed test rows, in that order.

Encode a sign by bit one for a positive normal value and bit zero for a negative value.
At a branch testing row $t$, store its bit-one child as the first $\mathsf{CC}$ child, appending $t$, and its bit-zero child as the second, appending $-t$.
Thus each child list is bitwise its parent list with the required signed test appended; this child-field order is separate from lexicographic order.
Every internal node has two children, so the number of nodes is twice the number of leaves minus one, and is at most twice the number of representatives minus one.

A node's stored prefix is the longest common prefix of the words below it, but the cone printed at that node tests only the bits at its proper ancestors' branch positions.
The other bits of the stored prefix are skipped.
We show that these skipped signs nevertheless hold weakly throughout the printed cone.
For a fixed rational point \(x\), use numerical induction on node depth with the following invariant: if \(x\) belongs to a node's printed cone, then every sign in its full stored prefix holds weakly at \(x\).
The root prefix is empty, so the base case is immediate.

Consider a child whose cone contains \(x\).
Its parent cone also contains \(x\), so the induction hypothesis applies to this same point.

Choose a strict retained representative \(z\) among the child's descendants.
For each originally nonzero \(h_j\cdot x\), choose \(\varepsilon>0\) smaller than \( |h_j\cdot x|/(2(1+|h_j\cdot z|))\), using any positive bound if this list is empty.
The point \(x+\varepsilon z\) preserves all originally nonzero global signs and is strict on every row: where \(h_j\cdot x=0\), strictness follows from \(h_j\cdot z\ne0\).

The parent's prefix signs become strict, because they hold weakly at \(x\) and strictly at \(z\); the child's appended branch sign becomes strict for the same reason.
Its represented word therefore belongs to the child's descendant interval.
An opposite nonzero skipped sign at \(x\) would be preserved by the perturbation, contradicting the common prefix of that interval.
This proves the child's full weak prefix.

The induction formula has only bounded number quantifiers over nodes and row indices, with \(x\) and the preformed arrays fixed.
It never applies its induction hypothesis to the perturbed point; completeness of the represented strict words supplies that point's word separately.
A second generic perturbation is unnecessary, since the chosen descendant representative is already strict on every listed row.

A leaf's complete word supplies weak signs for every nonzero first-layer row and hence a first activation pattern.
That pattern selects the relevant second-layer expansions from the listed pool; their weak signs give the second activation pattern.
Assign bit zero to identically zero forms.
Each sign inequality is valid on its printed cone.

On adjoining \(x_a\ge x_i\) for every \(i\), the already proved rational correctness of \(Q\) makes both signs of its output minus \(x_a\) valid.
Apply the conic alternative to each such row.
The separating case contradicts that validity; its selected nonnegative sparse coefficients therefore give the exact multiplier fields of \Cref{rc:sign-identities,rc:winner-identities}.
Omitted entries are explicit zero numerators over denominator one.
This constructs the current certificate schema, with its actual child-list and multiplier equations.

All these chosen direct graphs have \(b\)-local output bounds: the actual dimensions, compact references, scalar term counts and headers depend only on \(b\) and \(Q\), rather than on the ambient \(n\).
These bounds are explicit; write \(\bar b=b+1\) and use the comparisons of \Cref{ir:packets}.

For \(b=2\) the block of \Cref{ir:small} has three neurons in each hidden layer and entries in \(\{0,1,-1\}\), so every bound below holds for it a fortiori.
For \(b\ge3\), the block \(Q\) has first-layer width \(\mathfrak{w}_1=b^2+b\le\bar b^2\), and its second layer holds the dictionary features, one for each root of arity \(b\) and ternary tag, below \(P_2(b^2)P_2(b)P_2(2b)\), and at most \(\bar b^2\) copied hinges, so its width is \(\mathfrak{w}_2<P_2(3\bar b^2)\).

Let \(L_Q\) be the largest component length of a rational entry of \(Q\).
Its first-layer entries lie in \(\{0,1,-1\}\), since its first-layer rows are coordinate-sign and oriented pair hinges.
Its copy entries are integers of magnitude at most two, and its inner second-layer coefficients have magnitude at most \(\mathfrak{m}\).

An output coefficient is a sum, over fewer than \(P_2(b^2)\) subset chains, of products of at most \(b\) factors, each an integer numerator of magnitude at most \(N_T\mathfrak{m}(b+1)D_0<P_2(42\bar b^3+2\bar b+2+\bar b+b^2)\le P_2(43\bar b^3)\) or the factor \(D_0\le P_2(b^2)\), over \(D_0^b\le P_2(b^3)\), and the final linear rows add sums of at most \(b\) such coefficients.
So \(L_Q\le45\bar b^4+\bar b+2\le46\bar b^4\).

The listed rows number \(\mathfrak{q}\le \mathfrak{w}_1+\mathfrak{w}_2 P_2(\mathfrak{w}_1)+b<P_2(5\bar b^2)\).
An expanded row is a sum of at most \(\mathfrak{w}_1\) products of an entry of length at most \(L_Q\) with an integer in \(\{0,1,-1\}\), so the clearing bound of \Cref{ir:arithmetic} gives normal components of length at most \(L_N=4(\bar b^2+1)(L_Q+8)\le400\bar b^6\).
A target is a sum of at most \(\mathfrak{w}_2\) products of such rows with output coefficients, so its components have length at most \(4(\mathfrak{w}_2+1)(L_Q+L_N+6)<P_2(3\bar b^2+6\bar b+12)\le P_2(7\bar b^2)\).

Every other rational produced for the certificate, namely a cofactor ray, a representative, a Gram entry, a Gram solution, a dual row, an update matrix, a deletion-word product and a final coordinate, is obtained from normals and targets by at most twenty successive operations.
Each operation is a binary product or difference, a clearing of at most \(\bar b^2\) summands, a solve of order at most \(b\), or a matrix word of at most \(b+1\) factors of dimension at most \(b+1\).

The longest such chain is the deletion argument of the conic alternative: Gram entries take two operations, Gram solutions three, projections and dual rows five, residuals six and update matrices seven, the improvement numerator eight, the improved point eleven, its image under the deletion word twelve, and its coordinates and minimum-ratio tests at most sixteen; every other listed rational occurs at an earlier stage of one of these chains.

Each operation maps a length cap \(L\) to at most \(1000\bar b^5(L+1)\), by the displayed formulas \(2L+3\), \(4(\mathfrak{n}+1)(L+2)\), \(\mathsf{Solve}\) and \(\mathsf{Word}_{\Q}\).
Number induction on the number of operations then gives at most \((1000\bar b^5)^{20}(P_2(7\bar b^2)+20)\le P_2(7\bar b^2+100\bar b+201)\le P_2(80\bar b^2)\).

The rays number \(R_0=2(\mathfrak{q}+b)^{b-1}<P_2(1+\bar b(5\bar b^2+1))\le P_2(6\bar b^3)\), representative index tuples fewer than \(P_2(6\bar b^4)\), and trie nodes fewer than \(P_2(6\bar b^4+1)\), which also bounds every cone-list length \(r\).
Projection supports number at most \((r+b+1)^b<P_2(7\bar b^5)\), and, over all leaves and the at most \(\mathfrak{w}_1+\mathfrak{w}_2+2b<P_2(3\bar b^2+1)\) identity rows, fewer than \(P_2(10\bar b^5)\).
Deletion words and coordinate walks are below \(P_2(\bar b^2+\bar b)\) each, sorting and duplicate removal use fewer than \(P_2(12\bar b^4+2)\) pairs, and hit words number at most \((2\mathfrak{q}+2)^b<P_2(6\bar b^3)\).
A product of any of these domains with one deletion word or walk stays below \(P_2(12\bar b^5)\).
So every padded domain of this subsection is below \(P_2(w_C(b))\), with \(w_C(b)=12(b+1)^5\).

The certificate \(d_b\) has six array levels.
Its row counts are below \(P_2(6\bar b^4+2)\), including the paired row indices \(\langle i,k\rangle\le(i+k+2)^2\) of its matrices.
Indeed, a cone list has one test row for each proper internal ancestor and these test distinct bit positions, so \(r\le\mathfrak{q}<P_2(5\bar b^2)\).

Hence every matrix index is below \(P_2(5\bar b^2+1)\), so \(i+k+2\le2P_2(5\bar b^2+1)\) and \((2P_2(5\bar b^2+1))^2=P_2(10\bar b^2+4)<P_2(6\bar b^4+2)\).
Its innermost strings, namely integer components, unary fields and bit patterns, have length below \(P_2(80\bar b^2)\), \(P_2(6\bar b^4+2)\) and \(P_2(3\bar b^2)\) respectively.
All these bounds are at most \(P_2(11\bar b^4)\), so the nesting bound of \Cref{ir:packets} gives \(|d_b|<P_2(64(11\bar b^4+4))\le P_2(w_d(b))\), with \(w_d(b)=768(b+1)^4\).

The network record of \(Q\) has fewer than eight array levels, as counted in the proof of \Cref{uc:balanced}.
Its row counts are at most its parameter count \(M\) plus one, where \(M\le\bar b^3+(\bar b^2+2)\mathfrak{w}_2+1<P_2(3\bar b^2)\) by the sharper width bound \(\mathfrak{w}_2<P_2(b^2+3b)+\bar b^2\).
Its innermost strings, the integer components of length at most \(46\bar b^4\) and the unary fields of length at most \(M\), have length below \(P_2(3\bar b^2)\).
So \(|Q|<P_2(128(3\bar b^2+4))\le P_2(w_Q(b))\), with \(w_Q(b)=512(b+1)^2\).
The packed output uses these fields directly; it does not truncate a rational sum already evaluated over the ambient width.
All auxiliary chosen graphs have polynomial bounds under the corresponding count guard.

\subsection{Direct Block Supply and the Depth Inequality}
\label{ir:packets}

This subsection first collects the power-table comparisons and the allocation exponents used by the earlier subsections.
It then fixes the arity guard under which certified blocks are supplied, proves \Cref{ir:small}, and proves \Cref{ir:rate}.
Every exponent in the allocation is an explicit polynomial in \(\bar b\).

Write \(\bar b=b+1\ge3\), and collect, with proofs, the elementary comparisons that the earlier subsections use.
Let \(X\) be a table of binary integers with \(X(0)=1\) and \(X(j+1)=X(j)r\) for an integer \(r\ge2\): the binary power table \(P_2\), with \(r=2\), or the interval-product powers \(R_g^j\), with \(r=R_g\).
Number induction on \(j\) gives \(X(j)\ge1\), and number induction on \(a'\) gives \(X(a)X(a')=X(a+a')\) whenever \(a+a'\) is within the index bound of the table; with \(X(a')\ge1\) this gives monotonicity, \(X(a)\le X(a+a')\).

For \(P_2\), monotonicity gives \(P_2(a)+P_2(a')\le2P_2(\max(a,a'))=P_2(\max(a,a')+1)\).
Number induction on \(j\), with \(U\) as in \Cref{rc:selection}, from \(U(1)=P_2(0)\) and \(U(j+2)=U(j+1)+1\le2P_2(j)=P_2(j+1)\), gives \(U(j+1)\le P_2(j)\).
Hence \(\bar b\le P_2(\bar b)\); \(U(2j)=2U(j)\le2P_2(j-1)=P_2(j)\) for \(j\ge1\), in particular \(2b\le P_2(b)\); and \(\bar b^k\le P_2(k\bar b)\).
We call these the power-table comparisons.

A canonical integer of magnitude below \(P_2(\lambda)\) has length at most \(\lambda+1\), and one of length \(\lambda\) has magnitude below \(P_2(\lambda)\), by its definition in \Cref{sec:foundations}.
Every inequality between polynomials in \(\bar b\) with nonnegative coefficients displayed in \Cref{ir:direct,ir:readout,ir:tiers,ir:cones} holds for \(\bar b\ge3\) and is proved in \(\bm{\mathcal T}\) by substituting \(b=\bar b-1\) and \(\bar b=3+\bar b'\), expanding, and checking that every coefficient of the difference of the two sides, as a polynomial in \(\bar b'\), is nonnegative.

The nesting bound states that if an exact array has at most \(\delta\) levels, every row count is at most \(P_2(W)\), and every innermost string has length at most \(P_2(W)\), then its length is below \(P_2(2^\delta(W+4))\).
Indeed, by (Arr) and \(\langle v,B\rangle\le(v+B+2)^2\), an array of at most \(P_2(W)\) rows of length at most \(P_2(c)\), with \(c\ge W\), has length at most \((4P_2(c))^2=P_2(2c+4)\), and number induction on the level shows that level \(k\) has length at most \(P_2(c_k)\) with \(c_k=2^kW+4(2^k-1)\le2^k(W+4)\).
The explicit allocation exponents are
\[
w_D=24\bar b^3,\quad w_R=56\bar b^3,\quad w_{\mathrm{pt}}=300\bar b^3,\quad w_C=12\bar b^5,\quad w_Q=512\bar b^2,\quad w_d=768\bar b^4,
\]
where \(w_D\) comes from \Cref{ir:direct}, \(w_R\) and \(w_{\mathrm{pt}}\) come from \Cref{ir:tiers} for the readout and point domains, and \(w_C\), \(w_Q\) and \(w_d\) come from \Cref{ir:cones} for the cone domains, the block and the certificate.

The precision caps \(B_0\), \(B_{\mathrm{tr}}\), \(B_v\), \(B_z\), \(B_d\), \(\mathfrak{t}_b\) and \(B_T\) are bounded by polynomials in \(\bar b\) with nonnegative coefficients after the following replacements.
The maximum in \(B_{\mathrm{lin}}\) can be removed: whenever a local root is present, \(m\ge3\), and
\[
B_F\ge4(m+1)(B_d^0+2)>2B_d^0+3=C_1(B_d^0),
\qquad B_F>3.
\]
Thus \(B_{\mathrm{lin}}=C_3(B_F)\) before the remaining majorizations.

First write \(b+1=\bar b\), and bound the remaining \(b\), \(m\) and \(m+1\) by \(\bar b\).
Use \(\lceil\log_2(r+1)\rceil\le\bar b\) for \(r\le\bar b\) and \(\lceil\log_2R_g\rceil\le\bar b+2\).
In \(B_T\), use \(\lceil\log_2(N_T+1)\rceil\le42\bar b^3+1\) and \(\lceil\log_2(D_0+1)\rceil\le\bar b^2\).
For the remaining logarithms, use \(\lceil\log_2(t+1)\rceil\le t\).
In particular, \(\lceil\log_2(m(m+1)+1)\rceil\le\bar b^2\) and \(\lceil\log_2(b+2)\rceil\le\bar b\).
The conversion multiplier cap \(\mathfrak m=4\bar b^2\) of \Cref{ir:readout} gives \(\lceil\log_2(\mathfrak m+1)\rceil\le4\bar b^2\).
Drop the negative constants in \(C_k\).

The coefficient sum of a polynomial with nonnegative coefficients is its value at one.
\Cref{tab:ir-precision} records sufficient bounds on degree and coefficient sum for the resulting majorants.
For example, the majorant of \(B_F\) expands to
\[
14+34\bar b+3\bar b^2+1600\bar b^5+8000\bar b^7.
\]
The majorants of \(B_{\mathrm{root}}\) and \(B_t^0\) then have values \(558642\) and \(558715\) at one.
Substituting these into the displayed formula for \(\mathsf{Word}_{\Z}\) gives the first table entry:
\[
10\left(11(558715+2)+(558642+2)+11+2\right)
=67045440.
\]
Applying the remaining sum and product formulas gives the other entries; their degrees follow by taking the maximum under addition and adding degrees under multiplication.

\begin{table}[!bp]
\caption{Bounds on the degrees and coefficient sums of the precision majorants.}
\label{tab:ir-precision}
\centering
\begin{tabular}{@{}lcrp{.38\linewidth}@{}}
\toprule
Cap & Degree at most & Coefficient sum at most & Defining formula\\
\midrule
\(B_0\) & \(12\) & 67,045,440 & \Cref{ir:direct}, \emph{Uniform translations and primitive roots}.\\[.4em]
\(B_{\mathrm{tr}}\) & \(12\) & 67,045,484 & \Cref{ir:direct}, \emph{Closed term and vertex keys}.\\[.4em]
\(B_v\) & \(14\) & 8,045,458,050 & \Cref{ir:direct}, \emph{Closed term and vertex keys}.\\[.4em]
\(B_z\) & \(18\) & 8,045,458,052,210 & \Cref{ir:tiers}, direction precision bounds.\\[.4em]
\(B_d\) & \(18\) & 8,045,525,102,575 & \Cref{ir:tiers}, direction precision bounds.\\[.4em]
\(\mathfrak t_b\) & \(4\) & 46 & \Cref{ir:tiers}, direction precision bounds.\\[.4em]
\(B_T\) & \(3\) & 64 & \Cref{ir:readout}, coefficient precision bound.\\
\bottomrule
\end{tabular}
\end{table}

Every entry has degree at most eighteen and coefficient sum below \(2^{44}\).
Since each power \(\bar b^i\), \(0\le i\le18\), is at most \(\bar b^{18}\), each cap is at most
\[
2^{44}\bar b^{18}\le P_2(44+18\bar b)\le P_2(w_D).
\]
The first comparison uses \(\bar b\le P_2(\bar b)\); the second uses \(44+18\bar b\le24\bar b^3\) for \(\bar b\ge3\).
Likewise, the cone precisions \(400\bar b^6\) and \(P_2(80\bar b^2)\) are at most \(P_2(w_C)\): use \(400<2^9\), \(9+6\bar b\le12\bar b^5\) and \(80\bar b^2\le12\bar b^5\).

Put \(A=771\) and \(r_\ast=5\), and set \(\Gamma_b=2^{A(b+1)^{r_\ast}}\).
Then \(w+3\le A\bar b^{r_\ast}\) for each displayed exponent \(w\), since each has degree at most five and coefficient at most \(768\), so \(\Gamma_b\) dominates every padded key, minor, word and certificate count guard and every precision cap.
In particular, for the block \(Q\) and certificate \(d_b\) that \Cref{ir:small} supplies, \(U(b+2|Q|+2|d_b|+4)\le\Gamma_b\), by the computation in its proof.

The bounds above concern actual output strings, including scalar precision, dense zeros, headers and references.
All scalar calculations use their actual \(b\)-derived summand ranges before repacking.
The block-supply statement of \Cref{ir:small} packages the constructions of \Cref{ir:direct,ir:readout,ir:tiers,ir:cones} as two total functions with an explicit size bound.

\paragraph{\emph{Supply under the guard.}}
We now prove the block-supply statement of \Cref{ir:small} from the direct construction and its allocation bounds.

\begin{restatedtheorem}[\Cref{ir:small}]
For the fixed values \(A=771,r_\ast=5\), there are chosen total \(\mathsf{FTC}^0\) functions \(Q_{\mathrm{dir}}(n,b),D_{\mathrm{dir}}(n,b)\) such that \(\bm{\mathcal T}\) proves the following.
If \(2\le b\le n\) and the guarded comparison \(\Gamma_b\le U(n)\) holds, they return a homogeneous rational two-hidden-layer block \(Q\) and a certificate \(d_b\) with
\[
\mathsf{CC}(b,Q,d_b),\qquad U(b+2|Q|+2|d_b|+4)\le\Gamma_b\le U(n),\qquad b+2|Q|+2|d_b|+4\le n.
\]
Off the guard they return empty strings.
\end{restatedtheorem}

\begin{proof}[Proof of \Cref{ir:small}]
We first fix the guard.
For \(2\le b\le n\), the binary integer \(\Gamma_b\) is the entry \(P_2(A(b+1)^{r_\ast})\) of the binary power table of \Cref{rc:selection}, formed through the index bound \(A(n+1)^{r_\ast}\), so its length is at most a fixed polynomial in unary \(n\), even when the comparison \(\Gamma_b\le U(n)\) below fails.
The guard is the signed integer comparison \(\Gamma_b\le U(n)\), with \(U\) as in \Cref{rc:selection}.
Both functions are defined by a case distinction on this guard and on \(2\le b\le n\), and return empty strings off it.

On the guard, the guarded conversion of \Cref{rc:selection} with cap \(n\) gives a number \(e_b\le n\) with \(U(e_b)=\Gamma_b\).
Let \(w_1,\ldots,w_6\) be the explicit allocation exponents \(w_D,w_R,w_{\mathrm{pt}},w_C,w_Q,w_d\) at \(b\), namely \(24(b+1)^3\), \(56(b+1)^3\), \(300(b+1)^3\), \(12(b+1)^5\), \(512(b+1)^2\) and \(768(b+1)^4\) of \Cref{ir:packets}.
With these exponents, \Cref{ir:direct,ir:readout,ir:tiers,ir:cones} allocate their padded key domains, count guards and output bounds as binary powers \(P_2(w_k(b))\).
Take \(A=771\) and \(r_\ast=5\), as in \Cref{ir:packets}, so that \(w_k(b)+3\le A(b+1)^{r_\ast}\) for every \(1\le k\le6\) and every \(b\ge2\).

Each of these inequalities is between number terms with nonnegative coefficients, and \(\bm{\mathcal T}\) proves it from the semiring and order laws of the number sort by comparing monomials.
Monotonicity of the power table, proved in \Cref{ir:packets}, gives \(P_2(w_k(b))\le\Gamma_b\le U(n)\).
The guarded conversion therefore turns each of these bounds into a number at most \(n\), and only these converted numbers size arrays.
Every auxiliary function checks shape, denominator and range guards before indexing or inversion; invalid candidates have their stated zero defaults.

We now construct the local tables, all binary-lineage arrays, the face closure and generic directions, complete operator words, integer relation and subset-chain arrays, and the literal block.
Each stage is a fixed aggregate of the previously defined arithmetic, comparison, minimization and reconstruction functions.
There is no iteration of a function on the arity.
Each array is formed by comprehension below a converted bound, so the composite is a total \(\mathsf{FTC}^0\) function by the closure properties of \Cref{fact:tc-interface}, and its defining equations hold by construction.

\emph{Rational correctness.}
For a rational \(b\)-vector \(x\), let \(Q(x)\) be the output entry of the ordinary history \(\mathsf{Hist}_2(Q,x)\) of \Cref{rc:cc-lemma}.
For \(b\ge3\), \Cref{ir:tiers} proves \(Q(x)=_{\Q}\MAX_b(x)\) in \(\bm{\mathcal T}\) with \(n\), \(b\) and the string \(x\) as free parameters, under the guard; its point and value arrays are bounded by a fixed polynomial in \(n+|x|\), because their counts are among the guarded domains.
Generalization on \(x\) proves $Q(x)=_{\Q}\MAX_b(x)$ for every valid rational $b$-vector $x$, under the arity and padding guard.

For \(b=3\), the arity induction of \Cref{ir:tiers} consists of its base case alone, which starts with the zero low array of \Cref{ir:paired} and uses no smaller-arity expression.
For \(b=2\), use first rows \(e_1-e_0,e_0,-e_0\), second matrix \(I_3\), output \((1,1,-1)\), and zero biases.
Its value is \(\relu(x_1-x_0)+\relu(x_0)-\relu(-x_0)=x_0+\relu(x_1-x_0)=\max(x_0,x_1)\), by the case distinctions on the signs of \(x_0\) and \(x_1-x_0\) in the ordered field, and \(\max(x_0,x_1)=\MAX_2(x)\) by \Cref{uc:label-max}.

\emph{The certificate.}
The construction of \Cref{ir:cones}, applied to \((b,Q)\), forms the node array, the cone lists, the leaf activation patterns and, for every leaf \(\nu\) and every identity row \(\varrho\) of \Cref{rc:sign-identities,rc:winner-identities}, the output of the conic alternative for the target row computed by the checker of \Cref{rc:cones}.

That output is either nonnegative coefficients on at most \(b\) independent rows of the relevant list, or a rational point \(x_{\nu,\varrho}\) with \(g\cdot x_{\nu,\varrho}\ge0\) for every row \(g\) of that list and a negative value at the target.
We prove the admissible formula, with \(\nu\) and \(\varrho\) as free number variables and the arrays as parameters, stating that the alternative returns coefficients; generalization then covers all leaves and rows, and no induction on \(\nu\) or \(\varrho\) is used.

Suppose, for contradiction, that the alternative returns the point \(x=x_{\nu,\varrho}\).
Then \(x\) lies in the printed cone of the leaf \(\nu\), so the weak-prefix induction of \Cref{ir:cones}, with \(x\) as a parameter, shows that every sign of the leaf's complete word holds weakly at \(x\).

If \(\varrho\) is the sign row of a first-layer neuron \(i\), its target is \((2\chi_i-1)A_i\), where \(\chi_i\) is the leaf's first-layer activation bit for neuron \(i\) and \(A_i\) is row \(i\) of the first-layer matrix of \(Q\), unrelated to the constant \(A\).
If \(A_i=0\), the target value is zero; otherwise the word's bit for \(A_i\) is \(\chi_i\), so the target value is nonnegative; either case contradicts negativity.
If \(\varrho\) is the sign row of a second-layer neuron \(j\), the checker's row \(\gamma_j=\sum_iB_{ji}\chi_iA_i\) is rationally equal to the listed expanded row \(\gamma_j(\chi)\), by congruence of bounded sums, and the same argument applies with its bit \(\psi_j\).

If \(\varrho\) is a winner row for a coordinate \(a\), then \(x\) also satisfies \(x_a\ge x_i\) for every \(i\).
Recall the coefficient rows \(\beta_i=\chi_iA_i\), \(\delta_j=\psi_j\gamma_j\) and \(f=\sum_jc_j\delta_j\) that the checker of \Cref{rc:cones} computes, where \(c\) is the output row of \(Q\).
The weak signs give \(\relu(A_i\cdot x)=\beta_i\cdot x\) for every \(i\) and \(\relu(\gamma_j\cdot x)=\delta_j\cdot x\) for every \(j\), and the adjacent equations of \(\mathsf{Hist}_2(Q,x)\), with distributivity and the exchange of bounded double sums, give \(Q(x)=_{\Q}f\cdot x\), as in the proof of \Cref{rc:cc-lemma}.
By \Cref{uc:label-max}, \(\MAX_b(x)=_{\Q}x_a\), and rational correctness at this \(x\) gives \(f\cdot x=_{\Q}x_a\).
Hence both winner targets \(\pm(f-e_a)\) have value zero at \(x\), which contradicts negativity.
So every leaf and every row receives nonnegative coefficients \(\lambda_k\) on a support \(I\) of rows with \(\sum_{k\in I}\lambda_kg_k\) rationally equal to the target.

Store these coefficients in the multiplier fields at the positions of \(I\), with explicit zero numerators over denominator one elsewhere.
The checker's sum over all positions is rationally equal to the sum over \(I\), by the injective reindexing equation of \Cref{tc:parallel-arithmetic} applied to the inclusion of \(I\), since the omitted entries are zero; this gives every identity of \Cref{rc:sign-identities,rc:winner-identities} as a rational-value equality, as \(\mathsf{CC}\) requires.
The internal child lists are formed by appending the signed test row to the parent list, so the child identities hold bitwise by exact reconstruction (Arr), and the tags, unary indices, distinct in-range children, empty root list and dimensions hold by construction.
Thus \(\mathsf{CC}(b,Q,d_b)\) holds.

\emph{The size bound.}
The block \(Q\) and the certificate \(d_b\) are formed by comprehension below converted bounds \(P_2(w_Q(b))\) and \(P_2(w_d(b))\) among the \(P_2(w_k(b))\), so \(U(|Q|)\le P_2(w_Q(b))\) and \(U(|d_b|)\le P_2(w_d(b))\).
Put \(N_b=A(b+1)^{r_\ast}\).

Each of \(U(b)\le P_2(b)\), \(P_2(w_Q(b))\), \(P_2(w_d(b))\) and \(U(4)\le P_2(2)\) is at most \(P_2(N_b-3)\), by monotonicity and \(N_b\ge b+3\ge5\), so \(U(b)+2P_2(w_Q(b))+2P_2(w_d(b))+U(4)\le6P_2(N_b-3)\le P_2(N_b)=\Gamma_b\).
Additivity of \(U\) gives \(U(b+2|Q|+2|d_b|+4)\le\Gamma_b=U(e_b)\), and strict monotonicity of \(U\) gives \(b+2|Q|+2|d_b|+4\le e_b\le n\).

Under the guard all static output fields have the stated \(b\)-local bound.
The primitive arithmetic graphs retain their fixed polynomial computation-witness bounds in the actual argument lengths; after substitution these are polynomial in ambient \(n\), and in \(n+|x|\) for the point arrays.
A \(b\)-local output bound does not assert that arbitrary ambient padding of an auxiliary witness is also \(b\)-local.
This proves the lemma.
\end{proof}

It remains to prove the depth bound of \Cref{ir:rate}.

\begin{proof}[Proof of \Cref{ir:rate}]
We first bound \(k(n)\) on every branch, and then, when \(L_2\) is large, use \Cref{ir:small} to obtain a large admitted arity.
Use the direct arity selector of \Cref{rc:selection} and the literal compiler of \Cref{rc:compiler}.
On every branch \(k(n)\le2(L_1+1)\).

In block mode, write \(B\) for the selected arity and \(h\) for the least exponent of \Cref{rc:compiler} with \(P_B(h)\ge U(n)\), so that \(k(n)=2h\).
The prefix equations and \(B\ge2\) prove \(P_B(j)\ge P_2(j)\) by numerical induction on \(j\), starting at one and multiplying by the respective bases.
Moreover \(L_1+1\le n\): \(L_1\le n\) by its definition, and \(L_1\ne n\) since \(P_2(n)\ge U(n+1)>U(n)\).
Since \(P_B(L_1+1)\ge P_2(L_1+1)>U(n)\) and \(L_1+1\) is an index of the table, least height gives \(h\le L_1+1\).

On the fallback branch \(k(n)=L(n)\), which by \Cref{uc:powers} is the least \(i\le n\) with \(P_2(i)\ge U(n)\), where \(P_2(i)\) is IMUL on \(i\) copies of two, the same term as in \Cref{rc:selection}.
Hence \(P_2(L_1+1)>U(n)\) applies to this minimization.
Since \(L_1+1\le n\), the index \(L_1+1\) lies in the range of this minimization.
Minimality therefore gives \(L(n)\le L_1+1\le2(L_1+1)\).

Choose a fixed integer \(a\) with \(A\le2^a\), and set
\[
C_0=2(a+r_\ast)+2.
\]
If \(L_2\ge C_0\), put \(q=\lfloor L_2/(2r_\ast)\rfloor\).
Then \(L_2\ge2r_\ast+2\) gives \(q\ge1\), and \(q\le L_2\), so the binary power satisfies
\[
0<P_2(q)\le P_2(L_2)\le U(L_1)\le U(n).
\]
Apply the guarded conversion of \Cref{rc:selection}: let \(b_0\le n\) be the least number with \(P_2(q)\le U(b_0)\).
It satisfies \(U(b_0)=P_2(q)\), hence \(b_0=2^q\) and \(b_0\ge2\).
The quotient equation \(L_2=2r_\ast q+t\), \(0\le t<2r_\ast\), gives \(L_2<2r_\ast q+2r_\ast\le4r_\ast q\), since \(q\ge1\).
Thus \(q\ge L_2/(4r_\ast)\), and
\[
A(b_0+1)^{r_\ast}\le2^{a+r_\ast(q+1)}\le2^{L_2}\le L_1.
\]
The middle comparison uses \(r_\ast q\le L_2/2\) and \(a+r_\ast\le L_2/2\).
Hence \(\Gamma_{b_0}=P_2(A(b_0+1)^{r_\ast})\le P_2(L_1)\le U(n)\), by monotonicity of the power table and \(P_2(\ell(n))\le U(n)\) from \Cref{rc:selection}.
This arity is admitted by \Cref{ir:small}, so the greatest-arity selector has \(B\ge b_0\).

The least height \(h\) of block mode satisfies
\[
2^{q(h-1)}\le B^{h-1}<n<2^{L_1+1},\qquad q(h-1)\le L_1.
\]
To extract the exponent inequality, first form binary-string powers through the polynomial numerical cap \(n^2+n\).
Product-prefix regrouping gives \((2^q)^j=2^{qj}\); strict power monotonicity then gives the displayed numerical inequality without converting a large exponential magnitude.
Since \(k=2h\),
\[
kL_2\le8r_\ast L_1+2L_2\le(8r_\ast +2)L_1,
\qquad k(L_2+1)\le(8r_\ast +4)(L_1+1).
\]

If \(L_2<C_0\), integrality gives \(L_2+1\le C_0\), so the all-branch bound yields \(k(L_2+1)\le2C_0(L_1+1)\).
Thus the single standard constant \(C_\ast=\max(8r_\ast +4,2C_0)\) works for every \(n\ge2\), including zero iterated-log cases.
\end{proof}
\space

\section{Translation Cases Left as Exercises}
\label{sec:translation-cases}

This appendix writes out the cases of the proof of \cite[Theorem~7.57]{CookNguyen} that Cook and Nguyen leave as exercises \cite[Exercises~7.27, 7.58 and~7.59]{CookNguyen}.
The two-sorted sequent calculus $\mathrm{LK}^2$ uses the propositional rules specified in \Cref{sec:proofs}, with number and string quantifiers and equality axioms.
Its quantifier rules have the four patterns specified in \Cref{sec:proofs}, now with variables of either arithmetic sort.
The substituting expression is a number term for a number variable, and a string variable or the empty-string constant for a string variable.
A fresh variable is absent from the conclusion; substitutions preserve sorts and avoid capture.

The system $\mathrm{LK}^2$-$\tilde{\mathbf V}^1$ has the term-substitution instances of the $\text{2-BASIC}$, $\Sigma^B_0$ comprehension and equality axioms as its non-logical axioms.
It also has the induction rule
\[
\frac{\Gamma,A(b)\Rightarrow A(b+1),\Delta}
{\Gamma,A(0)\Rightarrow A(t),\Delta},
\]
where the number variable $b$ is absent from the conclusion and $A$ has the single-string form $\exists X\le s\,\varphi$ with $\varphi\in\Sigma^B_0$.
This is the system of \cite[Definitions~6.39 and~6.40 and Corollary~6.43]{CookNguyen}.
An anchored proof has every cut formula in a non-logical axiom or as $A(0)$ or $A(t)$ in an induction instance.

A parameter variable is a variable that is free in the endsequent.
In free variable normal form, a variable can disappear from a sequent only as the fresh variable of $\forall$-right, $\exists$-left or the induction rule; it is then not a parameter, and every nonparameter free variable serves exactly once as such a fresh variable.
This is the two-sorted induction version of \cite[Definition~2.39 and Sections~4D.1 and~6D.2]{CookNguyen}.
The translation proof starts from an anchored derivation in this normal form and translates it sequent by sequent.
The other steps of that proof are used as given by Cook and Nguyen, and this appendix does not reprove them.

We use the rules of $\mathrm{LK}^2$ and of $\mathbf G$ as defined in \cite[Sections~2A.1, 2B.3, 4D and~7C]{CookNguyen}, with their convention that free and bound variables are distinct symbols.
In particular, the propositional axioms, connective rules, weakening, exchange and contraction are exactly the rules specified in \Cref{sec:proofs}.

\subsection{Conventions}

An assignment $\vec a$ to a list of free variables gives each number variable a value in $\N$ and each string variable a length in $\N$.
For a number term $s$ whose variables are assigned, $\operatorname{val}_{\vec a}(s)$ is its value when each number variable takes its assigned value and each $|\gamma|$ takes the length assigned to $\gamma$.

A string variable $\gamma$ of length $n$ has the Boolean variables $\mathsf p^\gamma_0,\ldots,\mathsf p^\gamma_{n-2}$, and none when $n\le1$.
For $v=\operatorname{val}_{\vec a}(s)$, the atom $\gamma(s)$ translates to $\mathsf p^\gamma_v$ if $v<n-1$, to $\top$ if $v=n-1$, and to $\bot$ if $v\ge n$.
The empty-string constant $\emptyset$ of \cite[Section~4D.1]{CookNguyen} is translated as a string of length $0$, so that $\emptyset(s)$ translates to $\bot$ and $|\emptyset|$ has value $0$.
A numerical atom translates to $\top$ or $\bot$ according to its truth value at the assigned values, and the constants $\top$ and $\bot$ translate to themselves.

For a formula $\varphi$ and an assignment $\vec a$ that gives a value or a length to every free variable of $\varphi$, we write $\varphi[\vec a]$ for the translation of $\varphi$ under $\vec a$, with the simplification fixed below.
We write $\vec a,x\mapsto v$ for the assignment that extends $\vec a$ by giving the variable $x$ the value $v$ if $x$ is a number variable and the length $v$ if $x$ is a string variable.

Every string equality, including an atom $\gamma=\gamma$ that \cite[Section~7B.1]{CookNguyen} translates directly to $\top$, is translated through its defining formula, as \cite[Section~7B.3]{CookNguyen} does for the equality axioms.
This uniform choice is the one under which \Cref{lem:tr-substitution} holds.

Under this choice an instance $\Rightarrow\gamma=\gamma$ of the equality axiom E1$''$ \cite[Definition~4.26]{CookNguyen}, with $\gamma$ of length $n$, translates to $\Rightarrow\top$ when $n\le1$ and otherwise to the sequent whose one formula is the conjunction over $j\le n-2$ of $(\neg\mathsf p^\gamma_j\vee\mathsf p^\gamma_j)\wedge(\neg\mathsf p^\gamma_j\vee\mathsf p^\gamma_j)$.
Each conjunct comes from two copies of the axiom $\mathsf p^\gamma_j\Rightarrow\mathsf p^\gamma_j$ by $\neg$-right, an exchange and $\vee$-right in each copy and one $\wedge$-right.
Then $n-2$ further applications of $\wedge$-right give the whole sequent.
Hence this sequent has a tree-like cut-free derivation with at most $10n$ sequents.
The instance $\Rightarrow\emptyset=\emptyset$ of E1$''$ and the axiom $\Rightarrow|\emptyset|=0$ of \cite[Section~4D.1]{CookNguyen} both translate to $\Rightarrow\top$.

The implication $A\supset B$ abbreviates $\neg A\vee B$.
The bounded quantifiers $\forall w\le t\,\psi$ and $\exists w\le t\,\psi$ abbreviate $\forall w\,(w\le t\supset\psi)$ and $\exists w\,(w\le t\wedge\psi)$, with $|w|\le t$ in place of $w\le t$ when $w$ is a string variable \cite[Definition~4.13]{CookNguyen}.
Here $s<t$ abbreviates $s\le t\wedge\neg s=t$ \cite[Section~7B.1]{CookNguyen}.
For a number variable $w$, we let the strict bound $\forall w<t\,\psi$ or $\exists w<t\,\psi$ abbreviate $\forall w\le t\,(w<t\supset\psi)$ or $\exists w\le t\,(w<t\wedge\psi)$.
Under this convention the translation of a strict bound is the conjunction or disjunction of the instances below the value of $t$, as stated in \cite[Section~7B.1]{CookNguyen}, since the instance at that value is $\top$ or $\bot$ and is omitted.

The translation simplifies logical constants as in \cite[Section~7B.1, and Section~7E, displays~(148) and~(149)]{CookNguyen}, and we fix the simplification as follows.
The negation of $\top$ is $\bot$, and the negation of $\bot$ is $\top$.
A conjunction with a conjunct $\bot$ is $\bot$, conjuncts $\top$ are omitted, and an empty conjunction is $\top$.
A disjunction with a disjunct $\top$ is $\top$, disjuncts $\bot$ are omitted, and an empty disjunction is $\bot$.
A block of Boolean quantifiers in front of $\top$ or $\bot$ is omitted.
Iterated conjunctions and disjunctions over a finite index set are taken in increasing index order and bracketed to the right, as in \cite[Section~2A.1]{CookNguyen}.

Let $\Phi$ be $\forall w\le t\,\psi(w)$ or $\exists w\le t\,\psi(w)$, where $w$ is a number variable or a string variable, and let $\vec a$ assign the free variables of $\Phi$.
Put $r=\operatorname{val}_{\vec a}(t)$.
For $i\le r$, the instance $P_i$ is the translation of $\psi(w)$ under $\vec a$ extended by $w\mapsto i$.
For a string variable $w$, the instance $P_i$ may contain $\mathsf p^w_0,\ldots,\mathsf p^w_{i-2}$.
Let $\mathcal I=\{i\le r: P_i\notin\{\top,\bot\}\}$.
Let $\vec{\mathsf p}$ be the list $\mathsf p^w_0,\ldots,\mathsf p^w_{r-2}$ for a string variable $w$, and the empty list for a number variable $w$.

Then $(\forall w\le t\,\psi)[\vec a]$ is $\bot$ if some $P_i$ is $\bot$, is $\top$ if every $P_i$ is $\top$, and is $\forall\vec{\mathsf p}\,\bigwedge_{i\in \mathcal I}P_i$ otherwise.
Dually, $(\exists w\le t\,\psi)[\vec a]$ is $\top$ if some $P_i$ is $\top$, is $\bot$ if every $P_i$ is $\bot$, and is $\exists\vec{\mathsf p}\,\bigvee_{i\in \mathcal I}P_i$ otherwise.

For a sequent $S$ and an assignment $\vec a$ to its free variables, $S[\vec a]$ is the axiom $\Rightarrow\top$ if some antecedent formula translates to $\bot$ or some succedent formula translates to $\top$.
Otherwise $S[\vec a]$ lists the translations of the formulas of $S$ in order, omitting antecedent formulas $\top$ and succedent formulas $\bot$ \cite[Section~7B.3]{CookNguyen}.

\begin{lemma}[Substitution]
\label{lem:tr-substitution}
Let $\psi(x)$ be a bounded formula and $s$ a number term all of whose variables are free variables \cite[Section~4D]{CookNguyen}, let $\vec a$ assign the other free variables of $\psi(x)$ and the variables of $s$, and let $v=\operatorname{val}_{\vec a}(s)$.
Then $\psi(s)[\vec a]$ is $\psi(x)[\vec a,x\mapsto v]$.
Let $\psi(Z)$ be a bounded formula, let $\gamma$ be a free string variable other than $Z$, and let $\vec a$ assign the other free variables of $\psi(Z)$, giving $\gamma$ the length $n$ if $\gamma$ occurs in $\psi(Z)$.
Then $\psi(\gamma)[\vec a,\gamma\mapsto n]$ is $\psi(Z)[\vec a,Z\mapsto n]$ with every free occurrence of $\mathsf p^Z_j$ renamed $\mathsf p^\gamma_j$.
The same holds with $\emptyset$ in place of $\gamma$ and $n=0$, with no renaming.
\end{lemma}
\begin{proof}
For a number term $u(x)$, induction on $u$ shows $\operatorname{val}_{\vec a}(u(s))=\operatorname{val}_{\vec a,x\mapsto v}(u(x))$.
Both sides apply the same function symbols to equal values, and the occurrences of $x$ have value $v$ on the right while $s$ has value $v$ on the left.
In the second statement $|\gamma|$ and $|Z|$ both have value $n$, and in the third $|\emptyset|$ has value $0$.

We argue by induction on the formula obtained from $\psi$ by replacing every string equality with its defining formula.
By the convention above this formula has the same translation as $\psi$, and the induction is simultaneous for all assignments.
A numerical atom translates to the constant determined by the values of its terms, and these agree by the statement for terms.
A membership atom translates according to the value of its index term and the length of its string, and these agree.

In the second statement, $Z(u)$ and $\gamma(u)$ therefore translate to $\mathsf p^Z_j$ and $\mathsf p^\gamma_j$ for the same $j$, or to the same constant.
In the third statement, $Z(u)$ under length $0$ and $\emptyset(u)$ both translate to $\bot$.
Replacing a string equality with its defining formula commutes with the substitution, so string equalities need no separate case.

For a connective, the translation, with its simplification, is determined by the translations of the immediate subformulas.
Renaming maps each constant to itself, so it commutes with simplification.
For a bounded quantifier $Q y\le t'\,\chi(y)$, the variable $y$ does not occur in $t'$ \cite[Definition~4.13]{CookNguyen}, and the values of $t'$ agree by the statement for terms.
The variable $\gamma$ and the variables of $s$ are free and $y$ is bound, so $y$ is distinct from $\gamma$ and does not occur in $s$ \cite[Section~4D]{CookNguyen}.

Suppose first that $y$ is the substituted variable $x$ or $Z$.
Then the substituted variable has no free occurrence in $Q y\le t'\,\chi(y)$, so the substitution leaves this formula unchanged, and its translation does not depend on the value or length assigned to the substituted variable.
For $y=Z$, every $\mathsf p^Z_j$ in that translation is bound by its block of Boolean quantifiers, since the instance for $Z\mapsto i$ contains $\mathsf p^Z_j$ only for $j\le i-2$.
So the renaming of free occurrences changes nothing, and both sides are the translation of the same formula under assignments that agree on its free variables.

Suppose next that $y$ is distinct from the substituted variable.
The instances agree by the induction hypothesis for $\chi(y)$ under the assignments extended by $y\mapsto i$, which leave the value of $s$ equal to $v$, since $y$ does not occur in $s$.
The renaming does not touch the bound Boolean variables $\mathsf p^y_j$, since $y$ is distinct from $Z$, and it creates no capture, since $y$ is distinct from $\gamma$.
\end{proof}

\subsection{The Propositional Lemma for Comprehension Axioms}

This subsection proves the propositional lemma that supplies the comprehension axiom cases.
\begin{lemma}[{\cite[Exercise~7.58]{CookNguyen}}]
\label{lem:tr-comp-tails}
Let $A_0,\ldots,A_\ell$ be quantifier-free formulas, where $\ell\ge0$.
Put $N_k=\bigwedge_{i=k}^{\ell}\neg A_i$ for $0\le k\le\ell$, put $D_k=A_k\wedge N_{k+1}$ for $0\le k<\ell$, and put $D_\ell=A_\ell$.
Let $L=\sum_{i=0}^{\ell}(|A_i|+4)$, where $|A|$ is the number of symbols of $A$.
The sequent
\[
\Rightarrow N_0,D_0,D_1,\ldots,D_\ell
\]
has a tree-like cut-free derivation in the propositional rules of $\mathbf G$.
The derivation has at most $(\ell+1)(\ell+3)^3$ sequents, each with at most $\ell+3$ formulas of at most $L$ symbols each.
It is computed from $A_0,\ldots,A_\ell$ in time polynomial in $L$.
\end{lemma}
\begin{proof}
Each $N_k$ and each $D_k$ has at most $L$ symbols, since each conjunct $\neg A_i$ contributes $|A_i|+1$ symbols and its conjunction sign and brackets three more.
For $0\le k\le\ell$, write $\Theta_k$ for the list $D_0,\ldots,D_{k-1},A_k,\ldots,A_\ell$.
Thus $\Theta_0$ is $A_0,\ldots,A_\ell$, and $\Theta_\ell$ is $D_0,\ldots,D_\ell$, since $D_\ell=A_\ell$.
We derive $\Rightarrow\Theta_k,N_0$ for $k=0,\ldots,\ell$ in turn, replacing $A_k$ by $D_k$ at each step, using the two constructions below.

First, let $\Xi$ be a list of $\ell+1$ formulas that contains $A_j$.
The axiom $A_j\Rightarrow A_j$ and $\neg$-right give $\Rightarrow A_j,\neg A_j$.
Then $\ell$ weakenings add the other formulas of $\Xi$ at the end, and at most $(\ell+2)(\ell+1)/2$ exchanges give $\Rightarrow\Xi,\neg A_j$.
This takes at most $(\ell+3)^2$ sequents.

Second, let $\Xi$ be a list of $\ell+1$ formulas that contains $A_k,\ldots,A_\ell$.
We derive $\Rightarrow\Xi,N_j$ for $j=\ell,\ell-1,\ldots,k$ in turn.
For $j=\ell$ this is $\Rightarrow\Xi,\neg A_\ell$, derived as above.
For $j<\ell$, $\wedge$-right applied to $\Rightarrow\Xi,\neg A_j$ and $\Rightarrow\Xi,N_{j+1}$ gives $\Rightarrow\Xi,N_j$, since $N_j=\neg A_j\wedge N_{j+1}$.
This takes at most $(\ell+1)(\ell+3)^2+\ell\le(\ell+2)(\ell+3)^2$ sequents.

The second construction with $\Xi=\Theta_0$ and $k=0$ derives $\Rightarrow\Theta_0,N_0$.
Let $k<\ell$, and suppose that $\Rightarrow\Theta_k,N_0$ has been derived.
Let $\Xi_k$ be the list $D_0,\ldots,D_{k-1},A_{k+1},\ldots,A_\ell,N_0$, which has $\ell+1$ formulas and contains $A_{k+1},\ldots,A_\ell$.
At most $\ell-k+1$ exchanges turn $\Rightarrow\Theta_k,N_0$ into $\Rightarrow\Xi_k,A_k$.
The second construction derives $\Rightarrow\Xi_k,N_{k+1}$.
From these two sequents $\wedge$-right gives $\Rightarrow\Xi_k,D_k$, since $D_k=A_k\wedge N_{k+1}$.
At most $\ell-k+1$ exchanges turn it into $\Rightarrow\Theta_{k+1},N_0$.
This step adds at most $(\ell+2)(\ell+3)^2+2\ell+3\le(\ell+3)^3$ sequents.
Finally $\ell+1$ exchanges turn $\Rightarrow\Theta_\ell,N_0$ into $\Rightarrow N_0,D_0,\ldots,D_\ell$.

Each derived sequent is used once, so the derivation is tree-like, and it has no cut.
It has at most $(\ell+2)(\ell+3)^2+\ell(\ell+3)^3+\ell+1\le(\ell+1)(\ell+3)^3$ sequents.
Every sequent has at most $\ell+3$ formulas, each of which is some $A_k$, $\neg A_k$, $N_k$ or $D_k$, and so has at most $L$ symbols.
The derivation is written down step by step from $A_0,\ldots,A_\ell$ in time polynomial in $L$.
\end{proof}

The derivation of \Cref{lem:tr-comp-tails} has no cut, so it is a $\mathbf G_1^\star$ derivation.

\subsection{The Bounded Quantifier Cases}

This subsection treats the four quantifier rules when the principal formula is bounded.
\begin{lemma}[The quantifier rules]
\label{lem:tr-quantifier-cases}
Let $S$ follow from $S_1$ by a quantifier rule of $\mathrm{LK}^2$ whose principal formula $\Phi$ is $\forall w\le t\,\psi(w)$ or $\exists w\le t\,\psi(w)$, where $w$ is a number variable or a string variable.
Suppose that, for every assignment $\vec b$ to the free variables of $S_1$, a tree-like $\mathbf G_1^\star$ derivation $\pi_1(\vec b)$ of $S_1[\vec b]$ is given.
Let $\vec a$ assign the free variables of $S$, and put $r=\operatorname{val}_{\vec a}(t)$.
Then $S[\vec a]$ has a tree-like $\mathbf G_1^\star$ derivation built from the derivations $\pi_1(\vec b)$ for at most $r+1$ assignments $\vec b$, each used once, and at most $4r+4$ further inferences, none of which is a cut.

Every added sequent has at most as many symbols as $S[\vec a]$.
For the fixed rule instance, the assignments $\vec b$ and the added inferences are computed from $\vec a$ in time polynomial in the values and lengths that $\vec a$ assigns.
\end{lemma}
\begin{proof}
Let $A_1(s)$ be $s\le t\supset\psi(s)$ when $\Phi$ is universal and $s\le t\wedge\psi(s)$ when $\Phi$ is existential, with $|s|\le t$ for a string variable $w$.
The rule has one of the following four forms, in which $\Lambda$ and $\Pi$ are lists of side formulas.
\[
\frac{\Lambda\Rightarrow\Pi,A_1(b)}{\Lambda\Rightarrow\Pi,\Phi}
\qquad
\frac{A_1(s),\Lambda\Rightarrow\Pi}{\Phi,\Lambda\Rightarrow\Pi}
\qquad
\frac{\Lambda\Rightarrow\Pi,A_1(s)}{\Lambda\Rightarrow\Pi,\Phi}
\qquad
\frac{A_1(b),\Lambda\Rightarrow\Pi}{\Phi,\Lambda\Rightarrow\Pi}
\]
These are $\forall$-right, $\forall$-left, $\exists$-right and $\exists$-left, with $\Phi$ universal in the first two and existential in the last two.
The formula $A_1(b)$ or $A_1(s)$ in the premise is the auxiliary formula, and $\Phi$ is the principal formula.
In $\forall$-right and $\exists$-left the eigenvariable $b$ does not occur in $S$.
In $\forall$-left and $\exists$-right, $s$ is a number term when $w$ is a number variable, and a string variable or $\emptyset$ when $w$ is a string variable.
For a string variable $w$, write $\gamma$ for the string variable $b$ or $s$.

For $i\le r$, let $\vec a_i$ extend $\vec a$ by $b\mapsto i$.
Let $\vec a'$ extend $\vec a$ by the value $0$, or the length $0$, for each free variable of $S_1$ that does not occur in $S$.
Let $P_i$, $\mathcal I$ and $\vec{\mathsf p}$ be the data of $\Phi$ under $\vec a$.
For a string variable $w$, let $P^\gamma_i$ be $P_i$ with every free occurrence of $\mathsf p^w_j$ renamed $\mathsf p^\gamma_j$; otherwise, and when $s$ is $\emptyset$, put $P^\gamma_i=P_i$.

The formulas of $\Lambda$ and $\Pi$ contain neither $b$ nor any variable assigned by $\vec a'$ but not by $\vec a$.
So their translations under $\vec a_i$ and under $\vec a'$ are their translations under $\vec a$.
If some formula of $\Lambda$ translates to $\bot$ or some formula of $\Pi$ translates to $\top$, then $S[\vec a]$ is the axiom $\Rightarrow\top$.
Assume otherwise, and let $\Lambda'$ and $\Pi'$ be the translations of $\Lambda$ and $\Pi$ with antecedent formulas $\top$ and succedent formulas $\bot$ omitted.

We compute the translation of the auxiliary formula.
Let $v$ be the value, or the length, of $s$ under $\vec a'$; for the eigenvariable $b$ under $\vec a_i$, put $v=i$.
The term $t$ occurs in $S$, so it contains neither $b$ nor a variable outside $S$, and its value is $r$.
Hence $s\le t$, or $|s|\le t$, translates to $\top$ if $v\le r$ and to $\bot$ if $v>r$.
When $v\le r$, \Cref{lem:tr-substitution} shows that $\psi(s)$ translates to $P^\gamma_v$.
For a string variable $\gamma$ that occurs in $\psi(w)$, this uses that $\gamma$ is assigned the length $v$.

By the simplification rules, the auxiliary formula of a universal rule translates to $P^\gamma_v$ if $v\le r$ and to $\top$ if $v>r$.
The auxiliary formula of an existential rule translates to $P^\gamma_v$ if $v\le r$ and to $\bot$ if $v>r$.

When $\mathcal I$ is nonempty, write $\mathcal I=\{i_1<\cdots<i_m\}$.
Put $C_m=P_{i_m}$ and $C_j=P_{i_j}\wedge C_{j+1}$ for $j<m$, and put $E_m=P_{i_m}$ and $E_j=P_{i_j}\vee E_{j+1}$ for $j<m$.
Thus $C_1=\bigwedge_{i\in \mathcal I}P_i$ and $E_1=\bigvee_{i\in \mathcal I}P_i$.
Write $C^\gamma_j$ and $E^\gamma_j$ for the renamed formulas.

In $\forall$-left and $\exists$-right with $v\le r$, let $\sigma$ be the following substitution at the free occurrences.
It replaces $\mathsf p^w_j$ by $\mathsf p^\gamma_j$ when $j\le v-2$, and by $\bot$ when $v-1\le j\le r-2$.
When $w$ is a number variable or $s$ is $\emptyset$, every $\mathsf p^w_j$ is replaced by $\bot$, and the list is empty for a number variable.
Since $\mathsf p^w_j$ occurs free in $P_v$ only for $j\le v-2$, we have $P_v\sigma=P^\gamma_v$.

\emph{$\forall$-right.}
Suppose that some $P_i$ is $\bot$.
Then $\Phi[\vec a]=\bot$ is omitted from the succedent, and $S[\vec a]$ is $\Lambda'\Rightarrow\Pi'$.
The auxiliary formula translates under $\vec a_i$ to $P^\gamma_i=\bot$, so $S_1[\vec a_i]$ is the same sequent, and $\pi_1(\vec a_i)$ derives $S[\vec a]$.
If every $P_i$ is $\top$, then $S[\vec a]$ is the axiom $\Rightarrow\top$.

Otherwise $\mathcal I$ is nonempty, and $S_1[\vec a_i]$ is $\Lambda'\Rightarrow\Pi',P^\gamma_i$ for each $i\in \mathcal I$.
For $j=m-1,\ldots,1$ in turn, apply $\wedge$-right to $S_1[\vec a_{i_j}]$ and $\Lambda'\Rightarrow\Pi',C^\gamma_{j+1}$, which gives $\Lambda'\Rightarrow\Pi',C^\gamma_j$.
For a number variable, $C^\gamma_1=C_1$, and $\Lambda'\Rightarrow\Pi',C_1$ is $S[\vec a]$.
For a string variable, apply $\forall$-right for $j=r-2,\ldots,0$, each time with eigenvariable $\mathsf p^\gamma_j$, which is replaced by the bound variable $\mathsf p^w_j$.
The eigenvariable condition holds because $\gamma$ occurs in neither $\Lambda$ nor $\Pi$, so $\mathsf p^\gamma_j$ occurs in neither $\Lambda'$ nor $\Pi'$.
After the step for $j$, only the variables $\mathsf p^\gamma_l$ with $l<j$ remain free in the principal formula.
The result is $\Lambda'\Rightarrow\Pi',\forall\vec{\mathsf p}\,C_1$, which is $S[\vec a]$.
At most $(m-1)+r\le2r$ inferences were added.

\emph{$\forall$-left.}
If some $P_i$ is $\bot$, then $\Phi[\vec a]=\bot$ lies in the antecedent, and $S[\vec a]$ is the axiom $\Rightarrow\top$.
Suppose that every $P_i$ is $\top$.
Then $\Phi[\vec a]=\top$ is omitted, and $S[\vec a]$ is $\Lambda'\Rightarrow\Pi'$.
The auxiliary formula translates under $\vec a'$ to $\top$ whether $v\le r$ or $v>r$, so $S_1[\vec a']$ is the same sequent.
Otherwise $\Phi[\vec a]=\forall\vec{\mathsf p}\,C_1$.
No $P_i$ is $\bot$ here, so $P_v$ is not $\bot$ when $v\le r$.
If $v>r$ or $P_v=\top$, then $S_1[\vec a']$ is $\Lambda'\Rightarrow\Pi'$, and one weakening gives $S[\vec a]$.

Otherwise $v=i_h$ for some $h$, and $S_1[\vec a']$ is $P^\gamma_v,\Lambda'\Rightarrow\Pi'$, which is $P_v\sigma,\Lambda'\Rightarrow\Pi'$.
If $h<m$, a weakening by $C_{h+1}\sigma$, an exchange and $\wedge$-left give $C_h\sigma,\Lambda'\Rightarrow\Pi'$.
For $j=h-1,\ldots,1$, a weakening by $P_{i_j}\sigma$ and $\wedge$-left give $C_j\sigma,\Lambda'\Rightarrow\Pi'$.
For a string variable, apply $\forall$-left for $j=r-2,\ldots,0$, introducing $\forall\mathsf p^w_j$ with the quantifier-free instance $\sigma(\mathsf p^w_j)$.
This gives $\forall\vec{\mathsf p}\,C_1,\Lambda'\Rightarrow\Pi'$, which is $S[\vec a]$.
The substituted formulas are free Boolean variables or $\bot$, so no variable is captured.
At most $3+2(h-1)+r\le3r+3$ inferences were added.

\emph{$\exists$-right.}
If some $P_i$ is $\top$, then $\Phi[\vec a]=\top$ lies in the succedent, and $S[\vec a]$ is the axiom $\Rightarrow\top$.
Suppose that every $P_i$ is $\bot$.
Then $\Phi[\vec a]=\bot$ is omitted, and $S[\vec a]$ is $\Lambda'\Rightarrow\Pi'$.
The auxiliary formula translates under $\vec a'$ to $\bot$, so $S_1[\vec a']$ is the same sequent.
Otherwise $\Phi[\vec a]=\exists\vec{\mathsf p}\,E_1$.
No $P_i$ is $\top$ here, so $P_v$ is not $\top$ when $v\le r$.
If $v>r$ or $P_v=\bot$, then $S_1[\vec a']$ is $\Lambda'\Rightarrow\Pi'$, and one weakening gives $S[\vec a]$.

Otherwise $v=i_h$ for some $h$, and $S_1[\vec a']$ is $\Lambda'\Rightarrow\Pi',P_v\sigma$.
If $h<m$, a weakening by $E_{h+1}\sigma$ and $\vee$-right give $\Lambda'\Rightarrow\Pi',E_h\sigma$.
For $j=h-1,\ldots,1$, a weakening by $P_{i_j}\sigma$, an exchange and $\vee$-right give $\Lambda'\Rightarrow\Pi',E_j\sigma$.
For a string variable, apply $\exists$-right for $j=r-2,\ldots,0$ with the quantifier-free witness $\sigma(\mathsf p^w_j)$.
This gives $\Lambda'\Rightarrow\Pi',\exists\vec{\mathsf p}\,E_1$, which is $S[\vec a]$.
At most $2+3(h-1)+r\le4r+2$ inferences were added.

\emph{$\exists$-left.}
Suppose that some $P_i$ is $\top$.
Then $\Phi[\vec a]=\top$ is omitted from the antecedent, and $S[\vec a]$ is $\Lambda'\Rightarrow\Pi'$.
The auxiliary formula translates under $\vec a_i$ to $P^\gamma_i=\top$, so $S_1[\vec a_i]$ is the same sequent.

If every $P_i$ is $\bot$, then $S[\vec a]$ is the axiom $\Rightarrow\top$.
Otherwise $\mathcal I$ is nonempty, and $S_1[\vec a_i]$ is $P^\gamma_i,\Lambda'\Rightarrow\Pi'$ for each $i\in \mathcal I$.
For $j=m-1,\ldots,1$, $\vee$-left applied to $S_1[\vec a_{i_j}]$ and $E^\gamma_{j+1},\Lambda'\Rightarrow\Pi'$ gives $E^\gamma_j,\Lambda'\Rightarrow\Pi'$.
For a number variable this is $S[\vec a]$.
For a string variable, apply $\exists$-left for $j=r-2,\ldots,0$ with eigenvariable $\mathsf p^\gamma_j$, which is allowed for the reason given in the $\forall$-right case.
This gives $\exists\vec{\mathsf p}\,E_1,\Lambda'\Rightarrow\Pi'$, which is $S[\vec a]$.
At most $2r$ inferences were added.

In every case each derivation $\pi_1(\vec b)$ is used at most once, so the result is tree-like.
Its cuts are those of the derivations $\pi_1(\vec b)$, so it is a $\mathbf G_1^\star$ derivation.
Each added sequent is $S[\vec a]$, or it is obtained from $S[\vec a]$ by replacing the translation of $\Phi$ with at most two renamed or instantiated subformulas of that translation, which together have fewer symbols than that translation.
Renaming and instantiation replace a Boolean variable by a Boolean variable or by $\bot$, so each added sequent has at most as many symbols as $S[\vec a]$.

The formulas of $S$ and $S_1$ are fixed, so the translation of any of them under an assignment is computed by expanding each bounded quantifier into its instances up to the value of its bound, in time polynomial in the values and lengths assigned.
This applies to the instances $P_i$ for $i\le r$, to the translations of $\Lambda$ and $\Pi$, and to the value $v$, and $r=\operatorname{val}_{\vec a}(t)$ is itself bounded by a polynomial in the values and lengths assigned by $\vec a$.
Hence the assignments $\vec a_i$ and $\vec a'$, the index set $\mathcal I$, the substitution $\sigma$ and the added inferences are computed from $\vec a$ in time polynomial in the values and lengths that $\vec a$ assigns.
\end{proof}

Taking $w$ to be a number variable gives the cases of \cite[Exercise~7.27]{CookNguyen}, together with the $\forall$-right case proved there.
Taking $w$ to be a string variable gives the cases of \cite[Exercise~7.59]{CookNguyen}, together with the $\exists$ cases proved there for their main subcases.
\space

\section{Depth Two for the Maximum}
\label{app:depth-two}

This appendix concerns networks with two hidden layers for $\MAX$ when the weights are unrestricted rationals, and the arithmetic needed to prove them correct.
Here a theory is called polynomial-bounded when it extends $\mathbf V^0$, has an axiomatization by bounded formulas, and proves base-language polynomial bounds for every function in its language.
A theory is sound when its axioms are true in the standard model under the intended interpretation of its symbols.
The three conditions defining a polynomial-bounded theory are the hypotheses of the two-sorted Parikh bound, \Cref{d2:parikh}, which bounds the length of a string witness for a bounded existence assertion that such a theory proves.

The appendix gives three results, in this order.
First, $\mathbf{VTC}^0$ proves a padded depth-two collapse: whenever the number sort contains a number $n\ge2^{771(b+1)^5}$, the direct block of \Cref{ir:appendix} for the arity $b$ is a certified two-hidden-layer network whose rational correctness is proved by the theory (\Cref{d2:padded}).
Second, one instance of exponentiation, the axiom $\mathrm{Exp}_A$, removes the padding (\Cref{d2:exp}).
Third, if a sound polynomial-bounded theory proves a bounded depth-two existence assertion, in the sense of \Cref{d2:section}, then $\MAX_n$ has networks of depth at most two and polynomial total size (\Cref{d2:poly-size,d2:all-bounded}).
The end of \Cref{d2:general-bound} shows that a proof of the unpadded assertion of \Cref{d2:exp} in such a theory would have the same consequence.
Between the second and third results, \Cref{d2:constructions} compares the three known depth-two constructions.

\subsection{The Padded Collapse}
\label{d2:padded-section}

Recall that $\bm{\mathcal T}$ is the conservative extension of $\mathbf{VTC}^0$ by the chosen $\mathsf{FTC}^0$ functions of \Cref{uc:section}.
We use the fixed standard integers $A=771$ and $r_\ast=5$ of \Cref{ir:packets}, the functions $P_2$ and $U$ of \Cref{uc:powers}, and the chosen functions $Q_{\mathrm{dir}},D_{\mathrm{dir}}$ of \Cref{ir:small}.
For every number $b$ put $\Gamma_b=P_2(A(b+1)^{r_\ast})$, a binary integer; for every number $n$ with $2\le b\le n$ it equals the comparison string of \Cref{rc:selection}.
The integer $A$ is unrelated to the first matrix of a block.
Write $\mathsf{Vec}(b,x)$ for validity of the rational $b$-vector code of \Cref{sec:foundations}.
We also use the history checker $\mathsf{LEval}$ of \Cref{sec:foundations}, the output $\operatorname{out}(Z)$ of a history from \Cref{rc:body}, the certificate checker $\mathsf{CC}$ of \Cref{rc:cones}, and the ordinary two-hidden-layer history function $\mathsf{Hist}_2$ of \Cref{rc:cc-lemma}.
For a number $b$ and strings $Q,d$, define
\begin{equation}
\begin{aligned}
\mathsf{DMax}(b,Q,d)\equiv{}&\Net(Q,b)\wedge\Depth(Q,2)\wedge\neg\Depth(Q,1)\wedge\mathsf{CC}(b,Q,d)\\
&{}\wedge\forall x\,\left(\mathsf{Vec}(b,x)\rightarrow\exists Z\,\mathsf{LEval}(Q,b,x,Z)\right)\\
&{}\wedge\forall x\,\forall Z\,\left(\mathsf{Vec}(b,x)\wedge\mathsf{LEval}(Q,b,x,Z)\rightarrow\operatorname{out}(Z)=_{\Q}\MAX_b(x)\right).
\end{aligned}
\label{d2:dmax}
\end{equation}
The first line says that $Q$ is a valid network on $b$ inputs with exactly two hidden layers whose cone certificate $d$ is accepted.
Since $\mathsf{CC}$ requires the valid block of \Cref{rc:cones}, the hidden widths of $Q$ are then positive and its biases are zero.
The other two lines say that $Q$ has an evaluation history at every valid rational input and that every such history outputs the maximum.
Their input and history quantifiers are unbounded.

\begin{corollary}[The padded depth-two collapse]
\label{d2:padded}
The theory $\bm{\mathcal T}$ proves the following for all numbers $n$ and $b$ with $b\ge2$ and $\Gamma_b\le U(n)$, where $Q=Q_{\mathrm{dir}}(n,b)$ and $d=D_{\mathrm{dir}}(n,b)$.
\begin{enumerate}[label=\textup{(\roman*)}]
\item $b<n$ and $b+2|Q|+2|d|+4\le n$.
\item $\Net(Q,b)\wedge\Depth(Q,2)\wedge\neg\Depth(Q,1)$ and $\mathsf{CC}(b,Q,d)$ hold, both hidden widths of $Q$ are positive, and every bias of $Q$ has rational value zero.
\item For every valid rational $b$-vector $x$, $\mathsf{LEval}(Q,b,x,\mathsf{Hist}_2(Q,x))$ holds, and every string $Z$ with $\mathsf{LEval}(Q,b,x,Z)$ satisfies $\operatorname{out}(Z)=_{\Q}\MAX_b(x)$.
\end{enumerate}
In particular, $\bm{\mathcal T}$ proves
\begin{equation}
\forall n\,\forall b\,\left(b\ge2\wedge\Gamma_b\le U(n)\rightarrow\exists Q\,\exists d\,\left(|Q|\le n\wedge|d|\le n\wedge\mathsf{DMax}(b,Q,d)\right)\right).
\label{d2:padded-sentence}
\end{equation}
Replacing, in \Cref{d2:padded-sentence}, each occurrence of $\Net$, $\Depth$, $\mathsf{CC}$, $\mathsf{Vec}$, $\mathsf{LEval}$, the output comparison $\operatorname{out}(Z)=_{\Q}\MAX_b(x)$ and the guard $\Gamma_b\le U(n)$ by its $\Delta^B_1$ definition gives a theorem of $\mathbf{VTC}^0$.
In the standard model, for every $b\ge2$ and every $n\ge2^{A(b+1)^{r_\ast}}$, the network $Q_{\mathrm{dir}}(n,b)$ has exactly two hidden layers and zero biases and computes $\MAX_b$ on all of $\R^b$.
\end{corollary}
\begin{proof}
Argue in $\bm{\mathcal T}$, and assume $b\ge2$ and $\Gamma_b\le U(n)$.
For (i), form the tables of $U$ and $P_2$ through $A(b+1)^{r_\ast}+n+1$ by aggregate closure, with $n$ and $b$ as parameters, as in \Cref{uc:powers}.
In these tables the power-table comparisons of \Cref{ir:packets} give $U(b+1)\le P_2(b)$.
Monotonicity of $P_2$, also proved there, together with $b\le A(b+1)^{r_\ast}$ gives $P_2(b)\le\Gamma_b$.
Hence $U(b+1)\le U(n)$, and strict monotonicity of $U$ (\Cref{rc:selection}) gives $b<n$.
The hypotheses $2\le b\le n$ and $\Gamma_b\le U(n)$ of \Cref{ir:small} therefore hold, and that lemma gives $\mathsf{CC}(b,Q,d)$ and $b+2|Q|+2|d|+4\le n$.

For (ii), \Cref{ir:small} returns a homogeneous rational two-hidden-layer block, and $\mathsf{CC}(b,Q,d)$ requires the valid block of \Cref{rc:cones}.
Such a block has exactly $b$ inputs, two hidden layers of positive widths and zero biases, and it is stored in the common network record of \Cref{sec:foundations}.
This gives $\Net(Q,b)$, and the recorded hidden depth two gives $\Depth(Q,2)$ and $\neg\Depth(Q,1)$.

For (iii), let $x$ be a valid rational $b$-vector.
By (ii), $Q$ is a valid rational network of hidden depth two and input dimension $b$, so \Cref{rc:cc-lemma}, applied to $Q$, $x$ and $d$, gives $\mathsf{LEval}(Q,b,x,\mathsf{Hist}_2(Q,x))$ and $\operatorname{out}(\mathsf{Hist}_2(Q,x))=_{\Q}\MAX_b(x)$.
Let $Z$ be any string with $\mathsf{LEval}(Q,b,x,Z)$.
As in the proof of \Cref{rc:upper}, value uniqueness of local histories (paragraph ``Local histories and fixed-depth evaluation'' of \Cref{uc:section}) gives $\operatorname{out}(Z)=_{\Q}\operatorname{out}(\mathsf{Hist}_2(Q,x))=_{\Q}\MAX_b(x)$.

The bounds of (i) give $|Q|\le n$ and $|d|\le n$.
Items (ii) and (iii), with the history $\mathsf{Hist}_2(Q,x)$ as the witness for the existential history quantifier, give $\mathsf{DMax}(b,Q,d)$.
Generalization on $n$ and $b$ proves \Cref{d2:padded-sentence}.
The replacement of the chosen symbols by their $\Delta^B_1$ definitions gives a sentence of the original language that $\bm{\mathcal T}$ proves, and $\mathbf{VTC}^0$ and $\bm{\mathcal T}$ prove the same sentences of the original language (\Cref{uc:section}).
Either the $\Sigma^B_1$ or the $\Pi^B_1$ form of each definition may be used, since $\bm{\mathcal T}$ proves them equivalent; both forms are bounded formulas.

For the standard model, the intended interpretation of the chosen functions satisfies $\bm{\mathcal T}$.
There $P_2(i)$ has value $2^i$ and $U(v)$ has value $v$ (\Cref{uc:powers}), so the guard $\Gamma_b\le U(n)$ holds exactly when $n\ge2^{A(b+1)^{r_\ast}}$.
Item (ii) then holds in the standard model, and the first sentence of \Cref{rc:cc-lemma} gives $Q(x)=\MAX_b(x)$ for every real $x$.
\end{proof}

The number $n$ only makes room for the block.
The functions $Q_{\mathrm{dir}},D_{\mathrm{dir}}$ are $\mathsf{FTC}^0$ functions, so their outputs have length polynomial in the number inputs, while the block for the arity $b$ has length exponential in a polynomial of $b$.
For $b\ge3$ its first hidden layer has width $b^2+b$, its second hidden layer has width below $2^{b^2+3b}+(b+1)^2$, and its code has length below $2^{512(b+1)^2}$ (\Cref{ir:cones}); for $b=2$ it has three neurons in each hidden layer (\Cref{ir:small}).

Three limits of \Cref{d2:padded} should be kept in view.
It is a statement about $Q_{\mathrm{dir}}(n,b)$ for each admissible $n$; we do not prove that this string is the same for two admissible values of $n$, although its size bounds depend on $b$ alone.
Inside the theory, correctness is proved for valid rational inputs; correctness on all of $\R^b$ is the standard-model sentence of the corollary.
Finally, the corollary gives the unbounded formula $\mathsf{DMax}$.
We do not derive for these blocks the bounded base-language formula $\mathsf{BMax}_2$ of \Cref{d2:base-language}, which checks correctness through the computation tableau of a fixed machine; this paper does not relate that tableau to $\mathsf{LEval}$ inside the theory.

\subsection{Exponentiation and the Unpadded Collapse}
\label{d2:exp-section}

The padding hypothesis of \Cref{d2:padded} asks the number sort to contain a number at least $2^{A(b+1)^{r_\ast}}$.
In the $\mathbf{VTC}^0$ form of \Cref{d2:padded-sentence}, each occurrence of $\Net$, $\Depth$, $\mathsf{CC}$, $\mathsf{Vec}$, $\mathsf{LEval}$, the output comparison and the guard is replaced by its $\Delta^B_1$ definition, as in \Cref{d2:padded}.
Write $\mathsf{Pad}(n,b)$ for the bounded base-language formula that this replacement puts in place of the guard $\Gamma_b\le U(n)$, and $\mathsf{DMax}^0(b,Q,d)$ for the base-language formula that it puts in place of $\mathsf{DMax}(b,Q,d)$.
Let $\mathrm{Exp}_A$ be the base-language sentence
\[
\forall b\,\exists n\,\mathsf{Pad}(n,b).
\]
In the standard model $\mathsf{Pad}(n,b)$ says $n\ge2^{A(b+1)^{r_\ast}}$, so $\mathrm{Exp}_A$ is true there; it asserts, for each $b$, the existence of a number at least $2^{A(b+1)^{r_\ast}}$, an instance of the totality of exponentiation.
Its existential number quantifier is unbounded.

\begin{corollary}[The unpadded collapse under exponentiation]
\label{d2:exp}
The theory $\mathbf{VTC}^0+\mathrm{Exp}_A$ proves
\[
\forall b\,\left(b\ge2\rightarrow\exists Q\,\exists d\,\mathsf{DMax}^0(b,Q,d)\right).
\]
In the standard model the witnesses can be taken to be $Q_{\mathrm{dir}}(n,b)$ and $D_{\mathrm{dir}}(n,b)$ for any $n\ge2^{A(b+1)^{r_\ast}}$, so each witness network has exactly two hidden layers and zero biases and computes $\MAX_b$ on all of $\R^b$.
\end{corollary}
\begin{proof}
By \Cref{d2:padded}, $\mathbf{VTC}^0$ proves
\[
\forall n\,\forall b\,\left(b\ge2\wedge\mathsf{Pad}(n,b)\rightarrow\exists Q\,\exists d\,\left(|Q|\le n\wedge|d|\le n\wedge\mathsf{DMax}^0(b,Q,d)\right)\right).
\]
Argue in $\mathbf{VTC}^0+\mathrm{Exp}_A$, and let $b\ge2$.
The axiom $\mathrm{Exp}_A$ gives a number $n$ with $\mathsf{Pad}(n,b)$.
Instantiating the displayed theorem at $n$ and $b$ gives strings $Q,d$ with $\mathsf{DMax}^0(b,Q,d)$, and generalization on $b$ proves the claim.
The standard-model sentence is the one of \Cref{d2:padded}.
\end{proof}

The proof uses only first-order logic on top of a theorem of $\mathbf{VTC}^0$, so it needs no conservativity of an extension of $\bm{\mathcal T}+\mathrm{Exp}_A$ over $\mathbf{VTC}^0+\mathrm{Exp}_A$.
The end of \Cref{d2:general-bound} shows that $\mathbf{VTC}^0$ alone does not prove $\mathrm{Exp}_A$, and that a proof of the conclusion of \Cref{d2:exp} in a sound polynomial-bounded theory would give polynomial-size depth-two networks.

\subsection{The Three Known Constructions}
\label{d2:constructions}

Three constructions of two-hidden-layer networks for $\MAX_n$ with unrestricted rational weights are known to us, and their correctness proofs differ in kind.
The blocks of \Cref{d2:padded} follow the finite integer-feature dictionary of Zimmermann~\cite[Definitions~3--5 and Theorem~7]{Zim26}.
Their correctness is proved in $\bm{\mathcal T}$ by the verification on fixed arrays of rational points of \Cref{ir:tiers} and by the cone certificates of \Cref{ir:cones}.

Brunck, Hertrich, Rue{\ss}, Stade and Yehudayoff give two proofs~\cite{BHRSY26}.
Here $\MAX_n$ is the support function of the standard simplex, the convex hull of the unit vectors, and a paramid is the convex hull of a unit vector and a parallelepiped.
Their direct proof writes $\MAX_n$ as a signed combination of support functions of paramids, modulo support functions of lower-dimensional polytopes~\cite[Theorem~25]{BHRSY26}.
They handle the lower-dimensional terms by induction.
The steps of the direct proof are finite identities apart from one imported property, the valuation property of support functions.
A product expansion and a bijection give a pointwise indicator identity~\cite[Lemma~23]{BHRSY26}, binomial identities evaluate the coefficients~\cite[Lemma~22 and Theorem~25]{BHRSY26}, and the valuation property, which they cite, turns indicator identities into support-function identities modulo lower-dimensional ones~\cite[Lemma~20]{BHRSY26}.
This proof is therefore a plausible candidate for a padded formalization in $\mathbf{VTC}^0$, as for our blocks, or an unpadded one in $\mathbf{VTC}^0+\mathrm{Exp}_A$; this paper carries out neither.
Such a formalization would also have to make the lower-dimensional remainder explicit, which~\cite{BHRSY26} does not list.
Their homological proof instead imports results on polytope algebras and does not write out a network.

Tallavarjula and Rastegar~\cite{TR26} prove their explicit network correct through Laplace transforms.
A rational-function identity is transferred to the pointwise identity by uniqueness of the Laplace transform~\cite[Lemma~10 and Theorem~11]{TR26}, which they prove with holomorphic functions and Fourier uniqueness.
We see no evident route for this step in $\mathbf{VTC}^0$ or in $\mathbf{VTC}^0+\mathrm{Exp}_A$.

None of the three constructions is known to have polynomial size.
Before identical units are merged, the network of~\cite{TR26} has $2+\sum_{r=1}^{n-1}(r-1)!(4^r-3^r)$ second-layer units~\cite[Section~6]{TR26}, a number between $(n-2)!$ and $2+(n-1)!4^{n-1}$ and hence $2^{\Theta(n\log n)}$; their Theorem~2 states this number as an upper bound on the second-layer width.
The top-level identity of~\cite[Theorem~25]{BHRSY26} alone is a signed sum of $n!2^{n-2}$ support functions of paramids, which is $2^{\Theta(n\log n)}$, and~\cite{BHRSY26} states no size bound.
For the blocks here we prove only the upper bounds recorded after \Cref{d2:padded}.


\subsection{Provability and Polynomial Total Size}
\label{d2:section}

\Cref{d2:padded,d2:exp} prove the depth-two collapse with padding in $\mathbf{VTC}^0$ and without padding in $\mathbf{VTC}^0+\mathrm{Exp}_A$.
This subsection derives polynomial total description length from bounded proofs of explicit network existence.
We make this consequence precise for the ordinary explicit codes fixed in \Cref{sec:foundations}.
No denominator restriction is imposed here.

Recall that a theory is polynomial-bounded if it extends $\mathbf V^0$, has an axiomatization by bounded formulas, and proves base-language polynomial bounds for every function in its language, as in \Cref{d2:parikh}.
It is sound when its axioms are true in the standard model under the intended interpretation of its symbols.
For $i\ge0$, $\mathbf V^i$ is the base two-sorted theory with comprehension for bounded formulas having $i$ alternating blocks of bounded string quantifiers, beginning existentially, followed by a matrix whose quantifiers are all bounded number quantifiers \cite[Definition~5.3]{CookNguyen06}.
\Cref{d2:poly-size} states the consequence for a threshold-arithmetic theory and a bounded correctness formula defined below.
\Cref{d2:all-bounded} formulates it in the base language and applies even to $\bigcup_{i<\omega}\mathbf V^i$.

\subsubsection{The Bounded Depth-Two Assertion}

Let $\mathbf{T}_2$ be the finite conservative extension of $\mathbf{VTC}^0$ by the chosen parsing, rational comparison and fixed-depth evaluation functions used in \Cref{tc:parallel-arithmetic}, with depth cap $2$.
These functions have provable polynomial output-length bounds and bounded defining axioms.
We use their intended interpretations.
In particular, $\mathsf{Eval}_2(M,x)$ is the guarded evaluator specified there, and equality of its rational result with a target means equality of rational values.

In the correctness formula \Cref{d2:bounded-max} below, the coordinate maximum is expressed by a bounded number formula: for some $1\le i\le n$, the output of the network equals $x_i$ and $x_j\le_{\Q}x_i$ for every $1\le j\le n$.
Thus the coordinate maximum introduces no real-sort or unbounded quantifier.
The network string $M$ of that formula lists all matrix entries, biases and the affine output row.
The string $M$ is an explicit network string, rather than a program for its entries.
The number $n$ denotes the input dimension, in the first sort, with its usual unary computational size convention.

Recall from \Cref{d2:padded-section} that $\mathsf{Vec}(n,x)$ denotes validity of the rational $n$-vector code of \Cref{sec:foundations}.
Using the fixed polynomial $p$ of \Cref{sec:foundations}, define
\begin{equation}
\begin{aligned}
\mathsf{ShMax}_2(n,M)\equiv{}&
\Net(M,n)\wedge\Depth(M,2) \\
&{}\wedge\forall x\le p(|M|)\,
\left(\mathsf{Vec}(n,x)\to
\mathsf{Eval}_2(M,x)=_{\Q}\MAX_n(x)\right).
\end{aligned}
\label{d2:bounded-max}
\end{equation}
Here $x\le t$ abbreviates the string-length bound $|x|\le t$ for a number term $t$.
The validity predicate $\Net(M,n)$ requires scalar output and rational parameters with positive denominators.
At every fixed depth, the chosen evaluator agrees with layer evaluation on valid codes, including the cases of zero or one hidden layer.
By \Cref{lem:short} and $p\ge p_0$, the standard-model meaning of \Cref{d2:bounded-max} is exactly all-real correctness of the valid network $M$.
The formula is bounded, including its universal string quantifier.
It is not asserted to be a $\Sigma^B_0$ formula.

In the following fact, a base-language number term is a number term built from $0,1,+,\cdot$, number variables and lengths $|Y|$ of string variables, and the bounded formula $\psi$ may contain bounded string quantifiers.
We cite a separately identified draft because its numbering differs from the September 2008 draft used elsewhere in this paper.
\begin{fact}[Two-sorted Parikh bound]
\label{d2:parikh}
Let $\mathbf{S}$ extend $\mathbf V^0$, be axiomatized by bounded formulas, and prove base-language polynomial bounds for every function in its language.
If $\psi(a,Y)$ is bounded and $\mathbf{S}\vdash\forall a\,\exists Y\,\psi(a,Y)$, then some base-language number term $t(a)$ satisfies
\[
\mathbf{S}\vdash\forall a\,\exists Y\,
(|Y|\le t(a)\wedge\psi(a,Y)).
\]
This is the one-number-input, one-string-witness specialization of \cite[Definition~5.24 and Theorem~5.25, p.~94]{CookNguyen06}.
\end{fact}

The base theory has a bounded axiomatization by \Cref{fact:tc-interface}: the counting-sequence axiom can bound its witness by the polynomial pairing term in its input length.
Together with the base-language polynomial bounds on the extra functions and their bounded defining axioms, this proves that $\mathbf{T}_2$ satisfies the hypotheses.
Conservativity alone would not establish them.
The occurrence of $|M|$ in the inner input bound in \Cref{d2:bounded-max} is permitted: $M$ is a free variable of that bounded formula.

Both size results below apply \Cref{d2:parikh} in the same way, so we isolate that step.
\begin{lemma}[Polynomial witness bounds]
\label{d2:poly-witness}
Let $\mathbf{S}$ satisfy the hypotheses of \Cref{d2:parikh}, and let $\varphi(n,M)$ be a bounded formula of its language whose only free variables are $n$ and $M$.
If $\mathbf{S}\vdash\forall n\ge1\,\exists M\,\varphi(n,M)$, then a polynomial $r$ with nonnegative integer coefficients satisfies
\[
\mathbf{S}\vdash\forall n\ge1\,\exists M\,(|M|\le r(n)\wedge\varphi(n,M)).
\]
\end{lemma}
\begin{proof}
Set $\psi(n,M)\equiv(n=0)\vee\varphi(n,M)$, a bounded formula.
For $n=0$ the empty string witnesses $\psi$, and for $n\ge1$ the hypothesis supplies a witness of $\varphi$.
Thus $\mathbf{S}$ proves $\forall n\,\exists M\,\psi(n,M)$.
\Cref{d2:parikh} gives a base-language number term $t(n)$ and an $\mathbf{S}$-proof of
\[
\forall n\,\exists M\,(|M|\le t(n)\wedge\psi(n,M)).
\]
Recall that base-language number terms are built from $0,1,+,\cdot$, number variables and lengths of string variables.
Here $n$ is the only free variable, so $t(n)$ is built from $n,0,1,+$ and $\cdot$.
Hence $t$ is itself a polynomial in $n$ with nonnegative integer coefficients, and we take $r=t$.
For $n\ge1$ the first disjunct of $\psi$ is false, which gives the displayed bounded existence assertion.
\end{proof}

The first application is to the theory $\mathbf{T}_2$ and the bounded depth-two correctness formula $\mathsf{ShMax}_2$ of \Cref{d2:bounded-max}.
\begin{proposition}[Depth-two provability forces polynomial total size]
\label{d2:poly-size}
If
\begin{equation}
\mathbf{T}_2\vdash\forall n\ge1\,\exists M\,
\mathsf{ShMax}_2(n,M),
\label{d2:collapse-premise}
\end{equation}
then a polynomial $r$ with nonnegative integer coefficients satisfies
\[
\mathbf{T}_2\vdash\forall n\ge1\,\exists M\,
(|M|\le r(n)\wedge\mathsf{ShMax}_2(n,M)).
\]
Consequently, in the standard model, $\MAX_n$ has rational networks of depth at most two and total explicit encoding length polynomial in $n$.
\end{proposition}
\begin{proof}
The formula $\mathsf{ShMax}_2(n,M)$ is bounded in the language of $\mathbf{T}_2$, its only free variables are $n$ and $M$, and $\mathbf{T}_2$ satisfies the hypotheses of \Cref{d2:parikh}, as shown before \Cref{d2:poly-witness}.
\Cref{d2:poly-witness}, applied to \Cref{d2:collapse-premise}, gives the polynomial $r$ and the displayed $\mathbf{T}_2$-proof.
The correctness predicate in this conclusion is the same formula $\mathsf{ShMax}_2(n,M)$, since \Cref{d2:poly-witness} leaves its bounded formula unchanged; in particular its input bound is still $p(|M|)$, not $p(r(n))$.
This form suffices for the standard-model argument below, because \Cref{lem:short} bounds the length of a short error by $p_0(|M|)$, a function of the length of the network itself.

The standard model satisfies the axioms and chosen definitions of $\mathbf{T}_2$.
It therefore contains, for each $n\ge1$, a valid network $M$ of length at most $r(n)$ satisfying \Cref{d2:bounded-max}.
If this network failed on a real input, \Cref{lem:short} would supply a rational error of length at most $p_0(|M|)\le p(|M|)$.
That error would contradict \Cref{d2:bounded-max}.
Thus the same $M$ computes $\MAX_n$ on all real inputs.
Its length counts the entire explicit representation, including every parameter and its precision, which proves the claimed total-size bound.
\end{proof}

This is a necessary size consequence of a threshold-arithmetic proof of the specified bounded collapse assertion.
It is not a polynomial-size construction supplied by the classical representation theorem.
The universal string block in \Cref{d2:bounded-max} also remains after Parikh bounding, so the witnessing theorem of \Cref{fact:tc-interface}, which needs a matrix without quantified strings, does not by this argument produce an $\mathsf{FTC}^0$ constructor.
A semantic formalization with additional sorts or unbounded quantifiers requires a proved reduction to \Cref{d2:collapse-premise} before this consequence can be applied.

The general size implication applies to theories beyond $\mathbf{VTC}^0$: \Cref{d2:all-bounded} includes every $\mathbf V^i$ and their union, using the base-language bounded correctness assertion.
The special role of $\mathbf{VTC}^0$ in the main duality is the threshold-computable construction and refutation guarantees, rather than an upper limit on Parikh bounding.
\space

\subsubsection{A Base-Language Formulation}
\label{d2:base-language}

To apply the size argument beyond threshold arithmetic, we express correctness in the base two-sorted language.
A bounded computation tableau replaces the threshold evaluator, so the formulation does not require its function symbols or totality axioms.

A bit machine here is a deterministic Turing machine with one tape, a finite alphabet and a finite set of states.
A configuration records the tape symbols, the unique head position and the state; a step changes the scanned symbol and state and moves the head by at most one cell.
The tape has a left end marker, and a halting configuration repeats unchanged, with a recorded output bit.

Fix a deterministic polynomial-time bit machine $A$ that, on strings $M,x$, reads the input dimension $k$ of $M$.
This machine is unrelated to the integer $A=771$ of \Cref{d2:padded-section}, to the first matrix $A$ of a block in \Cref{rc:cones} and to the subscript of $\mathrm{Exp}_A$.
It returns one when $x$ is not a valid rational $k$-vector.
Otherwise it returns one exactly when the value of the valid depth-at-most-two network $M$ at $x$ equals the coordinate maximum of $x$.
On malformed network codes it uses a fixed default.
The two fixed hidden layers give polynomial bit-length bounds for all intermediate rational values; ordinary integer arithmetic and a scan of the input coordinates give a polynomial running-time bound.

Put $u=|M|+|x|$, choose a polynomial time bound $T(u)$, and reserve $W(u)=u+T(u)+3$ tape cells for the delimited input $M,x$ and the computation.
Encode each cell by a fixed number of bits specifying its symbol and, at the unique head, its state.
A tableau is the exact array of these configuration rows for times $0,\ldots,T(u)$.
The initial row contains the input, end marker, blanks and initial head state.
At each later time, the code of a cell is determined by the preceding codes at that cell and its two neighbors through a fixed finite transition table, with fixed boundary markers.
Require exactly one head per row and a halting final row.
Every check quantifies only over the polynomially bounded time, cell and bit indices, so it is a bounded-number formula in the base language.

Choose a polynomial $s$ with nonnegative integer coefficients bounding the length of this exact tableau encoding, including its fixed array-pairing overhead.
The local transition predicate $\mathsf{Tr}_A(M,x,H)$, which says that the string $H$ is such a tableau of $A$ on $M,x$, is a bounded-number formula of the base two-sorted language.
It checks the initial configuration, the fixed machine's local transition rule at each time and tape position, and the terminal configuration.

Write $\mathsf{Acc}(H)$ for the assertion that the final output recorded in $H$ is one.
This formulation uses no division or threshold function as an additional axiom.
In the standard model every input has its deterministic tableau of length at most $s(|M|+|x|)$.

Let $\mathsf{Net}_2^0(n,M)$ be a fixed bounded formula of the base language that holds in the standard model exactly when $\Net(M,n)\wedge\Depth(M,2)$, that is, when $M$ is a valid rational network of input dimension $n$, scalar output and at most two hidden layers.
Such a formula exists.
Let $A'$ be a deterministic polynomial-time bit machine deciding this property of $M$ and $n$, with $n$ given in unary, and form $\mathsf{Tr}_{A'}$ for these inputs as $\mathsf{Tr}_A$ is formed above.
Let $s'$ be a polynomial with nonnegative integer coefficients bounding the length of its tableau on inputs with $n\le|M|$ as a function of $|M|$.
Then take $\mathsf{Net}_2^0(n,M)$ to be $n\le|M|\wedge\exists H\le s'(|M|)\,(\mathsf{Tr}_{A'}(M,n,H)\wedge\mathsf{Acc}(H))$, since every valid code has $n\le|M|$ (\Cref{sec:foundations}).

With the fixed polynomial $p$ of \Cref{sec:foundations}, put
\[
\mathsf{BMax}_2(n,M)\equiv{}\mathsf{Net}_2^0(n,M)
{}\wedge\forall x\le p(|M|)\,
\forall H\le s(|M|+|x|)\,
(\mathsf{Tr}_A(M,x,H)\to\mathsf{Acc}(H)).
\]
Both universal string quantifiers are length-bounded.
The formula $\mathsf{BMax}_2(n,M)$ is therefore bounded, but need not be $\Sigma^B_0$.
In the standard model it is equivalent to all-real correctness of $M$.
If $M$ is wrong at some real input, then by \Cref{lem:short} it is wrong at some short rational input, and the actual computation tableau $H$ at that input ends in output zero, so $\mathsf{Acc}(H)$ fails.
Conversely, for an all-real correct network every valid deterministic tableau $H$ satisfies $\mathsf{Acc}(H)$.

\subsubsection{Beyond Threshold Arithmetic}
\label{d2:general-bound}

The polynomial-size consequence is not restricted to threshold arithmetic.
The hypothesis needed by \Cref{d2:parikh} concerns bounds on the language and axioms, rather than a polynomial-time limit on the theory's reasoning.
In particular it applies to every finite level of the usual bounded-arithmetic hierarchy and to their union.
Recall from \Cref{d2:section} that $\mathbf V^i$ is the base two-sorted theory with bounded $\Sigma_i^B$ comprehension; $\Sigma_i^B$ permits $i$ alternating blocks of bounded string quantifiers, beginning existentially, followed by a bounded-number matrix.
\begin{fact}[Polynomial-bounded arithmetic theories]
\label{fact:hierarchy-bounded}
Every finite level $\mathbf V^i$ extends $\mathbf V^0$, has an axiomatization by bounded formulas and proves base-language polynomial bounds on its function symbols \cite[Definitions~5.3 and~5.24 and the observation following Definition~5.24]{CookNguyen06}.
The union $\bigcup_{i<\omega}\mathbf V^i$ in the same base language also has these properties.
\end{fact}
For the union, the function vocabulary is unchanged and the union of the bounded comprehension axioms is still an axiomatization by bounded formulas.
No fixed alternation bound on the entire axiom set is required by \Cref{d2:parikh}.
We can therefore apply that fact to the hierarchy as follows.
\begin{corollary}[Polynomial size from bounded arithmetic]
\label{d2:all-bounded}
Let $\mathbf{S}$ be a theory whose axioms are true in the standard two-sorted model under the intended interpretation of its symbols, and which satisfies the polynomial-bounded hypotheses of \Cref{d2:parikh}.
If
\[
\mathbf{S}\vdash\forall n\ge1\,\exists M\,\mathsf{BMax}_2(n,M),
\]
then $\MAX_n$ has rational depth-at-most-two networks of polynomial total explicit encoding length in $n$.
This applies in particular when $\mathbf{S}=\bigcup_{i<\omega}\mathbf V^i$ in the base two-sorted language.
\end{corollary}
\begin{proof}
The formula $\mathsf{BMax}_2(n,M)$ is bounded in the base language, and its only free variables are $n$ and $M$.
\Cref{d2:poly-witness} gives a polynomial $r$ with nonnegative integer coefficients and an $\mathbf{S}$-proof that every $n\ge1$ has a witness $M$ with $|M|\le r(n)$.
The axioms of $\mathbf{S}$ are true in the standard model, so such witnesses exist there.
Standard-model soundness and the short-error/tableau equivalence above give all-real correctness of these witnesses.
Their lengths count all entries and parameter precision.

For the stated union, \Cref{fact:hierarchy-bounded} verifies the hypotheses.
\end{proof}

This is a sufficient general criterion, not a claim of a uniquely strongest theory among all possible languages and axiom systems.
The bounded universal string quantifiers still prevent this argument from extracting a polynomial-time constructor.

\paragraph{\emph{Exponentiation and the barrier.}}
The axiom $\mathrm{Exp}_A$ of \Cref{d2:exp} is a $\forall\exists$ sentence whose existential number quantifier is unbounded.
The axiom is not provable in $\mathbf{VTC}^0$, and $\mathbf{VTC}^0+\mathrm{Exp}_A$ has no axiomatization by bounded base-language formulas, so \Cref{d2:parikh} does not apply to it and \Cref{d2:exp} does not conflict with \Cref{d2:poly-size,d2:all-bounded}.

For the first claim, that $\mathbf{VTC}^0$ does not prove $\mathrm{Exp}_A$, note that $\mathbf{VTC}^0$ satisfies the hypotheses of \Cref{d2:parikh} by \Cref{fact:tc-interface}, as for $\mathbf{T}_2$ in \Cref{d2:section}, and its axioms are true in the standard model.
Every number $m$ is the length of the string $\{i:i<m\}$, formed by $\Sigma^B_0$ comprehension, so a proof of $\forall b\,\exists n\,\mathsf{Pad}(n,b)$ would give a proof of $\forall b\,\exists Y\,\mathsf{Pad}(|Y|,b)$, whose matrix is bounded.
\Cref{d2:parikh} would then give a base-language number term $t(b)$ such that, in the standard model, every $b$ has a number $m\le t(b)$ with $m\ge2^{A(b+1)^{r_\ast}}$.
This fails for large $b$, because $t(b)$ has $b$ as its only variable and is therefore a polynomial in $b$, as in the proof of \Cref{d2:poly-witness}.

For the second claim, the same argument applies to any set of bounded base-language axioms extending $\mathbf V^0$ with the same theorems as $\mathbf{VTC}^0+\mathrm{Exp}_A$, since those theorems are true in the standard model and every function of the base language is polynomially bounded, as \Cref{d2:parikh} requires.
Such axioms would therefore not prove $\mathrm{Exp}_A$, although $\mathbf{VTC}^0+\mathrm{Exp}_A$ does, so no such axiomatization exists.

The size consequence also applies to the unpadded assertion of \Cref{d2:exp} itself, because its conjunct $\mathsf{CC}$ is bounded.
Let $\mathbf{S}$ be a theory in the base language whose axioms are true in the standard model and which satisfies the hypotheses of \Cref{d2:parikh}, and suppose that $\mathbf{S}$ proves $\forall b\,(b\ge2\rightarrow\exists Q\,\exists d\,\mathsf{DMax}^0(b,Q,d))$.
Let $\mathsf{CC}^0(b,Q,d)$ be the bounded base-language formula that replaces $\mathsf{CC}(b,Q,d)$ in $\mathsf{DMax}^0$; the theory $\bm{\mathcal T}$ proves it equivalent to $\mathsf{CC}(b,Q,d)$, so the two agree in the standard model.
Let $\varphi(b,Y)$ be the bounded formula stating that $b=1$, or that some strings $Q,d$ of length at most $|Y|$ are the two entries of $Y$ and satisfy $\mathsf{CC}^0(b,Q,d)$.
Here $Y$ is an exact tuple with empty row zero, as in the proof of \Cref{uc:duality}.

Since $\mathsf{CC}^0$ is a conjunct of $\mathsf{DMax}^0$, the hypothesis gives $\mathbf{S}\vdash\forall b\,(b\ge2\rightarrow\exists Q\,\exists d\,\mathsf{CC}^0(b,Q,d))$, and for $b=1$ the first disjunct of $\varphi$ holds.
Since $\mathbf{S}$ extends $\mathbf V^0$, it forms the array from $Q$ and $d$ by bounded comprehension, so it proves $\forall b\ge1\,\exists Y\,\varphi(b,Y)$.

\Cref{d2:poly-witness} gives a polynomial $r$ with $\mathbf{S}\vdash\forall b\ge1\,\exists Y\,(|Y|\le r(b)\wedge\varphi(b,Y))$.
In the standard model, for each $b\ge2$, the entries $Q,d$ of such a $Y$ satisfy $\mathsf{CC}(b,Q,d)$, and $|Q|\le r(b)$.
Since $\mathsf{CC}$ requires the valid block of \Cref{rc:cones}, the network $Q$ has exactly two hidden layers, and the first sentence of \Cref{rc:cc-lemma} shows that it computes $\MAX_b$ on all of $\R^b$.
Thus a proof of the conclusion of \Cref{d2:exp} in such a theory, for instance in $\mathbf{VTC}^0$, would give rational depth-two networks for $\MAX_n$ of polynomial total encoding length.
None of the three constructions of \Cref{d2:constructions} is known to have polynomial size.
Exponentiation therefore suffices for the unpadded collapse, while a proof of it in a sound polynomial-bounded theory would require depth-two networks of polynomial size, which are not known to exist.

\paragraph{\emph{What a weaker theory would give.}}
The particular exponent $A(b+1)^{r_\ast}$ in $\mathrm{Exp}_A$ does not matter.
Write $\mathrm{Exp}$ for the base-language sentence $\forall m\,\exists n\,\mathsf{Pad}_1(n,m)$, where $\mathsf{Pad}_1(n,m)$ replaces the comparison $P_2(m)\le U(n)$ by its $\Delta^B_1$ definition; in the standard model it says $n\ge2^m$, so $\mathrm{Exp}$ expresses the totality of $m\mapsto2^m$.
Since $\Gamma_b=P_2(A(b+1)^{r_\ast})$, the instance $m=A(b+1)^{r_\ast}$ of $\mathrm{Exp}$ is $\mathrm{Exp}_A$ at $b$.
Conversely, $\mathrm{Exp}_A$ at $b=m$ gives a number $n$ with $\Gamma_m\le U(n)$, and $m\le A(m+1)^{r_\ast}$ with the monotonicity of $P_2$ (\Cref{ir:packets}), in a table formed as in the proof of \Cref{d2:padded}, gives $P_2(m)\le\Gamma_m\le U(n)$.
These arguments are carried out in $\bm{\mathcal T}$, and both sentences are in the base language, so $\mathbf{VTC}^0$ proves $\mathrm{Exp}\leftrightarrow\mathrm{Exp}_A$.
So $\mathbf{VTC}^0+\mathrm{Exp}$ proves the unpadded collapse, and a proof of it in a theory that does not prove $\mathrm{Exp}$ would not follow from \Cref{d2:exp}.

For a sound polynomial-bounded theory, the size argument given before the paragraph \emph{What a weaker theory would give} shows what a proof of the unpadded assertion of \Cref{d2:exp} in that theory would give: rational depth-two networks for $\MAX_n$ of polynomial total encoding length.
Contrapositively, a superpolynomial lower bound on the total encoding length of rational depth-two networks for $\MAX_n$ would show that no sound polynomial-bounded theory proves the unpadded collapse.
For $\mathbf{VTC}^0$ itself, such a proof would also give a chosen $\mathsf{FTC}^0$ constructor of the certified networks.
Suppose that $\mathbf{VTC}^0$ proved the unpadded collapse.
The size argument, with $\mathbf{S}=\mathbf{VTC}^0$, then gives a polynomial $r$ such that $\mathbf{VTC}^0$ proves that every $b\ge2$ has strings $Q,d$ of length at most $r(b)$ with $\mathsf{CC}^0(b,Q,d)$.
Since $\bm{\mathcal T}$ proves $\mathsf{CC}^0$ equivalent to $\mathsf{CC}$, it proves $\forall b\ge1\,\exists Q,d\le r(b)\,(b=1\vee\mathsf{CC}(b,Q,d))$, which has the form of \Cref{uc:duality}(a) with the chosen $\mathsf{FTC}^0$ checker $b=1\vee\mathsf{CC}(b,Q,d)$.
That correspondence then gives a chosen $\mathsf{FTC}^0$ constructor of the certified networks, and \Cref{rc:cc-lemma} supplies the soundness of $\mathsf{CC}$.
\Cref{op:depth-two-theory} asks whether a theory that does not prove $\mathrm{Exp}$ can prove the unpadded collapse.

For the padded collapse, $\mathbf{VTC}^0$ is the natural floor for the formulation of \Cref{d2:padded}.
There the output equation of a history sums all second-layer units of the block, and these sums are handled with the threshold arithmetic of $\mathbf{VTC}^0$, which uses counting that $\mathbf V^0$ lacks.
A version in $\mathbf V^0$ would need a different formulation, for instance a padding large enough that every such sum has only polylogarithmically many terms in $n$; we do not know whether $\mathbf V^0$ proves the present formulation.
\space
\space

\end{document}